\documentclass[fleqn, final]{unmeethesis}
\usepackage{zj}
\usepackage{mdframed}
\usepackage[square,sort,comma,numbers]{natbib}
\usepackage{float}
\newcommand{\expect}[1]{\big\langle #1 \big\rangle}
\newcommand{\erf}[1]{Eq. (\ref{#1})}
\newcommand{\mbf}[1]{\mathbf{#1}}
\newcommand{\eff}{\text{eff}}

\begin{document}

\frontmatter


\title{Internal Spin Control, Squeezing and Decoherence in Ensembles of Alkali Atomic Spins}

\author{Leigh Morgan Norris}

\degreesubject{Ph.D., Physics}

\degree{Doctor of Philosophy \\ Physics}

\documenttype{Dissertation}

\previousdegrees{B.A., Carleton College, 2007}

\date{December 2014}

\maketitle

\makecopyright

\begin{dedication}
To my parents and grandparents
\end{dedication}

\begin{acknowledgments}
\vspace{2em}
First and foremost, I would like to thank my advisor Ivan Deutsch. Ivan has been a wonderful teacher and mentor, and has made a substantial contribution to my research. I would also like to thank my collaborators Ben Baragiola, Collin Trail, Poul Jessen, Enrique Monta$\tilde{\text{n}}$o and Pascal Mickelson. The material in this dissertation is the combined effort of many people. Before beginning graduate school at UNM, I was fortunate enough to take quantum mechanics with Arjendu Pattanayak at Carleton College. I would like to thank Arjendu for introducing me to quantum information and giving me the opportunity to work for him the summer after my junior year, during which I learned a great deal. I would also like to thank my professors at UNM, in particular Carl Caves and Andrew Landahl for their quantum information classes and Ivan Deutsch for his quantum mechanics and quantum optics classes. I draw on material from these courses on a routine basis.  Thanks to all of the CQuIC graduate students and postdocs, especially the current and former members of the Deutsch group, for exposing me to all sorts of topics in physics and for many engaging discussions. Thanks also to the Jessen group for explaining to me how experiments work and showing me how theories actually apply in the lab. Thank you to Vicky Bird and Alisa Gibson for helping me navigate the bureaucracy of graduate school and for their cheerfulness and patience.
 
Thank you to Tom for his support during my time in graduate school. I must also thank my good friends no longer at UNM, Krittika and Laura. Krittika was my trusty comrade at many a conference and retreat. We remained united even after getting lost in the woods of Northern Arizona and in the foothills of Boulder. Thanks to Laura for being my friend and always providing a sympathetic ear. 

Thank you to my parents and sister for being there for me when I've lived over a thousand miles away. Some combination of my cousin Jonathan, my aunts Ann and Meg and my mother Gwen has traveled all the way to New Mexico to visit me every summer while I have been in graduate school. Thank you for the summer adventures and the year-round support and friendship. Thanks also to my wonderful grandparents for inspiring me and being interested in even the most boring things I do.

\end{acknowledgments}

\maketitleabstract

 \begin{abstract}
 Large atomic ensembles interacting with light are one of the most promising platforms for quantum information processing. In the past decade, novel applications for these systems have emerged in quantum communication, quantum computing, and metrology.  Essential to all of these applications is the controllability of the atomic ensemble, which is facilitated by a strong coupling between the atoms and light. Non-classical spin squeezed states are a crucial step in attaining greater ensemble control. The degree of entanglement present in these states, furthermore, serves as a benchmark for the strength of the atom-light interaction. Outside the broader context of quantum information processing with atomic ensembles, spin squeezed states have applications in metrology, where their quantum correlations can be harnessed to improve the precision of magnetometers and atomic clocks.  
 
This dissertation focuses upon the production of spin squeezed states in large ensembles of cold trapped alkali atoms interacting with optical fields. While most treatments of spin squeezing consider only the case in which the ensemble is composed of two level systems or qubits, we utilize the entire ground manifold of an alkali atom with hyperfine spin $f\geq 1/2$, a qudit. Spin squeezing requires non-classical correlations between the constituent atomic spins, which are generated through the atoms' collective coupling to the light. Either through measurement or multiple interactions with the atoms, the light mediates an entangling interaction that produces quantum correlations. 

The spin squeezing treated in this dissertation ultimately originates from the coupling between the light and atoms. Conventional approaches of improving this squeezing have focused on increasing the optical density of the ensemble. The greater number of internal degrees of freedom and the controllability of the spin-$f$ ground hyperfine manifold enable novel methods of enhancing  squeezing. In particular, we find that state preparation using control of the internal hyperfine spin increases the entangling power of squeezing protocols when $f>1/2$. Post-processing of the ensemble using additional internal spin control converts this entanglement into metrologically useful spin squeezing. By employing a variation of the Holstein-Primakoff approximation, in which the collective spin observables of the atomic ensemble are treated as quadratures of a bosonic mode, we model entanglement generation, spin squeezing and the effects of internal spin control.

The Holstein-Primakoff formalism also enables us to take into account the decoherence of the ensemble due to optical pumping. While most works ignore or treat optical pumping phenomenologically, we employ a master equation derived from first principles.  Our analysis shows that state preparation and the hyperfine spin size have a substantial impact upon both the generation of spin squeezing and the decoherence of the ensemble. Through a numerical search, we determine state preparations that enhance squeezing protocols while remaining robust to optical pumping.

Finally, most work on spin squeezing in atomic ensembles has treated the light as a plane wave that couples identically to all atoms. In the final part of this dissertation, we go beyond the customary plane wave approximation on the light and employ focused paraxial beams, which are more efficiently mode matched to the radiation pattern of the atomic ensemble. The mathematical formalism and the internal spin control techniques that we applied in the plane wave case are generalized to accommodate the non-homogeneous paraxial probe. We find the optimal geometries of the atomic ensemble and the probe for mode matching and generation of spin squeezing.

   \clearpage 
 \end{abstract}

\tableofcontents
\listoffigures
\listoftables

\begin{glossary}{longeststring}
   \item[$\hat{f}_n$]
      internal spin
      
        \item[$\hat{F}_n$]
      collective spin
      
              \item[$N_A$]
      number of atoms in ensemble
      
               \item[$f$]
      internal spin quantum number
      
                  \item[$SCS$]
    spin coherent state
    
                \item[$SQL$]
      standard quantum limit
      
               \item[$\zeta_q$]
      $=2\Delta\hat{X}^2$, quadrature squeezing parameter
      
      \item[$\Delta\phi$]
   angular resolution of the collective spin or spin wave in a metrological application defined in Eqs. (\ref{AngRes}) and (\ref{angRes00})
       
       \item[$\zeta_m$]
       $=2N_Af\Delta F_z^2/\expect{\hat{F}_x}^2$, Wineland squeezing parameter
      
                      \item[$\hat{S}_n$]
      Stokes' component
      
             \item[$N_L$]
      number of photons per pulse of time $\Delta t$
      
      \item[$\chi$]
      $=g_f(\sigma_0/A)(\Gamma/6\Delta)$, Faraday rotation angle
      
      \item[$g_f$]
      Land\'{e} g-factor
      
        \item[$A$]
      beam area
      
       \item[$\sigma_0$]
      $=3\lambda^2/2\pi$, resonant cross section for a unit oscillator strength
      
      \item[$\Gamma$]
      transition linewidth
      
        \item[$\Delta$]
      detuning
      
          \item[$\Delta F_z^2$]
      projection noise
      
          \item[$\xi$]
      $=\chi^2N_L\Delta F_z^2$, collective spin coupling constant
      
      \item[$\ket{\uparrow}$]
      fiducial state
      
       \item[$\ket{\downarrow}$]
      coupled state
      
       \item[$\ket{\wr}$]
      transfer state
      
          \item[$\Sigma$]
      covariance matrix
      
         \item[$\sigma$]
      symplectic matrix
      
      \item[$\zeta_m^\uparrow$]
      $=2f(\Delta f_z^2)_\uparrow/\expect{\hat{f}_x}_\uparrow^2$ squeezing parameter of the internal spin prepared in the fiducial state

 \item[$\gamma_s$]
$=(N_L/\Delta t)\,(\sigma_0/A)\,(\Gamma^2/4\Delta^2)$, photon scattering rate

 \item[$\Gamma_\text{op}$]
$=2\gamma_s/3$, rate of optical pumping events

 \item[$C(\uparrow)$]
update of correlation term under optical pumping for fiducial state $\ket{\uparrow}$ defined in \erf{update}

 \item[$\ket{\widetilde{q}_{\uparrow}}$]
$ =\hat{W}_q\ket{\uparrow}$, unnormalized state to which the fiducial state is optically pumped

 \item[$\ket{\widetilde{q}_{\downarrow}}$]
$ =\hat{W}_q\ket{\downarrow}$, unnormalized state to which the coupled state is optically pumped

 \item[$N_\psi$]
$ =\sum_{i=1}^{N_A}\ket{\psi}\bra{\psi}_i$, number of atoms in internal spin state $\ket{\psi}$

 \item[$\Gamma_{\text{loss},\,\psi}$]
rate at which an atom in state $\ket{\psi}$ is lost to the other ground hyperfine manifold due to optical pumping, defined in \erf{lossPsi}

 \item[$\Gamma_{\text{loss}}$]
$=\Gamma_{\text{loss},\,\uparrow}+\Gamma_{\text{loss},\,\downarrow}$, total loss rate

 \item[$\Gamma_{\text{flip}}$]
rate of spin flips defined in \erf{eq::FlipRate}

 \item[$v(\uparrow)$]
 $=\sqrt{2(\Delta f_z^2)_\uparrow}$

 \item[$w(\uparrow)$]
 $=\sqrt{2(\Delta f_z^2)_\downarrow-2(\Delta f_z^2)_\uparrow}$
 
  \item[$\kappa$]
  $=\chi^2N_L/\Delta t$, measurement strength
 
  \item[$u_{pl}(\mathbf{r})$]
  Gaussian mode function
  
    \item[$\eta(\mathbf{r})$]
 density of atoms in the ensemble
 
     \item[$\beta_{pl}(\mathbf{r})$]
     $=u_{pl}^{*}(\mathbf{r})u_{00}(\mathbf{r})$
     
    \item[$\hat{F}_z^{pl}$]
          $=\sum_{i} \beta_{pl}(\mbf{r}_{\perp i}, z_i)\hat{f}_z^{(i)}$, spin wave
    
        \item[$N_{\text{eff}}^{pl\,(K)}$]
        $ =\int d^3\mbf{r} \, \eta (\mbf{r})\beta_{pl}(\mbf{r})^K$, effective atom number
        
         \item[$ O\!D_{\text{eff}}$]
        $=N_{\text{eff}}^{00\,(2)}\sigma_0/A$, effective optical density
        
            \item[$\zeta_{\text{para}}$] 
            $= 2f \frac{ \big(N^{(1)}_\eff \big)^2 }{N_\eff^{(2)}} \frac{\left(\Delta F_z^{00}\right)^2}{\expect{\hat{F}_x^{00}}^2}$, spin wave squeezing parameter
            
            \item[$\gamma_s(\mathbf{r})$] 
            spatially inhomogeneous scattering rate defined in \erf{Eq::LocalScatRate}

\end{glossary}

\mainmatter

\chapter{Introduction} \label{sec::Intro}
Since the advent of quantum mechanics, the quantum nature of the interaction between light and atoms has been an active area of study. With the emergence of quantum information science came novel applications for large atomic ensembles interacting with light, so-called atom-light interfaces.  For quantum information processing and quantum communication, these systems can form networks with light acting as a carrier of information and the ensemble acting as a quantum memory or repeater \cite{Fleischhauer2002,Polzik2004,Kimble2008,DLCZ,Kuzmich2004}. Collective degrees of freedom of the atomic ensemble are potential platforms for continuous variable quantum computing, in which light is utilized to perform gates and operations \cite{ContinuousVar}. Light can also mediate entangling interactions between the atoms in the ensemble, creating non-classical spin squeezed states for use in metrology \cite{KuzBig00,appel09,Takano2009,VulSqueezingClock,Koschorreck2010,TakTak05}. 

The controllability of the atomic ensemble is vital for all of these applications. Quantum memories and repeaters require long coherence times that permit the storage of information in the spin degrees of freedom without degradation. The ability to create states that are non-Gaussian in the collective spin of the atomic ensemble is essential for continuous variable quantum computing. Spin squeezed states require the generation of either interatomic entanglement or entanglement within the internal spin degrees of freedom of the atoms. In addition to having applications in quantum information processing, control of large atomic ensembles is of interest in fundamental physics. Creating non-classical spin states in mesoscopic systems can shed light on poorly understood phenomena like the quantum-classical boundary and many-body entanglement. 

The ability to produce spin squeezed states is a critical benchmark in the effort to control the collective spin of a large atomic ensemble. Spin squeezed states contain quantum correlations that reduce the variance of a spin component below the standard quantum limit (SQL).  These correlations, which can be produced by entanglement between the atoms in the ensemble, serve as a measure for the strength of the collective control achievable in the atom-light interface. Beyond quantum control, spin squeezed states are of practical interest in metrology, where their quantum correlations can be harnessed to improve the precision of atomic clocks \cite{Wineland94, VulSqueezingClock} and magnetometers \cite{KosMitSq, BudkerSq}. 
 
\section{Context of Dissertation}
This dissertation studies the generation of spin squeezed states in large ensembles of alkali atomic spins.   While ensembles of alkali atoms figure prominently in experimental demonstrations and theoretical investigations of spin squeezing \cite{KuzBig00,appel09,Takano2009,VulSqueezingClock,Koschorreck2010,TakTak05}, this dissertation focuses on a property of alkali atoms seldom studied in the context of spin squeezing, control over the internal hyperfine ground spin.  Ensembles of alkalis are unique platforms in that the spin of the hyperfine ground state is fully controllable \cite{ASmith13, MerkelControlPRA}. We seek to integrate control over the internal spins of the atoms that compose the ensemble with control over the collective spin. Squeezing of the collective spin and interatomic entanglement are generated by the Faraday interaction, which couples the collective spin to the polarization of the light. As we demonstrate in this dissertation, control over the internal spins of the alkalis has a surprising impact on both the strength of the Faraday interaction and the generation of the interatomic entanglement that contributes to spin squeezing. 
Specifically, we find that by preparing the internal spins in a state with large projection noise variance, we can maximize the resolution of the collective spin in a measurement mediated by the light. This increases the measurement backaction on the ensemble, a signature of the enhanced entanglement generation between the atoms. Through the application of subsequent control on the internal spins, this entanglement can be converted into metrologically useful spin squeezing. Although spin squeezing is a phenomenon involving the collective spin of the atomic ensemble, it is enhanced substantially by control over the internal spin.

Much of the work on spin squeezing in both alkali ensembles and other platforms has focused on the case in which the ensemble is composed of two-level systems or ``qubits". In an ensemble of alkalis, this corresponds to a hyperfine spin quantum number of $f=1/2$. This dissertation places particular emphasis on ensembles where the alkali atomic spins are ``qudits", i.e. $f\geq 1/2$. The higher spin case is especially interesting from the perspective of internal spin control because of the greater number of internal degrees of freedom. Whereas for  $f=1/2$ spin squeezing is achieved solely by the generation of interatomic entanglement, when  $f>1/2$ entanglement between the internal degrees of freedom creates squeezing of the internal spin. Internal spin squeezing can be combined with interatomic entanglement when $f>1/2$ for a substantial enhancement in the overall squeezing of the ensemble. 

Optical pumping due to spontaneous emission is the most significant source of decoherence in the atomic ensemble. In most previous work on spin squeezing, optical pumping is treated phenomenologically. We study the effects of optical pumping in the ensemble using a master equation derived from first principles in Ref. \cite{DeuJes09}. This treatment of optical pumping reveals a variety of interesting effects. Most relevant to our study is the substantial role played by the spin size, $f$, in the decay of spin squeezing due to optical pumping. Many of the optical pumping processes that damage spin squeezing are suppressed for larger $f$. Furthermore, ``transfers of coherence" that occur only when $f>1/2$ increase the robustness of interatomic entanglement. Even without internal spin squeezing, we find that alkali atoms with $f>1/2$ generate more squeezing than those with $f=1/2$ because they are less susceptible to decoherence.

Theoretical models of spin squeezing in large atomic ensembles often treat the light that couples to the atoms as a plane wave. In experimental implementations of squeezing, employing plane-like waves is undesirable because they are poorly mode matched to the radiation pattern of the ensemble.  The coupling between the light and the ensemble is enhanced by utilizing paraxial beams that match the paraxial radiation emitted by a spatially extended atomic ensemble. In this dissertation, we analyze the spin squeezing produced by a quantum nondemolition (QND) measurement of the ensemble's collective spin mediated by a paraxial beam, rather than a plane wave. Because the intensity of the paraxial beam is spatially varying, mode matching, spin squeezing and optical pumping are influenced by the geometries of the probe and the atomic ensemble. We the find optimal geometries of the ensemble and probe for both mode matching and spin squeezing in the presence of optical pumping. To our knowledge, this is the first treatment of optical pumping in a three-dimensional, inhomogeneous atom-light interface.

\section{Summary of Dissertation}
The remainder of this dissertation is organized as follows. Chapter \ref{sec::Squeezing} introduces spin squeezing and defines several important concepts, such as the collective spin and the standard quantum limit (SQL). We analyze the entanglement and quantum correlations that produce spin squeezing. Also discussed is the use of spin squeezed states in metrological applications, such as magnetometry. Special emphasis is placed on the role of the ensemble's mean spin in metrological applications, where it increases the resolvability of measurement outcomes.  We examine several methods of quantifying spin squeezing and introduce the Wineland squeezing parameter as a measure of spin squeezing and metrological usefulness. 

Chapter \ref{sec::ALinterface} examines the interaction between the light and the atomic ensemble in detail. From the AC-Stark Hamiltonian that couples the atomic spins and the polarization of the light, we derive the Faraday interaction. We show that the Faraday interaction creates entanglement between the light and ensemble that is enhanced when the ensemble is prepared in a state with large projection noise variance. The Holstein-Primakoff approximation on the light and the multilevel Holstein-Primakoff approximation on the atoms enable us to treat the combined system of the light and ensemble as a multimode Gaussian state on two effective modes. We outline properties of Gaussian states that are useful for modeling the dynamics of the light and ensemble, including the covariance matrix update formalism. 

In Chapter \ref{Sec:Protocols}, we introduce protocols that utilize the Faraday interaction to create spin squeezing in the atomic ensemble. In all of these protocols, the light mediates a nonlinear interaction in the ensemble's collective spin that creates interatomic entanglement. The strength of this interaction ultimately depends upon the initial entanglement between the light and atoms. The evolution of the light and ensemble under each protocol is modeled using the covariance matrix update formalism. Through this formalism, we show that the protocols create squeezing of a quadrature in a phase plane defined by the multilevel Holstein-Primakoff approximation. While this quadrature squeezing does not necessarily imply spin squeezing, it scales monotonically with interatomic entanglement. Through control over the internal spins of the atoms, we show how this entanglement is converted to metrologically relevant spin squeezing. Additional control can squeeze the internal spins of the atoms, further enhancing spin squeezing.

Chapter \ref{sec::OpticalPumping} introduces the primary source of decoherence in the system, optical pumping due to the spontaneous scattering of photons. Optical pumping, as we demonstrate, damages spin squeezing by destroying beneficial interatomic entanglement, increasing the collective spin variance and causing the mean spin to decay. Some of these damaging effects can be offset by internal spin control. The extent to which optical pumping damages spin squeezing depends substantially on the initial state preparation of the ensemble. We outline specific properties of state preparations that determine the ensemble's susceptibility to decoherence. For particular state preparations and for $f>1/2$, ``transfers of coherence" can preserve interatomic entanglement in the presence of optical pumping. 

Chapter \ref{sec::ModHPCovar} explores the dynamics of the ensemble and light as the system undergoes both squeezing interactions and decoherence due to optical pumping. Whereas we previously made the multilevel Holstein-Primakoff approximation on the ensemble by treating each atom as a qubit embedded in the higher dimensional hyperfine spin, preserving transfers of coherence requires that we treat the atoms as embedded qutrits.
We modify the multilevel Holstein-Primakoff approximation to accommodate the ensemble of embedded qutrits, which enables us to treat the ensemble as a Gaussian state on two modes. The effects of optical pumping can be expressed as an update on the covariance matrix of this Gaussian state, as can the coherent squeezing dynamics. Using the covariance matrix update formalism, we conduct a variety of numerical simulations to explore the influence of the spin size $f$ and the state preparation of the ensemble upon the achievable squeezing.  

The state preparation of the ensemble has substantial influence on the coherent generation of squeezing and upon decoherence of the ensemble due to optical pumping. In Chapter \ref{sec::Beyond}, we use optimal control techniques to determine the state preparations of the ensemble that maximize squeezing for multiple values of $f$. We specialize to the case of squeezing by quantum nondemolition (QND) measurement, which can be expressed in differential form. Each state preparation determines a unique set of coupled differential equations that give the evolution of the ensemble under both QND measurement and optical pumping. Utilizing a numerical search, we find the state preparations that maximize the spin squeezing determined by the solution to these differential equations.

In Chapter \ref{paraxial}, we extend our previous treatment of spin squeezing to a three-dimensional atom-light interface. We model the light as a paraxial Gaussian beam, which is more closely mode matched to the radiation pattern of the ensemble. The inhomogenous nature of the light and ensemble necessitates the introduction of ``spin waves", which are collective spin operators that take into account the non-uniform coupling between the light and ensemble.  QND measurement of a spin wave, in addition to decoherence of the ensemble from optical pumping and scattering of the light into transverse spatial modes outside of the probe, can be modeled through a system of coupled differential equations. We solve for the geometries of the probe and ensemble that maximize both spin squeezing and mode matching for $f=1/2$ and $f=4$.

Chapter \ref{conclusion} concludes the dissertation and discusses possible extensions of this work. 

\begin{table}[ht]
\centering
\begin{tabular}{c l }
\hline
Chapters & Publication \\ [0.5ex] 
\hline
&\\
4, 5, 6& L. M. Norris, C. M. Trail, P.S. Jessen and I. H. Deutsch, \\&\emph{Enhanced Squeezing of a Collective Spin via Control }\\&\emph{of Its Qudit Subsystems}, Phys. Rev. Lett. \textbf{109}, 173603 (2012).\\&\\
4,5,6,7&L. M. Norris, C. M. Trail, P.S. Jessen and I. H. Deutsch, \\&\emph{Spin Squeezing Ensembles of Qudits} (In preparation).\\&\\
8&B. Q. Baragiola, L. M. Norris, E. Monta$\tilde{\text{n}}$o, P. G. Mickelson, \\&P. S. Jessen and I. H. Deutsch, \emph{Three-dimensional light-matter} \\&\emph{interface for collective spin squeezing in atomic ensembles}, \\&Phys. Rev. A \textbf{89}, 033850 (2014).\\\\
\hline
\end{tabular}
\label{table:nonlin}
\caption{List of publications and the corresponding chapters of this dissertation.}
\end{table}

\chapter{Spin Squeezing}\label{sec::Squeezing}
Spin squeezed states contain quantum correlations that reduce the variance in a component of the collective spin below the standard quantum limit. In the pages that follow, we discuss precisely what is meant by the collective spin and the standard quantum limit. Additionally, we examine the interatomic entanglement and the types of quantum correlations that lead to spin squeezing. Because producing spin squeezed states is our focus, we introduce the Wineland squeezing parameter as a means of quantifying the metrologically useful spin squeezing that we create. The Wineland squeezing parameter is motivated by the application of spin squeezed states in magnetometry. Through the squeezing parameter, we outline some of the preliminary differences between generating spin squeezing in ensembles of ``qudits" versus ensembles of ``qubits".

\section{Collective Spin}
We study spin squeezing in a large ensemble of $N_A$ identical spins with quantum number $f$. Each spin $i$ in the ensemble has an angular momentum operator $\hat{\mathbf{f}}^{(i)}$ associated with its internal spin. Throughout this text, lowercase letters denote operators on the internal spins of the constituent spins that form the ensemble. We seek to squeeze the \emph{collective} spin of the ensemble, the components of which are the summation of the components of the internal spins,
\begin{align}
\hat{F}_{n}=\sum_{i=1}^{N_A}\hat{f}_n^{(i)},
\end{align}
where $n\in\{x,y,z\}$. 

A state of the ensemble with particular significance in metrology is the spin coherent state (SCS), the state in which all of the internal spins are polarized along the same spatial direction. The SCS along $x$, for instance, is the eigenstate of $\hat{F}_x$ with maximal spin eigenvalue,
\begin{align}
\ket{SCS(x)}=\ket{f,\,m_x=f}^{\otimes N_A}=\ket{F=N_Af,\,M_x=N_Af}.
\end{align}
While $\ket{SCS(x)}$ has no variance in $\hat{F}_x$, the variance of $\ket{SCS(x)}$ in the collective spin components orthogonal to $x$ defines the standard quantum limit (SQL) of the ensemble. For $N_A>>1$, the spin coherent state is approximately Gaussian distributed in  the collective spin components orthogonal to $x$. Consider the projection of $\ket{SCS(x)}$ onto the eigenstates of $\hat{F}_z$,
\begin{align}\label{eq::SCSBinomial}
\ket{SCS(x)}&=\frac{1}{2^{N_Af}}\sum_{M_z}\sqrt{\binom{2N_Af}{N_Af+M_z}}\ket{F=N_Af,\,M_z}\\\label{eq::SCSGauss}
&\approx (\pi N_Af)^{-1/4}\sum_{M_z}\text{exp}\left(-\frac{M_z^2}{2N_Af}\right)\ket{F=N_Af,\,M_z}.
 \end{align}
 The approximate equality in this expression arises from applying the central limit theorem to the binomial distribution in \erf{eq::SCSBinomial}. The probability of measuring the ensemble in the $\hat{F}_z$ eigenstate $\ket{F=N_Af,M_z}$ is given by the Gaussian probability density function 
 \begin{align}
 P(M_z)=\frac{1}{\sqrt{\pi N_Af}}\text{exp}\left(-\frac{M_z^2}{N_Af}\right).  
  \end{align}
The variance of $ P(M_z)$, $\Delta F_z^2=N_Af/2$, is the SQL, the smallest nonzero variance possible without quantum correlations.

\section{Spin Squeezing}\label{sec::SpinSqDef}
Spin squeezed states of the ensemble have quantum correlations either between the constituent spins or between the internal degrees of freedom of the constituent spins, permitting the variance of a collective spin component to fall below the SQL \cite{Wineland94, KitagawaUeda93}. For an ensemble symmetric under interchange of the constituent spins, this can be seen through a decomposition of the collective spin variance in terms of the internal spin components,
\begin{align}\label{variance}
\Delta F_z^2=N_A(N_A-1)\langle\Delta\hat{f}_z^{(i)}\Delta\hat{f}_z^{(j)}\rangle_{i\neq j}+N_A\Delta f_z^2.
\end{align}
The first term in \erf{variance} is proportional to the covariance between any two spins in the ensemble, while the second term depends upon the variance of the internal spin. For a spin coherent state of the ensemble along $x$, $\langle\Delta\hat{f}_z^{(i)}\Delta\hat{f}_z^{(j)}\rangle_{i\neq j}=0$ and $\Delta f_z^2=f/2$. Entanglement between the constituent spins can produce states with $\langle\Delta\hat{f}_z^{(i)}\Delta\hat{f}_z^{(j)}\rangle_{i\neq j}<0$, however, reducing the collective variance. Alternatively, $\Delta F_z^2$ can fall below the SQL if the variance of each internal spin is less than $f/2$ \cite{Chaudhury07, Fernholz08}. Either or both of these mechanisms produce states that are ``squeezed" in $\hat{F}_z$, having variance $\Delta F_z^2\leq N_Af/2$.  For an ensemble state with $\expect{\hat{F}_x}=N_Af$, the Heisenberg uncertainty principle mandates that the orthogonal collective spin components satisfy $\Delta F_y^2\Delta F_z^2\geq (N_Af/2)^2$.  Therefore, a state squeezed in $\hat{F}_z$ must be ``anti-squeezed" in $\hat{F}_y$, implying $\Delta F_y^2\geq N_Af/2$. 

\section{Quantifying Spin Squeezing}\label{sec::QuantSqueezing}

Squeezed states were first proposed in the context of 
light \cite{Stoler71,Lu72,Hollenhorst79}. Consider a mode of an optical field where the phase space position and momentum quadratures are defined as
\begin{align}
&\hat{X}=\frac{1}{\sqrt{2}}\left(\hat{a}^\dag+\hat{a}\right)\;\;\;\;\text{and}\\ 
&\hat{P}=\frac{i}{\sqrt{2}}\left(\hat{a}^\dag-\hat{a}\right).
\end{align}
The standard quantum limit is set by the phase space variance of a spin coherent state, given by $\Delta X^2=\Delta P^2=1/2$. A state is, thus, squeezed in the quadrature $\hat{X}$ if $\Delta X^2<1/2$. 
This squeezing is quantified by the ``quadrature squeezing parameter",
\begin{align}
\zeta_q=2\Delta X^2.
\end{align}
A state is squeezed when $\zeta_q<1$ with a smaller $\zeta_q$ indicating more squeezing. 

We could easily define an analogous parameter for the ensemble, scaling linearly with the variance of a collective spin component. For example, consider 
\begin{align}
\zeta=\frac{2}{N_Af}\Delta F_z^2. 
\end{align}
While $\zeta<1$ does indicate spin squeezing, this parameter has no dependence on the three dimensional structure of the collective spin. The three dimensional structure of the collective spin, as we shall see, is critical in metrological applications for spin squeezed states. 

\begin{figure}[H]
\centering
\includegraphics[scale=.38]{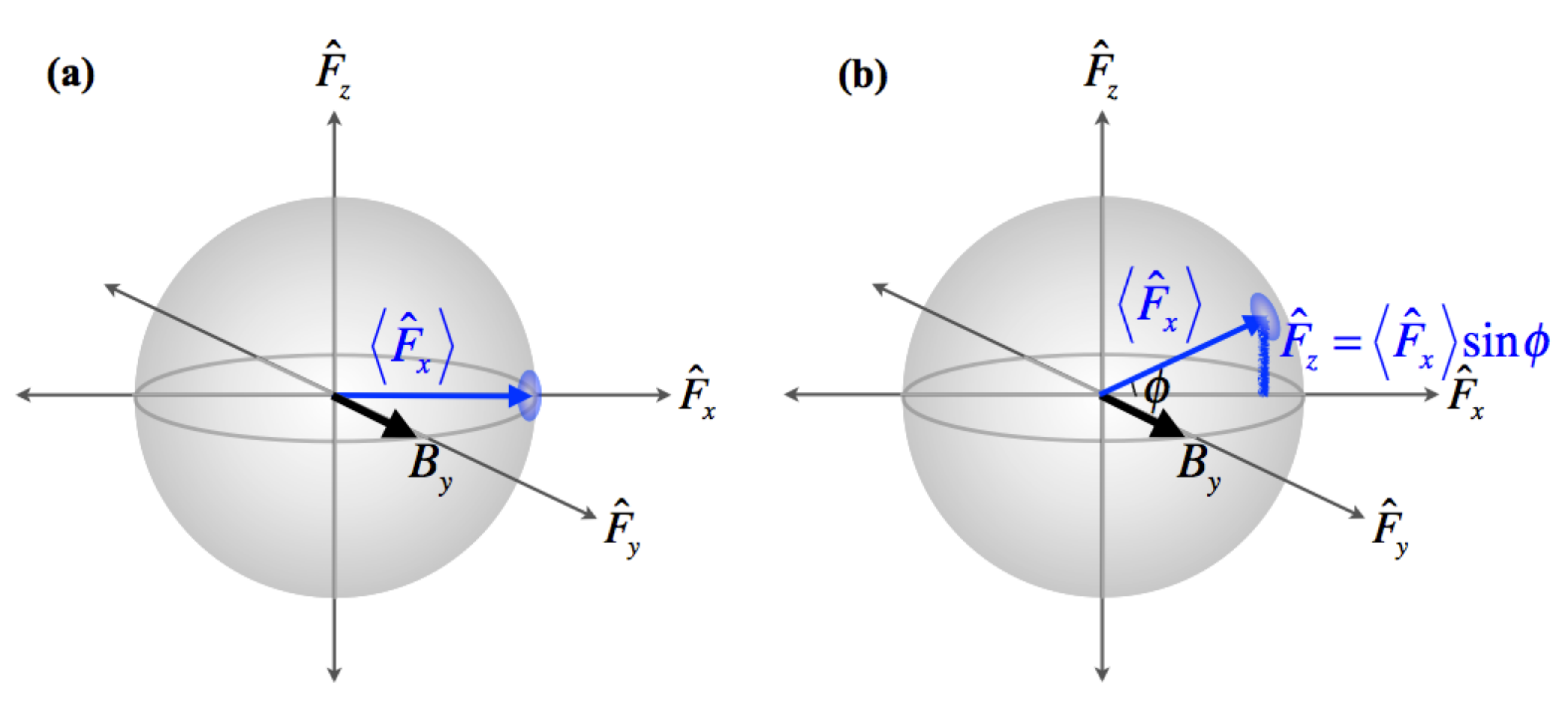}
\caption{(a) An atomic ensemble is prepared with the mean spin along $x$. (b) A magnetic field along $y$ of strength $B_y$ causes the spin of the ensemble to rotate about $y$ by an angle $\phi$. By measuring $\hat{F}_z$, the rotation angle, $\phi$, can be determined. The fundamental resolution is determined by the quantum uncertainty of the spin projection (projection noise), shown here as an uncertainty patch.}\label{fig::MeanSpin}
\end{figure}

Consider an example involving magnetometry. Suppose the ensemble is prepared in an initial state with the mean collective spin aligned along $x$, i.e. $\hat{\mathbf{F}}=\expect{\hat{F}_x}\mathbf{e}_x$, as depicted in Fig. \ref{fig::MeanSpin} (a).  If the ensemble is exposed to a magnetic field of unknown strength in the $y$-direction, given by $\mathbf{B}=B_y\mathbf{e}_y$, the collective spin will precess about $\hat{F}_y$.  The precession angle, $\phi=\gamma_g B_y \Delta t$, is proportional to the field strength along $y$, the gyromagnetic ratio $\gamma_g$ and the interaction time $\Delta t$. Our objective is to deduce the field strength, $B_y$, by measuring $\phi$. If $\phi$ is small, it can be determined by measuring $\hat{F}_z$, as
\begin{align}
\hat{F}_z=\expect{\hat{F}_x}\text{sin} \phi.
\end{align}
The variance of $\phi$ in the measurement of $\hat{F}_z$ is given by
\begin{align} \label{AngRes}
\Delta\phi^2=\frac{\Delta F_z^2}{\expect{\hat{F}_x}^2}.
\end{align} 
It is natural that the angular resolution should depend on the variance of the measured spin component, $\Delta F_z^2$. The angular resolution also depends on the ``mean spin" $\expect{\hat{F}_x}$, which explains why the three dimensional structure of the collective spin is essential for metrology. The mean spin, which is always a component orthogonal to the squeezed component, acts like a lever arm in the magnetometer, as shown in Fig. \ref{fig::MeanSpin}. A larger mean spin makes the measured displacement in $\hat{F}_z$ larger and more resolvable, leading to a more precise estimate of $\phi$. A squeezing parameter that takes into account the metrological usefulness of a squeezed state must depend on the mean spin. 

\begin{figure}[H]
\centering
\includegraphics[scale=.6]{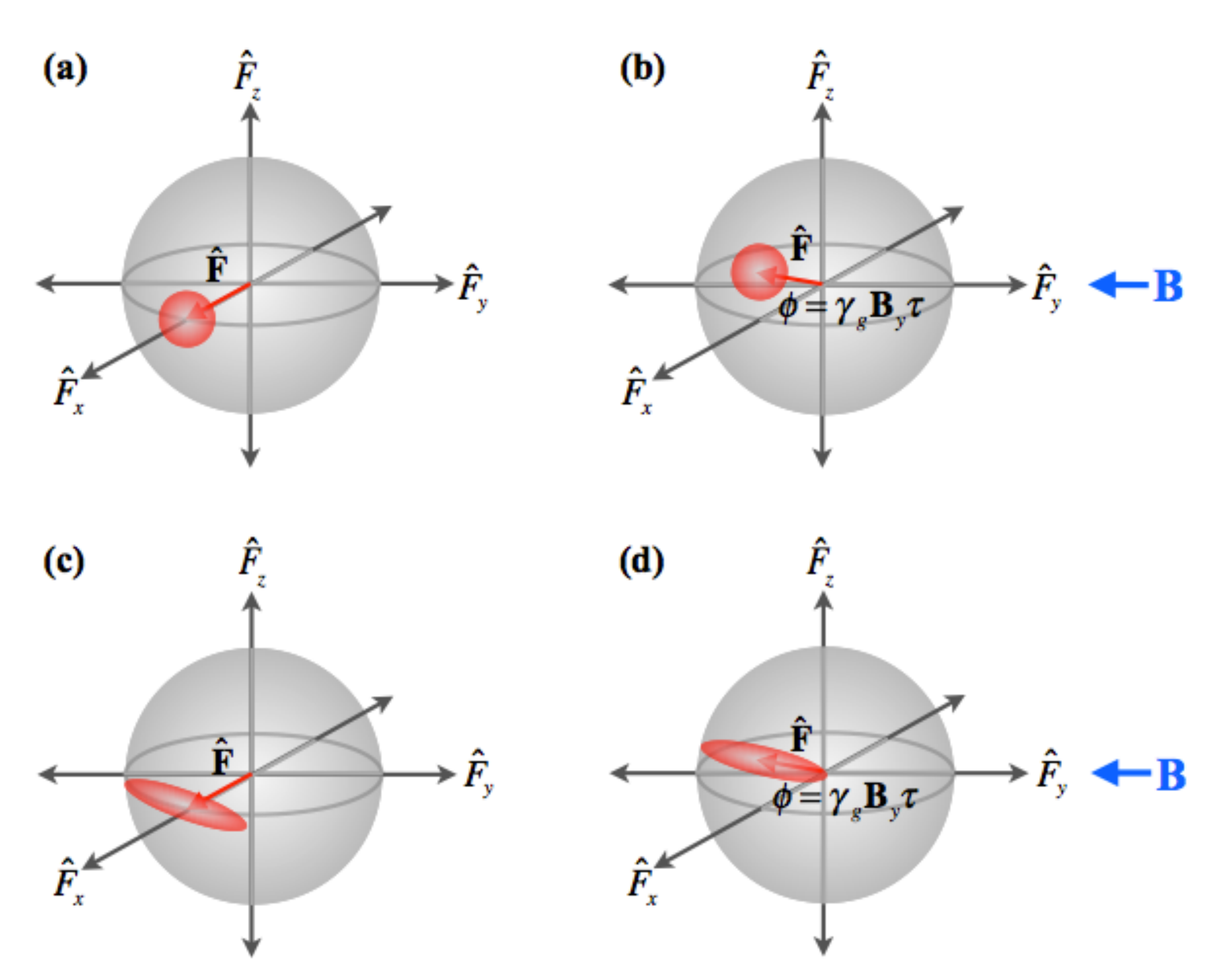}
\caption{Using an ensemble of spins to measure a magnetic field. (a) The ensemble is prepared in a spin coherent state along the $x$-axis with with transverse variance $\Delta F_{y,z}^2=N_Af/2$. (b) After the ensemble is subjected to a magnetic field along $y$, its mean spin rotates by an angle $\phi$ about $y$. The angle $\phi$  is deduced by measuring $\hat{F}_z$. The precision with which we can determine $\phi$ is, therefore, proportional to  $\Delta F_{z}^2=N_Af/2$. (c) The ensemble is prepared in a state squeezed in $\hat{F}_z$, i.e. with $\zeta_m<1$, and a mean spin along the $x$-axis. (d) After being subjected to a magnetic field along $y$, the mean spin of the ensemble rotates by an angle $\phi$ about $y$. Our ability to resolve the angle $\phi$ is enhanced over (b) because $\Delta F_z^2<N_Af/2$. }\label{fig::magnetometry}
\end{figure}

The previous example involving magnetometry is equivalent to Ramsey interferometry for atomic clocks when $f=1/2$. In the context of Ramsey interferometry, Wineland proposed a parameter that quantifies both squeezing and metrological usefulness \cite{Wineland94}. The Wineland squeezing parameter is defined as
\begin{align}\label{eq::SqParameter}
\zeta_m=\frac{(\Delta\phi^2)}{(\Delta\phi^2)_{SCS}}=\frac{2N_Af\Delta F_z^2}{\langle\hat{F}_x\rangle^2}.
\end{align}    
Here, $(\Delta\phi^2)_{SCS}$ is the angular resolution of a magnetometer that uses a spin coherent state and 
 $(\Delta\phi^2)$ is the angular resolution of a magnetometer that uses the state we wish to quantify. A state for which $\zeta_m<1$ improves the resolution of a magnetometer over a an SCS. Because the maximal value of $\expect{\hat{F}_x}$ is $N_Af$,  a state with $\zeta_m<1$ also has $\Delta F_z^2<SQL$. A smaller variance combined with a still sizable mean spin make a spin squeezed state metrologically useful. Fig. \ref{fig::magnetometry} compares a magnetometer using a SCS and one using a squeezed state. From this point onward in the text, we consider a squeezed state to be a state for which $\zeta_m<1$. 
 
\section{Collective Spin Squeezing and $f$}
The Wineland squeezing parameter sheds light on the relationship between collective spin squeezing and the internal spin size, $f$. By substituting \erf{variance} and $\expect{\hat{F}_x}=N_A\expect{\hat{f}_x}$ into \erf{eq::SqParameter}, the squeezing parameter becomes
\begin{align}
\zeta_m=\frac{2N_A\Delta F_z^2}{\expect{\hat{F}_x}^2}=\frac{2N_Af\expect{\Delta\hat{f}_z^{(i)}\Delta\hat{f}_z^{(j)}}_{i\neq j}}{\expect{\hat{f}_x^{(i)}}^2}+\zeta_m^{(i)}
\end{align}
for $N_A>>1$. As before, $\expect{\Delta\hat{f}_z^{(i)}\Delta\hat{f}_z^{(j)}}_{i\neq j}$ is the covariance between any two different spins in the ensemble. The last term on the right-hand side of this expression is the value of the Wineland squeezing parameter for any single spin in the ensemble, given by 
\begin{align}
\zeta_m^{(i)}=\frac{2f\Delta f_z^{(i)\,2}}{\expect{f_x^{(i)}}^2}.
\end{align}
Like the variance $\Delta F_z^2$, $\zeta_m$ can be reduced in two different ways. Entanglement between the atoms creates negative correlations for which $\expect{\Delta\hat{f}_z^{(i)}\Delta\hat{f}_z^{(j)}}_{i\neq j}<0$. Alternatively, the internal spins can be squeezed, causing $\zeta_m^{(i)}<1$.  This latter option exists only when $f>1/2$, however, which we can demonstrate by calculating the squeezing parameter of an arbitrary qubit. 

Consider a $f=1/2$ qubit in the state $\ket{\psi}=\text{cos}\theta\ket{\uparrow}+e^{i\phi}\text{sin}\theta\ket{\downarrow}$, where \\$\ket{\uparrow}$ denotes $\ket{f=1/2,m_z=1/2}$ and  $\ket{\downarrow}$ denotes $\ket{f=1/2,m_z=-1/2}$.  The variance and mean spin of this state are $\Delta f_z^2=\text{cos}^2\theta\text{sin}^2\theta$ and $\expect{\hat{f}_x}=\text{cos}\theta\text{sin}\theta\text{cos}\phi$, respectively. The squeezing parameter of a single qubit is, therefore,
\begin{align}
\zeta_m^{f=1/2}=\frac{1}{\text{cos}^2\phi}.
\end{align}
Because the minimum value of $\zeta_m^{f=1/2}$ is 1, a qubit cannot be squeezed. Thus, entanglement is solely responsible for spin squeezing in an ensemble of spins with $f=1/2$. When $f\geq 1/2$, on the other hand, both entanglement and squeezing of the internal spin generate spin squeezing.

\chapter{Atom-Light Interface}\label{sec::ALinterface}
Large atomic ensembles interacting with light or ``atom-light interfaces" are platforms for many spin squeezing protocols and experimental demonstrations of spin squeezing \cite{KuzBig00,appel09,Takano2009,VulSqueezingClock,Koschorreck2010,TakTak05}. Typically, the correlations between the atomic spins that create spin squeezing are generated through the atoms' mutual coupling to the light. In this chapter, we analyze the interaction between the light and atoms in detail. In particular,  we concentrate on the entanglement generated between the light and ensemble, which is essential for creating spin squeezing. We also develop the mathematical formalism necessary to model the effects of spin squeezing protocols on the light and ensemble. By applying variations of the Holstein-Primakoff approximation, we greatly simplify the representation of the light and ensemble states \cite{HP,MultilevelHP}. This enables us to utilize the properties of Gaussian states to describe the joint evolution of the light and ensemble.

\section{The Faraday Interaction}\label{sec::FaradayH}
\begin{figure}
\centering
\includegraphics[scale=.55]{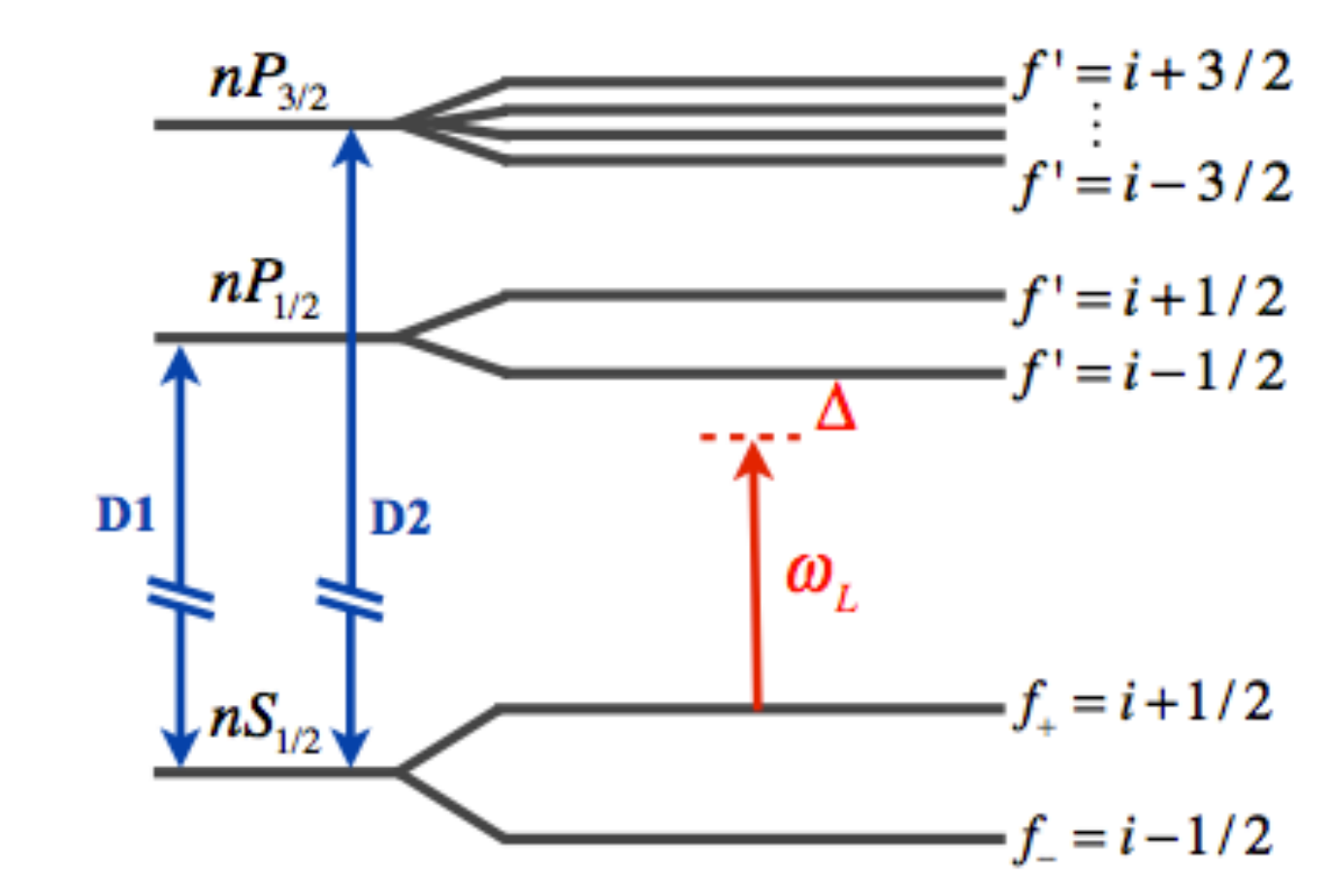}
\caption{Energy levels of an alkali atom with nuclear spin $i$ and ground hyperfine spins $f_+$ and $f_-$, both greater than $1/2$. Atoms in the ensemble are prepared in one of the ground hyperfine manifolds with spin $f$ equal to $f_+$ or $f_-$. The probe laser is detuned from the excited state hyperfine splitting of the D1 or D2 line. The spin quantum numbers of the excited hyperfine states are denoted $f'$.}
\label{fig::AlkaliLines}
\end{figure}

We study spin squeezing in an ensemble composed of $N_A>>1$ alkali atoms. The atoms, which are identically prepared in one of the ground hyperfine manifolds with spin $f$, interact with a probe laser far detuned from the D1 or D2 line, as depicted in Fig. \ref{fig::AlkaliLines}.  For a probe with sufficiently large detuning, $\Delta$, and weak intensity, the excited hyperfine manifolds of the atoms can be adiabatically eliminated. The resulting interaction couples the magnetic sublevels of the spin-$f$ ground hyperfine manifold to the polarization modes of the optical field. This is described by the ac-Stark Hamiltonian, which can be decomposed into irreducible tensor components as follows \cite{DeuJes09},   
\begin{align}\label{eq::ACstark}
&\hat{H}=\frac{\hbar\chi_0}{\Delta t}\sum_{i=1}^{N_A}[C^{(0)}\hat{S}_0+C^{(1)}\hat{f}_z^{(i)}\hat{S}_3\\
&+C^{(2)}((\hat{f}_x^{(i)2}-\hat{f}_y^{(i)2})\hat{S}_1/2-(3\hat{f}_z^{(i)\;2}-\hat{\mathbf{f}}^{(i)\;2})\hat{S}_0/6+(\hat{f}_x^{(i)}\hat{f}_y^{(i)}+\hat{f}_y^{(i)}\hat{f}_x^{(i)})\hat{S}_2/2)]\notag,
\end{align}
where $\chi_0=(\sigma_0\Gamma)/(2A\Delta)$, $\sigma_0$ is the resonant cross section for unit oscillator strength, $\Gamma$ is the excited state linewidth, $A$ is the cross sectional area of the probe and $\Delta t$ is the interaction time. Here, the light's direction of propagation is along $z$ and $\hat{\textbf{S}}$ is the quantized Stokes' vector of the light with components 
\begin{align}\label{Stokes}
&\hat{S}_0=\frac{1}{2}(\hat{a}_x^\dag\hat{a}_x+\hat{a}_{y}^\dag\hat{a}_{y})\;\;\;\;\;
\hat{S}_1=\frac{1}{2}(\hat{a}_x^\dag\hat{a}_{x}-\hat{a}_{y}^\dag\hat{a}_y)\\\notag
&\hat{S}_2=\frac{1}{2}(\hat{a}_{x}^\dag\hat{a}_y+\hat{a}_y^\dag\hat{a}_{x})\;\;\;\;\;
\hat{S}_3=\frac{1}{2i}(\hat{a}_{x}^\dag\hat{a}_{y}-\hat{a}_y^\dag\hat{a}_x),
\end{align}
where $x$ and $y$ denote orthogonal linearly polarized modes.  The Stokes' components 1 through 3 are analogous to angular momentum operators, satisfying the $su(2)$ commutation relations, $[\hat{S}_i,\hat{S}_j]=i\epsilon_{ijk}\hat{S}_k$.  

The Hamiltonian in \erf{eq::ACstark} is decomposed into scalar, vector and tensor terms with coefficients $C^{(0)}$,  $C^{(1)}$ and  $C^{(2)}$, respectively. The scalar term shifts the atomic energy levels by an amount proportional to the probe intensity. Because this shift is independent of the atomic state, the scalar term has no influence on the dynamics of the atom-light interface and can be discarded. The vector component, commonly known as the Faraday interaction, couples the circular polarization of the probe to the collective spin component in the direction of the light's propagation. This interaction generates atom-light entanglement, the crucial first step in many spin squeezing protocols. The tensor component, on the other hand, induces an undesirable birefringence on the light as well as nonlinear dynamics on the internal state of the atomic spins \cite{Smith03}. The magnitudes of the constants $C^{(K)}$ determine the contribution of each term to the dynamics. While the full form of the $C^{(K)}$ is given in \cite{DeuJes09}, for our purposes it suffices to know that each constant depends upon the detuning with $C^{(1)}\propto 1/\Delta$ and $C^{(2)}\propto 1/\Delta^2$ to leading order in $1/\Delta$.

Although it is smaller than the vector component by an order of $1/\Delta$, the tensor term is not negligible over the time scale during which spin squeezing occurs. The damaging birefringence effects of the tensor term are removed in the presence of a large bias magnetic field along the direction of the light's propagation \cite{Baragiola14}.  This averages to zero the coupling between the light and the atomic spin components transverse to $z$. This effect is evident when we transform into a frame rotating at the Larmor frequency of the bias, $\Omega_B$, via the unitary $\hat{U}_B(t)=\text{exp}(-i\Omega_Bt\hat{F}_z)$. The terms in the tensor component that depend upon the transverse internal spin components become
\begin{align}
&\expect{\hat{U}_B^\dag(t)\hat{f}_x^{(i)2}\hat{U}_B(t)}_t=(\hat{f}_x^{(i)2}+\hat{f}_y^{(i)2})/2\\
&\expect{\hat{U}_B^\dag(t)\hat{f}_y^{(i)2}\hat{U}_B(t)}_t=(\hat{f}_x^{(i)2}+\hat{f}_y^{(i)2})/2
\end{align}
and
\begin{align}
&\expect{\hat{U}_B^\dag(t)(\hat{f}_x^{(i)}\hat{f}_y^{(i)}+\hat{f}_y^{(i)}\hat{f}_x^{(i)})\hat{U}_B(t)}_t=0,
\end{align}
where $\expect{\cdot}_t$ denotes a time average. Transformed into the rotating frame and time averaged, the AC Stark Hamiltonian is
\begin{align}\label{eq::ACstarkRot}
&\hat{H}=\frac{\hbar\chi_0}{\Delta t}\sum_{i=1}^{N_A}[C^{(1)}\hat{f}_z^{(i)}\hat{S}_3
-C^{(2)}(3\hat{f}_z^{(i)\;2}-\hat{\mathbf{f}}^{(i)\;2})\hat{S}_0/6].
\end{align}
The residual tensor component of the Hamiltonian nonlinearly couples the internal spins of the atoms to the intensity of the probe. This coupling can, in principle, be eliminated with the internal spin control techniques covered in Sec. \ref{sec::IntControl}. In practice, one can also apply two probes  detuned from the D1 and D2 lines to cancel the residual tensor component without affecting the vector component \cite{Baragiola14}.

\begin{figure}
\centering
\includegraphics[scale=.4]{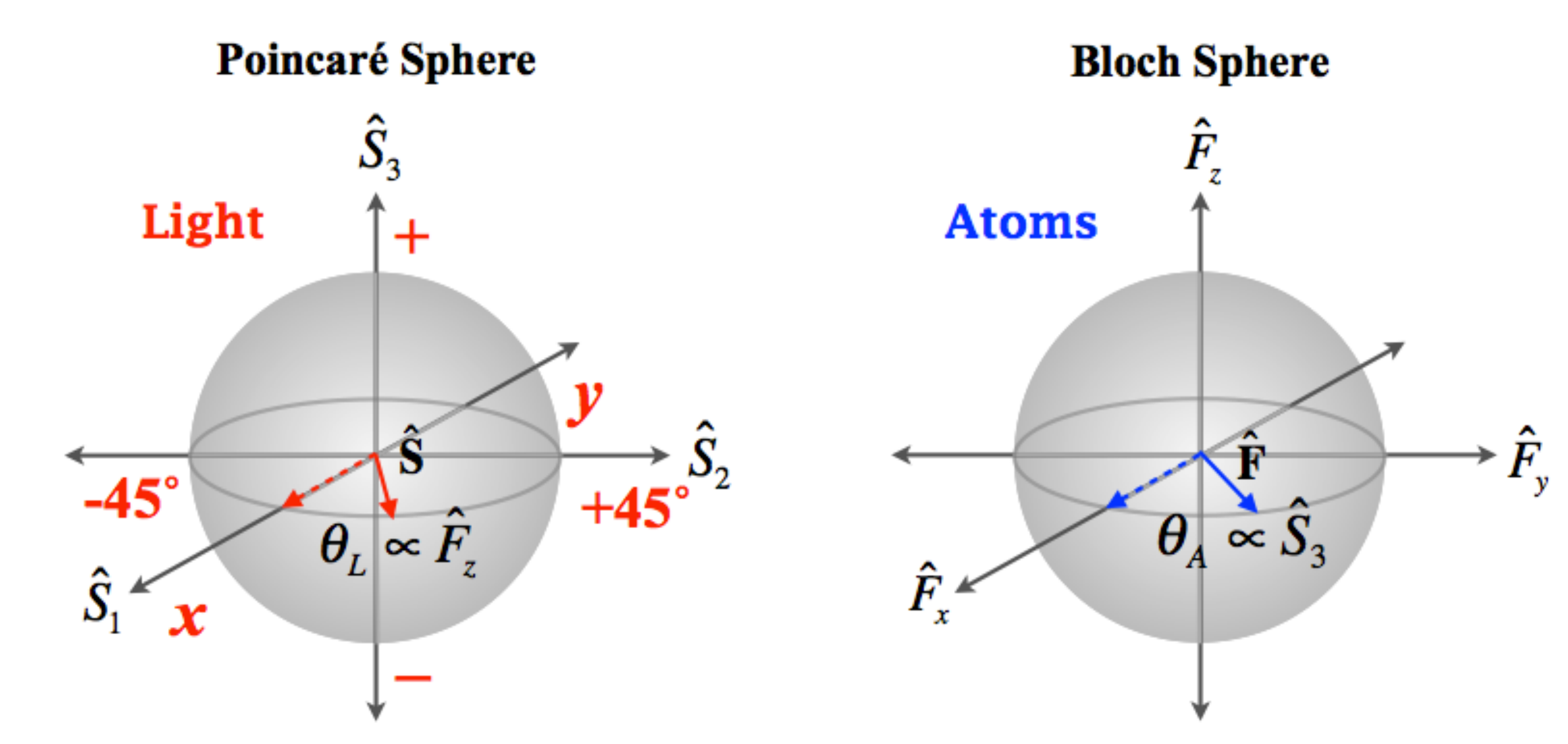}
\caption{The Poincar\'{e} sphere of the light (left) and the Bloch sphere of the atomic ensemble (right). The $x$, $y$ and $z$ axes of the Poincar\'{e} sphere represent $x/y$ linear polarization, diagonal/anti-diagonal linear polarization and right/left circular polarization, respectively. The Faraday interaction, given by the Hamiltonian $\hat{H}=\hbar\chi\hat{S}_3\hat{F}_z/\Delta t$, creates a rotation of the light's linear polarization in the Poincar\'{e} sphere about $\hat{S}_3$ by an angle $\theta_L$, which is proportional to the collective spin $\hat{F}_z$ of the ensemble. The Faraday interaction generates a similar rotation of the ensemble's collective spin about $\hat{F}_z$ by an angle $\theta_A$ proportional to the light's circular polarization, quantified by $\hat{S}_3$.   }\label{fig::FaradaySpheres}
\end{figure}

After removing the tensor term, the Faraday interaction is the dominant contribution to the Hamiltonian. We rewrite the Faraday Hamiltonian as 
\begin{align}\label{eq::FaradayDef}
\hat{H}=\frac{\hbar\chi}{\Delta t}\hat{S}_3\hat{F}_z,
\end{align}
where $\chi=g_f(\sigma_0/A)(\Gamma/6\Delta)$ is the Faraday rotation angle, $g_f$ is the Land\'{e} g-factor, $\Gamma$ is the linewidth of the transition, $A$ is the beam area and $\sigma_0=3\lambda^2/2\pi$ is the resonant cross section for a unit oscillator strength. The Faraday interaction is an effective spin-spin interaction that couples the polarization of the light to the collective spin of the ensemble. The linear polarization of the light rotates by an amount proportional to the $z$-component of the ensemble's collective spin. Similarly, the collective spin of the ensemble rotates about $z$ by an amount proportional to the light's circular polarization, quantified by $\hat{S}_3$.  Pictorially, these effects can be represented as rotations in the Poincar\'{e} and Bloch spheres as shown in Fig. \ref{fig::FaradaySpheres}. Note that the Faraday rotation angle, $\chi$, is proportional to the Land\'{e} g-factor, $g_f$. In the higher hyperfine ground manifold, the Land\'{e} g-factor is given by $g_f=f^{-1}$, while $g_f=(f+1)^{-1}$ in the lower manifold. As a consequence, the strength of the Faraday interaction decreases with increasing $f$ in both manifolds.

\section{Entanglement and the Faraday Interaction}\label{sec::EntangleFaraday}
The most important aspect of the Faraday interaction in relation to spin squeezing is the entanglement it generates between the light and the ensemble. Consider an initial state of the ensemble that can be decomposed in terms of the collective spin eigenstates of $\hat{F}_z$ as  $\ket{\Phi_A}=\sum_{M_z=-N_Af}^{N_Af}C(M_z)\ket{N_Af,M_z}$. For an initial state of the light $\ket{\Phi_L}$, the Faraday interaction produces the entangled state
\begin{align}\label{eq::FaradayState}
\ket{\Phi_{AL}}\!=e^{-i\chi\hat{S}_3\hat{F}_z}\ket{\Phi_L}\ket{\Phi_A}=\!\!\!\!\sum_{M_z=-N_Af}^{N_Af}\!\!\!\!C(M_z)\left(e^{-i\chi\hat{S}_3M_z}\ket{\Phi_L}\right)\ket{N_Af,M_z}.
\end{align}
Each $\hat{F}_z$ eigenstate of the ensemble with eigenvalue $M_z$ is coupled to a state of the light that has been rotated about $\hat{S}_3$ by an angle proportional to $M_z$. Measuring the rotation angle of the light provides information about the associated $\hat{F}_z$ eigenstate of the ensemble and vice versa, an indicator of entanglement. 

More quantitatively, the strength of the entanglement between the light and atoms can be determined by calculating the purity of the reduced density operator of the light after tracing out the ensemble. If the initial state of the ensemble consists of $N_A>>1$ separable identically prepared atomic spins, it is approximately Gaussian by the central limit theorem. The $C(M_z)$ coefficients in $\ket{\Phi_A}$ can, thus, be written as $C(M_z)=(2\pi\Delta F_z^2)^{-1/4}\text{exp}(-\frac{M_z^2}{4\Delta F_z^2})$. Consider an initial state of the light $\ket{N_L}_x$, where all photons are linearly polarized along $x$ and $N_L$ is the number of photons in a pulse of time $\Delta t$. We can take advantage of the $su(2)$ algebra of the Stokes' components to write the initial state as an effective spin eigenstate of $\hat{S}_1$, $\ket{\Phi_L}=\ket{N_L/2,M_1=N_L/2}$. Here, the effective total angular momentum is $N_L/2$ and $M_i$ denotes an eigenvalue of $\hat{S}_i$. In the basis of eigenstates of $\hat{S}_3$, $\ket{\Phi_L}=(\pi N_L/2)^{-1/4}\sum_{M_3=-N_L/2}^{N_L/2}\text{exp}\left(-\frac{M_3^{\;2}}{N_L}\right)\ket{N_L/2,M_3}$, analogous to a spin coherent state. In the limit of continuous $M_z$, the reduced density operator of the light is
\begin{align}
\hat{\rho}_L&=\text{Tr}_A(\ket{\Phi_{AL}}\bra{\Phi_{AL}})\\\notag
&=\sum_{M_3,\,M_3'=-N_L/2}^{N_L/2}\sqrt{\frac{2}{\pi N_L}}e^{-\frac{M_3^{2}+M_3'^{2}}{N_L}}
e^{-\frac{1}{2}(M_3-M_3')^2\Delta F_z^2\chi^2}\ket{N_L/2,M_3}\bra{N_L/2,M_3'}.
\end{align}
In the limit of continuous $M_3$ and $M_3'$, the purity is
\begin{align}\label{eq::purityXi}
\text{Tr}(\rho_L^2)=\frac{1}{\sqrt{1+\chi^2N_L\Delta F_z^2}}.
\end{align}
The purity decreases with the quantity
\begin{align}\label{eq::1stXi}
\xi=\chi^2N_L\Delta F_z^2,
\end{align}
 which we call the ``collective spin coupling constant".  A larger collective spin coupling constant, therefore, signifies greater entanglement between the light and ensemble.
  
While it seems natural that the entanglement between the light and ensemble should increase with $N_L$ and $\chi$, the presence of $\Delta F_z^2$ in the collective spin coupling constant is counterintuitive. The relationship between the ``projection noise" $\Delta F_z^2$ and entanglement is explained by considering a measurement of the ensemble's collective spin mediated by the Faraday interaction. In this form of measurement, the polarization of the light serves as a meter for the ensemble's collective spin. Recall that the Faraday interaction rotates the initial state of the light about $\hat{S}_3$ by an amount proportional to the projections of the ensemble's collective spin in  $\hat{F}_z$, as shown in \erf{eq::FaradayState}. If the ensemble is initially linearly polarized along $x$, $\expect{\hat{S}_2}=0$. By measuring the displacement of the light in $\hat{S}_2$, the rotation angle and corresponding $\hat{F}_z$ eigenstate of the ensemble can be deduced. Measuring the $\hat{S}_2$ Stokes component of the light is, thus, an indirect measurement of the ensemble's collective spin component $\hat{F}_z$.

\begin{figure}[H]
\centering
\includegraphics[scale=.45]{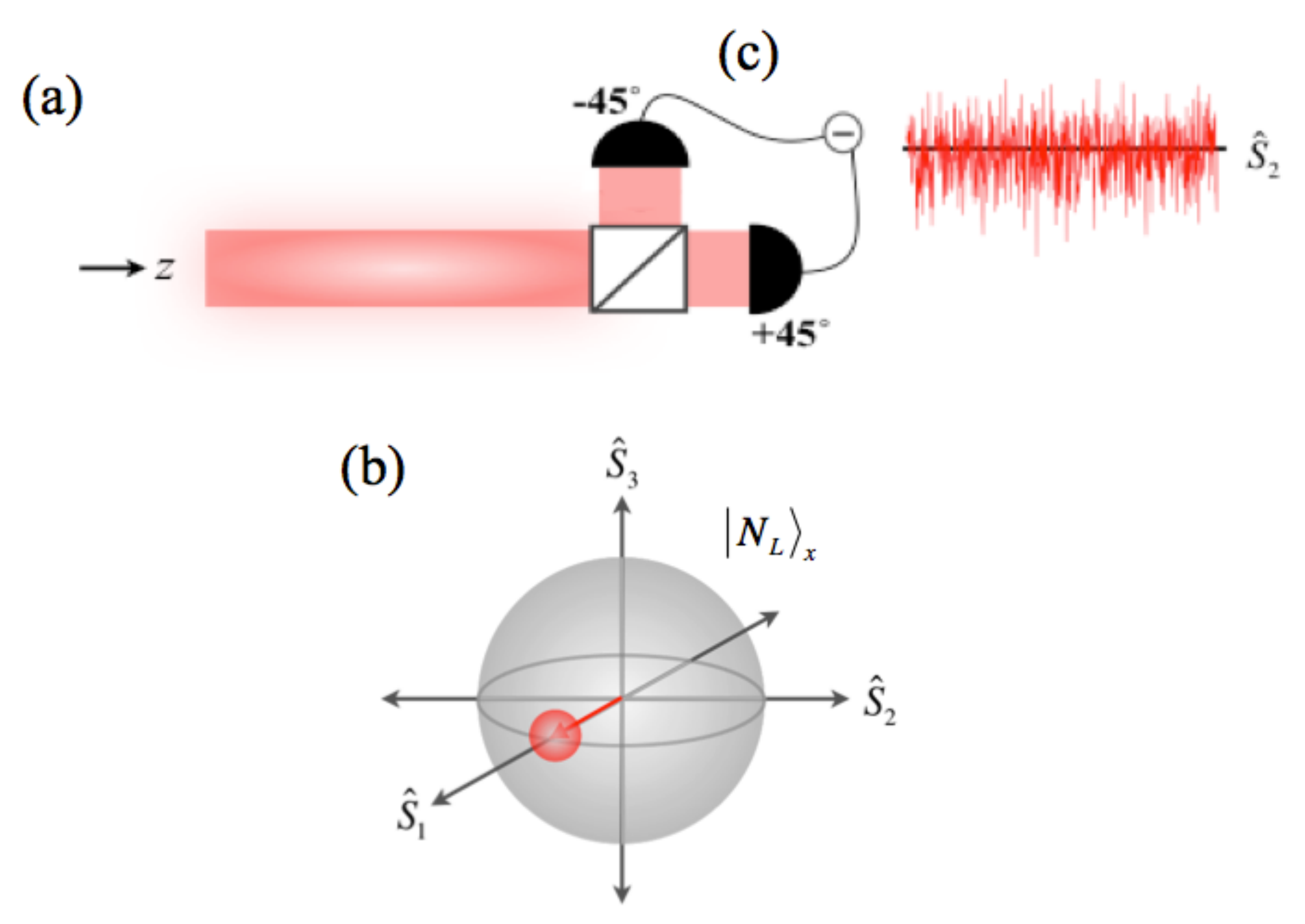}
\caption{The intrinsic shot noise of the light. (a) The light passes through the experimental apparatus, which contains no atoms. (b) The state of the light, $x$ polarized, is an eigenstate of $\hat{S}_1$. The light has variance in the Stokes' components $\hat{S}_2$ and $\hat{S}_3$, corresponding to the diagonal/anti-diagonal polarization and circular polarization of the light. This variance, given by $\Delta S_2^2=\Delta S_3^2=N_L/4$, is known as the shot noise. (c) Measuring the diagonal/anti-diagonal polarization gives a signal with mean zero, since $\expect{\hat{S}_2}=0$. The variance in the signal is the shot noise, $\Delta S_2^2$.   }\label{fig::QNDwoutAtoms}
\end{figure}

\begin{figure}[H]
\centering
\includegraphics[scale=.42]{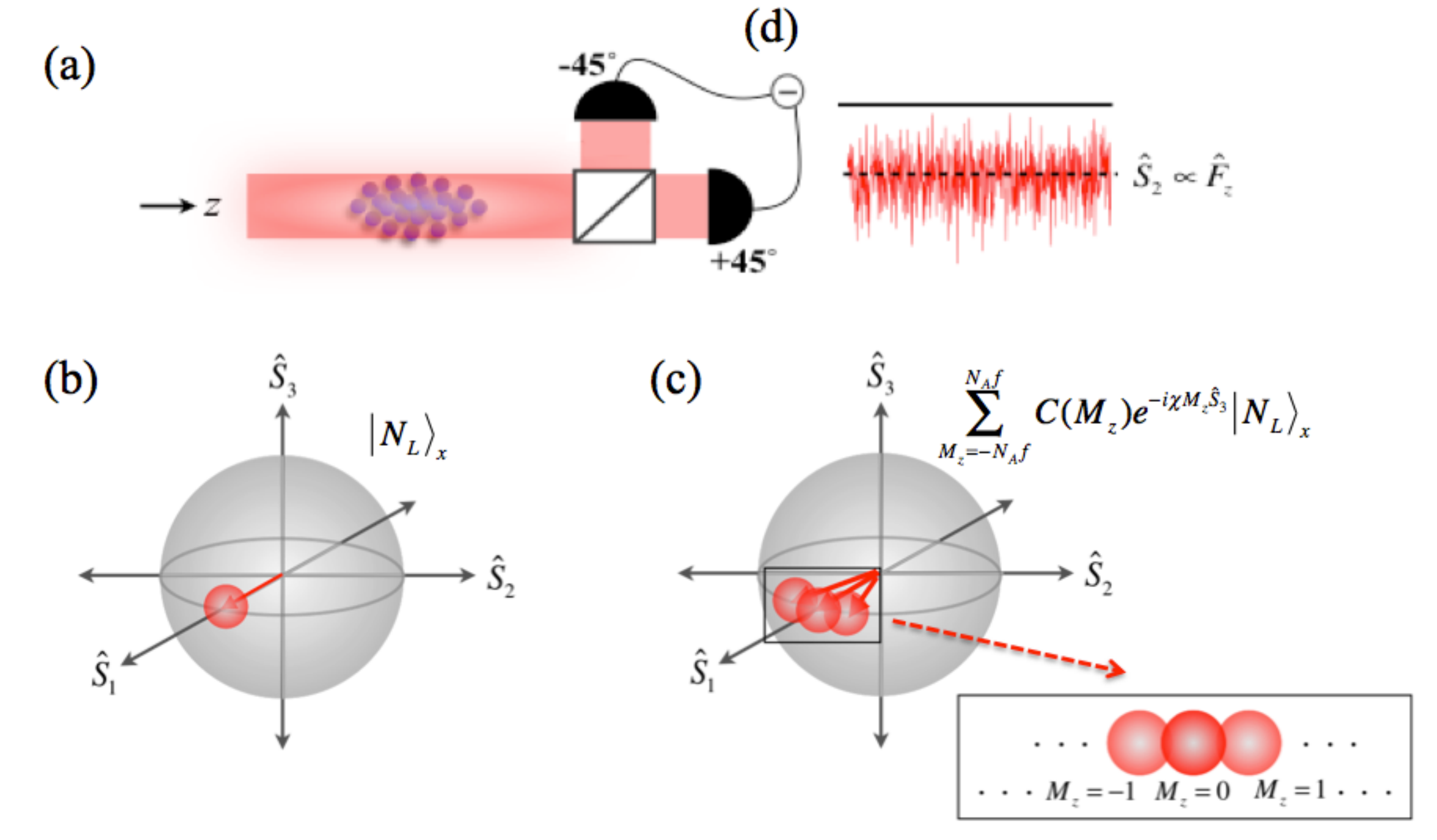}
\caption{The effect of atomic projection noise on atom-light entanglement. (a) The light passes through the experimental apparatus, where it interacts with the atoms. (b) The light, prior to interacting with the atoms, is in an eigenstate of $\hat{S}_1$. The variance in the transverse components, $\Delta S_2^2=\Delta S_3^2=N_L/4$, is the shot noise. (c) If the ensemble is prepared in an initial state $\ket{\Phi_A}=\sum_{M_z=-N_Af}^{N_Af}C(M_z)\ket{N_Af,M_z}$ with $\expect{\hat{F}_z}=0$, the Faraday interaction causes no net rotation. Instead, the Faraday interaction couples each $\hat{F}_z$ eigenstate with a state of the light that has been rotated about $\hat{S}_3$ by an angle proportional to the eigenvalue $M_z$, $\sum_{M_z=-N_Af}^{N_Af}C(M_z)e^{-i\chi M_z\hat{S}_3}\ket{N_L}_x\ket{N_Af,M_z}$. The spread in rotation angles about $\hat{S}_3$ increases with $\Delta F_z^2$ or the ``projection noise" of the ensemble. (d) When  $\hat{S}_2$ is measured by the polarimeter, the mean of the signal corresponds to the rotation angle about  $\hat{S}_3$, while the variance of the signal is the shot noise. The greater spread in rotation angles created by increased atomic projection noise makes the rotation angle more resolvable beneath the shot noise. Because the rotation angle corresponds to an $\hat{F}_z$ eigenstate of the ensemble, information about the ensemble's collective spin is also more resolvable with increased projection noise. A measurement of $\hat{S}_2$, thus, creates more measurement backaction in the ensemble, indicating greater atom-light entanglement.
}\label{fig::QNDwAtoms}
\end{figure}

To understand the role of the projection noise, we examine the measurement of $\hat{S}_2$ in greater detail. Consider  an apparatus used to measure $\hat{S}_2$ in absence of the atomic ensemble, depicted in Fig. \ref{fig::QNDwoutAtoms} (a) and (c). The light travels along $z$ until it encounters a polarizing beam splitter, whereupon half of the light is diverted to a photodetector that counts photons with $+45^\circ$ polarization and the other half is diverted to a photodetector that counts photons with $-45^\circ$ polarization. The subtracted intensities measured by the two photodetectors yields a measurement of $\hat{S}_2=\frac{1}{2}(\hat{a}_{+45^\circ}^\dag\hat{a}_{+45^\circ}-\hat{a}_{-45^\circ}^\dag\hat{a}_{-45^\circ})$. The variance in this measurement, $\Delta S_2^2$, is due to ``shot noise" arising from the vacuum fluctuations in the y-polarized mode. For an initial state of the light that is horizontally polarized, shown in Fig. \ref{fig::QNDwoutAtoms} (b), the shot noise variance is $\Delta S_2^2=N_L/4$. Figure \ref{fig::QNDwAtoms} (a) and (d) shows the same measurement apparatus, except that the light passes through the atomic ensemble en route to the photodetectors. If initial state of the light is $\ket{N_L}_x$, as depicted in Fig. \ref{fig::QNDwAtoms} (b), the state of the light after the Faraday interaction is shown schematically in Fig. \ref{fig::QNDwAtoms} (c). This state consists of a superposition of the states  $\ket{N_L}_x$, all of which have been rotated by an angle $\chi M_z$ about $\hat{S}_3$, corresponding to projections of the ensemble spin state in the $\hat{F}_z$ eigenbasis. For small rotation angles, $\chi M_z$, the displacement of each superposition in $\hat{S}_2$ is 
\begin{align}\label{eq::S2displacement}
\expect{\hat{S}_2}_{M_z}=(N_L/2)\text{sin}(\chi M_z)\approx N_L\chi M_z/2.
\end{align}
As shown in Fig. \ref{fig::QNDwAtoms} (d), the mean of the measurement signal from the photodetectors is equal to the displacement of one of the superpositions of $\ket{N_L}_x$. The variance of the signal over the time interval of the measurement is the shot noise variance of $\ket{N_L}_x$, which is once again $\Delta S_2^2=N_L/4$.

\begin{figure}[H]
\centering
\includegraphics[scale=.47]{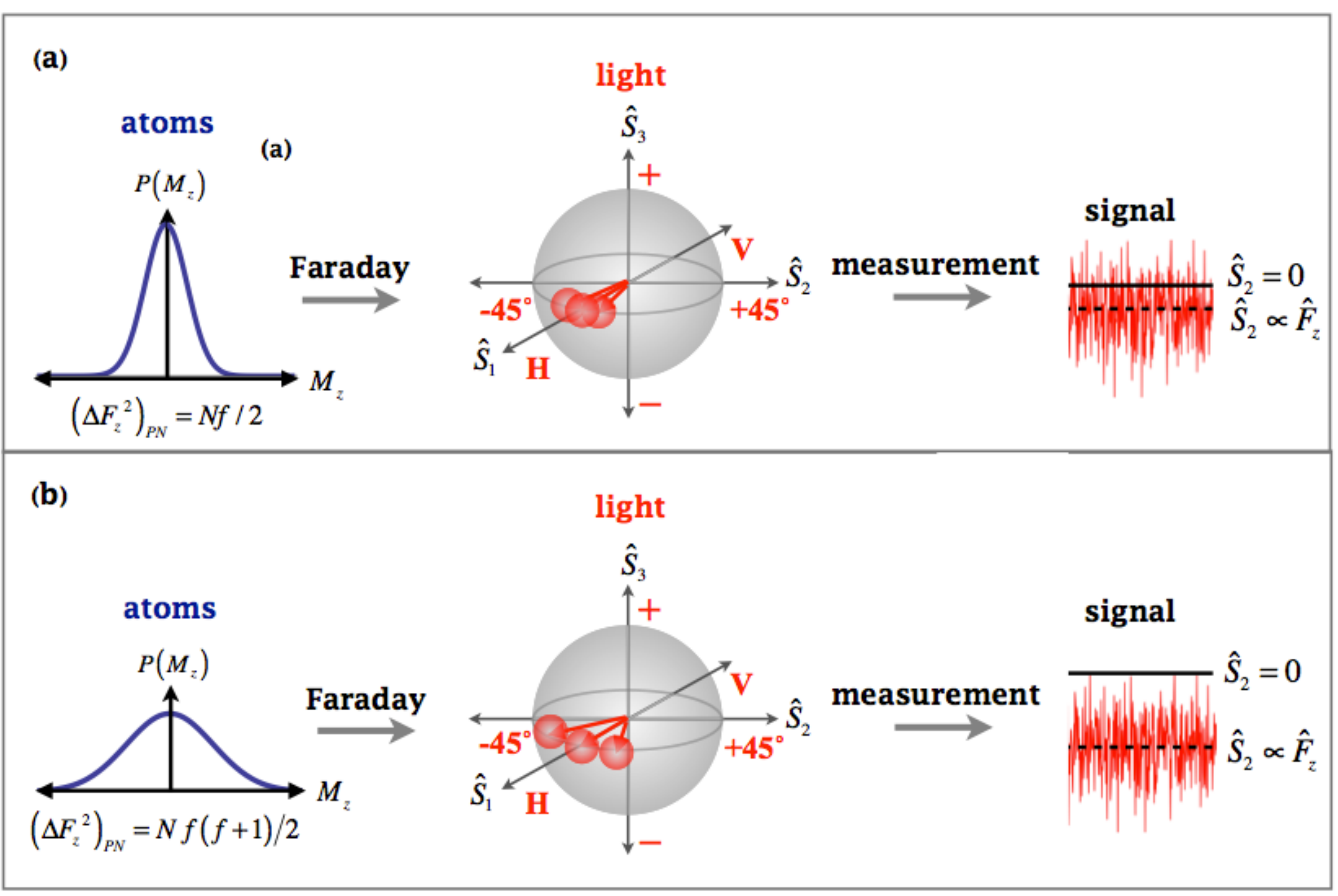}
\caption{The Faraday interaction increases the variance of the light's polarization by an amount proportional to the projection noise of the ensemble's collective spin along $\hat{F}_z$. (a) Variance induced on the light's polarization by an ensemble prepared in a spin coherent state, $\ket{f,\, m_x=f}^{\otimes N_A}$. The projection noise of the spin coherent state is $(\Delta F_z^2)_{PN}=N_Af/2$, the standard quantum limit. (b) Variance induced upon the light's polarization by an ensemble prepared in $\ket{f,\, m_x=0}^{\otimes N_A}$, a state with a larger projection noise than the spin coherent state. This enhanced projection noise, $(\Delta F_z^2)_{PN}=N_Af(f+1)/2$, induces more variance upon the light's polarization, increasing the resolvability of translations corresponding to eigenvalues of $\hat{F}_z$.}\label{fig::StatePreps2}
\end{figure}

Being able to deduce the value of $\hat{F}_z$ by measuring $\hat{S}_2$ requires that the value of the displacement in \erf{eq::S2displacement} is resolvable beneath the shot noise of the signal. The displacement is resolvable as long as $\expect{\hat{S}_2}_{M_z}>\Delta S_2$. The smallest separation between values of $\hat{F}_z$ that is resolvable beneath the shot noise of the light is, therefore,
\begin{align}
(\Delta F_z)_{SN}=\frac{1}{\chi\sqrt{N_L}}.
\end{align}
We refer to $(\Delta F_z^2)_{SN}$ as the shot noise resolution of a measurement of $\hat{F}_z$. Note that a large spread of measurement outcomes is more resolvable than a smaller one. It follows that different measurement outcomes of $\hat{F}_z$ are more resolvable in a measurement of the light when the ensemble has a large projection noise, $\Delta F_z^2$. As illustrated in Fig. \ref{fig::StatePreps2}, such an ensemble has a greater spread of $\hat{F}_z$ eigenstates, producing a greater spread of rotation angles in the state of the light after the Faraday interaction. This leads to larger relative displacements of $\hat{S}_2$ and greater resolution of $\hat{F}_z$. The ability to probe one system and obtain information about another is a signature of entanglement. Indeed, the increased resolution of $\hat{F}_z$ that results from a larger projection noise indicates greater entanglement between the light and atoms. For this reason, the collective spin coupling constant increases with the projection noise. The collective spin coupling constant from \erf{eq::1stXi} can alternatively be expressed as
\begin{align}\label{eq::2ndXI}
\xi=\frac{(\Delta F_z^2)_{PN}}{(\Delta F_z^2)_{SN}}=\chi^2N_L\Delta F_z^2,
\end{align}
where $(\Delta F_z^2)_{PN}$ denotes the projection noise. As demonstrated in \erf{eq::purityXi}, this is the key quantity in determining the entanglement between the light and ensemble. The resolvability of $\hat{F}_z$, and the entanglement generated between the light and ensemble, increases with the ratio of the projection noise to the shot noise resolution. 

\section{Holstein-Primakoff Approximations }\label{sec::HP}
Both the state of the light and the state of the atomic ensemble belong to Hilbert spaces with extremely large dimensions. The state of the light is specified by its two polarization modes, $x$ and $y$, and fixed excitation number $N_L$. In the Schwinger representation, this corresponds to a spin state with total angular momentum number $N_L/2$ and Hilbert space dimension $N_L+1$, where $N_L$ is on the order of $10^8$ for realistic parameters.  In the case of $f=1/2$, the collective spin state of the ensemble has total angular momentum quantum number $N_A/2$ and Hilbert space dimension $N_A+1$. Although this dimensionality is large, since $N_A \sim 10^6$, the situation is appreciably worse when $f>1/2$. Consider an ensemble state $\ket{\Psi}=\ket{\psi}^{\otimes N_A}$, where $\ket{\psi}$ is a state of the $2f+1$ dimensional internal spin. This state can be decomposed in terms of collective spin states with different total angular momentum quantum numbers, $F$, 
\begin{align}
\ket{\Psi}=\sum_F\sum_{M=-F}^FC_M^F\ket{F,M}.
\end{align}
In the case of $f=1/2$, all $C_M^F=0$ except when $F=N_A/2$. This symmetry enables the dimension of the Hilbert space to be reduced from $2^{N_A}$ to $N_A+1$. When $f>1/2$, $\ket{\Psi}$ has projections onto collective spin states with many different total angular momentum quantum numbers $F$. Reducing the dimensionality of the relevant Hilbert space below $(2f+1)^{N_A}$ requires determining the $C_M^F$, which is an open problem. Without analytic methods, we must instead rely upon an approximation. By applying variations of the Holstein-Primakoff (HP) approximation to both the light and the atomic ensemble, we can treat each system as a state on a single bosonic mode \cite{HP}.

\subsection{Holstein-Primakoff Transformation}
The Holstein-Primakoff transformation is a map between angular momentum operators, or any generating set of $su(2)$, and bosonic creation and annihilation operators.  Consider the angular momentum operators $\hat{J}_+$, $\hat{J}_-$ and $\hat{J}_z$. In terms of the creation and annihilation operators $\hat{a}^\dag$ and $\hat{a}$, these angular momentum operators can be expressed as
\begin{align}\label{eq::HPJplus}
\hat{J}_+=\hat{J}_y+i\hat{J}_z=\sqrt{2J}\sqrt{1-\frac{\hat{a}^\dag\hat{a}}{2J}}\hat{a}\\\label{eq::HPJminus}
\hat{J}_-=\hat{J}_y-i\hat{J}_z=\sqrt{2J}\hat{a}^\dag\sqrt{1-\frac{\hat{a}^\dag\hat{a}}{2J}}
\end{align}
and
\begin{align}\label{eq::HPJx}
\hat{J}_x=J-\hat{a}^\dag\hat{a},
\end{align}
where $J$ is the total angular momentum quantum number. 

The Holstein-Primakoff transformation is an exact correspondence. If we desire to reduce the dimensionality of the relevant Hilbert space associated with the angular momentum $\hat{\mathbf{J}}$, we must make an approximation. In cases where $J$ is very large, the expressions in Eqs. (\ref{eq::HPJplus}), (\ref{eq::HPJminus}) and (\ref{eq::HPJx}) become
\begin{align}
\hat{J}_+=\hat{J}_y+i\hat{J}_z\approx\sqrt{2J}\hat{a}\\
\hat{J}_-=\hat{J}_y-i\hat{J}_z\approx\sqrt{2J}\hat{a}^\dag
\end{align}
and
\begin{align}
\hat{J}_x\approx J.
\end{align}
This mapping is referred to as the Holstein-Primakoff approximation.  The operator $\hat{J}_x$ is treated as a classical quantity with no variance, while the operators $\hat{J}_+$ and $\hat{J}_-$ remain quantum. By making the Holstein-Primakoff approximation, we have restricted the Hilbert space to states where $\expect{\hat{J}_x}\approx J$.

\subsection{Holstein-Primakoff Approximation on the Light}
Because the Stokes' components satisfy the $su(2)$ commutation relations, we can write these operators in terms of creation and annihilation operators through the Holstein-Primakoff transformation. For the squeezing protocols we will later consider, the initial state of the light is linearly polarized along $x$, which is equivalent to the effective angular momentum eigenstate of $\hat{S}_1$ with maximal spin projection, $\ket{N_L/2, M_1=N_L/2}$. Since the effective total angular momentum of this state is $N_L/2>>1$, we can apply the Holstein-Primakoff approximation to the Stokes' components. The Stokes' component $\hat{S}_1$ becomes a classical quantity with 
\begin{align}\label{eq::S1NL}
\hat{S}_1=\frac{1}{2}(\hat{a}_x^\dag\hat{a}_x-\hat{a}_y^\dag\hat{a}_y)\approx N_L/2. 
\end{align}
The number of excitations  in the $x$ mode, likewise, becomes effectively classical with $\hat{a}_x^\dag\hat{a}_x\approx N_L$. The creation and annihilation operators on the $y$ mode, on the other hand, are non-classical, containing all of the quantum uncertainty associated with the light. The operators on the light that remain quantum depend upon $\hat{a}_y^\dag$ and $\hat{a}_y$ with
\begin{align}
\hat{S}_+=\hat{S}_2+i\hat{S}_3\approx\sqrt{N_L}\hat{a}_y
\end{align}
and
\begin{align}
\hat{S}_-=\hat{S}_2-i\hat{S}_3\approx\sqrt{N_L}\hat{a}_y^\dag.
\end{align}
The Stokes' components $\hat{S}_2$ and $\hat{S}_3$ become
\begin{align}\label{eq::S2Xy}
\hat{S}_2=\sqrt{\frac{N_L}{2}}\hat{X}_y
\end{align}
and
\begin{align}\label{eq::S3Py}
\hat{S}_3=\sqrt{\frac{N_L}{2}}\hat{P}_y,
\end{align}
where $\hat{X}_y=(\hat{a}_y^\dag+\hat{a}_y)/\sqrt{2}$ and $\hat{P}_y=i(\hat{a}_y^\dag-\hat{a}_y)/\sqrt{2}$ are the position and momentum quadratures on the mode $y$.

\begin{figure}[H]
\centering
\includegraphics[scale=.77]{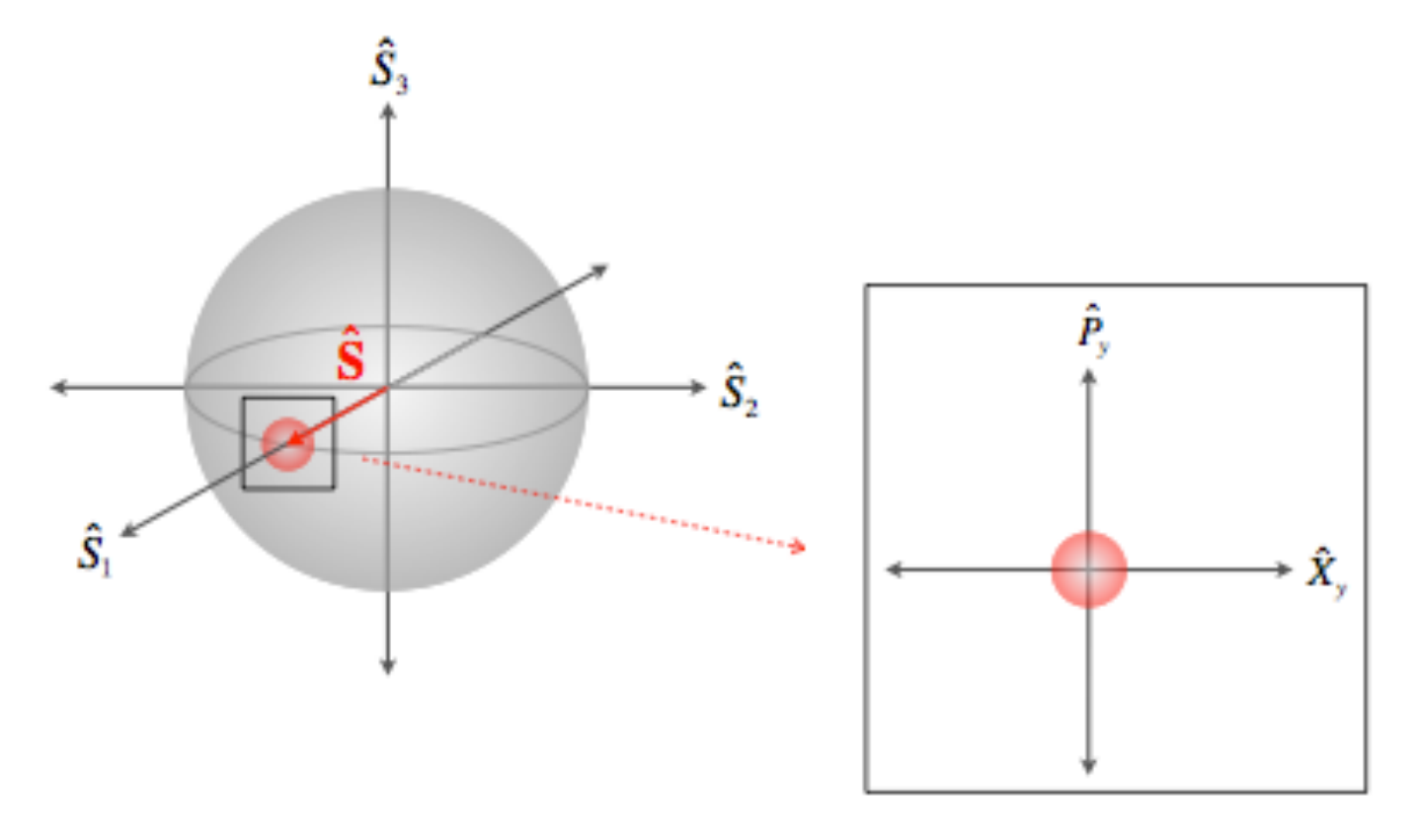}
\caption{The Holstein-Primakoff approximation on the light. The light, completely polarized along $x$, is in the initial state $\ket{N_L}_x$ for which $\hat{S}_1\approx N_L/2$. In this regime, the uncertainties of the transverse Stokes' components $\Delta S_2=\Delta S_3=\sqrt{N_L}/2$ are much smaller than $\hat{S}_1$. The state of the light, thus, occupies a small locally flat region of the Poincar\'{e} sphere. In this region, the Stokes' components $\hat{S}_2$ and $\hat{S}_3$ are well approximated by the position and momentum quadratures $\hat{X}_y$ and $\hat{P}_y$.}\label{fig::HPlight}
\end{figure}

Interestingly, we can arrive at the same approximation directly from the definition of the Stokes' components. Again, we take the initial state of the light to be linearly polarized along $x$ so that $\langle\hat{S}_1\rangle=N_L/2$. Because $\expect{\hat{S}_1}$ is large relative  to the uncertainties of the orthogonal Stokes' components, given by $\Delta S_2=\Delta S_3=\sqrt{N_L}/2$, the state of the light  is confined to a small locally flat region of the Poincar\'{e} sphere as shown in Fig. \ref{fig::HPlight}. In this region, the number of $x$-polarized photons remains approximately constant at $N_L$, implying $\hat{a}_x^\dag\approx\hat{a}_x\approx\sqrt{N_L}$. Under this approximation, the Stokes' vector components in \erf{Stokes} become
\begin{align}
\hat{S}_1&\approx N_L/2\\
\hat{S}_2&\approx\frac{\sqrt{N_L}}{2}(\hat{a}_y+\hat{a}_y^\dag)=\sqrt{\frac{N_L}{2}}\hat{X}_y
\end{align}
and
\begin{align}
\hat{S}_3&\approx\frac{\sqrt{N_L}}{2i}(\hat{a}_y-\hat{a}_y^\dag)=\sqrt{\frac{N_L}{2}}\hat{P}_y, 
\end{align}
equivalent to Eqs. (\ref{eq::S1NL}), (\ref{eq::S2Xy}) and (\ref{eq::S3Py}).

Regardless of how we obtain the Holstein-Primakoff approximation, the result is that the state of the light is specified by a single bosonic mode $y$. The initial state of the light, completely polarized along $x$, corresponds to the vacuum state, $\ket{0}_y$. So long as the light undergoes weak interactions that do not lead to large displacements on the Poincar\'{e} sphere, implying that the light remains in a region where $\expect{\hat{S}_1}\approx N_L/2$, its state remains well approximated by a single mode. The Faraday interaction fits this criteria, as $\chi<<1$ for realistic parameters.

\subsection{Holstein-Primakoff Approximation on the Atomic \\Ensemble}\label{sec::HPEnsemble}
In a manner similar to the light, we can greatly simplify the state of the atomic ensemble through the Holstein-Primakoff approximation. We first consider a spin coherent state of the ensemble along $x$ for $f=1/2$, which is given by \\$\ket{N_A/2,M_x=N_A/2}=\ket{1/2,m_x=1/2}^{\otimes N_A}$. This is an exact analogy to the case in which the light is linearly polarized along $x$ with the angular momentum operators $\hat{F}_x$, $\hat{F}_y$ and $\hat{F}_z$ taking the place of the Stokes' components $\hat{S}_1$, $\hat{S}_3$ and $\hat{S}_2$, respectively. This parallel is made especially clear when we express the collective angular momentum operators of the ensemble in  the Schwinger representation. The Schwinger representation is another mapping from angular momentum operators to bosonic creation and annihilation operators. Unlike the Holstein-Primakoff transformation, however, the Schwinger representation expresses angular momentum operators in terms of two bosonic modes. Consider the collective spin operators acting on the $f=1/2$ atomic ensemble written in the basis of $\hat{f}_x$ eigenstates \\$\ket{\uparrow}=\ket{f=1/2,m_x=1/2}$ and $\ket{\downarrow}=i\ket{f=1/2,m_x=-1/2}$,
\begin{align}\label{eq::FxHalf}
\hat{F}_x=\frac{1}{2}\sum_{i=1}^{N_A}\left(\ket{\uparrow}\bra{\uparrow}_i-\ket{\downarrow}\bra{\downarrow}_i\right)\\\label{eq::FyHalf}
\hat{F}_y=\frac{i}{2}\sum_{i=1}^{N_A}\left(\ket{\downarrow}\bra{\uparrow}_i-\ket{\uparrow}\bra{\downarrow}_i\right)
\end{align}
and
\begin{align}\label{eq::FzHalf}
\hat{F}_z=\frac{1}{2}\sum_{i=1}^{N_A}\left(\ket{\downarrow}\bra{\uparrow}_i+\ket{\uparrow}\bra{\downarrow}_i\right).
\end{align}
The phase of the state $\ket{\downarrow}$ has been selected so that the collective spin operator $\hat{F}_z$ corresponds with the derivation in Sec. \ref{sec::MultiHPEnsemble}, which employs the multilevel Holstein-Primakoff approximation. In the expressions above, the term $\sum_{i=1}^{N_A}\ket{\downarrow}\bra{\uparrow}_i$ that occurs in Eqs. (\ref{eq::FxHalf}) and (\ref{eq::FyHalf}) can be viewed as annihilating an atom in the state $\ket{\uparrow}$ and creating an atom in the state $\ket{\downarrow}$. Likewise, the term $\sum_{i=1}^{N_A}\ket{\uparrow}\bra{\downarrow}_i$ creates an atom in the state $\ket{\uparrow}$ and annihilates an atom in the state $\ket{\downarrow}$. The terms $\sum_{i=1}^{N_A}\ket{\uparrow}\bra{\uparrow}_i$ and $\sum_{i=1}^{N_A}\ket{\downarrow}\bra{\downarrow}_i$ in \erf{eq::FzHalf} are number operators, quantifying the number of atoms in states $\ket{\uparrow}$ and $\ket{\downarrow}$, respectively.  The Schwinger representation recasts the collective spin operators in terms of creation and annihilation operators on the ``modes" $\uparrow$ and $\downarrow$,
\begin{align}\label{eq::FxHalf2}
\hat{F}_x=\frac{1}{2}\left(\hat{a}_{\uparrow}^\dag\hat{a}_{\uparrow}-\hat{a}_{\downarrow}^\dag\hat{a}_{\downarrow}\right)\\\label{eq::FyHalf2}
\hat{F}_y=\frac{i}{2}\left(\hat{a}_{\downarrow}^\dag\hat{a}_{\uparrow}-\hat{a}_{\uparrow}^\dag\hat{a}_{\downarrow}\right)
\end{align}
and
\begin{align}\label{eq::FzHalf2}
\hat{F}_z=\frac{1}{2}\left(\hat{a}_{\downarrow}^\dag\hat{a}_{\uparrow}+\hat{a}_{\uparrow}^\dag\hat{a}_{\downarrow}\right).
\end{align}
Note that these are identical to the Stokes' components with the modes $x$ and $y$ being replaced by the modes $\uparrow$ and $\downarrow$. 

\begin{figure}[H]
\centering
\includegraphics[scale=.42]{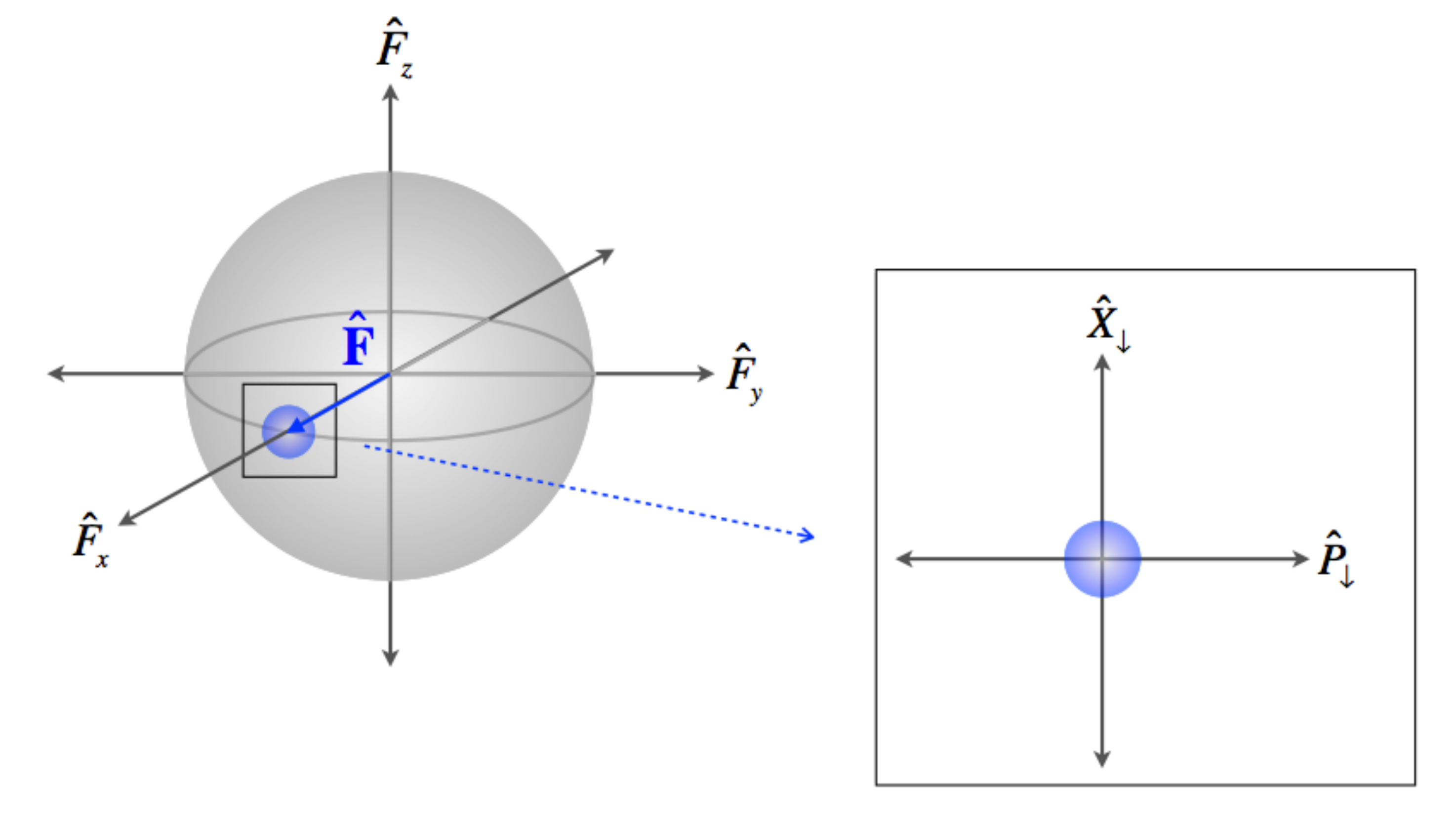}
\caption{The Holstein-Primakoff approximation on the ensemble prepared in a spin coherent state. The HP approximation is applied in the exact same manner as it was on the light in Fig. \ref{fig::HPlight}. Because the atoms are completely spin polarized along $x$, $\hat{F}_x\approx N_Af$. Because the transverse collective spin uncertainties $\Delta F_y$ and $\Delta F_z$ are so much smaller than $\hat{F}_x$, the state of the ensemble occupies a small, locally flat region of the Bloch sphere. The collective spin operators $\hat{F}_y$ and $\hat{F}_z$ are approximated as the position and momentum quadratures $\hat{P}_\downarrow$ and $\hat{X}_\downarrow$.}\label{fig::HPatoms}
\end{figure}

Because the ensemble prepared in a spin coherent state for  $f=1/2$  has total angular momentum $F=N_A/2>>1$, the Holstein-Primakoff approximation holds on the collective spin operators. The collective spins take a form identical to the Stokes' components under the Holstein-Primakoff approximation, 
\begin{align}
\hat{F}_x&\approx N_A/2\\
\hat{F}_y&\approx
\frac{\sqrt{N_A}}{2i}\big(\hat{a}_{\downarrow}-\hat{a}_{\downarrow}^\dag\big)=\sqrt{\frac{N_A}{2}}\hat{P}_{\downarrow}
\end{align}
and
\begin{align}
\hat{F}_z&\approx\frac{\sqrt{N_A}}{2}\big(\hat{a}_{\downarrow}+\hat{a}_{\downarrow}^\dag\big)=\sqrt{\frac{N_A}{2}}\hat{X}_{\downarrow}. 
\end{align}
The collective spin component $\hat{F}_x$ is treated classically, analogous to $\hat{S}_1$.  The $\uparrow$ mode is, likewise, the treated classically with $\hat{a}_{\uparrow}^\dag=\hat{a}_{\uparrow}=\sqrt{N_A}$, while the $\downarrow$ mode is quantum. The state of the ensemble is specified solely by the mode $\downarrow$. The initial spin coherent state with each atom in the state $\ket{\uparrow}$, is equivalent to the vacuum, $\ket{0}_{\downarrow}$.  As shown in Fig. \ref{fig::HPatoms}, the Holstein-Primakoff approximation can also be understood pictorially on the Bloch sphere similarly to the Poincar\'{e} sphere in the case of the light. When $\expect{\hat{F}_x}=N_A/2>>1$, it is much larger than the uncertainties of the transverse collective spin components, $\Delta F_y=\Delta F_z=\sqrt{N_A}/2$. The state of the ensemble is, thus, confined to a small region of the Bloch sphere that can be treated as a locally flat plane. In this region, the transverse collective spin components are well approximated by  quadratures on a single bosonic mode. 

\subsection{Multilevel Holstein-Primakoff Approximation on the Ensemble}\label{sec::MultiHPEnsemble}
For an ensemble with $f=1/2$ prepared in a spin coherent state, the Holstein-Primakoff approximation allows us to treat the state of the ensemble and all associated operators as being on a single bosonic mode. We would like to do this for a more general ensemble state where $f\geq1/2$, however.  This can be accomplished through the multilevel Holstein-Primakoff approximation of Kurucz and M\o lmer \cite{MultilevelHP}. Consider a separable initial state of the ensemble of the form $\ket{\Psi}=\ket{\uparrow}^{\otimes N_A}$, where each atom is identically prepared in some arbitrary state $\ket{\uparrow}$ in the spin-$f$ ground manifold. We should stress that, unlike in the previous section, $\ket{\uparrow}$ is not the state of a qubit, but the state of a $2f+1$ dimensional \textit{qudit}. To first order in $\chi$ when $\bra{\uparrow}\hat{f}_z\ket{\uparrow}$=\,0 and $\ket{\uparrow}$ is not an eigenstate of $\hat{f}_z$, the Faraday interaction maps each atom to an orthogonal state $\ket{\downarrow}$,
\begin{eqnarray}
(\mathbb{I}^{(j)}-i\chi\hat{S}_3\hat{f}_z^{(j)})\ket{\uparrow}^{(j)}=\ket{\uparrow}^{(j)}-i\chi\hat{S}_3\sqrt{(\Delta f_z^2)_\uparrow}\ket{\downarrow}^{(j)},
\end{eqnarray}   
where $(\Delta f_z^2)_\uparrow$ is shorthand for $\bra{\uparrow}(\Delta\hat{f}_z)^2\ket{\uparrow}$. For small time scales, therefore, we can approximate each atomic spin as a qubit embedded in the larger $2f+1$ dimensional hyperfine manifold.  The states defining our embedded qubit, $\ket{\uparrow}$ and $\ket{\downarrow}$, we refer to as the ``fiducial state" and the ``coupled state''.  In terms of the fiducial state, the coupled state is given by
\begin{align}\label{eq::coupledDef}
\ket{\downarrow}=\frac{\hat{f}_z\ket{\uparrow}}{\sqrt{(\Delta\hat{f}_z^2)_\uparrow}},
\end{align}
Here, we have assumed for simplicity that $\expect{\hat{f}_z}_\uparrow=\expect{\hat{f}_z}_\downarrow=0$, where $\expect{\hat{f}_z}_\psi$ is shorthand for $\bra{\psi}\hat{f}_z\ket{\psi}$. We treat the most general case, where the means of the fiducial and coupled states are not restricted to being zero, in Chapter \ref{sec::Beyond}. Also note that no coupled state exists when the fiducial state is an eigenstate of $\hat{f}_z$.

Restricting each atomic spin to the fiducial and coupled states, the collective spin $\hat{F}_z$ becomes 
\begin{align}\label{FzExpand}
\hat{F}_z=\sum_{i=1}^{N_A}\sqrt{(\Delta f_z^2)_\uparrow}(\ket{\uparrow}\bra{\downarrow}_i+\ket{\downarrow}\bra{\uparrow}_i).
\end{align}
Each atom is now an ``embedded qubit" in the larger $2f+1$ dimensional hyperfine manifold. We can use the same techniques that we employed in the case of the light and the $f=1/2$ ensemble.
The Schwinger representation enables us to express $\hat{F}_z$ in terms of creation and annihilation operators on a pair of oscillator modes $\uparrow$ and $\downarrow$,
\begin{align}\label{eq::Schwinger}
\hat{F}_z&\approx\sqrt{(\Delta f_z^2)_\uparrow}(\hat{a}_\uparrow^\dag\hat{a}_\downarrow+\hat{a}_\downarrow^\dag\hat{a}_\uparrow).
\end{align}
The operators $\hat{a}_{\uparrow(\downarrow)}^\dag$ and $\hat{a}_{\uparrow(\downarrow)}$ create and annihilate an atom in the state $\ket{\uparrow\!\!(\downarrow)}$. Because the number of atoms in the fiducial state remains approximately equal to $N_A>>1$ when $\chi<<1$, we can treat the $\uparrow$ mode as classical by taking  $\hat{a}_{\uparrow}^\dag\approx\hat{a}_{\uparrow}\approx\sqrt{N_A}$. Upon making this approximation, 
\begin{eqnarray}\label{eq::MeanZeroPosition}
\hat{F}_z\approx\sqrt{2N_A(\Delta f_z^2)_\uparrow}\hat{X}_{\downarrow}.
\end{eqnarray}
where $\hat{X}_{\downarrow}$ is the position quadrature on the $\downarrow$ mode. 

In terms of operators on the embedded qubit ensemble, the position quadrature is defined as
\begin{align}
\hat{X}_\downarrow&=\frac{1}{\sqrt{2}}(\hat{a}_\downarrow^\dag+\hat{a}_\downarrow)\notag\\
&\label{eq::position}
\approx\frac{1}{\sqrt{2N_A}}\sum_{i=1}^{N_A}(\ket{\uparrow}\bra{\downarrow}_i+\ket{\downarrow}\bra{\uparrow}_i).
\end{align}
While it is not related to a collective spin component in the same manner as $\hat{X}_{\downarrow}$, a momentum quadrature is similarly defined as an operator on the embedded qubit ensemble,
\begin{align}
\hat{P}_{\downarrow}&=\frac{i}{\sqrt{2}}(\hat{a}_\downarrow^\dag-\hat{a}_\downarrow)\notag\\
&\label{eq::momentum}
\approx\frac{i}{\sqrt{2N_A}}\sum_{i=1}^{N_A}(\ket{\downarrow}\bra{\uparrow}_i-\ket{\uparrow}\bra{\downarrow}_i).
\end{align}
In the limit where nearly all of the $N_A$ atoms remain in the fiducial state, the ensemble quadratures obey the canonical commutation relations,
\begin{align}
[\hat{X}_\downarrow,\hat{P}_\downarrow]=\frac{i}{N_A}\sum_{i=1}^{N_A}(\ket{\uparrow}\bra{\uparrow}_i-\ket{\downarrow}\bra{\downarrow}_i)\approx i.
\end{align}


After making the multilevel Holstein-Primakoff approximation, the state of the ensemble can be succinctly expressed in terms of a single oscillator mode. Similar to the light, the initial state of the ensemble with each atom prepared in $\ket{\uparrow}$ corresponds to the vacuum state, $\ket{0}_\downarrow$. The ensemble effectively resides in a phase plane defined by the position and momentum quadratures, $\hat{X}_\downarrow$ and $\hat{P}_\downarrow$. This approximation holds for a weak Faraday interaction that does not transfer atoms outside the embedded qubit appreciably.

\section{Phase Plane Faraday Interaction}\label{Sec::PhasePlaneFaraday}
After making the Holstein-Primakoff approximation on the light and the multilevel Holstein-Primakoff approximation on the ensemble, the Faraday interaction acts on the effective modes $y$ and $\downarrow$. By combining Eqs. (\ref{eq::FaradayDef}), (\ref{eq::S3Py}) and (\ref{eq::MeanZeroPosition}), we can write the Faraday Hamiltonian in terms of the quadratures on modes $y$ and $\downarrow$ as
\begin{align}\label{eq::HPfaraday}
\hat{H}=\frac{\hbar\chi}{\Delta t}\sqrt{N_LN_A(\Delta f_z^2)_\uparrow}\hat{P}_y\hat{X}_\downarrow.
\end{align}
This interaction generates entanglement between the light and atoms by coupling the modes $y$ and $\downarrow$.  In the phase plane picture, the Faraday interaction generates a translation of the light in $\hat{X}_y$ by an amount proportional to the position of the atoms, $\hat{X}_\downarrow$. Likewise, the Faraday interaction translates the state of the ensemble along $\hat{P}_\downarrow$ by an amount depending on the the momentum of the light, $\hat{P}_y$. Note that the strength of the coupling between the light and ensemble increases with the projection noise of the ensemble, given by $(\Delta F_z^2)=N_A(\Delta f_z^2)_\uparrow$. This is another analytic demonstration that greater ensemble projection noise leads to increased entanglement between the light and atoms.

\section{Internal Spin Control}\label{sec::IntControl}
The previous sections have demonstrated that the strength of the Faraday interaction and the resulting atom-light entanglement increase with the initial projection noise of the ensemble.  Because the initial projection noise is proportional to the variance of the fiducial state, it follows that we can enhance the Faraday interaction by preparing each atom in a fiducial state, $\ket{\uparrow}$, with larger internal spin variance, $(\Delta\hat{f}_z^2)_\uparrow$.  Through a combination of radio-frequency and microwave magnetic fields, the hyperfine ground state of an alkali atom is completely controllable \cite{MerkelControlPRA}. This control has been experimentally demonstrated in the $f=3$ and $f=4$ ground manifolds of cesium \cite{ASmith13}.  Controllability of the ground hyperfine spin enables us to apply any unitary transformation $\hat{u}$ to the space spanned by the magnetic sublevels of $f$. For the ensemble, control over the internal hyperfine spins of the atoms permits local unitary transformations of the form $\hat{u}^{\otimes N_A}$. Because of this restriction, the ensemble can be prepared in any state of the form $\ket{\uparrow}^{\otimes N_A}$.  Internal spin control also enables us to apply arbitrary rotations and a variety of other useful transformations, which will be detailed in later sections. 

\subsection{State Preparations}\label{sec::StatePreps}
To more concretely demonstrate how state preparation influences entanglement generation, we analyze the performance of squeezing protocols using three specific fiducial states. The results we present can be generalized to any fiducial state, however. The the fiducial states we consider have varying amounts of projection noise, leading to different degrees of atom-light entanglement and spin squeezing. In later chapters, we will also use these fiducial states to explore the impact of state preparation upon the decoherence of the ensemble. These studies demonstrate that state preparation greatly influences achievable spin squeezing.

The first state preparation we consider is the familiar spin coherent state ($SCS$) in which the fiducial state is the maximal spin projection along $x$,
\begin{align}
\ket{\uparrow_{SCS}}&=\ket{f,m_x=f}.    
\end{align}
The coupled state in the multilevel HP approximation is
\begin{align}
\ket{\downarrow_{SCS}}=i\ket{f,m_x=f-1},
\end{align}
since $\Delta\hat{f}_z\ket{f,m_x=f} =i\sqrt{f/2}\:\ket{f,m_x=f-1}$, where $\left(\Delta f_z^2\right)_{\uparrow_{SCS}}=f/2$. The SCS has the smallest initial projection noise of any state preparation we consider. As a consequence, the collective spin coupling constant, which is given by $\xi(\uparrow_{SCS})=\\\gamma\Delta t OD/(18f)$, decays the most markedly with $f$.

The ``cat" preparation, in which each atom is prepared in the fiducial state
\begin{align}
\ket{\uparrow_{\text{cat}}}&=\frac{1}{\sqrt{2}}\left(\ket{f,m_z=f}+\ket{f,m_z=-f}\right),
\end{align}
has the largest projection noise of any initially separable ensemble state with 
\\$(\Delta f_z^2)_{\uparrow_{\text{cat}}}=f^2$. From Eq. (\ref{eq::coupledDef}), the coupled state for the cat preparation
\begin{align}
\ket{\downarrow_{\text{cat}}}=\frac{1}{\sqrt{2}}\left(\ket{f,m_z=f}-\ket{f,m_z=-f}\right). 
\end{align}
Due to its sizable initial projection noise, the cat preparation exhibits the largest collective spin coupling constant, $\xi(\uparrow_{\text{cat}})=\gamma\Delta t OD/9$. Interestingly,  $\xi(\uparrow_{\text{cat}})$ is also independent of $f$.

Although the cat preparation generates the largest coherent interaction strength, it is extremely susceptible to decoherence, as our analysis will later show. Because of this, we consider an additional state preparation, ``$m_x=0$", with an intermediate projection noise between the $SCS$ and the cat. In the this preparation, each atom is prepared in the magnetic sublevel with zero spin projection along $x$,  
\begin{align}
\ket{\uparrow_0}=\ket{f,m_x=0}.
\end{align}
The variance of the fiducial state, $(\Delta f_z^2)_{\uparrow_0}=f(f+1)/2$, scales quadratically with $f$ like the cat state. By again using Eq. (\ref{eq::coupledDef}), we determine coupled state to be
\begin{align}\label{coupledMx0} 
\ket{\downarrow_0}=\frac{i}{\sqrt{2}}\left(\ket{f,m_z=-1}-\ket{f,m_z=1}\right). 
\end{align}
While the collective spin coupling constant $\xi(\uparrow_0)=\gamma\Delta t OD(f+1)/(18f)$ still decreases with $f$, it does so at much reduced rate compared to the $SCS$.

\section{Gaussian States}\label{sec::GaussianStates}
Making the HP approximation on the light and the multilevel HP approximation on the atoms enables us to treat both systems on equal footing. The states of the light and the atoms become states of bosonic modes, each initially prepared in the vacuum state. This fact is particularly significant, as the vacuum state is a Gaussian state in phase space.  Moreover, the interaction between the light and atomic ensemble preserves this Gaussianity to good approximation. Gaussian states, which have positive Gaussian-distributed Wigner functions, possess many useful properties that greatly simplify the description of the atom-light interface. Below we highlight several of the properties we will later utilize to characterize  light-mediated squeezing and decoherence of the atomic ensemble. For more comprehensive reviews of Gaussian states, see Refs. \cite{Giedke02, PlenioEisert, Wang07, Weedbrook12, Adesso14}. Throughout this section, we consider the most general Gaussian state on a set of bosonic modes $1,...,n$. Specializing to the case of the atom-light system requires that we consider only two such modes, one associated with the ensemble, labeled $\downarrow$, and one associated with the light, labeled $y$.

A Gaussian state $\hat{\rho}\,$ on modes $1,...,n$ is fully specified by the first and second order moments of the phase space quadratures $\hat{X}_1,\hat{P}_1,...,\hat{X}_n,\hat{P}_n$. Because we are primarily concerned with entanglement generation, we need only consider the variances and covariances of the quadratures, which contain the atom-light and interatomic correlations. Information regarding these correlations is stored in the $2n\times 2n$ covariance matrix, $\Sigma$, with elements 
\begin{align}
\Sigma_{ij}=\frac{\langle\Delta\hat{\textbf{d}}_i\Delta\hat{\textbf{d}}_j+\Delta\hat{\textbf{d}}_j\Delta\hat{\textbf{d}}_i\rangle}{2},
\end{align}
where $\hat{\textbf{d}}=\{\hat{X}_1,\hat{P}_1,...,\hat{X}_n,\hat{P}_n\}^T$ and $\Delta\hat{\textbf{d}}_i=\hat{\textbf{d}}_i-\langle\hat{\textbf{d}}_i\rangle$. All covariance matrices corresponding to physical states satisfy 
\begin{align}\label{continuousUncert}
\Sigma +\frac{i}{2}\sigma\geq 0,
\end{align}
where the matrix $\sigma$, known as the symplectic matrix, is defined in terms of the canonical commutation relations
\begin{align}
\sigma_{jk}=-i[\hat{d}_j,\hat{d}_k]
\end{align}
or
\begin{eqnarray}\label{eq::sympMatrix}
\sigma=\bigoplus_{i=1}^{n}\left(\begin{matrix} 0 & 1 \\ -1 & 0\end{matrix}\right).
\end{eqnarray}
As a consequence of \erf{continuousUncert}, the Heisenberg uncertainty relations are fulfilled on pairs of conjugate quadratures \cite{SymplecticGeo}.

Up to a translation in phase space, the Wigner function of a Gaussian state depends purely upon the covariance matrix, 
\begin{align}
W(\textbf{d})=\frac{1}{(2\pi)^n\sqrt{\text{det}(\Sigma)}}e^{-\frac{1}{2}\textbf{d}^T\Sigma^{-1}\textbf{d}},
\end{align}
where $\textbf{d}=\{X_1,P_1,...,X_n,P_n\}^T$. A vacuum state in all modes, such as the initial state of the atom-light system, has the covariance matrix
\begin{align}\label{eq::vacuumCov}
\Sigma_0=\bigoplus_{i=1}^{n}\left(\begin{matrix} 1/2 & 0 \\ 0 & 1/2\end{matrix}\right).
\end{align}
The covariance matrix $\Sigma_0$ generates a symmetric, Gaussian-distributed Wigner function
\begin{align}
W_0(\textbf{d})=\frac{1}{\pi^n}e^{-(X_1^2+P_1^2+...+X_n^2+P_n^2)}
\end{align}
with zero mean and a variance of 1/2 in all quadratures. As the light and ensemble evolve via a Faraday based squeezing protocol, the Gaussian character of the combined state is preserved. Because it stores all information regarding entanglement, we need only track the evolution of the covariance matrix. Below, we describe how the covariance matrix evolves as the corresponding Gaussian state undergoes unitary and dissipative dynamics. 

The evolution of $\hat{\rho}$ under any operation that preserves Gaussianity can be represented in the Heisenberg picture as a transformation of the covariance matrix. We first consider unitary maps that preserve Gaussianity, which are generated by Hamiltonians of the form 
\begin{align}\label{eq::GaussH}
\hat{H}=\sum_{i,j}h^{(i,j)}(\hat{\textbf{d}}_i\hat{\textbf{d}}_j+\hat{\textbf{d}}_j\hat{\textbf{d}}_i) 
\end{align}
for real $h^{(i,j)}$ \cite{PlenioEisert}. Associated with each Gaussian-preserving $\hat{H}$ is a \emph{symplectic map} $S_H$, which dictates the evolution of both the first order moments and the covariance matrix. Under a unitary map generated by $\hat{H}$, the central first order moments and the covariance matrix of $\hat{\rho}$ transform as
\begin{align}
\Delta\hat{\textbf{d}}'= S_H\Delta\hat{\textbf{d}}
\end{align}
and
\begin {align}
\Sigma'= S_H\Sigma S_H^{T}.
\end{align}  
All symplectic maps satisfy
\begin {align}\label{symplectic1}
S_H\sigma S_H^{T}=\sigma,
\end{align} 
which ensures that \erf{continuousUncert} is preserved on the output covariance matrix $\Sigma'$.

Unitary maps that preserve Gaussianity are a subset of the more general class of completely positive maps that preserve Gaussianity, so-called Gaussian channels. Associated with each Gaussian channel $\mathcal{E}$ acting on $\hat{\rho}\,$ is a matrix $M_\mathcal{E}$ and a symmetric, positive semidefinite matrix $N_\mathcal{E}$. Under $\mathcal{E}$, the evolution of the central first order moments and covariance matrix are given by 
\begin{align}
\Delta\hat{\textbf{d}}'= M_\mathcal{E}\Delta\hat{\textbf{d}}
\end{align}
and
\begin{align}
\Sigma'= M_\mathcal{E}\Sigma M_\mathcal{E}^{T}+N_\mathcal{E}.
\end{align}
Preserving \erf{continuousUncert} on $\Sigma'$ requires that $M_\mathcal{E}$ and $N_\mathcal{E}$ satisfy \cite{PlenioEisert}
\begin{align}\label{ChannelCond}
N_\mathcal{E}+\frac{i}{2}\sigma-\frac{i}{2}M_\mathcal{E}\sigma M_\mathcal{E}^T\geq 0.
\end{align}
The matrix $N_\mathcal{E}$, referred to as the noise component, increases the variance of the quadratures. Consequently, $\mathcal{E}$ is called a \emph{noise channel} in the case that $M_\mathcal{E}=\mathbb{I}$. When $N_\mathcal{E}=0$, condition \erf{ChannelCond} is equivalent to \erf{symplectic1} and $M_\mathcal{E}$ is a symplectic map. In the case where $N_\mathcal{E}\neq0$, $\mathcal{E}$ is a dissipative transformation. Gaussian channels of this sort frequently arise when $\hat{\rho}\,$ becomes entangled with another system that is subsequently traced out, such as an environment. 

In addition to Gaussian channels, operations that preserve Gaussianity include several forms of measurement, one of the most important being homodyne detection. Suppose that we wish to perform homodyne detection on the $n$th mode of the $n$-mode Gaussian state $\hat{\rho}$. The initial covariance matrix of $\hat{\rho}$ can be written in terms of the submatrices $A$, $B$ and $C$ as
\begin{align}
\Sigma=\left(\begin{matrix} A & C \\ C^T & B\end{matrix}\right),
\end{align} 
where $A$ is the covariance matrix of modes 1 through $n-1$, $B$ is the covariance matrix of mode $n$, and the matrix $C$ contains the covariances between modes 1 through $n-1$ and $n$. Consider a homodyne measurement of $\hat{X}_n$, which we will denote by $h[\hat{X}_n]$. In a homodyne measurement with perfect efficiency, the state of the $n$th mode is projected onto an eigenstate of $\hat{X}_n$ with no variance.  Although the $n$th mode is no longer Guassian, the state remains Gaussian on modes $1,...,\,n-1$ and has the covariance matrix 
\begin{eqnarray}\label{HCovariance}
\Sigma'=h[{\hat{X}_n}](\Sigma)=A-C(\mathbb{P} B\mathbb{P})^{-1}C^T,
\end{eqnarray} 
where $\mathbb{P}=\text{diag}(1,0)$ and the inverse symbol denotes the Moore-Penrose pseudoinverse \cite{PlenioEisert}. As evidenced in \erf{HCovariance}, the Gaussian state on modes 1 through $n$ evolves deterministically, independent of the measurement outcome of $\hat{X}_n$.

\chapter{Squeezing Protocols} \label{Sec:Protocols}

As discussed in Sec. \ref{sec::EntangleFaraday}, the Faraday interaction creates entanglement between the light and atoms with strength given by the collective spin coupling constant, $\xi$. It is entanglement between the atoms, however, that creates spin squeezing.  In squeezing protocols that utilize the Faraday interaction, a mode of the light acts as a quantum data bus, inducing entanglement through its mutual coupling to all atoms.  Atom-light entanglement can be converted into interatomic entanglement through measurement of the light, which subjects the ensemble to measurement backaction \cite{KuzMan98,Koschorreck2010,Takano2009}. In other protocols, the light mediates an effective nonlinear interaction on the ensemble \cite{TakTak05,TraDeu10}. The commonality in all protocols is that spin squeezing ultimately depends upon the strength of the initial entanglement between the light and atoms. 

Using the covariance matrix update rules of Gaussian states, we can describe these squeezing protocols in a simple, compact manner.  We represent the ensemble and light collectively as a multimode Guassian state with $\hat{\textbf{d}}=(\hat{X}_\downarrow,\hat{P}_\downarrow,\hat{X}_y,\hat{P}_y)^T$, where the initial vacuum state of the system has the covariance matrix given in \erf{eq::vacuumCov}. Because the protocols produce an ensemble state that is squeezed in the $\hat{X}_\downarrow$-$\hat{P}_\downarrow$ phase plane, we employ the quadrature squeezing parameter, $\zeta_q$, to quantify squeezing. The relationship between the Wineland spin squeezing parameter, $\zeta_m$, and the quadrature squeezing parameter, $\zeta_q$, will be explored in Sec. \ref{Sec::SqParameters}. 

\section{Quantum Nondemolition (QND) Measurement}\label{sec::QNDmeas}
\begin{figure}[H]
\centering
\includegraphics[scale=.35]{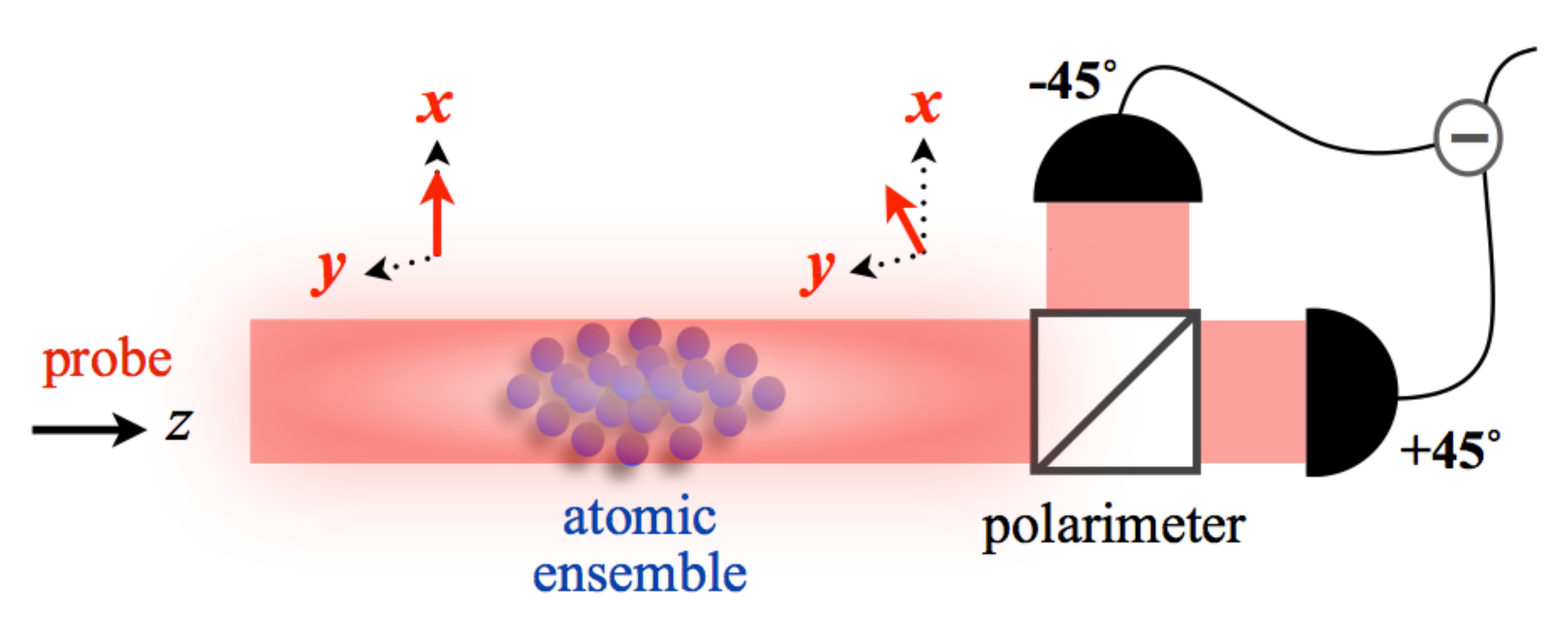}
\caption{Schematic of the experimental setup for the QND measurement protocol. Horizontally polarized probe light passes through the ensemble, causing a rotation of its linear polarization. The magnitude of this rotation, given by the value of the quadrature $\hat{X}_y$, is proportional to the collective spin of the ensemble, $\hat{F}_z$, or the quadrature, $\hat{X}_\downarrow$, in the generalized Holstein-Primakoff approximation.  A balanced polarimeter at $\pm45^\circ$ measures the quadrature $\hat{X}_y$, an effective measurement of $\hat{F}_z$. This causes squeezing of the ensemble along $\hat{F}_z$ by measurement backaction. In the generalized Holstein-Primakoff picture, this is equivalent to phase plane squeezing of the quadrature $\hat{X}_\downarrow$.  }
\label{fig::QND}
\end{figure}

Quantum nondemolition measurement was one of the earliest techniques used to create spin squeezing in atomic ensembles \cite{KuzMan98,Koschorreck2010,Takano2009}. A schematic of this protocol is shown in Fig. \ref{fig::QND}.  Squeezing by QND measurement employs the procedure described in Sec. \ref{sec::EntangleFaraday}, in which the collective spin component $\hat{F}_z$ of the ensemble is determined through a measurement of the Stokes' component $\hat{S}_2$ of the light. Because we have made the Holstein-Primakoff approximation, phase space quadratures take the place of the collective spin component $\hat{F}_z$ and the Stokes' component $\hat{S}_2$. In the phase plane picture, the position quadrature $\hat{X}_\downarrow$ of the ensemble is inferred through a measurement of the position quadrature $\hat{X}_y$ of the light. 

The light and atoms are first entangled by the Faraday interaction, the action of which over a small time $\Delta t$ is described by the symplectic matrix
\begin{align}\label{eq::FaradayS}
S_F(\Delta t)=\left(\begin{matrix} 1 & 0 & 0 & 0 \\ 0 &1& 0 & -\sqrt{\xi}\\ \sqrt{\xi} & 0&1& 0\\0 & 0 & 0 & 1\end{matrix}\right).
\end{align}
Equation (\ref{eq::FaradayS}) follows from the evolution of the quadratures in the Heisenberg picture. The atom-light entanglement is evidenced in the transformed quadratures,
\begin{align}\label{firstpass}
\left(\begin{matrix}\hat{X}_{\downarrow}(\Delta t)\\ \hat{P}_{\downarrow}(\Delta t)\\\hat{X}_{y}(\Delta t)\\\hat{P}_{y}(\Delta t)\end{matrix}\right)=
\left(\begin{matrix}\hat{X}_{\downarrow}(0)\\ \hat{P}_{\downarrow}(0)-\sqrt{\xi}\hat{P}_{y}(0)\\\hat{X}_{y}(0)+\sqrt{\xi}\hat{X}_{\downarrow}(0)\\\hat{P}_{y}(0)\end{matrix}\right).
\end{align}
Note that the position quadrature of the light, $\hat{X}_y$, contains information about the position quadrature of the ensemble, $\hat{X}_\downarrow$. A measurement of the light by the balanced polarimeter in Fig. \ref{fig::QND} behaves like a homodyne measurement with the probe light acting as a local oscillator. Performing homodyne detection on $\hat{X}_y$ is an effective measurement of the ensemble's $\hat{X}_\downarrow$, inducing backaction on the ensemble that reduces the variance of $\hat{X}_\downarrow$. For an arbitrary covariance matrix at time $t$, this protocol is described by the update 
\begin{align}\label{eq::QNDupdate}
\Sigma_{\downarrow}(t+\Delta t)=h[\hat{X}_y]\left(S_F(\Delta t)\Sigma(t) S_F^{T}(\Delta t)\right).
\end{align}
For the initial vacuum state of the atom-light system, the covariance matrix of the ensemble resulting from the QND measurement protocol is 
\begin{align}\label{eq::QNDcov}
\Sigma_{\downarrow}(\Delta t)
=\frac{1}{2}\left(\begin{matrix}(1+\xi)^{-1}& 0\\ 0 &1+\xi\end{matrix}\right).
\end{align}
The squeezing of the position quadrature $\hat{X}_\downarrow$, which follows from \erf{eq::QNDcov}, is
\begin{align}
\zeta_q=\frac{1}{1+\xi}.
\end{align}
Note that  $\zeta_q$ decreases with increasing $\xi$. This signifies that squeezing, like the atom-light entanglement, improves with a larger collective spin coupling constant.

\section{Double Pass Protocols}\label{sec::DP}
\begin{figure}
\centering
\includegraphics[scale=.5]{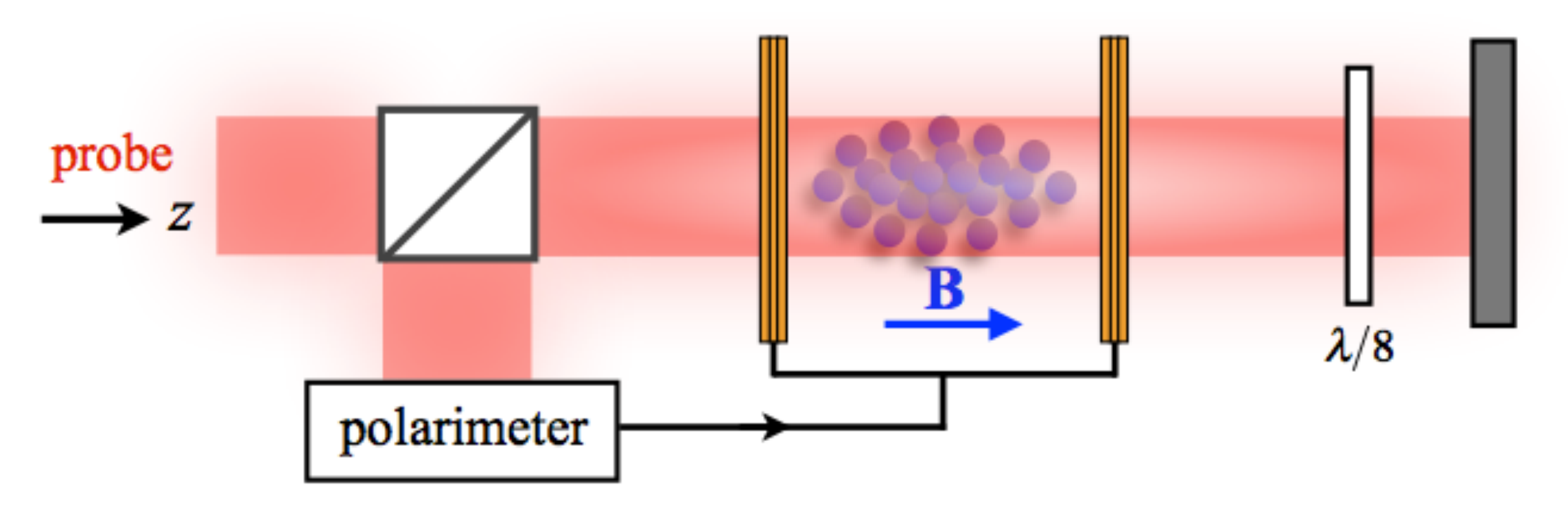}
\caption{Schematic of the experimental setup for the double pass protocols. Upon first entering the apparatus, light passes through the ensemble and undergoes Faraday rotation. The light is then reflected back towards the ensemble, passing twice through a $\lambda/8$ wave plate. The light then passes through the ensemble a second time, once again undergoing Faraday rotation. In the quantum eraser protocol, the light's polarization is measured upon exiting the ensemble. A magnetic field is applied to the ensemble along $z$ conditioned upon the measurement value, rotating the collective spin. }\label{fig::DPsetup}
\end{figure}

Generating spin squeezing through a double pass geometry, in which the light and atoms interact twice via the Faraday interaction, was first proposed in \cite{TakTak05}. Over the two passes, the light mediates an effective atom-atom interaction that generates interatomic entanglement and squeezing. This process is depicted in Fig. \ref{fig::DPsetup}. 

\begin{figure}
\centering
\includegraphics[scale=.4]{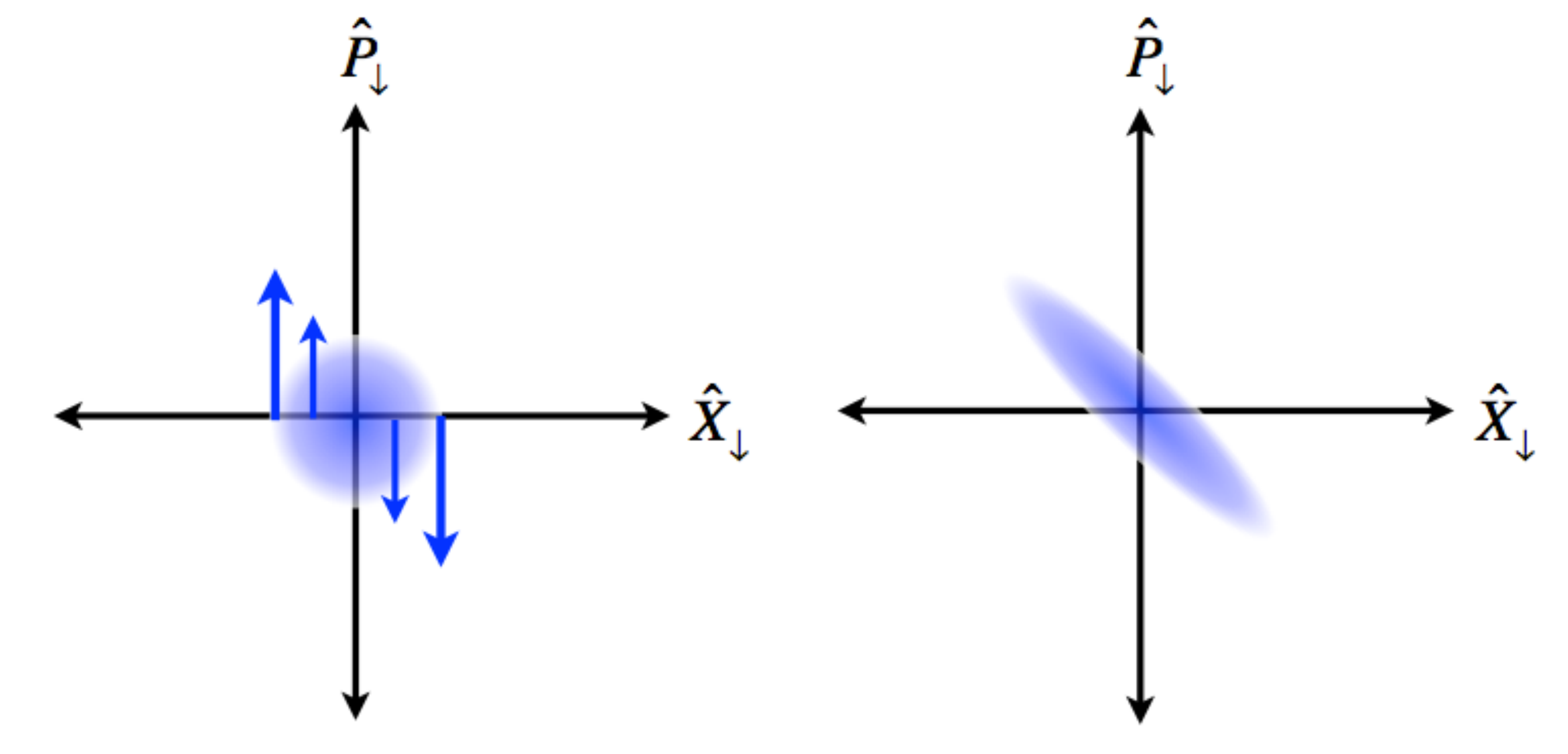}
\caption{The effect of the one-axis twisting interaction, $e^{i\xi\hat{X}_\downarrow^2/2}$, in the phase plane of the ensemble. One-axis twisting imparts a translation of the ensemble state along $\hat{P}_\downarrow$ by an amount proportional to the state's displacement in $\hat{X}_\downarrow$. This shearing creates squeezing in the $\hat{X}_\downarrow$-$\hat{P}_\downarrow$ phase plane.}\label{fig::Twisting}
\end{figure}

The light and ensemble are initially entangled by the Faraday interaction, transforming the quadratures as in Eq. (\ref{firstpass}). Recall that after the first pass, the position quadrature of the light $\hat{X}_y$ contains information about the position quadrature of the ensemble $\hat{X}_\downarrow$.  After the first pass, the light proceeds twice through a $\lambda/8$ wave plate, rotating the $\hat{X}_y$ and $\hat{P}_y$ quadratures by $\pi/2$,
\begin{align}
\left(\begin{matrix}\hat{X}_{\downarrow}(\Delta t)\\ \hat{P}_{\downarrow}(\Delta t)\\\hat{X}_{y}(\Delta t)\\\hat{P}_{y}(\Delta t)\end{matrix}\right)=
\left(\begin{matrix}\hat{X}_{\downarrow}(0)\\ \hat{P}_{\downarrow}(0)-\sqrt{\xi}\hat{P}_{y}(0)\\-\hat{P}_{y}(0)\\\hat{X}_{y}(0)+\sqrt{\xi}\hat{X}_{\downarrow}(0)\end{matrix}\right).
\end{align} 
The light passes through the ensemble a second time with the $\hat{P}_y $ quadrature containing information about $\hat{X}_\downarrow$. The Faraday interaction then couples $\hat{P}_y $ to $\hat{X}_\downarrow$, creating an effective $\hat{X}_\downarrow^2$ interaction. The nonlinear nature of this interaction is alternatively illustrated by decomposing the protocol into unitary operators, 
\begin{align}\label{Udp}
\hat{U}_{DP}=\hat{U}_{\text{Faraday}}\hat{U}_{\frac{\lambda}{8}}\hat{U}_{\frac{\lambda}{8}}\hat{U}_{\text{Faraday}}=e^{i\sqrt{\xi}(\hat{X}_y-\hat{P}_y)\hat{X}_\downarrow}e^{i\frac{\xi}{2}\hat{X}_\downarrow^2}.
\end{align}
The first exponential on the right hand side of \erf{Udp} does not create squeezing. Rather, it indicates the residual entanglement between the light and atoms after the second pass. The second exponential, however, is a well studied spin squeezing interaction known as one-axis twisting \cite{KitagawaUeda93}.  One axis-twisting is a shearing interaction that creates squeezing in the $\hat{X}_\downarrow$-$\hat{P}_{\downarrow}$ phase plane as depicted in Fig. \ref{fig::Twisting}.

For an arbitrary covariance matrix at time $t$, the double pass protocol is described by the update 
\begin{align}
\Sigma_{\downarrow}(t+2\Delta t)=\text{Tr}_y\left(S_F(\Delta t)R_{\frac{\lambda}{4}}S_F(\Delta t)\Sigma(t)S_F^T(\Delta t)R_{\frac{\lambda}{4}}^TS_F^T(\Delta t)\right).
\end{align}
Here, $\text{Tr}_y$ denotes taking the partial trace over the light, which is performed by discarding all entries of the covariance matrix involving operators on the $y$ mode. After tracing out the light, we are left with the ``reduced" covariance matrix of the ensemble, $\Sigma_{\downarrow}$. The eigenvalues of $\Sigma_{\downarrow}$ are the variances of the squeezed and anti-squeezed quadratures in phase space. The reduced density matrix can be diagonalized by performing a phase space rotation via a symplectic rotation matrix
\begin{align}
R(\theta)=\left(\begin{matrix}\text{cos}\theta&-\text{sin}\theta\\
\text{sin}\theta&\text{cos}\theta\end{matrix}\right).
\end{align}
Applying the double pass protocol to the initial vacuum state and rotating the resulting reduced covariance matrix by an angle
\begin{align}
\phi=-\text{tan}^{-1}\left(\frac{2\xi}{2\xi+\xi^2+\xi\sqrt{\xi^2+4\xi+8}}\right)
\end{align} 
yields
\begin{align}
&R(\phi)\Sigma_{\downarrow}R(\phi)^T=\\&\frac{1}{4}\left(\begin{matrix}2+2\xi+\xi^2-\xi\sqrt{\xi^2+4\xi+8}&0\\
0&2+2\xi+\xi^2+\xi\sqrt{\xi^2+4\xi+8}\end{matrix}\right).
\end{align}
This indicates that the double pass protocol produces squeezing in the phase plane along the angle $\pi-\phi$. For $\xi>>1$, the squeezing parameter becomes 
\begin{align}
\zeta_q\approx\frac{2}{\xi}.
\end{align}
Although the double pass performs slightly worse than the QND measurement protocol, it does not require a high quantum efficiency measurement of the light. 
 
 \subsection{Quantum Eraser}
Although the ensemble is squeezed after the double pass protocol, the squeezing is limited because the light and atoms are still entangled after the second pass. This entanglement is evidenced in \erf{Udp} and in the quadrature evolution after the second pass,     
\begin{align}\label{eq::doublepassQuads}
\left(\begin{matrix}\hat{X}_{\downarrow}(2\Delta t)\\ \hat{P}_{\downarrow}(2\Delta t)\\\hat{X}_{y}(2\Delta t)\\\hat{P}_{y}(2\Delta t)\end{matrix}\right)=
\left(\begin{matrix}\hat{X}_{\downarrow}(0)\\ \hat{P}_{\downarrow}(0)-\sqrt{\xi}\hat{P}_{y}(0)-\sqrt{\xi}\hat{X}_{y}(0)-\xi\hat{X}_{\downarrow}(0)\\ -\hat{P}_{y}(0)+\sqrt{\xi}\hat{X}_{\downarrow}(0)\\ \hat{X}_{y}(0)+\sqrt{\xi}\hat{X}_{\downarrow}(0)\end{matrix}\right).
\end{align}
	Taking the partial trace over the light, in effect discarding the light while it is still entangled to the atoms, causes the ensemble to decohere and limits squeezing. This decoherence can be remedied with a quantum eraser, a measurement that disentangles two systems \cite{Scully91,TraDeu10}.  Homodyne detection on the $y$ mode projects the light into a pure state, separable from the ensemble.  Note from \erf{eq::doublepassQuads} that the quadrature $\hat{X}'_{y}=(\hat{X}_{y}-\hat{P}_{y})/\sqrt{2}$ contains no atomic information. A measurement of $\hat{X}'_{y}$, thus, disentangles the atoms from the light without causing measurement backaction on the ensemble \cite{TraDeu10}. The Kraus operator that evolves the ensemble conditioned on a measurement value $X_y'$ of $\hat{X}'_{y}$ is given by
\begin{align}
\bra{X_y'}\hat{U}_{DP}\ket{0}_y=e^{i\sqrt{2\xi}X_y'\hat{X}_\downarrow}e^{i\frac{\xi}{2}\hat{X}_\downarrow^2}\bra{X_y'}0\rangle_y.
\end{align}
We see that the net effect of the measurement is a translation of the ensemble along $\hat{P}_\downarrow$. This translation can be cancelled by applying a magnetic field along the $z$ axis, resulting in a pure one-axis twisting interaction. In the covariance matrix update picture, the Takeuchi protocol with the quantum eraser is given by 
\begin{align}\label{eq::QE}
\Sigma_{\downarrow}(t+2\Delta t)=h[{\hat{X}'_y}]\left(S_F(\Delta t)R_{\frac{\lambda}{4}}S_F(\Delta t)\Sigma(t)S_F^T(\Delta t)R_{\frac{\lambda}{4}}^TS_F^T(\Delta t)\right).
\end{align}
By applying \erf{eq::QE} to the covariance matrix of the initial vacuum state, we find the resultant squeezing.   For $\xi>>1$,
\begin{align}
\zeta_q\approx\frac{1}{\xi^2}.
\end{align}
This quadratic scaling with $\xi$ is a substantial improvement over the double pass alone.

\subsection{Phase-matching}
In eliminating an important source of decoherence on the ensemble, the quantum eraser generates a pure one-axis twisting interaction. The spin squeezing generated by the double pass protocol can be enhanced even further, however. Decomposing the one-axis twisting interaction in terms of creation and annihilation operators on the $\downarrow$ mode yields
\begin{align}\label{eq::twisting}
e^{i\frac{\xi}{2}\hat{X}_\downarrow^2}=\text{exp}\left(i\frac{\xi}{4}(\hat{a}_\downarrow^{\dag 2}+\hat{a}_\downarrow^{2})+i\frac{\xi}{2}\hat{a}_\downarrow^{\dag}\hat{a}_\downarrow+i\frac{\xi}{4}\right).
\end{align}
The first term on the right hand side is a Bogoliubov transformation, a pure squeezing interaction. The second term is a rotation in the $\hat{X}_\downarrow$-$\hat{P}_\downarrow$ phase plane, while the last term is a phase that is irrelevant to the dynamics. Due to the rotation term, the pure squeezing generated by the Bogoliubov transformation takes place along a variable axis. By eliminating this rotation or ``phase-matching", we can ensure that squeezing occurs along a consistent axis \cite{TraDeu10}. To achieve this, we apply the double pass with the quantum eraser over a very small interaction time $2\Delta t/n$ followed by a rotation in the phase plane by an angle $\xi/2n$, which can be generated with internal spin control.   Through a Trotter expansion, we alternatingly apply the double pass plus quantum eraser and the rotation over infinitesimal time steps. This procedure generates a Bogoliubov transformation without extraneous rotation, 
\begin{align}\label{bogoliubov}
\hat{U}_{PM}&=\text{lim}_{n\rightarrow\infty}\left(e^{-i\frac{\xi}{2n}\hat{a}_\downarrow^\dag\hat{a}_\downarrow}e^{i\frac{\xi}{2n}\hat{X}_\downarrow^2}\right)^n=e^{i\frac{\xi}{4}(\hat{a}_\downarrow^{\dag2}+\hat{a}_\downarrow^2)}.
\end{align}

The phase matching procedure can also be implemented as a symplectic transformation on the covariance matrix.  The one-axis twisting unitary in \erf{eq::twisting} applied over a small interaction time $2\Delta t/n$ becomes the unitary transformation
\begin{align}\label{eq::SmallTwisting}
\hat{U}_n=e^{i\frac{\xi}{2n}\hat{X}_\downarrow^2}.
\end{align}
Because this unitary is generated by a Hamiltonian of the form in \erf{eq::GaussH}, it preserves Gaussianity and acts upon the covariance matrix via a symplectic map, $S_n$.
To first order in $\xi/n$, the singular value decomposition of $S_n$ is
\begin{align}
S_n=R(\theta_-) D R(-\theta_+)
\end{align}
where
\begin{align}
D=\left(\begin{matrix}e^{-\frac{\xi}{2n}}& 0\\ 0& e^{\frac{\xi}{2n}}\end{matrix}\right)
\end{align}
and the angles $\theta_{\pm}\approx\pi/4\pm\frac{\xi}{4n}$. Alternating the one-axis twisting interaction and rotations of $\theta_+-\theta_-\approx\xi/(2n)$, creates pure squeezing along a consistent axis,
\begin{align}\label{eq::phasematchingTheta}
...R(\theta_+-\theta_-)S_{1}^{(n)}R(\theta_+-\theta_-)S_{1}^{(n)}=R(\theta_+) D^n R(-\theta_-).
\end{align}
Up to a rotation in phase space, the resulting symplectic map is
\begin{align}\label{exponentialSq}
 D^n=\left(\begin{matrix}e^{-\frac{\xi}{2}}& 0\\ 0& e^{\frac{\xi}{2}}\end{matrix}\right).
\end{align}
Applying this symplectic map to the vacuum covariance matrix produces squeezing that scales exponentially with $\xi$, 
\begin{align}
\zeta_q=e^{-\xi}.
\end{align}
Through a combination of feedback and internal spin controls, the scaling of spin squeezing with the collective spin coupling constant is substantially enhanced.

\section{Quadrature Squeezing vs. Spin Squeezing}\label{Sec::SqParameters}
The spin squeezing protocols discussed in Sections \ref{sec::QNDmeas} and \ref{sec::DP} create squeezing in the phase plane defined by the ensemble quadratures $\hat{X}_\downarrow$ and $\hat{P}_\downarrow$. When the fiducial state is a spin coherent state, this is equivalent to squeezing a component of the collective spin. For more general fiducial states, however, the relationship between squeezing in the phase plane and squeezing of a collective spin component is not immediately clear. To shed light on this, we revisit equations (\ref{eq::position}) and (\ref{eq::momentum}), which show that the quadratures have an alternative interpretation as operators on the ensemble of embedded qubits. Recall that each embedded qubit $j$ is defined on a basis consisting of the fiducial and coupled states, $\ket{\uparrow}_j$ and $\ket{\downarrow}_j$, within the $2f+1$ dimensional hyperfine spin of each atom $j$. Consider the  collective spin components of the embedded qubit ensemble given by 
\begin{align}
&\hat{\Sigma}_x=\frac{1}{2}\Sigma_{j=1}^N\hat{\sigma}_x^{(j)} \;\;\;\text{and}  
\\ &\hat{\Sigma}_y=\frac{1}{2}\Sigma_{j=1}^N\hat{\sigma}_y^{(j)}, 
\end{align}
where 
\begin{align}
&\hat{\sigma}_x^{(j)}=\ket{\downarrow}\bra{\uparrow}_j+\ket{\uparrow}\bra{\downarrow}_j\;\;\;\; \text{and}\\ 
&\hat{\sigma}_y^{(j)}=i(\ket{\downarrow}\bra{\uparrow}_j-\ket{\uparrow}\bra{\downarrow}_j) 
\end{align}
are the Pauli spin operators on a qubit $j$ with basis states $\ket{\uparrow}_j$ and $\ket{\downarrow}_j$. From equations (\ref{eq::position}) and (\ref{eq::momentum}),
\begin{align}
&\hat{X}_\downarrow\approx\sqrt{2}\hat{\Sigma}_x \;\;\;\text{and}  
\\ &\hat{P}_\downarrow\approx\sqrt{2}\hat{\Sigma}_y. 
\end{align}
Squeezing in the phase plane is, thus, equivalent to squeezing the collective spin of the embedded qubit ensemble in a plane defined by $\hat{\Sigma}_x$ and $\hat{\Sigma}_y$. 

Squeezing of the embedded qubit ensemble does not necessarily imply squeezing of the atomic ensemble composed of spin-$f$ qudits, however. While $\hat{\Sigma}_x$ is proportional to $\hat{F}_z$ under the multilevel HP approximation,  $\hat{\Sigma}_y$ does not in general relate to a collective spin component of the qudit ensemble when $f\geq1/2$. Squeezing in the plane defined by  $\hat{\Sigma}_x$ and $\hat{\Sigma}_y$, nonetheless, reveals valuable information about the state of the qudit ensemble. Recall that because a qubit cannot be internally squeezed, spin squeezing in an ensemble of qubits is the product of entanglement alone. Squeezing in the $\hat{\Sigma}_x$-$\hat{\Sigma}_y$ plane, which is synonymous with squeezing in the $\hat{X}_\downarrow$-$\hat{P}_\downarrow$ phase plane, implies entanglement between the atoms in the ensemble. We can, furthermore, view the squeezing parameter $\zeta_q$, as a measure of this interatomic entanglement. While the squeezing protocols outlined in Sections \ref{sec::QNDmeas} and \ref{sec::DP}  do not necessarily squeeze a component of the collective spin $\hat{F}_n$, they generate entanglement between the atoms in the ensemble. 

While $\zeta_q<1$ indicates the presence of  interatomic entanglement, the question remains as to how exactly $\zeta_q$ relates to the metrological spin squeezing parameter, $\zeta_m$. In terms of the phase plane squeezing parameter, we can express the metrological spin squeezing parameter as
\begin{align}\label{parameters}
\zeta_m=\zeta_m^\uparrow\zeta_q,
\end{align}  
where $\zeta_m^\uparrow$ is the metrological squeezing parameter of a single atomic spin prepared in the fiducial state,
\begin{align}
\zeta_m^\uparrow=\frac{2f\left(\Delta f_z^2\right)_\uparrow}{\langle\hat{f}_x\rangle^2}.
\end{align} 
From \erf{parameters}, it is clear that when the fiducial state is anti-squeezed, i.e. $\zeta_m^\uparrow>1$, squeezing of the collective spin cannot  occur unless there is a sufficiently high degree of squeezing in the phase plane or, equivalently, interatomic entanglement. To achieve metrologically relevant spin squeezing $\zeta_q$ must be very small, specifically $\zeta_q<(\zeta_m^\uparrow)^{-1}$. 

\section{Post-Processing}\label{sec::postprocessing}
Our protocol for enhancing the Faraday interaction relies on preparing the atoms in fiducial states, such as $\ket{\uparrow_\text{cat}}$ and $\ket{\uparrow_0}$, which increases projection noise variance of the ensemble thereby maximizing the resolvability of a collective spin projection in a measurement of the light. Increasing the projection noise seems antithetical to the goal of squeezing. Indeed, equation (\ref{parameters}) appears to suggest that we strengthen the Faraday interaction, and the resulting interatomic entanglement, at the expense of metrologically relevant squeezing. Through internal spin control, however, we can convert this enhanced interatomic entanglement into metrologically relevant squeezing. 

Note that when the ensemble is prepared in a spin coherent state, \\$\bra{\uparrow_{SCS}}\hat{f}_x\ket{\uparrow_{SCS}}=f$ and $\bra{\uparrow_{SCS}}(\Delta\hat{f}_z)^2\ket{\uparrow_{SCS}}=f/2$, implying $\zeta_m^{\uparrow_{SCS}}=1$.  By \erf{parameters}, the phase plane squeezing and metrologically relevant spin squeezing are equivalent, i.e. $\zeta_m=\zeta_q$. We can take advantage of this relationship by mapping the squeezing created in the $\hat{X}_{\downarrow}$-$\hat{P}_{\downarrow}$ phase plane into the $\hat{X}_{\downarrow_{SCS}}$-$\hat{P}_{\downarrow_{SCS}}$ phase plane. Using internal spin control, we can generate the partial isometry
\begin{align}\label{Uscs}
\hat{U}_{SCS}=\bigotimes_{i=1}^{N_A}\left(\ket{\uparrow_{SCS}}\bra{\uparrow}_i+\ket{\downarrow_{SCS}}\bra{\downarrow}_i\right).
\end{align}
On each atom $i$,  $\hat{U}_{SCS}$ maps an arbitrary fiducial state to $\ket{\uparrow_{SCS}}_i$ and an arbitrary coupled state to $\ket{\downarrow_{SCS}}_i$. The phase plane quadratures are transformed as
\begin{align}
&\hat{U}_{SCS}^\dag\hat{X}_\downarrow\hat{U}_{SCS}=\hat{X}_{\downarrow_{SCS}}\;\;\;\;\text{and}\\
&\hat{U}_{SCS}^\dag\hat{P}_\downarrow\hat{U}_{SCS}=\hat{P}_{\downarrow_{SCS}}.
\end{align}
An ensemble state $\ket{\zeta}$ that is squeezed in the  $\hat{X}_{\downarrow}$-$\hat{P}_{\downarrow}$ phase plane will be mapped to a state $\ket{\zeta'}=\hat{U}_{SCS}\ket{\zeta}$ that is squeezed in the $\hat{X}_{\downarrow_{SCS}}$-$\hat{P}_{\downarrow_{SCS}}$ phase plane. We can see this more clearly by looking at the quadrature variances,
\begin{align}
&\bra{\zeta}(\Delta\hat{X}_\downarrow)^2\ket{\zeta}=\bra{\zeta'}(\Delta\hat{X}_{\downarrow_{SCS}})^2\ket{\zeta'}\;\;\;\;\text{and}\label{eq::UscsVarX}\\
&\bra{\zeta}(\Delta\hat{P}_\downarrow)^2\ket{\zeta}=\bra{\zeta'}(\Delta\hat{P}_{\downarrow_{SCS}})^2\ket{\zeta'}.\label{eq::UscsVarP}
\end{align}
As a result of Eqs. (\ref{eq::UscsVarX}) and  (\ref{eq::UscsVarP}), the phase plane squeezing parameter $\zeta_q$ is preserved by $\hat{U}_{SCS}$. The metrological squeezing parameter of the fiducial state, however, is not preserved,
\begin{align}
\zeta_m^\uparrow\rightarrow\frac{2f\bra{\uparrow}\hat{U}_{SCS}^\dag(\Delta\hat{f}_z)^2\hat{U}_{SCS}\ket{\uparrow}}{\bra{\uparrow}\hat{U}_{SCS}^\dag\hat{f}_x\hat{U}_{SCS}\ket{\uparrow}^2}=\zeta_m^{\uparrow_{SCS}}=1.
\end{align}
The metrologically relevant spin squeezing parameter, consequently, transforms as:
\begin{align}
\zeta_m=\zeta_m^\uparrow\zeta_q\rightarrow\zeta_m^{\uparrow_{SCS}}\zeta_q=\zeta_q.
\end{align}
The partial isometry $\hat{U}_{SCS}$ converts squeezing in the phase plane into metrologically relevant spin squeezing. 

\begin{figure}[H]
\centering
\includegraphics[scale=.8]{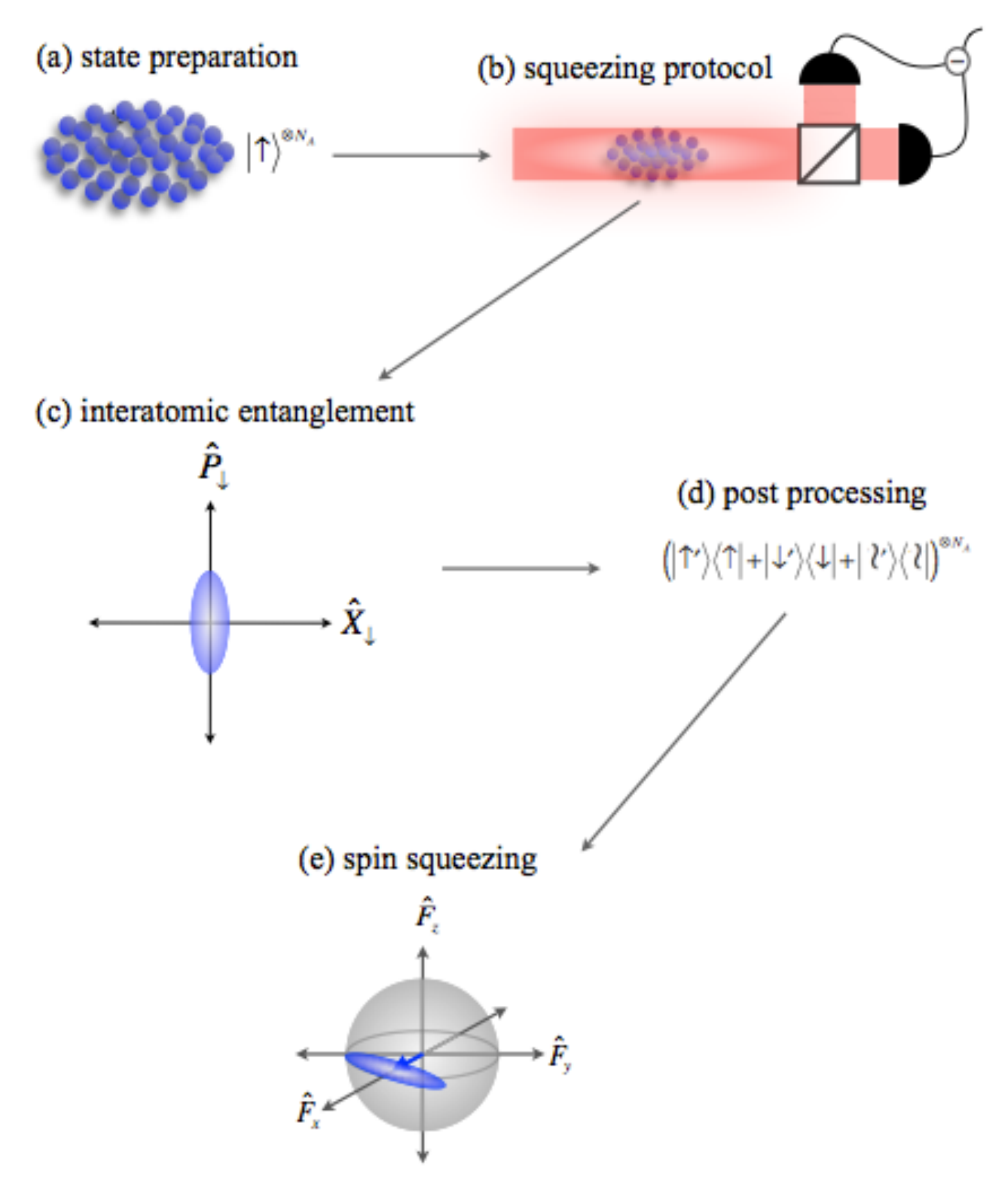}
\caption{Using internal spin control to enhance spin squeezing. (a) Internal spin control is first used to prepare each atom in the fiducial state, $\ket{\uparrow}$, creating the ensemble state $\ket{\uparrow}^{\otimes N_A}$. The fiducial state is chosen so that the ensemble has a larger value of projection noise, $\Delta F_z^2=N_A (\Delta f_z^2)_\uparrow$. (b) A Faraday spin squeezing protocol is applied to the ensemble. Interatomic entanglement is created through the coupling of the light and ensemble. The coupling between the light and ensemble and the interatomic entanglement that results increases with the initial projection noise. (c) The interatomic entanglement generated by the squeezing protocol creates squeezing in the phase plane of the HP quadratures $\hat{X}_\downarrow$ and $\hat{P}_\downarrow$. The greater the squeezing, the greater the interatomic entanglement. (d) By applying a partial isometry, the squeezing in the phase plane is converted into squeezing of the collective spin. (e) The end result is a spin squeezed state of the ensemble with $\zeta_m<1$.
}\label{fig::SqScheme}
\end{figure}

By combining state preparation, a Faraday-based squeezing protocol and ``post-processing" in the form of a partial isometry, we can create metrologically relevant spin squeezing that increases with the initial projection noise variance of the ensemble \cite{NorDeu12}. The steps of this procedure are shown in Fig. \ref{fig::SqScheme}.
The enhanced entanglement created by preparing the ensemble in a state with larger projection noise does not come at the expense of metrologically relevant squeezing. Indeed, through this series of controls, squeezing produced by the Faraday interaction is increased substantially. For QND measurement, in particular, one can obtain more squeezing than what would seem possible given the shot noise resolution.

 \subsection{Internal Spin Squeezing}\label{sec::IntSpinSqueeze}
While the partial isometry in \erf{Uscs} enables us to convert interatomic entanglement into metrologically relevant spin squeezing, it is not the optimal map. A partial isometry that maps a fiducial state to an internally spin squeezed state $\ket{\uparrow_S}$, for which $\zeta_m^{\uparrow_S}<1$, generates an even larger amount of metrologically relevant squeezing. 

As the squeezed state $\ket{\uparrow_S}$, we choose the Yurke state \cite{Combes05,YurkeState}, which is defined for integer $f$ as
\begin{equation}\label{eq::defYur}
\ket{\uparrow_{\text{y}}} \equiv  \frac{\sin\alpha}{\sqrt{2}} \ket{f, m_z=1} +\cos\alpha  \ket{f, m_z=0} +\frac{\sin\alpha}{\sqrt{2}} \ket{f, m_z=-1}.
\end{equation}
The Yuke state is maximally squeezed as $\alpha\rightarrow 0$, meaning that $\zeta_m^{\uparrow_\text{y}}$ reaches the Heisenberg limit of $1/f$. For arbitrary $\alpha$, the squeezing parameter of the Yurke state is
\begin{align}
\zeta_m^{\uparrow_\text{y}}= \frac{1}{(f+1)\text{cos}^2\alpha}.
\end{align}
The state that couples to the Yurke state in the multilevel HP approximation is
\begin{equation}
\ket{\downarrow_{\text{y}}}=\frac{1}{\sqrt{2}}(\ket{f, m_z=1}-\ket{f, m_z=-1}),
\end{equation}
since $\Delta\hat{f}_z\ket{\uparrow_{\text{y}}}=\text{sin}\alpha\ket{\downarrow_{\text{y}}}$ and $(\Delta\hat{f}_z^2)_{\uparrow_\text{y}}=\text{sin}^2\alpha$. 

While the original Yurke state given in \erf{eq::defYur} is only defined for integer $f$, we introduce a ``half-integer Yurke state" for half-integer $f$,
\begin{align}
 \ket{\uparrow_\text{hy}}\!=\!\frac{\text{sin}\alpha}{\sqrt{2}}\ket{f,m_z=3/2}\!+\!\text{cos}\alpha\ket{f,m_z=1/2}
 \!+\!\frac{\text{sin}\alpha}{\sqrt{2}}\ket{f,m_z=-1/2}.
 \end{align}
 The squeezing parameter of the half-integer Yurke state is
  \begin{align}
 \zeta_m^{\uparrow_\text{hy}}=\frac{4f}{\text{cos}^2\alpha(\sqrt{(f+3/2)(f-1/2)}+f+1/2)^2}.
  \end{align}
Although the squeezing parameter of the half-integer Yurke state has a more complicated dependence on $f$ than the squeezing parameter of the Yurke state, it is also maximally squeezed, scaling roughly with $1/f$ as $\alpha\rightarrow 0$. The state coupled to the half-integer Yurke state in the multilevel HP approximation is 
 \begin{align}
 \ket{\downarrow_{\text{hy}}}=\frac{1}{\sqrt{2}}(\ket{f, m_z=3/2}-\ket{f, m_z=-1/2}),
 \end{align}
as $\Delta\hat{f}_z\ket{\uparrow_{\text{hy}}}=\text{sin}\alpha\ket{\downarrow_{\text{hy}}}$ and $(\Delta\hat{f}_z^2)_{\uparrow_\text{hy}}=\text{sin}^2\alpha$. 

To enhance metrologically relevant squeezing through squeezing of the internal spin, we take an approach similar to post-processing with the spin coherent state. Using internal spin control, we generate the partial isometry
\begin{align}
\hat{U}_{\text{y}}=\bigotimes_{i=1}^{N_A}\Bigg\{\begin{matrix}\left(\ket{\uparrow_\text{y}}\bra{\uparrow}_i+\ket{\downarrow_{\text{y}}}\bra{\downarrow}_i\right)&\text{integer $f$}\\
\left(\ket{\uparrow_\text{hy}}\bra{\uparrow}_i+\ket{\downarrow_{\text{hy}}}\bra{\downarrow}_i\right)&\text{half-integer $f$}\end{matrix}\;\;\;.
\end{align}
This partial isometry maps squeezing from the $\hat{X}_\downarrow$- $\hat{P}_\downarrow$ phase plane to the $\hat{X}_{\downarrow_\text{y}}$- $\hat{P}_{\downarrow_\text{y}}$ phase plane for integer $f$ and to the $\hat{X}_{\downarrow_\text{hy}}$- $\hat{P}_{\downarrow_\text{hy}}$ phase plane for half-integer $f$.
Like $\hat{U}_{SCS}$, the partial isometry $\hat{U}_{\text{y}}$ preserves both the quadrature variances and the phase plane squeezing parameter, $\zeta_q$. The metrological squeezing parameter is transformed, however, 
\begin{align}
\zeta_m=\zeta_m^\uparrow\zeta_q\rightarrow\Bigg\{\begin{matrix}\zeta_m^{\uparrow_\text{y}}\zeta_q=\frac{\zeta_q}{\text{cos}^2\alpha(f+1)}&\text{integer $f$}\\
\zeta_m^{\uparrow_\text{hy}}\zeta_q=\frac{4f\zeta_q}{\text{cos}^2\alpha(\sqrt{(f+3/2)(f-1/2)}+f+1/2)^2}&\text{half-integer $f$}\end{matrix}\;\;\;.
\end{align}
When $\alpha= 0$, the internal spin squeezing produces a multiplicative enhancement on the order of $1/f$ compared to the partial isometry $\hat{U}_{\text{SCS}}$. Also note that the gains produced by internal spin squeezing increase with $f$.  
 
 \chapter{Decoherence Due to Optical Pumping}\label{sec::OpticalPumping}
In the previous chapter, we demonstrated that internal spin control in the form of state preparation and post-processing can enhance the performance of Faraday-based squeezing protocols.  While the gains are significant, achievable spin squeezing is ultimately limited by the various sources of decoherence in the system. Under realistic experimental conditions, optical pumping of the atoms due to spontaneous scattering of photons is the most significant source of decoherence.  Because the rate of photon scattering increases with the entangling strength of the Faraday interaction, any attempt to squeeze the ensemble inevitably produces decoherence. Preparation of the ensemble with internal spin control adds another layer of complexity, as the effects of optical pumping are state dependent. Achievable spin squeezing is limited by tradeoffs between the increased entanglement generation and the increased susceptibility to decoherence that come with the choice of a fiducial state. 

Although this chapter concerns the effect of optical pumping upon the atoms in the ensemble, it should be noted that optical pumping arising from spontaneous photon scattering produces decoherence of the light as well. Absorption followed by spontaneous emission causes the light to be diffusely scattered outside the spatial mode of the probe, meaning that photons are lost. This notwithstanding, the number of photons per pulse is much greater than the number of atoms. Additionally, the atomic ensemble continually interacts with fresh pulses of light. While the effects of spontaneous photon scattering accumulate on the ensemble, this is not true of the light. For these reasons, loss of photons due to diffuse scattering has minimal impact and can be ignored.     

\section{Fundamentals of Optical Pumping}
\begin{figure}
\centering
\includegraphics[scale=.35]{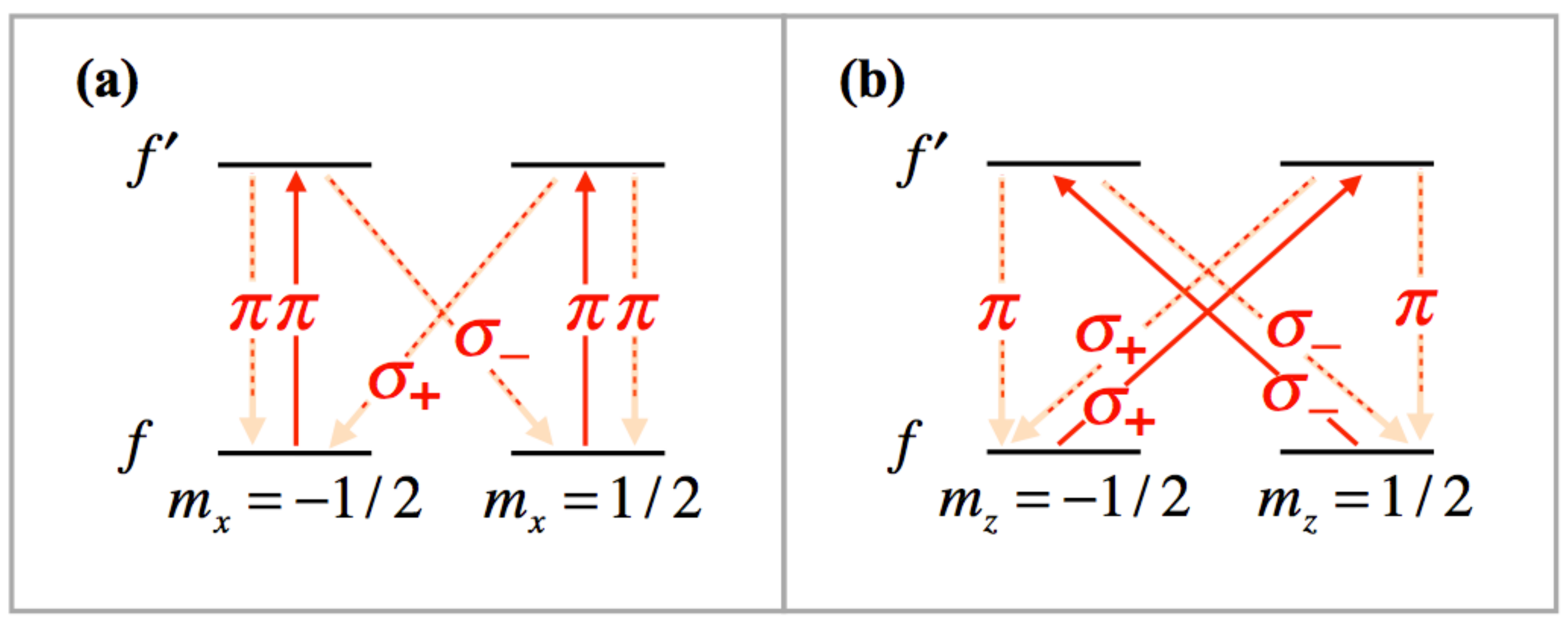}
\caption{Optical pumping in the case of $f=1/2$ for an atom being driven on a $f=1/2$ to $f'=1/2$ transition. The probe is linearly polarized along $x$. Solid lines represent absorption of a photon from the probe and dashed lines represent spontaneous emission of a photon. In (a), the quantization axis of the atom is parallel to the polarization of the probe, meaning that the probe is $\pi$ polarized. In (b), the  quantization axis of the atom is orthogonal to the polarization of the probe, meaning that the probe is a superposition of $\sigma_+$ and $\sigma_-$ light. }
\label{fig::OPqubit}
\end{figure}
To develop an intuition for the effects of optical pumping, we first examine a simple case for $f=1/2$. Depicted in  Fig. \ref{fig::OPqubit} is an atom with ground state angular momentum $f=1/2$ being driven on a transition to an excited state with angular momentum $f'=1/2$. Initially, the atom is in the state $\ket{f=1/2,m}$, where the spin projection number is $m=\pm1/2$. The probe photons driving the atom carry angular momentum with spin $s=1$. When the atom absorbs a probe photon, it is excited to the $f'=1/2$ manifold and its spin projection will change by an amount $\Delta m$ depending upon the polarization of the probe relative to the quantization axis of the atomic spin. If the probe is right circularly polarized with respect to the atomic quantization axis, which is called $\sigma_+$ light, $\Delta m=1$. Conversely, if the probe is left circularly polarized with respect to the atomic quantization axis, which is referred to as $\sigma_-$ light, $\Delta m=-1$. When the probe is linearly polarized along the quantization axis, which is called $\pi$ light, the spin projection of the atom does not change, i.e. $\Delta m=0$. 

Figure \ref{fig::OPqubit} depicts two possible orientations of the atomic quantization axis with respect to the probe polarization. In Fig.  \ref{fig::OPqubit} (a), the atom is quantized along the $x$ axis, which is parallel to the linear polarization of the probe. This $\pi$ light induces the transitions, $\ket{f=1/2,m=\pm1/2}\rightarrow\ket{f'=1/2,m'=\pm1/2}$.  In Fig.  \ref{fig::OPqubit} (b), the atom is quantized along $z$, which is perpendicular to the linear polarization of the probe. With respect to the quantization axis, the light's linear polarization along $x$ can be expressed in the basis of right and left circular polarization as $\mathbf{e}_x=(\mathbf{e}_L-\mathbf{e}_R)/\sqrt{2}$. This makes the probe a superposition of $\sigma_+$ and $\sigma_-$ light, which causes the transitions $\ket{f=1/2,m=\pm1/2}\rightarrow\ket{f'=1/2,m'=\mp1/2}$. 

After absorbing a probe photon and being excited up to the $f'$ manifold, the atom returns to the ground state by spontaneously emitting a photon. Due to conservation of angular momentum, spontaneous emission can change the spin projection of the atom. Like absorption, this change depends on the polarization of the emitted photon with respect to the atomic quantization axis. Spontaneous emission of a $\sigma_+$ photon changes the spin projection by $\Delta m=-1$, which induces the transition \\$\ket{f'=1/2,1/2}\rightarrow\ket{f=1/2,-1/2}$ in Figs.  \ref{fig::OPqubit} (a) and (b). When the atom in Figs.  \ref{fig::OPqubit} (a) and (b) spontaneously emits a $\sigma_-$ photon, changing the spin projection by $\Delta m=1$, it undergoes the transition  $\ket{f'=1/2,-1/2}\!\rightarrow\!\ket{f=1/2,1/2}$. Spontaneous emission of a $\pi$ photon does not change the spin projection number. In Figs.  \ref{fig::OPqubit} (a) and (b), spontaneous emission of a $\pi$ photon induces the transitions $\ket{f'=1/2,\pm1/2}\rightarrow\ket{f=1/2,\pm1/2}$. From Figs.  \ref{fig::OPqubit} (a) and (b), it is evident that the total change in the spin projection of the atom from one cycle of absorption and emission depends on the polarizations of both the absorbed and emitted photons. 

\section{Optical Pumping in the Higher Spin Alkali}
\begin{figure}[H]
\centering
\includegraphics[scale=.38]{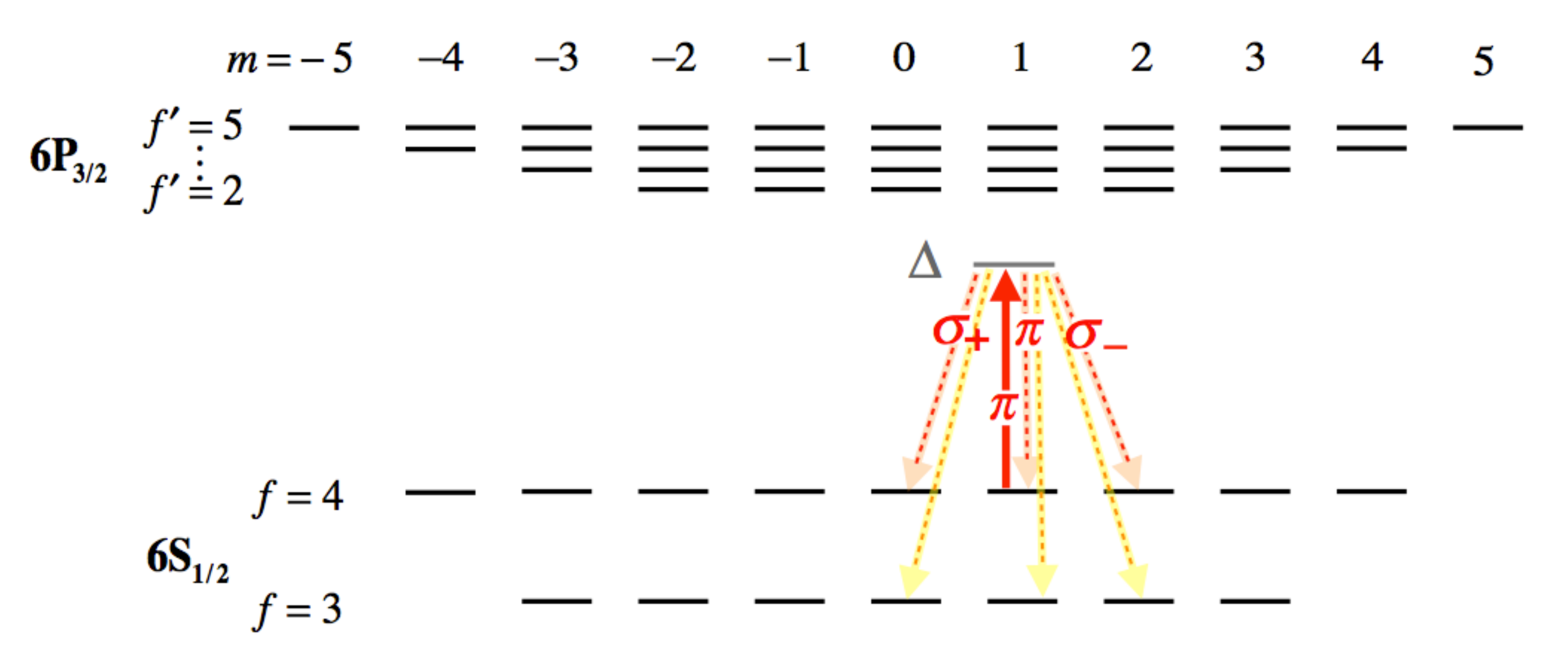}
\caption{An alkali atom is prepared in the ground hyperfine manifold with $f=4$. The quantization axis of the atom is parallel to the linear polarization of the probe. The solid line represents absorption of a photon from the probe and the dashed lines represent spontaneous emission of a photon. The atom absorbs a $\pi$ photon and transitions to the excited hyperfine multiplet. By spontaneously emitting a $\sigma_+$, $\sigma_-$ or $\pi$ photon, the atom transitions back to one of the ground hyperfine manifolds. Whether the atom is optically pumped into $f=4$ or $f_-=3$ depends on the color of the emitted photon.}
\label{fig::f4pi}
\end{figure}

\begin{figure}[H]
\centering
\includegraphics[scale=.38]{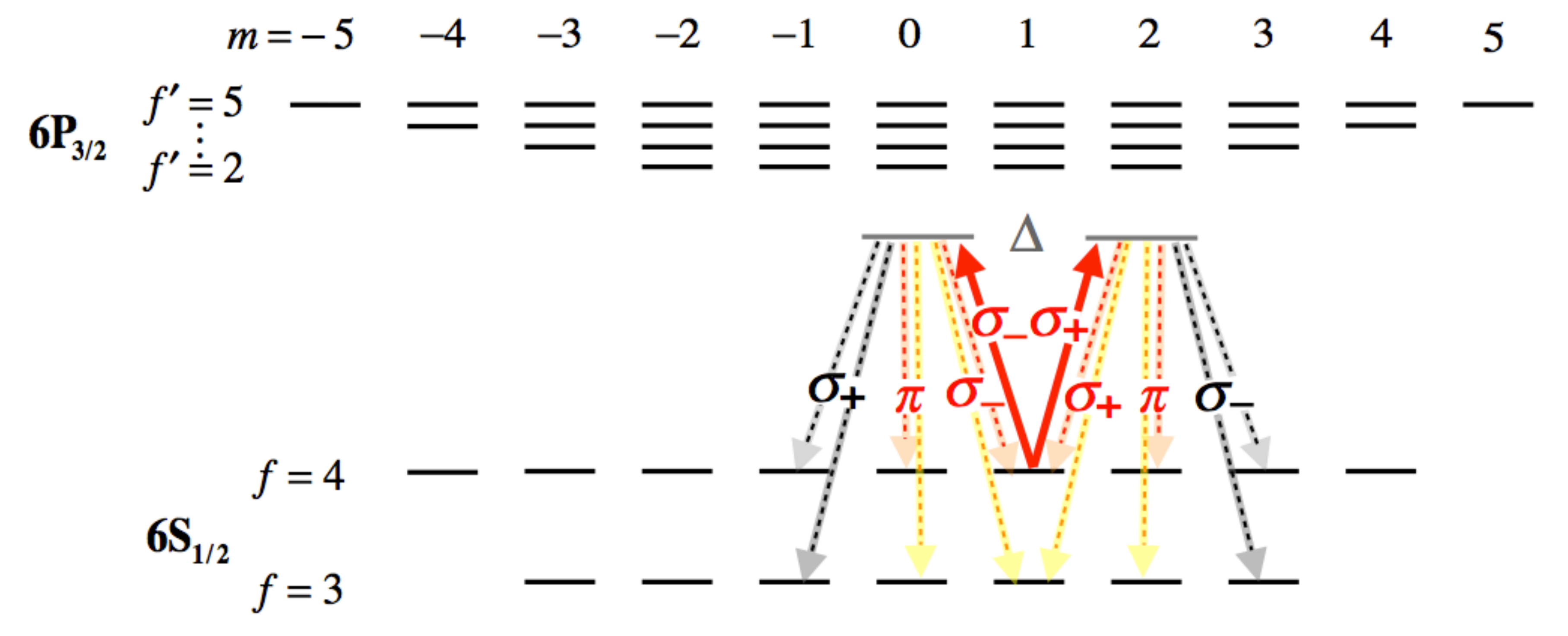}
\caption{An alkali atom is prepared in the ground hyperfine manifold with $f=4$. The quantization axis of the atom is perpendicular to the linear polarization of the probe, making the light a superposition of $\sigma_+$ and $\sigma_-$. Solid lines represent absorption of a photon from the probe and dashed lines represent spontaneous emission of a photon. The grey dashed lines denote transitions with negligibly small probability due to destructive interference. The atom absorbs the superposition of $\sigma_+$ and $\sigma_-$ and transitions to the excited hyperfine multiplet. By spontaneously emitting a $\sigma_+$, $\sigma_-$ or $\pi$ photon, the atom transitions back to one of the ground hyperfine manifolds. The atom can be optically pumped into $f=4$ or $f_-=3$, depending on the color of the emitted photon. }
\label{fig::f4sigma}
\end{figure}

Optical pumping affects higher spin alkali atoms in a similar fashion. As in the $f=1/2$ case, absorption of a $\sigma_\pm$ or $\pi$ photon changes the spin projection of the atom by $\Delta m=\pm 1$ or  $\Delta m=0$. Likewise, emission of a $\sigma_\pm$ or $\pi$ photon alters the spin projection by $\Delta m=\mp 1$ or  $\Delta m=0$. There are several important differences, however. One of the most significant differences is due to the presence of the second ground hyperfine manifold when $f>1/2$. Recall that when $f>1/2$, an alkali atom has two ground hyperfine manifolds with spins $f_\pm=i\pm1/2$, where $i$ is the nuclear spin. The ground hyperfine manifold with spin $f$, in which we prepare the atoms, can be either $f_+$ or $f_-$. For reference, the energy levels of an alkali are depicted in  Fig. \ref{fig::AlkaliLines}. Like $f=1/2$, a higher spin alkali atom initially in the $f$ manifold transitions to a higher energy level when it absorbs a probe photon.  Depending on the color of the photon it spontaneously emits, however, the atom might be optically pumped either back into the spin-$f$ manifold or into the other ground hyperfine manifold. When the detuning of the probe is sufficiently small compared to the ground state hyperfine splitting, atoms pumped into the other ground manifold are far off resonance, meaning that they are effectively lost from the system. For higher spin alkalis, optical pumping reduces the number of atoms in the ensemble in addition to changing the states of the atoms.  Examples of optical pumping when the atom is prepared in a ground hyperfine manifold with $f=4$ are shown in Figs. \ref{fig::f4pi} and \ref{fig::f4sigma}.

\subsection{Optical Pumping Master Equation}
For an alkali atom prepared in the spin-$f$ ground hyperfine manifold interacting with a probe linearly polarized along $x$ and detuned from the D1 or D2 line, optical pumping is approximately described by the master equation \cite{DeuJes09}  
\begin{align}\label{eq::Masteri}
\frac{d\hat{\rho}^{(i)}}{dt}\Big|_{\text{op}}=-\frac{2\gamma_s}{9}\hat{\rho}^{(i)}+\frac{g_f^2\gamma_s}{9}\left(\hat{f}_y^{(i)}\hat{\rho}^{(i)}\hat{f}_y^{(i)}+\hat{f}_z^{(i)}\hat{\rho}^{(i)}\hat{f}_z^{(i)}\right).
\end{align}
Here, the photon scattering rate is $\gamma_s=(N_L/\Delta t)\,(\sigma_0/A)\,(\Gamma^2/4\Delta^2)$. This equation neglects the tensor effects in optical pumping when the detuning is large compared to the excited state hyperfine splitting, but small compared to the fine structure splitting.  The internal spin operators $\hat{f}_y^{(i)}$ and $\hat{f}_z^{(i)}$ act on the spin-$f$ manifold. The effect of loss into the other ground manifold is taken into account by the fact that this master equation is not trace preserving when $f>1/2$. Although this master equation acts only on the $f$ manifold and, consequently, does not track the state of the atom in the other ground manifold, the probability of finding the atom in the $f$ manifold decreases with time due to loss. This is sufficient to determine the evolution of the ensemble observables that are relevant to spin squeezing since they are composed of sums of internal spin operators on the $f$ manifold.

To better understand how the master equation describes optical pumping, we decompose it into the form
\begin{align}\label{MostGenMasteri}
\frac{d\hat{\rho}^{(i)}}{dt}\Big|_{\text{op}}=-\Gamma_{\text{op}}\hat{\rho}^{(i)}+\sum_q\hat{W}_q^{(i)}\hat{\rho}^{(i)}\hat{W}_q^{(i)\dag}.
\end{align}
Here, $\Gamma_{\text{op}}=2\gamma_s/3$ is the total rate of optical pumping events and the $\hat{W}_q^{(i)}$ are jump operators. The jump operators update the state of atom in the $f$ manifold conditioned upon the polarization of the emitted photon. The index $q$ can take the value $+$, $0$ or $-$, representing the emission of a $\sigma_+$, $\pi$ or $\sigma_-$ photon, respectively. Assuming the detuning is large compared to the excited state hyperfine splitting, the exact form of the jump operators depends upon the quantization of the atomic spin relative to the probe polarization. For example, consider an atom quantized parallel to the linear polarization of the probe along $x$. The jump operators in this instance are approximately  \cite{DeuJes09}
\begin{align}\label{piPlus}
&\hat{W}_{+}^{(i)}=\frac{g_f}{3}\sqrt{\frac{\gamma_s}{2}}\hat{f}_{-}^{(i)},\\\label{pi0}
&\hat{W}_{0}=\frac{2\sqrt{\gamma_{s}}}{3}\mathbb{I}^{(i)}
\end{align}
and
\begin{align}\label{piMinus}
&\hat{W}_{-}^{(i)}=\frac{g_f}{3}\sqrt{\frac{\gamma_s}{2}}\hat{f}_{+}^{(i)},
\end{align}
which correspond to absorption of a $\pi$ photon and emission of a $\sigma_+$, $\pi$ and $\sigma_-$ photon, respectively. By referring to Fig. \ref{fig::f4pi}, we see that each of these jump operators updates the state of the atom how one would expect. The jump operators $\hat{W}_\pm^{(i)}$ change the spin projection by $\Delta m=\mp 1$ and the jump operator $\hat{W}_0^{(i)}$ leaves the spin projection unchanged. Alternatively, we can consider an atom quantized along $z$, perpendicular to the linear polarization of the probe along $x$. For a quantization axis along $z$ in the limit of large detuning, the jump operators are approximately
\begin{align}\label{sigmaPlus}
&\hat{W}_{+}^{(i)}=\sqrt{\frac{\gamma_s}{2}}\left(\frac{2}{3}\hat{\mathbb{I}}^{(i)}-\frac{g_f}{3}\hat{f}_z^{(i)}\right),  \\\label{sigma0}
&\hat{W}_0=\frac{g_f\sqrt{\gamma_{s}}}{3}\hat{f}_y^{(i)}
\end{align} 
and
\begin{align}\label{sigmaMinus}
&\hat{W}_{-}=\sqrt{\frac{\gamma_s}{2}}\left(\frac{2}{3}\hat{\mathbb{I}}^{(i)}+\frac{g_f}{3}\hat{f}_z^{(i)}\right),
\end{align} 
corresponding to the absorption of a photon in a superposition of $\sigma_+$ and $\sigma_-$   followed by the emission of a $\sigma_+$, $\pi$ and $\sigma_-$ photon, respectively. By referring to Fig. \ref{fig::f4sigma}, we can verify the action of these jump operators. When the detuning is significantly larger than the excited state hyperfine splitting, interference between transitions to the different spins $f'$ in the excited hyperfine multiplet eliminates changes in the spin projection of $\Delta m \pm 2 $. The jump operators $\hat{W}_{\pm}^{(i)}$, therefore, effectively describe absorption of a $\sigma_\pm$ photon followed by emission of of a $\sigma_\pm$ photon. These transitions leave the atomic state unchanged, as indicated by the jump operators. The jump operator $\hat{W}_0$ leaves the atom in a superposition of states with $\Delta m \pm 1$, as it should from Fig. \ref{fig::f4sigma}.

Ultimately, the polarization of the probe should dictate the effects of optical pumping and not the quantization axis we choose for the atomic spin. If we take the jump operators defined in Eqs. (\ref{piPlus})-(\ref{piMinus}) for a parallel quantization axis and substitute them into the general master equation in \erf{MostGenMasteri}, we will obtain the master equation in \erf{eq::Masteri}. Likewise, if we take the jump operators defined in Eqs. (\ref{sigmaPlus})-(\ref{sigmaMinus}) for a perpendicular quantization axis and substitute them into the general master equation in \erf{MostGenMasteri}, we will again obtain the master equation in \erf{eq::Masteri}. The quantization axis of the atom does not have any effect on optical pumping. We are, thus, free to choose the quantization axis that is more convenient for a particular atomic state or problem at hand.
  
 \subsection{Ensemble Master Equation}
The master equation describing the evolution of a single atom under optical pumping  in \erf{eq::Masteri} is generalized to describe the ensemble of $N_A$ atoms as follows,
 \begin{align}\label{eq::MasterEnsem}
\frac{d\hat{\rho}}{dt}\Big|_{\text{op}}=-\frac{2\gamma_s}{9}\sum_{i=1}^{N_A}\mathbb{I}^{(i)}\hat{\rho}+\frac{g_f^2\gamma_s}{9}\sum_{i=1}^{N_A}\left(\hat{f}_y^{(i)}\hat{\rho}\hat{f}_y^{(i)}+\hat{f}_z^{(i)}\hat{\rho}\hat{f}_z^{(i)}\right).
\end{align}
 This equation gives the evolution of the ensemble state, $\hat{\rho}$, as each atom, $i$, undergoes optical pumping independently. For an ensemble identical under interchange of atoms, taking the index $i$ in \erf{eq::MasterEnsem} to $k$ instead of $N_A$ gives the evolution of any group of $k$ atoms in the ensemble,
 \begin{align}\label{eq::Masterk}
\frac{d\hat{\rho}^{(1, ..., k)}}{dt}\Big|_{\text{op}}=&-\frac{2\gamma_s}{9}\sum_{i=1}^{k}\mathbb{I}^{(i)}\hat{\rho}^{(1, ..., k)}\\\notag&+\frac{g_f^2\gamma_s}{9}\sum_{i=1}^{k}\left(\hat{f}_y^{(i)}\hat{\rho}^{(1, ..., k)}\hat{f}_y^{(i)}+\hat{f}_z^{(i)}\hat{\rho}^{(1, ..., k)}\hat{f}_z^{(i)}\right).
\end{align}
Because the master equation is not trace preserving, care must be taken in deriving equations of motion for ensemble observables from \erf{eq::MasterEnsem}. The equation of motion for a $k$th order correlation function $\expect{\hat{o}^{(1)}...\hat{o}^{(k)}}$, for example, follows directly from $\hat{\rho}^{(1, ..., k)}$ in \erf{eq::Masterk},
\begin{align}\label{eq::kCorrelation}
\frac{d}{dt}\expect{\hat{o}^{(1)}&...\hat{o}^{(k)}}=\text{Tr}\Big(\hat{\rho}^{(1, ..., k)}\hat{o}^{(1)}...\hat{o}^{(k)}\Big) \\\notag
=&-\frac{2k\gamma_s}{9}\expect{\hat{o}^{(1)}...\hat{o}^{(k)}}
+\frac{g_f^2\gamma_s}{9}\sum_{i=1}^{k}\Big\langle\left(\hat{f}_y^{(i)}\hat{o}^{(i)}\hat{f}_y^{(i)}+\hat{f}_z^{(i)}\hat{o}^{(i)}\hat{f}_z^{(i)}\right)
\Pi_{j\neq i}\hat{o}^{(j)}\Big\rangle.
\end{align}

\section{Effects of Optical Pumping}\label{Sec::OPEvents}
Given a basic understanding of optical pumping, we now examine how it affects the ensemble state and the observables relevant to spin squeezing. Optical pumping has several harmful effects, which include a decay of the negative correlations that create spin squeezing, a decay of the mean spin and an increase in the collective spin variance that we are trying to squeeze. We explore optical pumping on an ensemble initially prepared in $\ket{\uparrow}^{\otimes N_A}$ with fiducial state $\ket{\uparrow}$ and coupled state $\ket{\downarrow}$. To avoid specifying a quantization axis for the atomic spins, we employ a master equation of the form
\begin{align}\label{MostGenMasterk}
\frac{d\hat{\rho}^{(1, ..., k)}}{dt}\Big|_{\text{op}}=-\Gamma_{\text{op}}\sum_{i=1}^{k}\mathbb{I}^{(i)}\hat{\rho}^{(1, ..., k)}+\sum_{i=1}^{k}\sum_q\hat{W}_q^{(i)}\hat{\rho}^{(1, ..., k)}\hat{W}_q^{(i)\dag},
\end{align}
where the $\hat{W}_q^{(i)}$ are arbitrary jump operators and $\hat{\rho}^{(1, ..., k)}$ is the density operator of any $k$ atoms.

\subsection{Growth of the Collective Spin Projection Variance}\label{sec::VarianceOP}
To understand the impact of optical pumping on the collective spin variance, we first study the interatomic entanglement that causes this variance to become ``squeezed". If the Bogoliubov transformation in \erf{bogoliubov} is rotated by $\pi/4$, it creates pure squeezing of the quadrature $\hat{X}_\downarrow$, which is proportional to $\hat{F}_z$ under the multilevel HP approximation.  To lowest order in $\xi/N_A<<1$, the spin squeezed state created by the rotated Bogoliubov transformation is 
\begin{align}
\ket{\psi}_\text{sq}\approx\ket{\uparrow}^{\otimes N_A}-\frac{\xi}{4N_A}\sum_{i\neq j}\ket{\downarrow_i\downarrow_j}\ket{\uparrow}_{\neq i, j}^{\otimes\left(N_A-2\right)}.
\label{SqueezedState}
\end{align}
This is an entangled state where every two atoms $i$ and $j$ are pairwise correlated. The density operator corresponding to this squeezed state is
\begin{align}\label{densityOp}
\hat{\rho}_{\text{sq}}\approx&\left(\ket{\uparrow}\bra{\uparrow}\right)^{\otimes N_A}\\\notag&-
\frac{\xi}{4N_A}\sum_{i\neq j}(\ket{\uparrow_i\uparrow_j}\bra{\downarrow_i\downarrow_j}+\ket{\downarrow_i\downarrow_j}\bra{\uparrow_i\uparrow_j})\left(\ket{\uparrow}\bra{\uparrow}\right)_{\neq i,j}^{\otimes N_A-2}.
\end{align}
Note that $\hat{\rho}_{\text{sq}}$ contains the term
\begin{align}\label{eq::pairwiseCoherence}
\hat{c}(i,j)=\left(\ket{\uparrow_i\uparrow_j}\bra{\downarrow_i\downarrow_j}+\ket{\downarrow_i\downarrow_j}\bra{\uparrow_i\uparrow_j}\right),
\end{align}
which is the coherence between two pairwise entangled atoms. To understand how pairwise entanglement influences spin squeezing, we can calculate the correlation term in \erf{variance} from $\hat{c}(i,j)$,
\begin{align}
\langle\Delta\hat{f}_z^{(i)}\Delta\hat{f}_z^{(j)}\rangle_{i\neq j}\approx-\frac{\xi}{2N_A}\text{Tr}(\hat{c}(i,j)\Delta\hat{f}_z^{(i)}\Delta\hat{f}_z^{(j)})
=-\xi(\Delta f_z^2)_\uparrow/N_A.
\label{negCorr}
\end{align}
Recall that collective spin squeezing requires this correlation term to be negative in absence of internal spin squeezing. Equation (\ref{negCorr}) demonstrates that negative correlations are fed by coherences of entangled, pairwise correlated states. Note also that the magnitude of the correlation term is proportional to the variance of the fiducial state.

Pairwise correlations decay at a rate dependent upon the choice of fiducial state, making some state preparations more robust than others. To see this, we calculate the evolution of the correlation term for the initial squeezed state in \erf{densityOp} over a small time step $\Delta t$ using the master equation in \erf{MostGenMasterk},
\begin{align}\label{corrDecay}
\langle\Delta\hat{f}_z^{(i)}\Delta&\hat{f}_z^{(j)}(\Delta t)\rangle_{i\neq j}\approx\left(1-2\Gamma_\text{op}\Delta t\right)\langle\Delta\hat{f}_z^{(i)}\Delta\hat{f}_z^{(j)}(0)\rangle_{i\neq j}\\\notag&-\frac{\xi}{N_A}\sqrt{(\Delta f_z^2)_\uparrow}\Delta t
\sum_q(\bra{\uparrow}\hat{W}_q^{\dag}\Delta\hat{f}_z\hat{W}_q\ket{\downarrow}+\bra{\downarrow}\hat{W}_q^{\dag}\Delta\hat{f}_z\hat{W}_q\ket{\uparrow}).
\end{align}
The first term in this equation represents a decay of the initial correlation term due to all optical pumping events. The second term, which updates the correlation term based upon the state of the atoms optically pumped back into the $f$ manifold, is proportional to the quantity
\begin{align}\label{update}
C(\uparrow)&=\sqrt{(\Delta f_z^2)_\uparrow}\sum_q\left(\bra{\uparrow}\hat{W}_q^{\dag}\Delta\hat{f}_z\hat{W}_q\ket{\downarrow}+\bra{\downarrow}\hat{W}_q^{\dag}\Delta\hat{f}_z\hat{W}_q\ket{\uparrow}\right)\\
&=2\sqrt{(\Delta f_z^2)_\uparrow}\sum_q\text{Re}\left[\bra{\widetilde{q}_\uparrow}\Delta\hat{f}_z\ket{\widetilde{q}_\downarrow}\right].
\end{align}
Here, $\ket{\widetilde{q}_\uparrow}=\hat{W}_q\ket{\uparrow}$ and $\ket{\widetilde{q}_\downarrow}=\hat{W}_q\ket{\downarrow}$ are the unnormalized states to which the fiducial and coupled states are mapped after an optical pumping event. If the fiducial and coupled states are mapped to states such that $C(\uparrow)>0$, negative correlations are maintained by optical pumping. If $C(\uparrow)<0$, on the other hand, optical pumping produces positive correlations. Consequently, squeezing decays more rapidly for fiducial states where $C(\uparrow)<0$. Pairwise correlations also decay due to loss into the other ground hyperfine manifold. Even if optical pumping back into the $f$ manifold preserves negative correlations, the total rate of optical pumping events is larger than the rate of replenishment into the $f$ manifold, signifying a net loss of atoms and a decay of negative correlations.  

As the ensemble undergoes optical pumping, the collective variance $\Delta F_z^2$ is affected both by the decay of negative correlations and by the variances of the states in the $f$ manifold to which the atoms are optically pumped. The variance of the initial squeezed state in \erf{densityOp} after a small time step $\Delta t$ is 
\begin{align}\label{SmallTvar}
\Delta F_z^2(\Delta t)\approx&\left(1-\Gamma_\text{op}\Delta t\right)\Delta F_z^2(0)+
\Gamma_\text{op} N_A\xi\Delta t(\Delta f_z^2)_\uparrow\notag\\&-
N_A\xi\Delta tC(\uparrow)+
N_A\Delta t\sum_q\bra{\widetilde{q}_\uparrow}(\Delta\hat{f}_z)^2\ket{\widetilde{q}_\uparrow}.
\end{align}
The first and second terms in \erf{SmallTvar} are the decay of the collective variance and the negative correlations, respectively, due to all optical pumping events. The third term is the familiar update on the correlation term, while the fourth term is a ``noise injection" that results from optical pumping back into the $f$ manifold. To a first order approximation, the noise term only depends upon the variance of the state $\ket{\widetilde{q}_\uparrow}$ to which the fiducial state is mapped. This occurs because the population of atoms in the fiducial state remains much larger than the  population of atoms in the coupled state. As in the case of the correlation term, loss into the other ground hyperfine manifold causes the collective variance to decay. The behavior of the collective variance under optical pumping ultimately depends upon competition between the noise injection, the decay of the correlation term and loss. 

\subsection{Decay of Mean Spin}
In addition to increasing the variance of the collective spin, optical pumping causes the mean spin to decay. Recall that the spin squeezing parameter in \erf{eq::SqParameter} is inversely proportional to the square of the mean spin, $\expect{\hat{F}_x}$.  A smaller mean spin, thus, results in less metrologically useful spin squeezing. To good approximation when $N_A>>1$ and $\chi<<1$, the mean spin is given by
\begin{align}\label{eq::FxPop}
\expect{\hat{F}_x}=\expect{\hat{f}_x}_\uparrow N_\uparrow
+\expect{\hat{f}_x}_\downarrow N_\downarrow.
\end{align}
Here, $N_\psi=\sum_{i=1}^{N_A}\ket{\psi}\bra{\psi}_i$ is a ``population" quantifying the number of atoms in state $\ket{\psi}$. Because the ensemble begins in an eigenstate of the populations, $N_\uparrow$ and $N_\downarrow$ have no initial variance. While $N_\uparrow$ and $N_\downarrow$ accumulate variance through optical pumping,  this variance influences the covariances of the ensemble observables as a second order effect in the scattering rate, $\gamma_s$. We, thus, approximate the populations as c-numbers.

Equation (\ref{eq::FxPop}) enables us to track the decay of the mean spin through the populations. The evolution of the populations under optical pumping is governed primarily by two processes, loss and ``spin flips".  Loss into the other ground hyperfine manifold is a familiar concept. Spin flips, on the other hand, occur when an atom is optically pumped from the fiducial state to the coupled state. While an atom could also be flipped from the coupled state to the fiducial state in principle, this process does not contribute appreciably to the dynamics since the ensemble is prepared with every atom in the fiducial state.  The rate at which an atom in state $\ket{\psi}$, where $\psi\in\{\uparrow,\downarrow\}$, is lost into the other ground hyperfine manifold is 
\begin{align}\label{lossPsi}
\Gamma_{\text{loss},\psi}=\Gamma_{\text{op}}-\sum_q|\bra{\psi}\hat{W}_q\ket{\psi}|^2.
\end{align}
The rate of spin flips from $\ket{\uparrow}$ to $\ket{\downarrow}$ is given by
\begin{align}\label{eq::FlipRate}
\Gamma_{\text{flip}}=\sum_q|\bra{\downarrow}\hat{W}_q\ket{\uparrow}|^2.
\end{align}
In terms of the loss and spin flip rates, the evolution of the mean spin over a small time step $\Delta t$ is given by
\begin{align}\label{eq::meanSpinEvol}
\expect{\hat{F}_x(\Delta t)}=&\expect{\hat{f}_x}_\uparrow\hat{N}_\uparrow(\Delta t)
+\expect{\hat{f}_x}_\downarrow\hat{N}_\downarrow(\Delta t)\\\notag
\approx&\expect{\hat{f}_x}_\uparrow(1-\Gamma_{\text{loss},\uparrow}\Delta t)\hat{N}_\uparrow(0)+
\\&\expect{\hat{f}_x}_\downarrow\big[(1-\Gamma_{\text{loss},\downarrow}\Delta t)\hat{N}_\downarrow(0)
+\Gamma_\text{flip}\Delta t\hat{N}_\uparrow(0)\big].\notag
\end{align}
From \erf{eq::meanSpinEvol}, it is evident that loss causes the mean spin to decay. Because loss also causes the collective variance to decay, however, damage to spin squeezing is much reduced. In general, spin flips are more damaging to spin squeezing. Spin flips destroy negative correlations at a rate depending upon $C(\uparrow)$. 
Because an atom returns to the $f$ manifold after a spin flip, there is no reduction in the collective variance. The extent to which spin flips cause the mean spin to decay depends upon the values of $\expect{\hat{f}_x}_{\uparrow'}$ and $\expect{\hat{f}_x}_{\downarrow'}$, where $\ket{\uparrow'}$ and $\ket{\downarrow'}$ are the states to which the fiducial and coupled states are mapped after post-processing. The partial isometries discussed in Sec. \ref{sec::postprocessing} map the atoms to fiducial and coupled states for which $\expect{\hat{f}_x}_{\uparrow'}>\expect{\hat{f}_x}_{\downarrow'}$, ensuring that every spin flip event reduces the mean spin. 

\section{Internal Spin Control and Optical Pumping}\label{sec::ControlOP}

To make the multilevel Holstein-Primakoff approximation, we modeled the atoms as an ensemble of embedded qubits composed of the fiducial and coupled states. As we have seen, optical pumping back into the $f$ manifold takes the fiducial and coupled states to the states $\ket{\widetilde{q}_\uparrow}$ and $\ket{\widetilde{q}_\downarrow}$, respectively, which may not be contained in the subspace spanned by the embedded qubit. Through internal spin control, however, we have the option of eliminating atoms that have been pumped outside the embedded qubit subspace. Using microwave pulses, which induce transitions between the ground hyperfine manifolds of alkali atoms, we can generate a map that takes all states in the $f$ manifold to the other ground hyperfine manifold, except for the fiducial and coupled states. For example, suppose that our ensemble has been prepared in a spin coherent state, where the fiducial and coupled states are $\ket{\uparrow_{SCS}}=\ket{f,m_x=f}$ and $\ket{\downarrow_{SCS}}=i\ket{f,m_x=f-1}$. Also suppose that the $f$ manifold is the ground hyperfine manifold with the larger spin quantum number, the spin quantum number of the other ground hyperfine manifold being $f-1$. Through internal spin control, we can generate the unitary map
\begin{align}
\hat{U}_-=&\bigotimes_{i=1}^{N_A}\!\Big(\ket{f,m_x=f}\bra{f,m_x=f}_i\!+\!\ket{f,m_x=f-1}\bra{f,m_x=f-1}_i\\
&+\sum_{m=-f}^{f-2}\ket{f-1,m_x=m+1}\bra{f,m_x=m}_i\Big).  
\end{align}
The unitary $\hat{U}_-$ has no effect upon the embedded qubit, but maps all other states in the $f$ manifold to the ground manifold with spin $f-1$. After applying $\hat{U}_-$ or a similar unitary operation on the internal spin,  all optical pumping events outside the embedded qubit are equivalent to loss. 

In addition to ensuring the validity of the embedded qubit approximation, this control has a substantial impact on the evolution of the projection noise. Recall that the noise injection in \erf{SmallTvar} increases with the variance of $\ket{\widetilde{q}_\uparrow}$. If a component of $\ket{\widetilde{q}_\uparrow}$ is contained outside the subspace spanned by the embedded qubit, mapping all states outside the embedded qubit subspace to the other ground manifold will reduce the noise injection. If $\ket{\widetilde{q}_\uparrow}$ lies entirely outside of the embedded qubit subspace, its contribution to the noise injection will be completely eliminated. Consequently, the only states contributing to the noise injection are the fiducial and coupled state. The amount that the noise injection contributes to the overall collective spin variance ultimately depends upon the magnitudes of $\bra{\uparrow'}(\Delta\hat{f}_z)^2\ket{\uparrow'}$ and $\bra{\downarrow'}(\Delta\hat{f}_z)^2\ket{\downarrow'}$, where $\ket{\uparrow'}$ and $\ket{\downarrow'}$ are the states to which the fiducial and coupled states are mapped after post-processing.

\section{A Detailed Look at The SCS}
\begin{figure}[H]
\centering
\includegraphics[scale=.6]{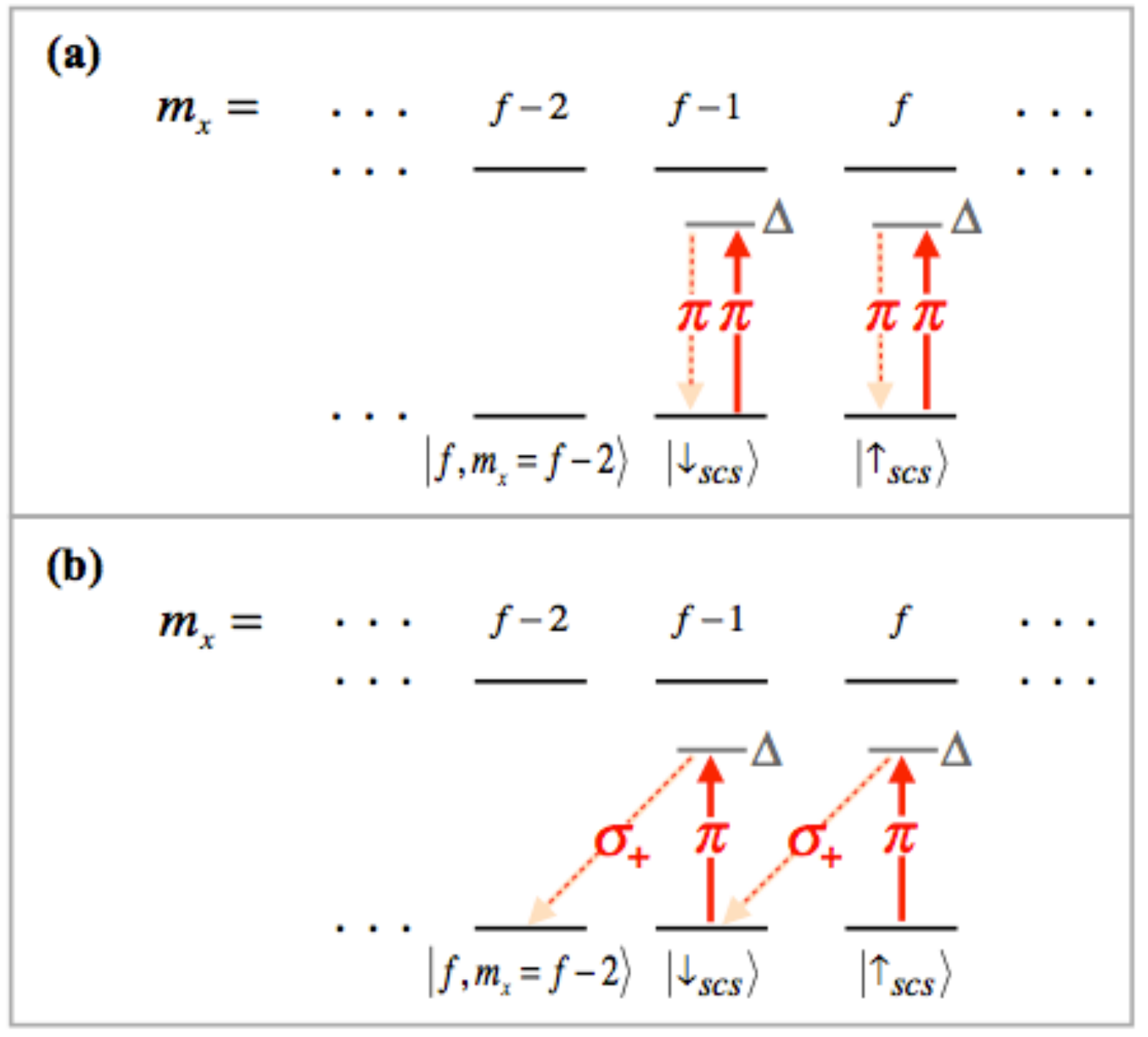}
\caption{Optical pumping processes permitted in the $SCS$ preparation. In (a) and (b), the polarization of the probe is parallel to the quantization axis of the atomic spin, meaning that it is $\pi$ polarized. Fig. (a) shows the absorption of a $\pi$ polarized photon followed by the emission of a $\pi$ polarized photon. This process does not change the spin state of the atom. Fig. (b) shows the absorption of a $\pi$ polarized photon followed by the emission of a $\sigma_+$ polarized photon, causing the projection of the atom's angular momentum along $x$ to decrease by 1. Not pictured is the absorption of a $\pi$ polarized photon and the emission of a $\sigma_-$ photon. This process is prohibited when the atom remains in the fiducial state, the state with maximal spin projection along $x$. Although the atom can emit a $\sigma_-$ photon after it has been optically pumped into the coupled state, this is a second order effect. Consequently,  the absorption of a $\pi$ polarized photon and the emission of a $\sigma_-$ photon does not contribute appreciably to the dynamics when the ensemble is prepared in the $SCS$ preparation.}
\label{fig::SCSpi}
\end{figure}

To apply the ideas of the previous sections, we now examine the effects of optical pumping on the SCS preparation in detail. For conceptual purposes, we neglect the bias magnetic field along the $z$-axis. Optical pumping in the presence of the bias magnetic field will be treated in Chapter \ref{sec::ModHPCovar}.  Because the fiducial state of the SCS preparation is $\ket{\uparrow_{SCS}}=\ket{f,m_x=f}$, it is natural to take the quantization axis of the atomic spin to be along $x$, parallel to the probe polarization. The optical pumping processes permitted in this configuration are depicted in Fig. \ref{fig::SCSpi} (a) and (b).   Note that when an atom absorbs a $\pi$ photon and then returns to the $f$ manifold by emitting a $\pi$ photon, its state is unchanged. Therefore, we concentrate on the two other optical pumping processes within the $f$ manifold: (1) the atom absorbs a $\pi$ polarized photon from the probe and emits a $\sigma_+$ photon and (2) the atom absorbs a $\pi$ polarized photon from the probe and emits a $\sigma_-$ photon. These optical pumping processes are described by the jump operators $\hat{W}_\pm$ in Eqs. (\ref{piPlus}) and (\ref{piMinus}). Because the expected number of atoms in the fiducial state remains large and an atom in the magnetic sublevel $m_x=f$ cannot emit a  $\sigma_-$ photon, process (2) does not contribute appreciably to the decoherence of the atomic state. Process (1), thus, 
is the dominant effect for the SCS.

Process (1) is a spin flip, sending $\ket{\uparrow_{SCS}}=\ket{f,m_x=f}$ to $\ket{\downarrow_{SCS}}=\\i\ket{f,m_x=f-1}$  up to a global phase, as depicted in Fig. \ref{fig::SCSpi} (b). While the fiducial state ``flips" to the coupled state, the coupled state is sent to another state, $\ket{f,m_x=f-2}$, up to a global phase. Using internal spin control, we could eliminate this state by sending it to the other ground manifold. However, an examination of the pairwise entanglement that creates spin squeezing reveals that this is not the optimal thing to do. Consider how the squeezed state in \erf{densityOp} for an initial fiducial state $\ket{\uparrow_{SCS}}$ transforms when one of the atoms undergoes process (1), 
\begin{align}\label{eq::NegCorTransferCoherence}
\hat{W}_+^{(1)}&\hat{\rho}_\text{sq}\hat{W}_+^{(1) \dag}\propto2f\ket{\downarrow_{SCS}}\bra{\downarrow_{SCS}}_1\left(\ket{\uparrow_{SCS}}\bra{\uparrow_{SCS}}\right)^{\otimes (N_A-1)}_{\neq 1}\\\notag&-\!
\frac{\xi\sqrt{f(2f-1)}}{N_A}\!\!\sum_{j\neq 1}\!(\ket{\downarrow_{SCS}}_1\ket{\uparrow_{SCS}}_j\bra{f-2}_1\bra{\downarrow_{SCS}}_j\!+\!\text{h.c.})\\\notag&\;\;\;\times\left(\ket{\uparrow_{SCS}}\bra{\uparrow_{SCS}}\right)_{\neq 1,j}^{\otimes N_A-2}
\\\notag&-\!
\frac{\xi f}{2N_A}\ket{\downarrow_{SCS}}\bra{\downarrow_{SCS}}_1\!\!\sum_{i\neq j,\;i,j\neq 1}\!(\ket{\uparrow_{SCS}}_i\ket{\uparrow_{SCS}}_j\bra{\downarrow_{SCS}}_i\bra{\downarrow_{SCS}}_j\!+\!\text{h.c.})\\\notag&\;\;\;\times\left(\ket{\uparrow_{SCS}}\bra{\uparrow_{SCS}}\right)_{\neq 1,i,j}^{\otimes N_A-3}.
\end{align}
Here, $\ket{f-2}$ is shorthand for $\ket{f,m_x=f-2}$. In the state above, atom 1 has been ``flipped". From this state, we can calculate the contribution of atom 1 to the negative correlations in \erf{variance},
\begin{align}\notag
\langle\Delta\hat{f}_z^{(j)}\Delta\hat{f}_z^{(1)}\rangle_{j\neq 1}=&-\frac{\xi\sqrt{f(2f-1)}}{N_A}\bra{f-2}_1\bra{\downarrow_{SCS}}_j\Delta\hat{f}_z^{(j)}\Delta\hat{f}_z^{(1)}\ket{\downarrow_{SCS}}_1\ket{\uparrow_{SCS}}_j\\&\;\;\;\;+\text{c.c.}\\
=&-\frac{\xi f(2f-1)}{N_A}\label{transferCorr2}.
\end{align}
After the optical pumping event, pairwise coherences between $\ket{\downarrow_{SCS}}$ and \\$\ket{f-2}$ take the place of pairwise coherences between $\ket{\uparrow_{SCS}}$ and $\ket{\downarrow_{SCS}}$, feeding the negative correlations that generate spin squeezing. By mapping $\ket{f-2}$ to the other ground hyperfine manifold, we destroy negative correlations.

To understand this another way, we can write $\hat{f}_z$ in the basis of eigenstates of  $\hat{f}_x$ as
\begin{align}\label{fzSCStransfer}
\hat{f}_z=&\sqrt{\frac{f}{2}}\Big(i\ket{f-1}\bra{f}-i\ket{f}\bra{f-1}\Big)\\\notag
&+\sqrt{\frac{2f-1}{2}}\Big(i\ket{f-2}\bra{f-1}-i\ket{f-1}\bra{f-2}\Big).
\end{align}
The collective spin $\hat{F}_z$ becomes
\begin{align}
\hat{F}_z=\sqrt{\frac{f}{2}}\hat{\Sigma}_{f-1}+\sqrt{\frac{2f-1}{2}}\hat{\Sigma}_{f-2},
\end{align}
where $\hat{\Sigma}_{f-1}=\sum_{j=1}^{N_A}(i\ket{f-1}\bra{f}_j-i\ket{f}\bra{f-1}_j)$ and $\hat{\Sigma}_{f-2}=\\\sum_{j=1}^{N_A}(i\ket{f-2}\bra{f-1}_j-i\ket{f-1}\bra{f-2}_j)$. Note that $\hat{\Sigma}_{f-1}$ is related to the position quadrature $\hat{X}_\downarrow$ since $\ket{\uparrow_{SCS}}=\ket{f}$ and $\ket{\downarrow_{SCS}}=i\ket{f-1}$. In terms of the operators $\hat{\Sigma}_{f-1}$ and $\hat{\Sigma}_{f-2}$, the collective spin variance is given by
\begin{align}
\Delta F_z^2=&\frac{f}{2}\Delta \Sigma_{f-1}^2+\frac{\sqrt{f(2f-1)}}{2}\expect{\Delta\hat{\Sigma}_{f-1}\Delta\hat{\Sigma}_{f-2}+\Delta\hat{\Sigma}_{f-2}\Delta\hat{\Sigma}_{f-1}}\\\notag&+\frac{2f-1}{2}\Delta \Sigma_{f-2}^2.
\end{align}
The second term in this expression, which is the covariance between the operators  $\hat{\Sigma}_{f-1}$ and $\hat{\Sigma}_{f-2}$, contains negative correlations that contribute to spin squeezing. These correlations are fed by pairwise coherences in of the form in \erf{eq::NegCorTransferCoherence} between atoms 1 and $j$. If we eliminate the state $\ket{f-2}$, the second term disappears and the variance becomes
\begin{align}
\Delta F_z^2=&\frac{f}{2}\Delta \Sigma_{f-1}^2+\frac{2f-1}{2}N_{\downarrow_{SCS}}.
\end{align}
When $\ket{f-2}$ is absent, negative correlations that reduce the variance are lost. Though it may seem counterintuitive, preserving $\ket{f-2}$ in addition to the fiducial and coupled states produces more spin squeezing.

\section{Transfers of Coherence}\label{sec::TransferState}
While internal spin control can reduce the noise injection and allow us to continue modeling our atoms as embedded qubits, eliminating all states besides the fiducial and coupled states is not always advantageous, as shown in the previous section. In the case of the SCS preparation,  we saw that preserving the state $\ket{f,m_x=f-2}$ maintains negative correlations that create spin squeezing. We call a state such as $\ket{f,m_x=f-2}$, which contributes to negative correlations in the presence of optical pumping, a ``transfer state". In this section, we explore the existence of transfer states for arbitrary state preparations. 

We first follow a procedure similar to Sec. \ref{sec::MultiHPEnsemble} and expand $\hat{{f}}_z$ in terms of the fiducial state, the coupled state and a third state, $\ket{\wr}$, orthogonal to both the fiducial and the coupled states, 
\begin{align}\notag
&\hat{f}_z\approx(\ket{\uparrow}\bra{\uparrow}+\ket{\downarrow}\bra{\downarrow}+\ket{\wr}\bra{\wr})\hat{f}_z(\ket{\uparrow}\bra{\uparrow}+\ket{\downarrow}\bra{\downarrow}+\ket{\wr}\bra{\wr})
\\\label{newFz}=&\sqrt{(\Delta f_z^2)_\uparrow}\left(\ket{\uparrow}\bra{\downarrow}\!+\!\ket{\downarrow}\bra{\uparrow}\right)
+\sqrt{(\Delta f_z^2)_\downarrow\!-\!(\Delta f_z^2)_\uparrow}\left(\ket{\downarrow}\bra{\wr}\!+\!\ket{\wr}\bra{\downarrow}\right).
\end{align}
Note that by the definition of the coupled state, $\bra{\wr}\hat{f}_z\ket{\uparrow}=\sqrt{(\Delta f_z^2)_\uparrow}\bra{\wr}\downarrow\rangle=0$, which explains the absence of terms coupling $\ket{\uparrow}$ and $\ket{\wr}$  in \erf{newFz}. For simplicity, we have also assumed that $\expect{\hat{f}_z}_\uparrow=\expect{\hat{f}_z}_\downarrow=\expect{\hat{f}_z}_\wr=0$.  States that satisfy this criteria, such as the SCS, cat and $m_x=0$ state preparations, are natural to consider for spin squeezing. This assumption will be relaxed in Chapter \ref{sec::Beyond}.  

The third orthogonal state in \erf{newFz}, which we denoted $\ket{\wr}$, is the transfer state. Although we will demonstrate this more rigorously below, compare \erf{newFz} to the expansion of $\hat{f}_z$ for the SCS preparation in \erf{fzSCStransfer}. The state $-\ket{f,m_x=f-2}$ in \erf{fzSCStransfer}, which is the transfer state for the SCS preparation, takes the place of $\ket{\wr}$. From \erf{newFz}, we can determine a general expression for the transfer state in terms of the coupled and fiducial states. When $\sqrt{(\Delta f_z^2)_\downarrow\!-\!(\Delta f_z^2)_\uparrow}\neq 0$, the transfer state is given by
\begin{align}\label{CoherenceState}
\ket{\wr}=\frac{1}{\sqrt{(\Delta f_z^2)_\downarrow\!-\!(\Delta f_z^2)_\uparrow}}\left(\hat{f}_z\ket{\downarrow}-\sqrt{(\Delta f_z^2)_\uparrow}\ket{\uparrow}\right).
\end{align}

As discussed in Sec. \ref{sec::VarianceOP}, the behavior of negative correlations under optical pumping is governed by the quantity $C(\uparrow)$, given in \erf{update}.  Recall that when $C(\uparrow)>0$, negative correlations are preserved.  Substituting \erf{newFz} into \erf{update} yields  an expression for $C(\uparrow)$ in terms of the fiducial, coupled and transfer states,
\begin{align}\label{expandedC}
C(\uparrow)=&(\Delta f_z^2)_\uparrow\sum_q\text{Re}[\langle\widetilde{q}_\uparrow|\uparrow\rangle\langle\downarrow|\widetilde{q}_\downarrow\rangle+\langle\widetilde{q}_\uparrow|\downarrow\rangle\langle\uparrow|\widetilde{q}_\downarrow\rangle]
\\\notag&+\sqrt{(\Delta f_z^2)_\uparrow((\Delta f_z^2)_\downarrow-(\Delta f_z^2)_\uparrow)}\sum_q\text{Re}[\langle\widetilde{q}_\uparrow|\downarrow\rangle\langle\wr|\widetilde{q}_\downarrow\rangle+\langle\widetilde{q}_\uparrow|\wr\rangle\langle\downarrow|\widetilde{q}_\downarrow\rangle].
\end{align}
The first term in this expression depends upon coherences between $\ket{\uparrow}$ and $\ket{\downarrow}$. The second term, which depends upon coherences between $\ket{\downarrow}$ and $\ket{\wr}$, determines the influence of the transfer state on negative correlations.

To gain insight, we analyze the second term in \erf{expandedC} in greater detail. The components of the second term for each $q$ are proportional to 
 \begin{align}\label{transferTerm}
 C_{\wr}(\uparrow,q)=\text{Re}\left[\langle q_\uparrow|\downarrow\rangle\langle\wr|q_\downarrow\rangle+\langle q_\uparrow|\wr\rangle\langle\downarrow| q_\downarrow\rangle\right],
 \end{align}
where $\ket{q_\uparrow}$ and $\ket{q_\downarrow}$ are the normalized versions of 
$\ket{\widetilde{q}_\uparrow}=\hat{W}_q\ket{\uparrow}$ and \\$\ket{\widetilde{q}_\downarrow}=\hat{W}_q\ket{\downarrow}$. We first consider two cases in which $C_{\wr}(\uparrow,q)$ is maximal,
 \begin{align}
 &(1) \;\;\;\langle q_\uparrow|\downarrow\rangle\langle\wr|q_\downarrow\rangle=1 \;\;\;\text{and}\;\;\; \langle q_\uparrow|\wr\rangle\langle\downarrow| q_\downarrow\rangle=0
  \end{align}
  and
  \begin{align}
&(2)  \;\;\;\langle q_\uparrow|\downarrow\rangle\langle\wr|q_\downarrow\rangle=0 \;\;\;\text{and}\;\;\; \langle q_\uparrow|\wr\rangle\langle\downarrow| q_\downarrow\rangle=1. 
 \end{align}
Recall that the noise injection in \erf{SmallTvar} is proportional to $\bra{q_\uparrow}(\Delta\hat{f}_z)^2\ket{q_\uparrow}$, the variance of the state to which the fiducial state is optically pumped. In case (2), the fiducial state is mapped to the transfer state. As a consequence, the transfer state contributes to the noise injection. Although preserving $\ket{\wr}$ maximizes $C(\uparrow)$ and the magnitude of the correlation term, this can be offset by a large injection of noise. Whether or not it is beneficial to retain the transfer state depends on the strength of its contribution to the negative correlations relative to the noise injection. If the contribution to the noise injection is larger, it is optimal to eliminate the noise injection by mapping the transfer state to the other hyperfine manifold. The situation is different for case (1), in which the fiducial state is mapped to the coupled state and the coupled state to the transfer state.  In this case, the noise injection is proportional to $\bra{\downarrow}(\Delta\hat{f}_z)^2\ket{\downarrow}$. Even if the noise injection is larger than the magnitude of the correlation update term, eliminating the coupled state through microwave control would destroy all beneficial pairwise correlations between $\ket{\uparrow}$ and  $\ket{\downarrow}$. It is instead advantageous to maximize $C(\uparrow)$ by retaining $\ket{\wr}$. This is the case in the SCS preparation.

From  Eqs. (\ref{SmallTvar})  and (\ref{expandedC}), we can determine the exact contribution of the transfer state to the negative correlations and to the noise injection. However, \erf{SmallTvar} presupposes that the ensemble was initially in a pure, highly squeezed state. In practice, we apply our squeezing protocols to states which are not squeezed at the initial time. As a consequence, the magnitude of the correlation update term is nearly always smaller than the noise injection. In this case, any contribution that the transfer state makes to the noise injection should be eliminated by mapping  the transfer state to the other ground manifold. Preserving the transfer state is beneficial when the following conditions hold: 
\begin{align}\label{eq::TofUp}
&(1)\; T(\uparrow)=\sum_q\text{Re}\left[\langle q_\uparrow|\downarrow\rangle\langle\wr| q_\downarrow\rangle\right]>0
\end{align}
and
\begin{align}
\label{eq::up2squiggle}
&(2)\; N(\uparrow)=\sum_q|\langle q_\uparrow|\wr\rangle|^2=0.
\end{align}
The first condition ensures that the transfer state contributes positively to $C(\uparrow)$, while the second condition guarantees that the transfer state does not contribute to the noise injection. There is a deeper physical meaning to condition (1), however. Recall from \erf{negCorr} that negative correlations are fed by pairwise coherences  between the fiducial and coupled states of the form $\ket{\uparrow_i\uparrow_j}\bra{\downarrow_i\downarrow_j}+\text{h.c.}$. Condition (1) being satisfied indicates that an optical pumping process transforms these coherences from $\ket{\uparrow_i\uparrow_j}\bra{\downarrow_i\downarrow_j}+\text{h.c.}$ to $\ket{\uparrow_i\downarrow_j}\bra{\downarrow_i\wr_j}+\text{h.c.}$. When $ T(\uparrow)>0$,     these transformed coherences feed negative correlations and generate spin squeezing. For this reason, we call this optical pumping process a ``transfer of coherence" \cite{CohenTannoudji75}. We can take advantage of a transfer of coherence, if one exists, by preserving the transfer state $\ket{\wr}$ and mapping all other states outside the embedded qubit to the other ground manifold.

\chapter{Optical Pumping and the Covariance Matrix Update Formalism}\label{sec::ModHPCovar} 
Having closely examined optical pumping, the principle source of decoherence in our system, we now seek to explore its effect on the achievable spin squeezing. As discussed in the previous chapter, retaining the transfer state in addition to the fiducial and coupled states can preserve negative correlations that would otherwise be lost to optical pumping. In cases where preserving the transfer state is beneficial, we model each atom as an embedded qutrit consisting of the fiducial, coupled and transfer states. Whereas the relevant ensemble observables in the embedded qubit case were the quadratures $\hat{X}_\downarrow$ and $\hat{P}_\downarrow$,  an expanded number of ensemble observables is required when we treat the atoms as embedded qutrits. Working in the Heisenberg picture, we determine the evolution of the means and covariances of these  observables under both coherent squeezing dynamics and decoherence from optical pumping.  For particular fiducial states, the multilevel Holstein-Primakoff approximation can be revised to accommodate the transfer state. When this approximation holds, the ensemble is a Gaussian state on two oscillator modes. This enables us to utilize the Gaussian formalism of Sec. \ref{sec::GaussianStates} and the squeezing protocols of Chapter \ref{Sec:Protocols} to track the evolution of the ensemble and light through the covariance matrix. 

\section{New Ensemble Observables}\label{sec::NewObs}
Preserving the transfer state in addition to the coupled and fiducial states produces an ensemble where each atomic spin is modeled as an embedded qutrit with basis states $\ket{\uparrow}$, $\ket{\downarrow}$ and $\ket{\wr}$. Operators on the qutrit state of a single atomic spin can be decomposed in terms of an operator basis   
\begin{align}\label{eq::qutritOp1}
\hat{x}_{\downarrow\uparrow}&=\frac{1}{\sqrt{2}}\left(\ket{\downarrow}\bra{\uparrow}+\ket{\uparrow}\bra{\downarrow}\right)\\
\hat{y}_{\downarrow\uparrow}&=\frac{i}{\sqrt{2}}\left(\ket{\downarrow}\bra{\uparrow}-\ket{\uparrow}\bra{\downarrow}\right)\\
\hat{x}_{\wr\downarrow}&=\frac{1}{\sqrt{2}}\left(\ket{\wr}\bra{\downarrow}+\ket{\downarrow}\bra{\wr}\right)\\
\hat{y}_{\wr\downarrow}&=\frac{i}{\sqrt{2}}\left(\ket{\wr}\bra{\downarrow}-\ket{\downarrow}\bra{\wr}\right)\\
\hat{x}_{\uparrow\wr}&=\frac{1}{\sqrt{2}}\left(\ket{\uparrow}\bra{\wr}+\ket{\wr}\bra{\uparrow}\right)\\
\hat{y}_{\uparrow\wr}&=\frac{i}{\sqrt{2}}\left(\ket{\uparrow}\bra{\wr}-\ket{\wr}\bra{\uparrow}\right)\\
\hat{n}_{\uparrow}&=\ket{\uparrow}\bra{\uparrow}\\
\hat{n}_\downarrow&=\ket{\downarrow}\bra{\downarrow}
\end{align}
and
\begin{align}\label{eq::qutritOp9}
\hat{n}_\wr&=\ket{\wr}\bra{\wr}.
\end{align}
Note that the operators $\hat{x}_{ij}$ and $\hat{y}_{ij}$ are Pauli spin operators expressed in the basis $\{\ket{i},\ket{j}\}$, but with a different normalization convention. By summing these operators over all atoms, we obtain a basis for collective operators on the ensemble of embedded qutrits,
\begin{align}
\hat{X}_{\downarrow\uparrow}&=\sum_{i=1}^{N_A}\hat{x}_{\downarrow\uparrow}^{(i)}=\frac{1}{\sqrt{2}}\sum_i\left(\ket{\downarrow}\bra{\uparrow}_i+\ket{\uparrow}\bra{\downarrow}_i\right)\label{eq::Xdownup}\\
\hat{Y}_{\downarrow\uparrow}&=\sum_{i=1}^{N_A}\hat{p}_{\downarrow\uparrow}^{(i)}=\frac{i}{\sqrt{2}}\sum_{i=1}^{N_A}\left(\ket{\downarrow}\bra{\uparrow}_i-\ket{\uparrow}\bra{\downarrow}_i\right)\\
\hat{X}_{\wr\downarrow}&=\sum_{i=1}^{N_A}\hat{x}_{\wr\downarrow}^{(i)}=\frac{1}{\sqrt{2}}\sum_{i=1}^{N_A}\left(\ket{\wr}\bra{\downarrow}_i+\ket{\downarrow}\bra{\wr}_i\right)\\
\hat{Y}_{\wr\downarrow}&=\sum_{i=1}^{N_A}\hat{p}_{\wr\downarrow}^{(i)}=\frac{i}{\sqrt{2}}\sum_{i=1}^{N_A}\left(\ket{\wr}\bra{\downarrow}_i-\ket{\downarrow}\bra{\wr}_i\right)\label{eq::Psquiggledown}\\
\hat{X}_{\uparrow\wr}&=\sum_{i=1}^{N_A}\hat{x}_{\uparrow\wr}^{(i)}=\frac{1}{\sqrt{2}}\sum_{i=1}^{N_A}\left(\ket{\uparrow}\bra{\wr}_i+\ket{\wr}\bra{\uparrow}_i\right)\\
\hat{Y}_{\uparrow\wr}&=\sum_{i=1}^{N_A}\hat{p}_{\uparrow\wr}^{(i)}=\sum_{i=1}^{N_A}\frac{i}{\sqrt{2}}\left(\ket{\uparrow}\bra{\wr}_i-\ket{\wr}\bra{\uparrow}_i\right)\label{eq::Sigmayupwr}\\
N_\uparrow&=\sum_{i=1}^{N_A}\hat{n}_{\uparrow}^{(i)}=\sum_{i=1}^{N_A}\ket{\uparrow}\bra{\uparrow}_i\label{eq::PopUp}\\
N_\downarrow&=\sum_{i=1}^{N_A}\hat{n}_\downarrow^{(i)}=\sum_{i=1}^{N_A}\ket{\downarrow}\bra{\downarrow}_i
\end{align}
and
\begin{align}
N_\wr&=\sum_{i=1}^{N_A}\hat{n}_\wr^{(i)}=\sum_{i=1}^{N_A}\ket{\wr}\bra{\wr}_i.\label{eq::PopWr}
\end{align}
The first six ensemble observables in Eqs. (\ref{eq::Xdownup})-(\ref{eq::Sigmayupwr}) are collective pseudo spin operators on the ensemble of embedded qutrits. Note that the first two pseudo spins, $\hat{X}_{\downarrow\uparrow}$ and $\hat{Y}_{\downarrow\uparrow}$, are related to the ensemble quadratures, $\hat{X}_\downarrow$ and $\hat{P}_\downarrow$. By writing  $\hat{X}_{\downarrow\uparrow}$ and $\hat{Y}_{\downarrow\uparrow}$ in the Schwinger representation and linearizing the operators in the $\uparrow$ mode, we obtained $\hat{X}_\downarrow$ and $\hat{P}_\downarrow$ in Sec. \ref{sec::HPEnsemble}. The final three ensemble observables in Eqs. (\ref{eq::PopUp})-(\ref{eq::PopWr}) are the familiar populations, which quantify the number of atoms in the fiducial, coupled and transfer states.

The central focus of this chapter is solving for the squeezing parameter as a function of time while the ensemble undergoes both spin squeezing and optical pumping. Recall that the metrological squeezing parameter in \erf{eq::SqParameter} depends upon the collective spin variance, the mean spin and the total number of atoms in the $f$ manifold. All of these quantities can be expressed in terms of the means and covariances of the unnormalized quadratures and the populations. In absence of post-processing via internal spin control, the collective spin variance, the mean spin and the total atom number are given by
 \begin{align}
 \Delta F_z^2&=v(\uparrow)^2\Delta\hat{X}_{\downarrow\uparrow}^{2}+2v(\uparrow)w(\uparrow)\expect{\Delta\hat{X}_{\downarrow\uparrow}\Delta\hat{X}_{\wr\downarrow}}_S
+w(\uparrow)^2\Delta\hat{X}_{\wr\downarrow}^{2},\label{eq::FzVarNewObs}\\
\expect{\hat{F}_x}&=\expect{\hat{f}_x}_\uparrow N_\uparrow+\expect{\hat{f}_x}_\downarrow N_\downarrow+\expect{\hat{f}_x}_\wr N_\wr
\label{eq::FxNewObs}
\end{align}
and
\begin{align}
N=N_\uparrow+ N_\downarrow+N_\wr.\label{eq::NNewObs}
\end{align}
Here, the functions $v(\uparrow)$ and $w(\uparrow)$ depend on the variances of the fiducial and coupled states,
\begin{align}\label{eq::vDef}
v(\uparrow)=\sqrt{2(\Delta f_z^2)_\uparrow} 
\end{align}
and 
\begin{align}\label{eq::wDef}
w(\uparrow)=\sqrt{2(\Delta f_z^2)_\downarrow-2(\Delta f_z^2)_\uparrow}. 
\end{align}
By tracking the collective pseudo spins and populations as a function of time, we can calculate the squeezing parameter.

\section{Equations of Motion Under Optical Pumping}
In order to determine the evolution of the correlation functions, we must describe their dynamics under optical pumping. We utilize the master equation description of optical pumping introduced in Chapter \ref{sec::OpticalPumping}.  In a realistic implementation of a squeezing protocol based on the Faraday interaction, one introduces a bias magnetic field in the direction of the light's propagation along $z$ to fix the quantization axis of the atoms. The bias field necessitates that we transform into a frame rotating at the Larmor frequency about $z$, as described in Sec. \ref{sec::FaradayH}. In the rotating frame, the the master equation in \erf{eq::MasterEnsem} becomes 
\begin{align}\label{eq::MasterRotating}
\frac{d\hat{\rho}}{dt}\Big|_{\text{op}}=&-\frac{2\gamma_s}{9}\hat{\rho}\sum_{i=1}\mathbb{I}^{(i)}+
\frac{g_f^2\gamma_s}{9}\sum_{i=1}\left(\hat{f}_z^{(i)}\hat{\rho}\hat{f}_z^{(i)}
+\frac{1}{2}\hat{f}_y^{(i)}\hat{\rho}\hat{f}_y^{(i)}+\frac{1}{2}\hat{f}_x^{(i)}\hat{\rho}\hat{f}_x^{(i)}\right)\\\label{eq::MasterD}
=&\gamma_s\sum_i\mathcal{D}^{(i)}(\hat{\rho}).
\end{align}
This master equation describes optical pumping of the ensemble resulting from the absorption of photons with equal probability of being polarized along $x$ or $y$.  

\subsection{Dynamics of First Order Moments}
To begin, we determine the equation of motion for first order collective operators under optical pumping. By summing \erf{eq::MasterRotating} over a single $i$, we obtain the equation of motion for the density matrix of an individual atom,
\begin{align}\label{eq::SingleMaster}
\frac{d\hat{\rho}^{(i)}}{dt}\Big|_{\text{op}}=\gamma_s\mathcal{D}^{(i)}(\hat{\rho}^{(i)}).
\end{align}
Evolution of a first order collective operator $\hat{O}=\sum_{i=1}^{N_A}\hat{o}^{(i)}$ follows from \erf{eq::SingleMaster},
\begin{align}\label{eq::FirstOrderEvol}
\frac{d}{dt}\hat{O}\Big|_{\text{op}}=\gamma_s\sum_{i=1}^{N_A}\mathcal{D}^{(i)}(\hat{o}^{(i)}).
\end{align}
This expression for $d\hat{O}/dt|_{\text{op}}$ can be decomposed in terms of the collective pseudo spins and populations as
\begin{align}\label{eq::FirstOrderEvolBasis1}
\frac{d}{dt}\hat{O}\Big|_{\text{op}}=&\gamma_s\text{Tr}(\mathcal{D}(\hat{o})\hat{x}_{\downarrow\uparrow})\hat{X}_{\downarrow\uparrow}+\gamma_s\text{Tr}(\mathcal{D}(\hat{o})\hat{p}_{\downarrow\uparrow})\hat{P}_{\downarrow\uparrow}+\gamma_s\text{Tr}(\mathcal{D}(\hat{o})\hat{x}_{\wr\downarrow})\hat{X}_{\wr\downarrow}\\\notag&+\gamma_s\text{Tr}(\mathcal{D}(\hat{o})\hat{p}_{\wr\downarrow})\hat{P}_{\wr\downarrow}+\gamma_s\text{Tr}(\mathcal{D}(\hat{o})\hat{x}_{\uparrow\wr})\hat{X}_{\uparrow\wr}+\gamma_s\text{Tr}(\mathcal{D}(\hat{o})\hat{p}_{\uparrow\wr})\hat{P}_{\uparrow\wr}\\\notag
&+\gamma_s\text{Tr}(\mathcal{D}(\hat{o})\hat{n}_\uparrow)N_\uparrow+\gamma_s\text{Tr}(\mathcal{D}(\hat{o})\hat{n}_\downarrow)N_\downarrow
+\gamma_s\text{Tr}(\mathcal{D}(\hat{o})\hat{n}_\wr)N_\wr.
\end{align}

Calculating the mean spin and total atom number requires knowledge of the populations in the fiducial, coupled and transfer states. From \erf{eq::FirstOrderEvolBasis1}, the evolution of a population $N_\phi$ is given by 
\begin{align}\label{eq::FirstOrderPopulations}
\frac{dN_\phi}{dt}\Big|_{\text{op}}=&\gamma_s\text{Tr}(\mathcal{D}(\hat{n}_\phi)\hat{x}_{\downarrow\uparrow})\expect{\hat{X}_{\downarrow\uparrow}}+\gamma_s\text{Tr}(\mathcal{D}(\hat{n}_\phi)\hat{p}_{\downarrow\uparrow})\expect{\hat{P}_{\downarrow\uparrow}}\\\notag&+\gamma_s\text{Tr}(\mathcal{D}(\hat{n}_\phi)\hat{x}_{\wr\downarrow})\expect{\hat{X}_{\wr\downarrow}}+\gamma_s\text{Tr}(\mathcal{D}(\hat{n}_\phi)\hat{p}_{\wr\downarrow})\expect{\hat{P}_{\wr\downarrow}}\\\notag&+\gamma_s\text{Tr}(\mathcal{D}(\hat{n}_\phi)\hat{x}_{\uparrow\wr})\expect{\hat{X}_{\uparrow\wr}}+\gamma_s\text{Tr}(\mathcal{D}(\hat{n}_\phi)\hat{p}_{\uparrow\wr})\expect{\hat{P}_{\uparrow\wr}}\\\notag
&+\gamma_s\text{Tr}(\mathcal{D}(\hat{n}_\phi)\hat{n}_\uparrow)N_\uparrow+\gamma_s\text{Tr}(\mathcal{D}(\hat{n}_\phi)\hat{n}_\downarrow)N_\downarrow
+\gamma_s\text{Tr}(\mathcal{D}(\hat{n}_\phi)\hat{n}_\wr)N_\wr,
\end{align}
for $\phi\in\{\uparrow,\,\downarrow,\,\wr\}$. For an ensemble prepared in the state $\ket{\uparrow}^{\otimes N_A}$, the means of the collective pseudo spins are initially zero. The population of atoms in the fiducial state, on the other hand, is extremely large with $N_\uparrow>>1$. Although the means of the pseudo spins can become nonzero if they are coupled to $N_\uparrow$ by their equations of motion, they remain much smaller than $N_\uparrow$. Because their influence upon the populations is second order in the scattering rate, the means of the pseudo spins can be eliminated from the equation of motion above, leaving
\begin{align}\label{eq::FirstOrderPopulations2}
\frac{dN_\phi}{dt}\Big|_{\text{op}}=\gamma_s\text{Tr}(\mathcal{D}(\hat{n}_\phi)\hat{n}_\uparrow)N_\uparrow+\gamma_s\text{Tr}(\mathcal{D}(\hat{n}_\phi)\hat{n}_\downarrow)N_\downarrow
+\gamma_s\text{Tr}(\mathcal{D}(\hat{n}_\phi)\hat{n}_\wr)N_\wr.
\end{align}
The result is a set of closed differential equations that couple the populations to one another.

\subsection{Dynamics of Second Order Moments}\label{sec::2ndOrderOP}
We next turn our attention to calculating the equation of motion for the second order collective operators. Specifically, we focus on the covariances between collective operators, which are essential for determining the collective spin variance.  Deriving these equations of motion requires the master equation describing the evolution of any two atoms in the ensemble, $i$ and $j$, 
\begin{align}\label{eq::rhoij}
\frac{d\hat{\rho}^{(i,j)}}{dt}\Big|_{\text{op}}=\gamma_s\mathcal{D}^{(i)}(\hat{\rho}^{(i,j)})+\gamma_s\mathcal{D}^{(j)}(\hat{\rho}^{(i,j)}).
\end{align}
This master equation follows from taking the sum in \erf{eq::MasterRotating} over the two indices $i$ and $j$. From $d\hat{\rho}^{(i,j)}/dt$ we obtain the evolution of  a second-order correlation function,
\begin{align}\label{eq::2ndCorFun}
\frac{d}{dt}\expect{\Delta\hat{o}^{(i)}\!\Delta\hat{a}^{(j)}}_{S,\,i\neq j}\Big|_{\text{op}}=&\gamma_s\expect{\Delta\mathcal{D}^{(i)}(\hat{o}^{(i)})\Delta\hat{a}^{(j)}}_{S,\,i\neq j}\\\notag&+\gamma_s\expect{\Delta\hat{o}^{(i)}\Delta\mathcal{D}^{(j)}(\hat{a}^{(j)})}_{S,\,i\neq j}.
\end{align}
Here and throughout, the notation $\expect{\Delta\hat{x}\Delta\hat{y}}_S$ on two operators $\hat{x}$ and $\hat{y}$ is shorthand for the covariance $\expect{\Delta\hat{x}\Delta\hat{y}+\Delta\hat{y}\Delta\hat{x}}/2$.
For two collective operators $\hat{O}=\sum_i\hat{o}^{(i)}$ and $\hat{A}=\sum_i\hat{a}^{(i)}$, the covariance depends on both first and second order correlation functions, i.e. both single-atom and two-atom expectation values,
\begin{align}
\expect{\Delta\hat{O}\Delta\hat{A}}_S=\sum_{i\neq j}\expect{\Delta\hat{o}^{(i)}\Delta\hat{a}^{(j)}}_S
+\sum_i\expect{\Delta\hat{o}^{(i)}\Delta\hat{a}^{(i)}}_S.
\end{align}
To obtain the evolution of this covariance under optical pumping, we employ the equations of motion for first and second order correlation functions given in Eqs. (\ref{eq::FirstOrderEvol}) and (\ref{eq::2ndCorFun}), which yield
\begin{align}\label{eq::CovarEvol}
\frac{d}{dt}\expect{\Delta\hat{O}\Delta\hat{A}}_S\Big|_{\text{op}}
=&\gamma_s\sum_{i=1}^{N_A}\expect{\Delta\mathcal{D}^{(i)}(\hat{o}^{(i)})\Delta\hat{A}}_S\\\notag
&+\gamma_s\sum_{j=1}^{N_A}\expect{\Delta\hat{O}\Delta\mathcal{D}^{(j)}(\hat{a}^{(j)})}_S\\\notag
&+\frac{\gamma_s}{2}\sum_{i=1}^{N_A}\expect{\mathcal{D}^{(i)}(\{\hat{o}^{(i)},\hat{a}^{(i)}\})}
-\frac{\gamma_s}{2}\sum_{i=1}^{N_A}\expect{\{\mathcal{D}^{(i)}(\hat{o}^{(i)}),\hat{a}^{(i)}\}}\\\notag
&-\frac{\gamma_s}{2}\sum_{i=1}^{N_A}\expect{\{\hat{o}^{(i)},\mathcal{D}^{(i)}(\hat{a}^{(i)})\}}.
\end{align}
We refer to the final three terms, collectively as the ``noise term". Note that the noise term is a first order collective operator, while the remaining terms of \erf{eq::CovarEvol} are second order. The contribution to the noise term from each atom is proportional to the superoperator
\begin{align}\label{eq::noiseSuperOp}
\mathcal{N}(\hat{o},\hat{a})=\frac{1}{2}\mathcal{D}(\{\hat{o},\hat{a}\})
-\frac{1}{2}\{\mathcal{D}(\hat{o}),\hat{a}\}
-\frac{1}{2}\{\hat{o},\mathcal{D}(\hat{a})\},
\end{align}
which acts upon the two internal spin operators $\hat{o}$ and $\hat{a}$. 

Utilizing the collective operator basis of the pseudo spins and populations, we can decompose the equation of motion for $\expect{\Delta\hat{O}\Delta\hat{A}}_S$ in a manner analogous to the equation of motion for the first order operator, $\hat{O}$. The evolution of $\expect{\Delta\hat{O}\Delta\hat{A}}_S$ becomes
\begin{align}\label{eq::CovarEvol2}
\frac{d}{dt}\expect{\Delta\hat{O}\Delta\hat{A}}_S\Big|_{\text{op}}
=&\gamma_s\sum_{\hat{X}\in\mathcal{S}}\text{Tr}(\mathcal{D}(\hat{o})\hat{x})\expect{\Delta\hat{X}\Delta\hat{A}}_S\\\notag
&+\gamma_s\sum_{\hat{X}\in\mathcal{S}}\text{Tr}(\mathcal{D}(\hat{a})\hat{x})\expect{\Delta\hat{O}\Delta\hat{X}}_S\\\notag
&+\gamma_s\text{Tr}(\mathcal{N}(\hat{o},\hat{a})\hat{n}_\uparrow)N_\uparrow
+\gamma_s\text{Tr}(\mathcal{N}(\hat{o},\hat{a})\hat{n}_\downarrow)N_\downarrow\\\notag
&+\gamma_s\text{Tr}(\mathcal{N}(\hat{o},\hat{a})\hat{n}_\wr)N_\wr
\end{align}
where $\hat{X}=\sum_{i=1}^{N_A}\hat{x}^{(i)}$ and $\mathcal{S}=\{\hat{X}_{\downarrow\uparrow},\hat{Y}_{\downarrow\uparrow},
\hat{X}_{\wr\downarrow},\hat{Y}_{\wr\downarrow},\hat{X}_{\uparrow\wr},\hat{Y}_{\uparrow\wr}\}$.
Because we treat the populations as c-numbers, $\Delta N_\uparrow\approx\Delta N_\downarrow\approx\Delta N_\wr\approx 0$. The covariances in the equation of motion for $\expect{\Delta\hat{O}\Delta\hat{A}}_S$ are, therefore, between the collective pseudo spins alone. As in the case of the equation of motion for the populations, the means of the collective pseudo spins remain small, enabling us to decompose the noise term as a sum of populations.

\section{Revising the Multilevel Holstein Primakoff \\Approximation}\label{sec::NewQuads}
In order to implement the squeezing protocols of Chapter \ref{Sec:Protocols}, we represented the ensemble as a Gaussian state on a single bosonic mode. This was accomplished through the multilevel Holstein Primakoff approximation, which relied upon describing each atom as an embedded qubit consisting of the fiducial and coupled states. As we have argued, retaining the transfer state in addition to the fiducial and coupled states can increase the robustness of a state preparation to optical pumping. In this section, we modify the multilevel Holstein-Primakoff approximation in order to preserve the transfer state. When the master equation describing optical pumping and the fiducial state satisfy certain properties, the ensemble becomes a Gaussian state on two effective collective spin modes. We adapt the squeezing protocols presented in Chapter \ref{Sec:Protocols}  to accommodate the state of the light and atomic ensemble,  which becomes a three mode Gaussian. Optical pumping can be expressed as a Gaussian channel upon the system covariance matrix, enabling us to combine coherent squeezing dynamics with dissipation.

\subsection{Effective Collective Spin Modes}\label{sec::EffModes}
In Sec. \ref{sec::NewObs}, we introduced the collective pseudo spin operators, which together with the populations, form a basis for operators on the ensemble of embedded qutrits. Of the collective pseudo spins defined in Eqs. (\ref{eq::Xdownup})-(\ref{eq::Sigmayupwr}), the operators $\hat{X}_{\downarrow\uparrow}$, $\hat{Y}_{\downarrow\uparrow}$, $\hat{X}_{\wr\downarrow}$ and $\hat{Y}_{\wr\downarrow}$ contain the coherences that generate spin squeezing. Recall that pairwise coherences between the fiducial and coupled states and between the coupled and transfer states create negative correlations that reduce $\Delta F_z^2$. Consider the commutator between a collective pseudo spin with subscript $\downarrow\uparrow$ and another with subscript $\wr\!\downarrow$. For $\hat{d}\in\{\hat{X},\hat{Y}\}$, this commutator is of the form
\begin{align}\label{eq::unnormDiffModes}
[\hat{d}_{\downarrow\uparrow},\hat{d}_{\wr\downarrow}]=\sum_{i=1}^{N_A}\left(C_{\uparrow\wr}\ket{\uparrow}\bra{\wr}_i+C_{\wr\uparrow}\ket{\wr}\bra{\uparrow}_i\right),
\end{align}
where $C_{\uparrow\wr}$ and $C_{\wr\uparrow}$ are complex constants. In cases where coherences between the fiducial and transfer states are negligible, the quadratures with subscripts $\downarrow\uparrow$ and $\wr\!\!\downarrow$ approximately commute like two bosonic modes.  

In general, coherences between the fiducial and transfer states are not negligible. As discussed in Sec. \ref{Sec::OPEvents}, squeezing protocols develop pairwise coherences between the fiducial coupled states of the form $\ket{\uparrow_i\uparrow_j}\bra{\downarrow_i\downarrow_j}+\text{h.c.}$. For certain fiducial states, optical pumping can transform these pairwise coherences into coherences of the form $\ket{\uparrow_i\uparrow_j}\bra{\downarrow_i\wr_j}+\text{h.c.}$. While $\expect{\ket{\wr}\bra{\uparrow}}=\expect{\ket{\uparrow}\bra{\wr}}=0$ when these pairwise coherences are present, second order moments involving coherences between the fiducial and transfer states are not necessarily zero. Consider, for example, the covariance $\expect{\Delta\hat{X}_{\downarrow\uparrow}\Delta\hat{X}_{\uparrow\wr}}_S$. The pairwise coherences $\ket{\uparrow_i\uparrow_j}\bra{\downarrow_i\wr_j}+\text{h.c.}$ contribute positively to this second order moment, since 
\begin{align}
\bra{\downarrow_i\wr_j}(\Delta\hat{X}_{\downarrow\uparrow}\Delta\hat{X}_{\uparrow\wr})_S\ket{\uparrow_i\uparrow_j}=\frac{1}{2}.
\end{align}
Consequently, we cannot neglect the value of the commutator in \erf{eq::unnormDiffModes} if the pairwise coherences $\ket{\uparrow_i\uparrow_j}\bra{\downarrow_i\wr_j}+\text{h.c.}$ are present.

In order for operators on the ``modes" $\downarrow\uparrow$ and $\wr\!\downarrow$ to commute, the fiducial state and the master equation describing optical pumping must satisfy certain properties. If the condition in \erf{eq::up2squiggle} holds, which guarantees that preserving the transfer state is beneficial for spin squeezing, optical pumping creates the pairwise coherences $\ket{\uparrow_i\uparrow_j}\bra{\downarrow_i\wr_j}+\text{h.c.}$ only when there exists a jump operator $\hat{W}_q$ such that both \\$\bra{\uparrow}\hat{W}_q\ket{\uparrow}\neq 0$ and $\bra{\downarrow}\hat{W}_q\ket{\wr}\neq 0$. Pairwise coherences between the fiducial and transfer states do not develop as long as
\begin{align}\label{eq::ModeCondition}
\sum_q\text{Re}[\bra{\uparrow}\hat{W}_q\ket{\uparrow}\bra{\downarrow}\hat{W}_q^{\dag}\ket{\wr}]=0.
\end{align}
For the remainder of this section, we consider only the case in which this condition holds. In Chapter \ref{sec::Beyond}, we treat the most general case in which pairwise coherences between the fiducial and transfer state are permitted to develop.

The condition in \erf{eq::ModeCondition} being satisfied implies that
\begin{align}
[\hat{d}_{\downarrow\uparrow},\hat{d}_{\wr\downarrow}]\approx 0.
\end{align}
The ensemble becomes a state on two effective collective spin modes, $\downarrow\uparrow$ and $\wr\!\downarrow$. The collective pseudo spins $\hat{X}_{\downarrow\uparrow}$, $\hat{Y}_{\downarrow\uparrow}$, $\hat{X}_{\wr\downarrow}$ and $\hat{Y}_{\wr\downarrow}$ are conjugate observables on these effective oscillator modes. There are several key differences between the collective pseudo spins and the quadratures $\hat{X}_\downarrow$ and $\hat{P}_\downarrow$ derived in Sec. \ref{sec::HPEnsemble}. First, the collective pseudo spins do not obey the canonical commutation relations. Instead,
\begin{align}\label{eq::commutator1}
[\hat{X}_{\downarrow\uparrow},\hat{Y}_{\downarrow\uparrow}]=i(N_\uparrow-N_\downarrow)
\end{align}
and
\begin{align}\label{eq::commutator2}
[\hat{X}_{\wr\downarrow},\hat{Y}_{\wr\downarrow}]=i(N_\downarrow-N_\wr).
\end{align}
Furthermore, because the populations $N_\uparrow$, $N_\downarrow$ and $N_\wr$ are not constant under optical pumping, the commutation relations are time-varying. From the commutators in equations (\ref{eq::commutator1}) and (\ref{eq::commutator2}), we can deduce the uncertainty relations on conjugate collective pseudo spins,
\begin{align}\label{eq::uncert1}
\Delta X_{\downarrow\uparrow}^{2}\Delta Y_{\downarrow\uparrow}^{2}\geq\frac{(N_\uparrow-N_\downarrow)^2}{4},
\end{align}
and
\begin{align}\label{eq::uncert2}
\Delta X_{\wr\downarrow}^{2}\Delta Y_{\wr\downarrow}^{2}\geq\frac{(N_\downarrow-N_\wr)^2}{4}.
\end{align}
Note that the lower bounds in the uncertainty relations are, likewise, time-varying. 

To place the atoms and the light on similar footing, we define ``collective pseudospins" on the $y$ mode of the light, 
\begin{align}
\hat{X}_y=\sqrt{\frac{N_L}{2}}(\hat{a}_y^\dag+\hat{a}_y)
\end{align}
and
\begin{align}
\hat{Y}_y=i\sqrt{\frac{N_L}{2}}(\hat{a}_y^\dag-\hat{a}_y).
\end{align}
These operators are equivalent to the HP quadratures of the light with a different normalization convention. The ``collective pseudospins" of the light satisfy the commutation relation
\begin{align}\label{eq::Commute3}
[\hat{X}_y,\hat{Y}_y]=iN_L
\end{align}
and the uncertainty relation
\begin{align}\label{eq::uncert3}
\Delta X_{y}^{2}\Delta Y_{y}^{2}\geq\frac{N_L^2}{4}.
\end{align}
Because the number of photons is approximately unchanged by spontaneous emission, the commutation and uncertainty relations are constant in time unlike those of the ensemble collective pseudospins.

The commutation relations of all observables on the light and ensemble can be expressed concisely as
\begin{align}
[\hat{\textbf{d}}_j,\hat{\textbf{d}}_k]=i(n\sigma)_{jk}
\end{align}
where $\hat{\textbf{d}}=\{\hat{X}_{\downarrow\uparrow},\hat{Y}_{\downarrow\uparrow},\hat{X}_{\wr\downarrow},\hat{Y}_{\wr\downarrow},\hat{X}_y,\hat{Y}_y\}^T$. Here, $\sigma$ is the symplectic matrix defined in \erf{eq::sympMatrix} and the matrix $n$ is given by
\begin{eqnarray}
n=\left(\begin{matrix} N_\uparrow-N_\downarrow& 0& 0& 0& 0& 0 \\ 0 & N_\uparrow-N_\downarrow&0&0& 0& 0
\\0&0&N_\downarrow-N_\wr&0& 0& 0\\0&0&0&N_\downarrow-N_\wr&0& 0\\0&0&0&0&N_L&0\\0&0&0&0&0&N_L
\end{matrix}\right).
\end{eqnarray}
For the collective pseudo spins, the matrix $n\sigma$ takes the place of the symplectic matrix, $\sigma$.

\subsection{Gaussianity}\label{sec::Gauss}
 In absence of optical pumping, the ensemble is a Gaussian state on a single mode, $\downarrow$. The previous section demonstrated that when optical pumping is taken into account, the ensemble can be approximated as a state on two collective spin modes, labeled by $\downarrow\uparrow$ and  $\wr\!\downarrow$. In this section, we show that the ensemble is Gaussian on these modes. The light and ensemble, thus, form a multimode Gaussian state that can be evolved via the covariance matrix update formalism. 

In our new definition of the ensemble modes, the mode $\downarrow\uparrow$ takes the place of the mode $\downarrow$. Like the mode  $\downarrow$, an excitation in $\downarrow\uparrow$ represents an atom taken from the fiducial state to the coupled state. Likewise, the initial state of the ensemble with each atom prepared in the fiducial state is equivalent to $\ket{0}_{\downarrow\uparrow}$, the vacuum state in the mode $\downarrow\uparrow$.  Similarly, excitations in the mode $\wr\!\downarrow$  correspond to an atom taken from the coupled state to the transfer state. The vacuum state in the mode $\wr\!\downarrow$ corresponds to an ensemble state with a comparatively large number of atoms in the coupled state and none in the transfer state. Due to the condition in \erf{eq::up2squiggle}, which prohibits optical pumping directly from the fiducial state to the transfer state, any population in the transfer state is the result of a second order optical pumping event. Because population in the coupled state accumulates due to spin flips, which are first order optical pumping events, the population of atoms in the coupled state is always substantially larger than the population of atoms in the transfer state. As soon as a spin flip event transfers population to the coupled state, the ensemble becomes a state on two modes. A spin flip event also transfers the beneficial coherences between the fiducial and coupled states into coherences between the coupled and transfer states. By evolving the density matrix in \erf{densityOp} under a spin flip, it can be seen that the coherences between the coupled and transfer states remain smaller than any population in the coupled state by an order of $\xi/N_A$.  This is also true of the coherences between the fiducial and coupled states, which are smaller than the population in the fiducial state by an order of $\xi/N_A$.

 In terms of the collective pseudo spins, the Wigner function of the initial vacuum state, $\ket{0}_{\downarrow\uparrow}\ket{0}_y$, of the light and the ensemble is
\begin{align}
W(X_{\downarrow\uparrow},Y_{\downarrow\uparrow},X_{y},Y_{y})=
\frac{1}{\pi^2N_AN_L}e^{-\frac{X_{\downarrow\uparrow}^{2}+Y_{\downarrow\uparrow}^{2}}{N_A}}
e^{-\frac{X_{y}^{2}+Y_{y}^{2}}{N_L}}.
\end{align}
From \erf{MostGenMasterk}, the evolution of the ensemble density matrix, $\hat{\rho}_A$, under optical pumping over a small time step $\Delta t$ is approximately
\begin{align}\label{eq::OPrho}
\hat{\rho}_A(\Delta t)\approx&\left(1-\Gamma_{\text{op}}\Delta tN_A+\Delta t N_A\sum_q|\bra{\uparrow}\hat{W}_q\ket{\uparrow}|^2\;\right)\ket{0}\bra{0}_{\downarrow\uparrow}\ket{0}\bra{0}_{\wr\downarrow}\\\notag
&+\Delta t\sqrt{N_A}\left(\sum_q\bra{\uparrow}\hat{W}_q\ket{\uparrow}\bra{\uparrow}\hat{W}_q^\dag\ket{\downarrow}\;\right)\ket{0}\bra{1}_{\downarrow\uparrow}\ket{0}\bra{0}_{\wr\downarrow}\\\notag
&+\Delta t\sqrt{N_A}\left(\sum_q\bra{\downarrow}\hat{W}_q\ket{\uparrow}\bra{\uparrow}\hat{W}_q^\dag\ket{\uparrow}\;\right)\ket{1}\bra{0}_{\downarrow\uparrow}\ket{0}\bra{0}_{\wr\downarrow}\\\notag
&+\Delta t\left(\sum_q|\bra{\downarrow}\hat{W}_q\ket{\uparrow}|^2\right)\left(\sum_{i=1}^{N_A}\ket{\downarrow}\bra{\downarrow}_i\ket{\uparrow}\bra{\uparrow}_{\neq i}^{\otimes N_A-1}\right)\ket{0}\bra{0}_{\wr\downarrow}.
\end{align}
In the expression above, $\ket{1}_{\downarrow\uparrow}=\sum_{i=1}^{N_A}\ket{\downarrow}_i\ket{\uparrow}_{\neq i}^{\otimes N_A-1}/\sqrt{N_A}$, which signifies the presence of an atom in the coupled state. After the first optical pumping event, whereupon population is transferred into the coupled state, the combined system of the ensemble and light becomes a state on three modes. Defining creation and annihilation operators on the ensemble modes as $\hat{a}_{mn}^\dag=\sum_{i=1}^{N_A}\ket{m}\bra{n}_i/\sqrt{N_A}$ and $\hat{a}_{mn}=\sum_{i=1}^{N_A}\ket{n}\bra{m}_i/\sqrt{N_A}$ for $m,\,n\in \{\uparrow,\downarrow,\wr\}$, we can calculate the Wigner function of the system from $\hat{\rho}_A(\Delta t)$. The Wigner function of the system after undergoing optical pumping for a time $\Delta t$ is approximately
\begin{align}
W_A(\mathbf{d})\approx W(X_{\downarrow\uparrow},Y_{\downarrow\uparrow})
\frac{1}{\pi N_\downarrow}e^{-\frac{X_{\wr\downarrow}^{2}+Y_{\wr\downarrow}^{2}}{N_\downarrow}}\frac{1}{\pi N_L}e^{-\frac{X_{y}^{2}+Y_{y}^{2}}{N_L}},
\end{align}
where $W(X_{\downarrow\uparrow},Y_{\downarrow\uparrow})$ is the Wigner function of the ensemble on the mode $\downarrow\uparrow$, given by
\begin{align}\label{eq::OPwigner}
W(X_{\downarrow\uparrow},Y_{\downarrow\uparrow})=&\frac{1}{\pi N_A}\Big(1-\Delta t N_A \Gamma_{\text{op}}+\Delta t N_A\sum_q|\bra{\uparrow}\hat{W}_q\ket{\uparrow}|^2\\\notag&\;\;\;\;\;\;\;\;\;\;+\Delta t N_A\sum_q|\bra{\downarrow}\hat{W}_q\ket{\uparrow}|^2\;\Big)e^{-\frac{X_{\downarrow\uparrow}^2+Y_{\downarrow\uparrow}^2}{N_A}}\\\notag
&+\frac{\Delta t 2^{3/2}}{\pi N_A}\text{Re}\Big(\bra{\uparrow}\hat{W}_q\ket{\uparrow}\bra{\uparrow}\hat{W}_q^\dag\ket{\downarrow}(X_{\downarrow\uparrow}+iY_{\downarrow\uparrow})\Big)e^{-\frac{X_{\downarrow\uparrow}^2+Y_{\downarrow\uparrow}^2}{N_A}}.
\end{align}
While the first term of $W(X_{\downarrow\uparrow},Y_{\downarrow\uparrow})$ is Gaussian, the second is non-Gaussian. Note, however, that the second term is smaller than the first by an order of $N_A$. 
Thus, to good approximation, the state of the ensemble is Gaussian on mode $\downarrow\uparrow$ after the first optical pumping event. The combined state of the light and ensemble becomes a multimode Gaussian on three modes.

\subsection{Covariance Matrix Update Formalism for \\Variable Atom Number}\label{sec::CovMarixOPUpdate}
Because the combined state of the ensemble and the light is a multimode Gaussian, we can adapt the formalism outlined in Sec. \ref{sec::GaussianStates}, which enables us to completely specify the state of the system by its covariance matrix. In terms of the collective pseudo spins, the covariance matrix is defined analogously to the covariance matrix in Sec. \ref{sec::GaussianStates}, with elements given by
\begin{align}\label{eq::newCovMatrix}
\widetilde{\Sigma}_{ij}=\left\langle\frac{\Delta\hat{\textbf{d}}_i\Delta\hat{\textbf{d}}_j+\Delta\hat{\textbf{d}}_j\Delta\hat{\textbf{d}}_i}{2}\right\rangle,
\end{align}
for  $\hat{\textbf{d}}=\{\hat{X}_{\downarrow\uparrow},\hat{Y}_{\downarrow\uparrow},\hat{X}_{\wr\downarrow},\hat{Y}_{\wr\downarrow},\hat{X}_y,\hat{Y}_y\}^T$ and $\Delta\hat{\textbf{d}}_i=\hat{\textbf{d}}_i-\langle\hat{\textbf{d}}_i\rangle$. 
The covariance matrix corresponds to a physical state provided that 
\begin{align}\label{covcondition2}
\widetilde{\Sigma} +\frac{i}{2}n\sigma \geq 0.
\end{align}
As a consequence of \erf{covcondition2},  the uncertainty relations in equations (\ref{eq::uncert1}), (\ref{eq::uncert2}) and (\ref{eq::uncert3}) are satisfied. From \erf{eq::newCovMatrix}, the covariance matrix of the initial vacuum state is  
\begin{align}
\widetilde{\Sigma}_0=\frac{1}{2}\left(\begin{matrix}N_A&0&0&0&0&0\\
0&N_A&0&0&0&0\\
0&0&0&0&0&0\\
0&0&0&0&0&0\\
0&0&0&0&N_L&0\\
0&0&0&0&0&N_L\end{matrix}\right).
\end{align}
Note that this covariance matrix satisfies \erf{covcondition2}, since the populations $N_\downarrow$ and $N_\wr$ are zero at the initial time.

Both unitary and dissipative evolution on the system can be expressed as update maps upon the covariance matrix.  Unitary transformations act upon the covariance matrix via a map $S$, similar to a symplectic transformation, 
\begin{align}
\widetilde{\Sigma}'=S\,\widetilde{\Sigma}\, S^{T}.
\end{align}
Rather than fulfilling the symplectic condition of \erf{symplectic1}, however, $S$ satisfies
\begin {eqnarray}
S\, n\sigma\, S^{T}=n\sigma,
\label{symplectic}
\end{eqnarray} 
which ensures that \erf{covcondition2} is preserved on the covariance matrix. Dissipative evolution of the system can be expressed as a Gaussian channel, transforming the covariance matrix as
\begin {eqnarray}\label{eq::UnormGaussChannel}
\widetilde{\Sigma}'= M\, \widetilde{\Sigma} \,M^{T}+N,
\end{eqnarray}
where $N$  is a positive semi-definite matrix.
Equation (\ref{covcondition2}) holds on $\widetilde{\Sigma}'$ provided that  
\begin {eqnarray}
N+\frac{i}{2}n\sigma-\frac{i}{2}M\, n\sigma\, M^T\geq 0.
\end{eqnarray}

The squeezing protocols of Sec. \ref{Sec:Protocols} are easily modified to act on the covariance matrix $\widetilde{\Sigma}$ with its alternative normalization and additional mode. Some of the symplectic matrices employed in the squeezing protocols must be adapted to the new normalization convention and their dimensionality increased to accommodate the  ensemble mode $\wr\!\downarrow$. For instance, the symplectic rotation matrix $R(\theta)$, which acts upon the ensemble covariance matrix in the double pass squeezing protocols, must be generalized to rotate both modes $\downarrow\uparrow$ and $\wr\!\downarrow$ by an angle $\theta$. The rotation matrix becomes
\begin{align}
R(\theta)=\left(\begin{matrix}\text{cos}\theta&-\text{sin}\theta&0&0\\
\text{sin}\theta&\text{cos}\theta&0&0\\
0&0&\text{cos}\theta&-\text{sin}\theta\\
0&0&\text{sin}\theta&\text{cos}\theta\end{matrix}\right).
\end{align}
The symplectic matrix corresponding to the Faraday interaction on modes $\downarrow\uparrow$, $\wr\!\downarrow$ and $y$ is also altered by the new normalization convention, taking the form \begin{align}
\widetilde{S}_{F}=\left(\begin{matrix} 1 & 0 & 0 & 0 & 0 & 0\\ 0 &1& 0 & 0 & 0 & -\sqrt{\xi_\uparrow}(N_\uparrow-N_\downarrow)\\ 0 & 0 & 1 & 0 & 0 & 0
 \\ 0 & 0 & 0 &1& 0 & -\sqrt{\xi_\downarrow}(N_\downarrow-N_\wr)\\ 
 \sqrt{\xi_\uparrow}N_L & 0&\sqrt{\xi_\downarrow}N_L& 0 & 1 &0\\0 & 0 & 0 & 0 & 0 & 1\end{matrix}\right),
\end{align}
where $\xi_\uparrow=\chi^2(\Delta f_z^2)_\uparrow$ and $\xi_\downarrow=\chi^2((\Delta f_z^2)_\downarrow-(\Delta f_z^2)_\uparrow)$. The procedure for homodyne measurement of a quadrature, outlined in \erf{HCovariance}, is independent of the  covariance matrix normalization.  Similarly unchanged is the method for taking the partial trace over a mode of the system.

\section{Covariance Matrix Update for Optical \\Pumping}\label{sec::covUpdate}
The master equation in \erf{eq::MasterRotating} can be described as a Gaussian channel acting upon the covariance matrix of the light and ensemble, taking the form of \erf{eq::UnormGaussChannel}. Because the effects of optical pumping due to spontaneous emission are negligible upon the light, the mean update matrix and noise matrix have the structure 
\begin{align}\label{eq::entireM}
M_\gamma=M_{A\gamma}\oplus\mathbb{I}_y 
\end{align}
and 
\begin{align}\label{eq::entireN}
N_\gamma=N_{A\gamma}\oplus\left(\begin{matrix}0&0\\0&0\end{matrix}\right)_y,
\end{align}
where $M_{A\gamma}$ and $N_{A\gamma}$ are matrices acting on the ensemble modes and the $y$ subscript denotes a matrix acting on the light.

We first determine the matrix $M_{A\gamma}$ through the equations of motion for the collective pseudo spins, which follow from the evolution of a first order moment under optical pumping given in \erf{eq::FirstOrderEvolBasis1}. Recall that first and second order moments involving coherences between the fiducial and transfer states can be neglected as long as $\sum_q\text{Re}[\bra{\uparrow}\hat{W}_q\ket{\uparrow}\bra{\downarrow}\hat{W}_q^{\dag}\ket{\wr}]=0$. For the master equation in \erf{eq::MasterRotating}, this condition is equivalent to $\bra{\uparrow}\mathcal{D}(\ket{\uparrow}\bra{\downarrow})\ket{\wr}= 0$.
When this condition is satisfied, we can discard the terms in \erf{eq::FirstOrderEvolBasis1} involving the quadratures $\hat{X}_{\uparrow\wr}$ and $\hat{Y}_{\uparrow\wr}$.  
After dispensing with $\hat{X}_{\uparrow\wr}$ and $\hat{Y}_{\uparrow\wr}$, we utilize \erf{eq::FirstOrderEvolBasis1} to calculate the evolution of the first order central moment, $\Delta\hat{O}=\hat{O}-\expect{\hat{O}}$.
The equation of motion for $\Delta\hat{O}$ is 
\begin{align}
\frac{d}{dt}\Delta\hat{O}|_{\text{op}}=&\;\gamma_s\text{Tr}\left(\mathcal{D}(\hat{o})\hat{x}_{\downarrow\uparrow}\right)\Delta\hat{X}_{\downarrow\uparrow}+
\gamma_s\text{Tr}\left(\mathcal{D}(\hat{o})\hat{p}_{\downarrow\uparrow}\right)\Delta\hat{Y}_{\downarrow\uparrow}\\\notag&+
\gamma_s\text{Tr}\left(\mathcal{D}(\hat{o})\hat{x}_{\wr\downarrow}\right)\Delta\hat{X}_{\wr\downarrow}
+\gamma_s\text{Tr}\left(\mathcal{D}(\hat{o})\hat{p}_{\wr\downarrow}\right)\Delta\hat{Y}_{\wr\downarrow}\\\notag&+
\gamma_s\text{Tr}\left(\mathcal{D}(\hat{o})\hat{n}_\uparrow\right)\Delta N_\uparrow
+\gamma_s\text{Tr}\left(\mathcal{D}(\hat{o})\hat{n}_\downarrow\right)\Delta N_\downarrow\\\notag&
+\gamma_s\text{Tr}\left(\mathcal{D}(\hat{o})\hat{n}_\wr\right)\Delta N_\wr.
\end{align}
Because we treat the populations as c-numbers, $\Delta N_\uparrow\approx \Delta N_\downarrow\approx \Delta N_\wr\approx0$. The resulting equation of motion for $\Delta\hat{O}$ is independent of the populations.
By integrating the equation of motion over a small time $\Delta t$, we obtain
\begin{align}\label{eq::CentralFirstIntegrate}
\Delta\hat{O}(\Delta t)\approx&\,\Delta\hat{O}(0)\\\notag&+\gamma_s\Delta t\,\text{Tr}\left(\mathcal{D}(\hat{o})\hat{x}_{\downarrow\uparrow}\right)\Delta\hat{X}_{\downarrow\uparrow}(0)+
\gamma_s\Delta t\, \text{Tr}\left(\mathcal{D}(\hat{o})\hat{y}_{\downarrow\uparrow}\right)\Delta\hat{Y}_{\downarrow\uparrow}(0)\\\notag&+
\gamma_s\Delta t\, \text{Tr}\left(\mathcal{D}(\hat{o})\hat{x}_{\wr\downarrow}\right)\Delta\hat{X}_{\wr\downarrow}(0)
+\gamma_s\Delta t\, \text{Tr}\left(\mathcal{D}(\hat{o})\hat{y}_{\wr\downarrow}\right)\Delta\hat{Y}_{\wr\downarrow}(0).
\end{align}
Taking $\hat{O}\in\{\hat{X}_{\downarrow\uparrow},\hat{Y}_{\downarrow\uparrow},\hat{X}_{\wr\downarrow},\hat{Y}_{\wr\downarrow}\}$ in \erf{eq::CentralFirstIntegrate} gives the us the update matrix for the central moments of the  quadratures,
\begin{align}\label{eq::Mgamma}
&M_{A\gamma}(\Delta t)=\mathbb{I}+\\\notag&\gamma_s\Delta t\left(\begin{matrix} 
\text{Tr}\left(\mathcal{D}(\hat{x}_{\downarrow\uparrow})\hat{x}_{\downarrow\uparrow}\right)&\text{Tr}\left(\mathcal{D}(\hat{x}_{\downarrow\uparrow})\hat{y}_{\downarrow\uparrow}\right)&\text{Tr}\left(\mathcal{D}(\hat{x}_{\downarrow\uparrow})\hat{x}_{\wr\downarrow}\right)&\text{Tr}\left(\mathcal{D}(\hat{x}_{\downarrow\uparrow})\hat{y}_{\wr\downarrow}\right)\\
\text{Tr}\left(\mathcal{D}(\hat{y}_{\downarrow\uparrow})\hat{x}_{\downarrow\uparrow})\right)&\text{Tr}\left(\mathcal{D}(\hat{y}_{\downarrow\uparrow})\hat{y}_{\downarrow\uparrow}\right)&\text{Tr}\left(\mathcal{D}(\hat{y}_{\downarrow\uparrow})\hat{x}_{\wr\downarrow}\right)&\text{Tr}\left(\mathcal{D}(\hat{y}_{\downarrow\uparrow})\hat{y}_{\wr\downarrow}\right)\\
\text{Tr}\left(\mathcal{D}(\hat{x}_{\wr\downarrow})\hat{x}_{\downarrow\uparrow}\right)&\text{Tr}\left(\mathcal{D}(\hat{x}_{\wr\downarrow})\hat{y}_{\downarrow\uparrow}\right)&\text{Tr}\left(\mathcal{D}(\hat{x}_{\wr\downarrow})\hat{x}_{\wr\downarrow}\right)&\text{Tr}\left(\mathcal{D}(\hat{x}_{\wr\downarrow})\hat{y}_{\wr\downarrow}\right)\\
\text{Tr}\left(\mathcal{D}(\hat{y}_{\wr\downarrow})\hat{x}_{\downarrow\uparrow}\right)&\text{Tr}\left(\mathcal{D}(\hat{y}_{\wr\downarrow})\hat{y}_{\downarrow\uparrow}\right)&\text{Tr}\left(\mathcal{D}(\hat{y}_{\wr\downarrow})\hat{x}_{\wr\downarrow}\right)&\text{Tr}\left(\mathcal{D}(\hat{y}_{\wr\downarrow})\hat{y}_{\wr\downarrow}\right)
\end{matrix}\right).
\end{align}
Update matrices for the different state preparations are given in Appendix \ref{sec::MgammaUps}. 

We now turn our attention to the noise matrix  $N_{A \gamma}$, which we derive from the equations of motion for the second order moments. Integrating \erf{eq::CovarEvol2} over a small time $\Delta t$ yields
\begin{align}\label{eq::CovarEvolInt}
\expect{\Delta\hat{O}\Delta\hat{A}}_S(\Delta t)
=&\expect{\Delta\hat{O}\Delta\hat{A}}_S(0)
+\gamma_s\Delta t\sum_{\hat{X}\in\mathcal{S}}\text{Tr}(\mathcal{D}(\hat{o})\hat{x})\expect{\Delta\hat{X}\Delta\hat{A}}_S\\\notag
&+\gamma_s\Delta t\sum_{\hat{X}\in\mathcal{S}}\text{Tr}(\mathcal{D}(\hat{a})\hat{x})\expect{\Delta\hat{O}\Delta\hat{X}}_S\\\notag
&+\gamma_s\Delta t\text{Tr}(\mathcal{N}(\hat{o},\hat{a})\hat{n}_\uparrow)N_\uparrow
+\gamma_s\Delta t\text{Tr}(\mathcal{N}(\hat{o},\hat{a})\hat{n}_\downarrow)N_\downarrow\\\notag
&+\gamma_s\Delta t\text{Tr}(\mathcal{N}(\hat{o},\hat{a})\hat{n}_\wr)N_\wr.
\end{align}
When $\hat{O}$ and $\hat{A}$ are collective pseudo spins, the first three terms of this equation arise from the action of $M_{A\gamma}$. Note that these terms are moments of second order collective operators. The noise matrix contains the final three terms, which are moments of first order collective operators. Because a second order scattering event is required to populate the transfer state, the population $N_\wr$ is negligible. The noise matrix becomes \pagebreak
\begin{align}
N_{A\gamma}(\Delta t)=\gamma_s\Delta t \sum_{\psi\in\{\uparrow,\downarrow\}}N_\psi\times
\end{align}
\vspace{-8 mm}
\begingroup\makeatletter\def\f@size{10}\check@mathfonts
\def\maketag@@@#1{\hbox{\m@th\large\normalfont#1}}
\begin{align}
\notag&
\left(\begin{matrix} 
\text{Tr}\left(\mathcal{N}(\hat{x}_{\downarrow\uparrow},\hat{x}_{\downarrow\uparrow})\hat{n}_\psi\right)&
\text{Tr}\left(\mathcal{N}(\hat{x}_{\downarrow\uparrow},\hat{p}_{\downarrow\uparrow})\hat{n}_\psi\right)&
\text{Tr}\left(\mathcal{N}(\hat{x}_{\downarrow\uparrow},\hat{x}_{\wr\downarrow})\hat{n}_\psi\right)&
\text{Tr}\left(\mathcal{N}(\hat{x}_{\downarrow\uparrow},\hat{p}_{\wr\downarrow})\hat{n}_\psi\right)\\
\text{Tr}\left(\mathcal{N}(\hat{p}_{\downarrow\uparrow},\hat{x}_{\downarrow\uparrow})\hat{n}_\psi\right)&
\text{Tr}\left(\mathcal{N}(\hat{p}_{\downarrow\uparrow},\hat{p}_{\downarrow\uparrow})\hat{n}_\psi\right)&
\text{Tr}\left(\mathcal{N}(\hat{p}_{\downarrow\uparrow},\hat{x}_{\wr\downarrow})\hat{n}_\psi\right)&
\text{Tr}\left(\mathcal{N}(\hat{p}_{\downarrow\uparrow},\hat{p}_{\wr\downarrow})\hat{n}_\psi\right)\\
\text{Tr}\left(\mathcal{N}(\hat{x}_{\wr\downarrow},\hat{x}_{\downarrow\uparrow})\hat{n}_\psi\right)&
\text{Tr}\left(\mathcal{N}(\hat{x}_{\wr\downarrow},\hat{p}_{\downarrow\uparrow})\hat{n}_\psi\right)&
\text{Tr}\left(\mathcal{N}(\hat{x}_{\wr\downarrow},\hat{x}_{\wr\downarrow})\hat{n}_\psi\right)&
\text{Tr}\left(\mathcal{N}(\hat{x}_{\wr\downarrow},\hat{p}_{\wr\downarrow})\hat{n}_\psi\right)\\
\text{Tr}\left(\mathcal{N}(\hat{p}_{\wr\downarrow},\hat{x}_{\downarrow\uparrow})\hat{n}_\psi\right)&
\text{Tr}\left(\mathcal{N}(\hat{p}_{\wr\downarrow},\hat{p}_{\downarrow\uparrow})\hat{n}_\psi\right)&
\text{Tr}\left(\mathcal{N}(\hat{p}_{\wr\downarrow},\hat{x}_{\wr\downarrow})\hat{n}_\psi\right)&
\text{Tr}\left(\mathcal{N}(\hat{p}_{\wr\downarrow},\hat{p}_{\wr\downarrow})\hat{n}_\psi\right)
\end{matrix}\right)
\end{align}
\endgroup
Noise matrices for the various state preparations are given in Appendix \ref{sec::MgammaUps}. 

Because the noise matrix, the mean spin and the total atom number all depend upon $N_\uparrow$ and $N_\downarrow$, we must derive a similar update matrix to evolve the populations in time. We focus only upon the populations of atoms in the fiducial and coupled states, since the population in the transfer state is negligible. Neglecting $N_\wr$ in \erf{eq::FirstOrderPopulations2} and integrating the equation of motion over a small time step $\Delta t$ yields
\begin{align}\label{eq::NumEvol3}
N_\psi(\Delta t)\approx N_\psi(0)+\gamma_s\Delta t\,\text{Tr}(\mathcal{D}(\hat{n}_\psi)\hat{n}_\uparrow)N_\uparrow(0)
+\gamma_s\Delta t\, \text{Tr}(\mathcal{D}(\hat{n}_\psi)\hat{n}_\downarrow)N_\downarrow(0).
\end{align}
Taking $\psi\in\{\uparrow,\downarrow\}$, we obtain the update matrix
\begin{align}\label{eq::popUpdate}
J_\gamma(\Delta t)=\mathbb{I}+\gamma_s\Delta t\left(\begin{matrix}
\text{Tr}(\mathcal{D}(\hat{n}_\uparrow)\hat{n}_\uparrow)&\text{Tr}(\mathcal{D}(\hat{n}_\uparrow)\hat{n}_\downarrow)\\
\text{Tr}(\mathcal{D}(\hat{n}_\downarrow)\hat{n}_\uparrow)&\text{Tr}(\mathcal{D}(\hat{n}_\downarrow)\hat{n}_\downarrow)
\end{matrix}\right).
\end{align}
The matrix $J_\gamma(\Delta t)$ evolves a vector of the populations forward in time by $\Delta t$,
\begin{align}
\left(\begin{matrix}N_\uparrow(t+\Delta t)\\N_\downarrow(t+\Delta t)\end{matrix}\right)
=J_\gamma(\Delta t)
\left(\begin{matrix}N_\uparrow(t)\\N_\downarrow(t)\end{matrix}\right).
\end{align}
Because the QND and double pass squeezing protocols have no appreciable effect upon the populations, $J_\gamma(\Delta t)$ is the only update matrix required to evolve $N_\uparrow$ and $N_\downarrow$.

\section{Simulating Gaussian Dynamics}\label{sec::GaussSim}
After deriving the covariance matrix and population updates for optical pumping, we can revisit the squeezing protocols of Chapter \ref{Sec:Protocols} to numerically evaluate their performance when decoherence is taken into account. To simulate both the coherent and dissipative evolution of the light and the atomic ensemble, we follow a procedure similar to Ref. \cite{MadMol}. By alternating over very small time increments the covariance matrix updates corresponding to a squeezing interaction and to optical pumping, we approximate the evolution of the system as it undergoes both processes simultaneously.

We first outline our approach for simulating the QND measurement protocol in the presence of optical pumping. Optical pumping acts on the covariance matrix by a Gaussian channel $\mathcal{P}$, which depends upon the mean update matrix  and the noise matrix,
\begin{align}
\widetilde{\Sigma}(t+\Delta t)&=\mathcal{P}[\widetilde{\Sigma}(t)]\\\notag
&=M_\gamma(\Delta t)\widetilde{\Sigma}(t)M_\gamma^T(\Delta t)+N_\gamma(\Delta t).
\end{align}
Here, $M_\gamma$ and $N_\gamma$ are the update matrices defined in Eqs. (\ref{eq::entireM}) and (\ref{eq::entireN}), acting on the covariance matrix of both the light and ensemble. The QND measurement protocol acts upon the covariance matrix by a map $\mathcal{Q}$,
\begin{align}
\widetilde{\Sigma}(t+\Delta t)&=\mathcal{Q}[\widetilde{\Sigma}(t)]
\\\notag&=h[\hat{X}_y](S_F(\Delta t)\widetilde{\Sigma}(t)S_F^T(\Delta t))
\oplus\widetilde{\Sigma}_{0y}.
\end{align}
This map differs from the update described in \erf{eq::QNDupdate} in that after the $\hat{X}_y$ quadrature is measured via homodyne detection, a vacuum state covariance matrix of the light is appended to the covariance matrix of the ensemble. This represents a fresh pulse of light entering the experimental apparatus. In the next iteration of the protocol, this light will interact with the ensemble and subsequently undergo homodyne detection. The covariance matrix is propagated forward in time by alternating the maps $\mathcal{P}$ and $\mathcal{Q}$ for some number of iterations $n$,
\begin{align}
\widetilde{\Sigma}(n\Delta t)=(\mathcal{Q}\cdot\mathcal{P})^n[\widetilde{\Sigma}(0)].
\end{align}
Through this procedure, we obtain the covariance matrix of the light and ensemble after the system undergoes QND measurement for a time $n\Delta t$.

We next consider the phase-matching protocol. The first step in the phase-matching protocol is implementing the one-axis twisting interaction, which can be combined with optical pumping to form a Gaussian channel $\mathcal {T}$,
\begin{align}
\widetilde{\Sigma}(t+2\Delta t)&=\mathcal{T}[\widetilde{\Sigma}(t)]
\\\notag&=h[\hat{X}'_y]\left(S_F(\Delta t)\mathcal{P}[R_{\frac{\lambda}{4}}S_F(\Delta t) \mathcal{P}[\widetilde{\Sigma}(t)]S_F^T(\Delta t)R_{\frac{\lambda}{4}}^T]S_F^T(\Delta t)\right)\oplus\!\widetilde{\Sigma}_{0y}
\end{align}
As in the map  $\mathcal{Q}$, a vacuum state covariance matrix of the light is appended to the covariance matrix of the ensemble after the quantum eraser, representing a fresh pulse of light. Similar to the coherent phase matching process depicted in \erf{eq::phasematchingTheta}, the one-axis twisting interaction is followed by a rotation of the ensemble given by the map $\mathcal{R}$,
\begin{align}
\mathcal{R}[\widetilde{\Sigma}(t)]=R(\theta_\text{opt})\widetilde{\Sigma}(t)R(\theta_\text{opt})^T.
\end{align}
For the coherent version of the phase matching protocol in \erf{eq::phasematchingTheta}, the ensemble is rotated by an angle $\xi/(2n)$ at each iteration. Here, the presence of noise from optical pumping changes the optimal angle for generating spin squeezing. A numerical optimization is used to determine the angle, $\theta_\text{opt}$, that maximizes squeezing at each iteration of the protocol. Like the case of QND measurement, we evolve the covariance matrix by alternating the maps $\mathcal{T}$ and $\mathcal{R}$ for some number of iterations $n$. This produces the covariance matrix at time $2n\Delta t$,
\begin{align}
\widetilde{\Sigma}(2n\Delta t)=(\mathcal{R}\cdot\mathcal{T})^n[\widetilde{\Sigma}(0)].
\end{align}
Note that, unlike the case of QND measurement, each iteration of the phase matching protocol takes a time $2\Delta t$ because of the double pass. 

In addition to the covariance matrix, we must also simulate the evolution of the populations. Because they evolve only by optical pumping, the populations at time $t=n\Delta t$ are given by
\begin{align}
\left(\begin{matrix}N_\uparrow(n\Delta t)\\N_\downarrow(n\Delta t)\end{matrix}\right)
=J_\gamma(\Delta t)^n
\left(\begin{matrix}N_\uparrow(0)\\N_\downarrow(0)\end{matrix}\right)
\end{align}
for both the QND measurement and phase matching  protocols. Here, $J_\gamma(\Delta t)$ is the update matrix given in \erf{eq::popUpdate}.

\section{Post-Processing Internal Spin Control}\label{sec::postprocessing2}
The final step in creating spin squeezing is post-processing via a partial isometry that acts identically on the internal spin state of each atom. In this section, we modify the partial isometries of Sections \ref{sec::postprocessing} and \ref{sec::IntSpinSqueeze}  to accommodate the transfer state. This enables us to derive an expression for the squeezing parameter following the application of a post-processing partial isometry. When combined with the numerical techniques of the previous section, this expression determines the amount of squeezing achievable in the atomic ensemble.

Consider an initial state preparation with fiducial, coupled and transfer states $\ket{\uparrow},\;\ket{\downarrow}$ and $\ket{\wr}$. We seek to map the ensemble to a final state preparation with fiducial, coupled and transfer states $\ket{\uparrow'},\;\ket{\downarrow'}$ and $\ket{\wr'}$. To preserve the mean spin and relevant correlations, a fiducial state must always be mapped to a fiducial state, a coupled state to a coupled state and a transfer state to a transfer state. The partial isometry that implements this map is given by 
\begin{align}\label{UArb}
\hat{U}_{\uparrow'}=\bigotimes_{i=1}^{N_A}\left(\ket{\uparrow'}\bra{\uparrow}_i+\ket{\downarrow'}\bra{\downarrow}_i+\ket{\wr'}\bra{\wr}_i\right).
\end{align}
This partial isometry leaves the total atom number invariant, but transforms the collective spin variance in \erf{eq::FzVarNewObs} as
 \begin{align}
 \Delta F_z^2&=v(\uparrow')^2\Delta\hat{X}_{\downarrow\uparrow}^{2}+2v(\uparrow')w(\uparrow')\expect{\Delta\hat{X}_{\downarrow\uparrow}\Delta\hat{X}_{\wr\downarrow}}_S
+w(\uparrow')^2\Delta\hat{X}_{\wr\downarrow}^{2},
\end{align}
where $w(\uparrow)$ and $v(\uparrow)$ are defined in Eqs. (\ref{eq::vDef}) and (\ref{eq::wDef}). In cases where it is not beneficial to preserve the transfer state, $\Delta\hat{X}_{\wr\downarrow}^{2}=N_\downarrow/2$ and $\expect{\Delta\hat{X}_{\downarrow\uparrow}\Delta\hat{X}_{\wr\downarrow}}_S=0$. The collective variance after the partial isometry is then 
 \begin{align}
  \Delta F_z^2&=v(\uparrow')^2\Delta\hat{X}_{\downarrow\uparrow}^{2}+w(\uparrow')^2N_\downarrow/2.
 \end{align}
The partial isometry also transforms the mean spin in \erf{eq::FxNewObs}, which is given by
 \begin{align}
\expect{\hat{F}_x}&=\expect{\hat{f}_x}_{\uparrow'} N_\uparrow+\expect{\hat{f}_x}_{\downarrow'} N_\downarrow+\expect{\hat{f}_x}_{\wr'} N_\wr.
\end{align}
 The final term in this expression is absent in cases where it is not beneficial to preserve the transfer state. With the transformed collective variance and mean spin, the metrological squeezing parameter becomes
\begin{align}\label{eq::unnormSqParam2}
\zeta_m=&2f(N_\uparrow+N_\downarrow+N_\wr)\times\\\notag
&\frac{v(\uparrow')^2\Delta\hat{X}_{\downarrow\uparrow}^{2}+2v(\uparrow')w(\uparrow')\expect{\Delta\hat{X}_{\downarrow\uparrow}\Delta\hat{X}_{\wr\downarrow}}_S
+w(\uparrow')^2\Delta\hat{X}_{\wr\downarrow}^{2}}{\left(\expect{\hat{f}_x}_{\uparrow'} N_\uparrow+\expect{\hat{f}_x}_{\downarrow'} N_\downarrow
+\expect{\hat{f}_x}_{\wr'} N_\wr\right)^2}
\end{align}
when the transfer state is preserved and
\begin{align}\label{eq::noTransfer}
\zeta_m=&2f(N_\uparrow+N_\downarrow)
\frac{v(\uparrow')^2\Delta\hat{X}_{\downarrow\uparrow}^{2}
+w(\uparrow')^2N_\downarrow/2}{\left(\expect{\hat{f}_x}_{\uparrow'} N_\uparrow+\expect{\hat{f}_x}_{\downarrow'} N_\downarrow\right)^2}
\end{align}
when the transfer state is eliminated.

The effect of post-processing upon the squeezing parameter is not as evident as it was in Sec. \ref{sec::postprocessing} in absence of decoherence. The squeezing parameter of \erf{parameters} can be recovered, however, by eliminating all except the leading order terms in Eqs. (\ref{eq::unnormSqParam2}) and (\ref{eq::noTransfer}). By preserving $\Delta\hat{X}_{\downarrow\uparrow}^{2}$ and $N_\uparrow$, we obtain 
\begin{align}\label{eq::unnormSqParam3}
\zeta_m \approx \frac{2fN_\uparrow v(\uparrow')^2\Delta\hat{X}_{\downarrow\uparrow}^{2}}{\left(\expect{\hat{f}_x}_{\uparrow'}N_\uparrow\right)^2}= \frac{4fN_\uparrow^2 (\Delta\hat{f}_z)_{\uparrow'}\Delta\hat{X}_\downarrow^2}{\left(\expect{\hat{f}_x}_{\uparrow'}N_\uparrow\right)^2}=\zeta_m^{\uparrow'}\zeta_q.
\end{align}
While $\zeta_m$ and this approximate expression differ in the presence of optical pumping, we see that post-processing improves squeezing by mapping the majority of atoms, which remain in the fiducial state $\ket{\uparrow}$, to a new fiducial state, $\ket{\uparrow'}$, with more internal spin squeezing.

In sections \ref{sec::postprocessing} and \ref{sec::IntSpinSqueeze}, we examined partial isometries that map from an arbitrary state preparation to the $SCS$, Yurke and half-integer Yurke state preparations. From \erf{eq::unnormSqParam2}, we can determine the squeezing parameters that result
when these partial isometries are modified to accommodate the transfer state. In cases where the transfer state is eminimated, we obtain the squeezing parameter by taking $\Delta\hat{X}_{\wr\downarrow}^{2}\rightarrow N_\downarrow/2$, $\expect{\Delta\hat{X}_{\downarrow\uparrow}\Delta\hat{X}_{\wr\downarrow}}_S\rightarrow0$ and $N_\wr\rightarrow 0$. When the transfer state is preserved, mapping to the SCS preparation requires that $\ket{\wr}$ is mapped to $\ket{\wr_{SCS}}=-\ket{f,m_x=f-2}$, the transfer state in the $SCS$ preparation. Implementing this partial isometry transforms the squeezing parameter as
\begin{align}\label{eq::SCSSqParamOP}
\zeta_m=&2f(N_\uparrow+N_\downarrow+N_\wr)\times\\\notag
&\frac{f\Delta\hat{X}_{\downarrow\uparrow}^{2}+2\sqrt{f(2f-1)}\expect{\Delta\hat{X}_{\downarrow\uparrow}\Delta\hat{X}_{\wr\downarrow}}_S+(2f-1)\Delta\hat{X}_{\wr\downarrow}^{2}}{(fN_\uparrow+(f-1)N_\downarrow+(f-2)N_\wr)^2}.
\end{align}
Mapping to the Yurke preparation similarly requires that $\ket{\wr}$ be mapped to
\begin{align}
\ket{\wr_\text{y}}=\frac{\text{cos}\alpha}{\sqrt{2}}\ket{f,m_z=1}-\text{sin}\alpha\ket{f,m_z=0}+\frac{\text{cos}\alpha}{\sqrt{2}}\ket{f,m_z=-1},
\end{align}
which is the transfer state in the Yurke preparation. The squeezing parameter that results from this partial isometry is
\begin{align}\label{eq::YurkeSqParamOP}
\zeta_m=&2f(N_\uparrow+N_\downarrow+N_\wr)\times\\\notag
&\frac{\text{sin}^2\alpha\Delta\hat{X}_{\downarrow\uparrow}^{2}+2\text{cos}\alpha\text{sin}\alpha\expect{\Delta\hat{X}_{\downarrow\uparrow}\Delta\hat{X}_{\wr\downarrow}}_S+\text{cos}^2\alpha\Delta\hat{X}_{\wr\downarrow}^{2}}{f(f+1)\text{cos}^2\alpha\;\text{sin}^2\alpha \;N_\uparrow^2}.
\end{align}
The transfer state in the half-integer Yurke preparation takes a similar form,
\begin{align}
\ket{\wr_\text{hy}}\!=\!\frac{\text{cos}\alpha}{\sqrt{2}}\ket{f,m_z=3/2}\!-\!\text{sin}\alpha\ket{f,m_z=1/2}\!+\!\frac{\text{cos}\alpha}{\sqrt{2}}\ket{f,m_z=-1/2}.
\end{align}
The partial isometry to the half-integer Yurke preparation produces the squeezing parameter 
\begin{align}\label{eq::YurkeLikeSqParamOP}
\zeta_m=&8f(N_\uparrow+N_\downarrow+N_\wr)\times\\\notag
&\frac{\text{sin}^2\alpha\Delta\hat{X}_{\downarrow\uparrow}^{2}+2\text{cos}\alpha\text{sin}\alpha\expect{\Delta\hat{X}_{\downarrow\uparrow}\Delta\hat{X}_{\wr\downarrow}}_S+\text{cos}^2\alpha\Delta\hat{X}_{\wr\downarrow}^{2}}
{(\sqrt{(f+3/2)(f-1/2)}+f+1/2)^2\;\text{sin}^2\alpha\;\text{cos}^2\alpha \;N_\uparrow^2}.
\end{align}
Recall that the squeezing of the Yurke and the half-integer Yurke states is maximal as $\alpha\rightarrow0$. Because the terms $\expect{\Delta\hat{X}_{\downarrow\uparrow}\Delta\hat{X}_{\wr\downarrow}}_S$ and $\Delta\hat{X}_{\wr\downarrow}^{\,2}$ are nonzero in the presence of optical pumping, however, the squeezing parameters in \erf{eq::YurkeSqParamOP} and \erf{eq::YurkeLikeSqParamOP} become infinite as $\alpha\rightarrow 0$. This occurs because atoms are optically pumped into the coupled states, which are infinitely anti-squeezed. In practice, when employing a partial isometry to the Yurke or half-integer Yurke state, we use a numerical search to determine the optimal nonzero $\alpha$. The optimal $\alpha$ maximizes the contribution of the squeezed fiducial state, while minimizing the contribution of the anti-squeezed coupled state.

\section{Results}\label{sec::HPResults}

The modified multilevel Holstein-Primakoff approximation along with the numerical methods introduced in this chapter enable us to determine the amount of squeezing achievable in the presence of optical pumping. The master equation describing optical pumping in the rotating frame and the SCS, cat, and $m_x=0\,$ state preparations satisfy the condition $\bra{\uparrow}\mathcal{D}(\ket{\uparrow}\bra{\downarrow})\ket{\wr}= 0$, ensuring that the modified multilevel Holstein-Primakoff approximation is valid. This allows us to quantitatively examine the influence of the fiducial state upon both coherent squeezing and decoherence of the ensemble. Using numerical techniques, we can also explore the impact of spin size and post-processing upon spin squeezing. In this section, we present a variety of numerical simulations showing achievable squeezing for various state preparations, spin sizes, post-processing partial isometries and squeezing protocols. In an effort to demonstrate what might be feasible in a laboratory setting, we have used experimentally realistic values of parameters relating to the light and atomic ensemble with $OD=300$, $N_A=10^6$, $N_L=3\times 10^8$, $\sigma_0/A=3\times 10^{-4}$ and $\Gamma/\Delta=10^{-3}$ for all simulations. 

\subsection{Testing the modified multilevel HP}
Before presenting numerical results relating to achievable squeezing, we test the validity of the modified multilevel Holstein-Primakoff approximation, which enabled us to treat the ensemble as a Gaussian state on two modes. For the case of $f=1$, the exact differential equations describing the evolution of the mean spin, the collective variance and the populations under QND measurement and optical pumping form a closed set that can be easily solved through standard numerical methods.
From Ref. \cite{MadMol}, the collective variance under continuous QND measurement of $\hat{F}_z$ obeys the nonlinear differential equation
\begin{align}
\frac{d}{dt}\Delta F_z^2\big|_\text{QND}=-\kappa(\Delta F_z^2)^2.
\end{align}
Here, $\kappa=\chi^2N_L/\Delta t$ is the ``measurement strength". Further details about the differential formulation of continuous QND measurement will be provided in Chapter \ref{sec::Beyond}. The equation of motion for the variance under both continuous QND measurement and optical pumping follows from the evolution of second order moments under optical pumping given in \erf{eq::CovarEvol},
\begin{align}\label{eq::NumVar}
\frac{d}{dt}\Delta F_z^2=-\kappa(\Delta F_z^2)^2-\frac{2\gamma_s}{9}\Delta F_z^2+\frac{\gamma_s}{9}\expect{\hat{N}_1+\hat{N}_0+\hat{N}_{-1}}.
\end{align}
Here, we have decomposed the noise term as a sum of the population operators $\hat{N}_1$, $\hat{N}_0$ and $\hat{N}_{-1}$, defined as $\hat{N}_m=\sum_{i=1}^{N_A}\ket{f,m_x=m}\bra{f,m_x=m}_i$. Unlike the multilevel HP approximation, we treat these populations as operators, rather than c-numbers. Because continuous QND measurement negligibly effects the populations and mean spin, their equations of motion depend soley on optical pumping. From the equation of motion for first order moments under optical pumping given in \erf{eq::FirstOrderEvol}, the populations and mean spin satisfy
\begin{align}
\frac{d}{dt}\expect{\hat{N}_1}=-\frac{\gamma_s}{9}\expect{\hat{N}_1}+\frac{\gamma_s}{18}\expect{\hat{N}_0},
\end{align}
\begin{align}
\frac{d}{dt}\expect{\hat{N}_0}=-\frac{2\gamma_s}{9}\expect{\hat{N}_0}+\frac{\gamma_s}{18}\expect{\hat{N}_1}+\frac{\gamma_s}{18}\expect{\hat{N}_{-1}},
\end{align}
\begin{align}
\frac{d}{dt}\expect{\hat{N}_{-1}}=-\frac{\gamma_s}{9}\expect{\hat{N}_{-1}}+\frac{\gamma_s}{18}\expect{\hat{N}_0}
\end{align}
and
\begin{align}\label{eq::NumMeanSpin}
\frac{d}{dt}\expect{\hat{F}_x}=-\frac{\gamma_s}{6}\expect{\hat{F}_x}.
\end{align}
The variance and populations form a closed set of differential equations that can be numerically solved, while the mean spin is an exponential that can be solved analytically.

\begin{figure}
 \centering
\includegraphics[scale=.4]{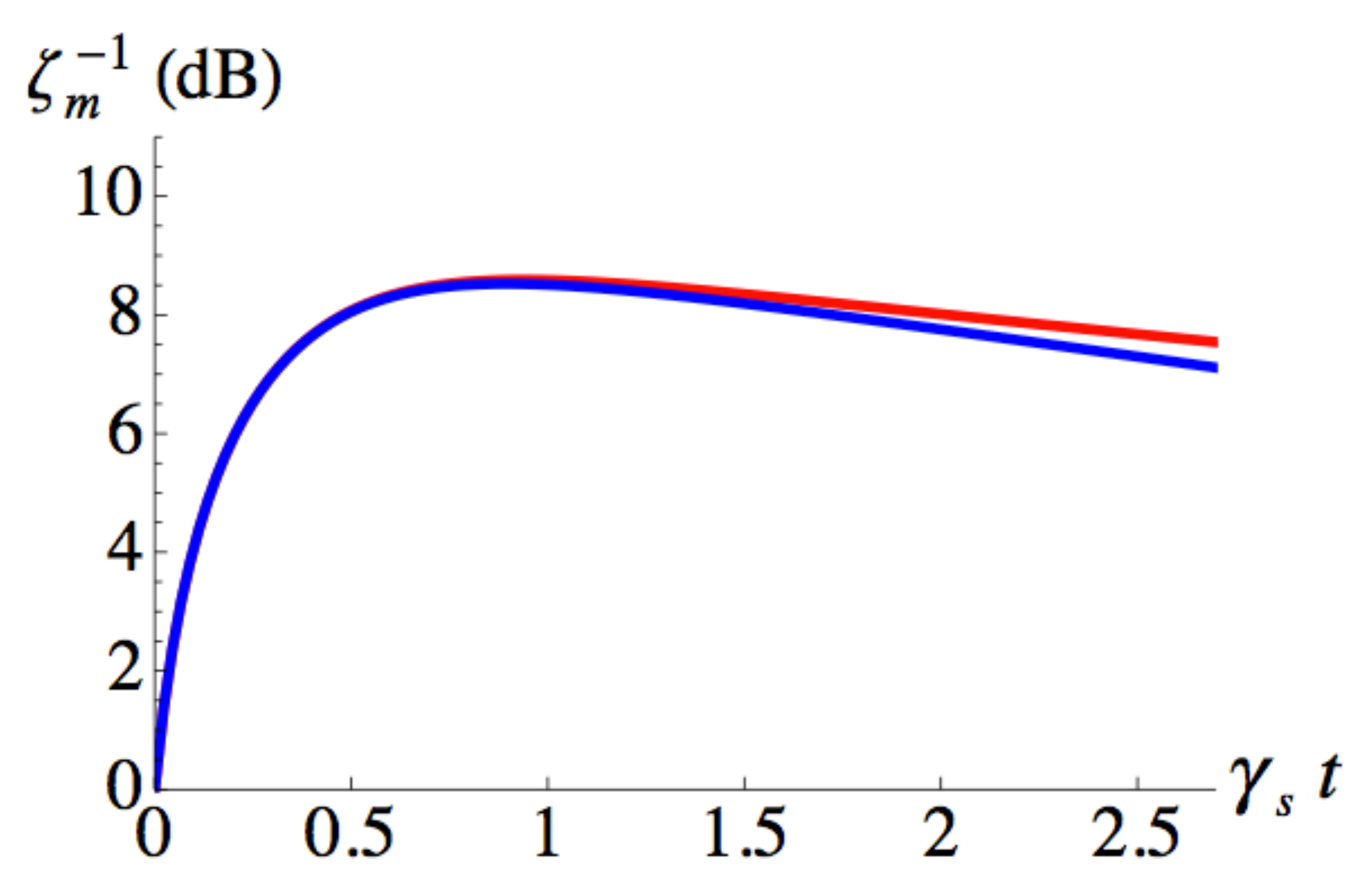}
\caption{Comparison of the modified multilevel Holstein-Primakoff approximation (red) with the numerical solution of the exact differential equations (blue) for the case of an ensemble with $f=1$ prepared in the $SCS$. The inverse of the metrological squeezing parameter is plotted against time in units of the scattering rate, $\gamma_s$. The maximal squeezing predicted by these models differs by .07dB. At the end of the plotted time interval, the models differ by .44 dB. }\label{fig::HPvsExact}
\end{figure}

In Fig. \ref{fig::HPvsExact}, we compare the squeezing generated by the solution of Eqs. (\ref{eq::NumVar}) through (\ref{eq::NumMeanSpin}) with the squeezing predicted by a simulation of the QND measurement protocol utilizing the modified multilevel Holstein-Primakoff approximation. We consider an ensemble of atoms with $f=1$, initially prepared in a spin coherent state. The solutions deviate very little in the time vicinity of peak squeezing with the peak squeezing predicted by both models differing only by .07 dB. From this plot, it is evident that neglecting the coherences between the fiducial and transfer states, treating $\downarrow\uparrow$ and $\wr\!\downarrow$ as two commuting modes and approximating the populations as c-numbers has little effect upon the predicted squeezing.

The case of $f=1$ is unique in that each atomic spin is an actual qutrit, rather than being a qutrit embedded in a higher dimensional qudit. For atoms with larger spin, the microwave internal spin control introduced in Sec. \ref{sec::ControlOP} can be utilized to map all states in the $f$-manifold to the other ground hyperfine manifold, with the exception of the fiducial, coupled and transfer states. When this control is applied, higher spin atoms are  in effect qutrits. The results in Fig. \ref{fig::HPvsExact} are, thus, indicative of the deviations that we expect between the two models for higher spins as well as for $f=1$.

\subsection{Performance of Squeezing Protocols for Different State Preparations}
The initial fiducial state of the ensemble has an enormous impact on the coherent generation of squeezing and decoherence due to optical pumping. The interatomic entanglement generated by the squeezing protocols increases with the variance of the fiducial state. Other properties of the fiducial state, such as the spin flip rate and atom loss rate, govern the robustness of the ensemble to optical pumping. The modified multilevel HP approximation and the numerical methods developed in this chapter allow us to assess the performance of the SCS, cat, and $m_x=0$ preparations in the presence of optical pumping. 

 \begin{figure}
 \centering
\includegraphics[scale=.35]{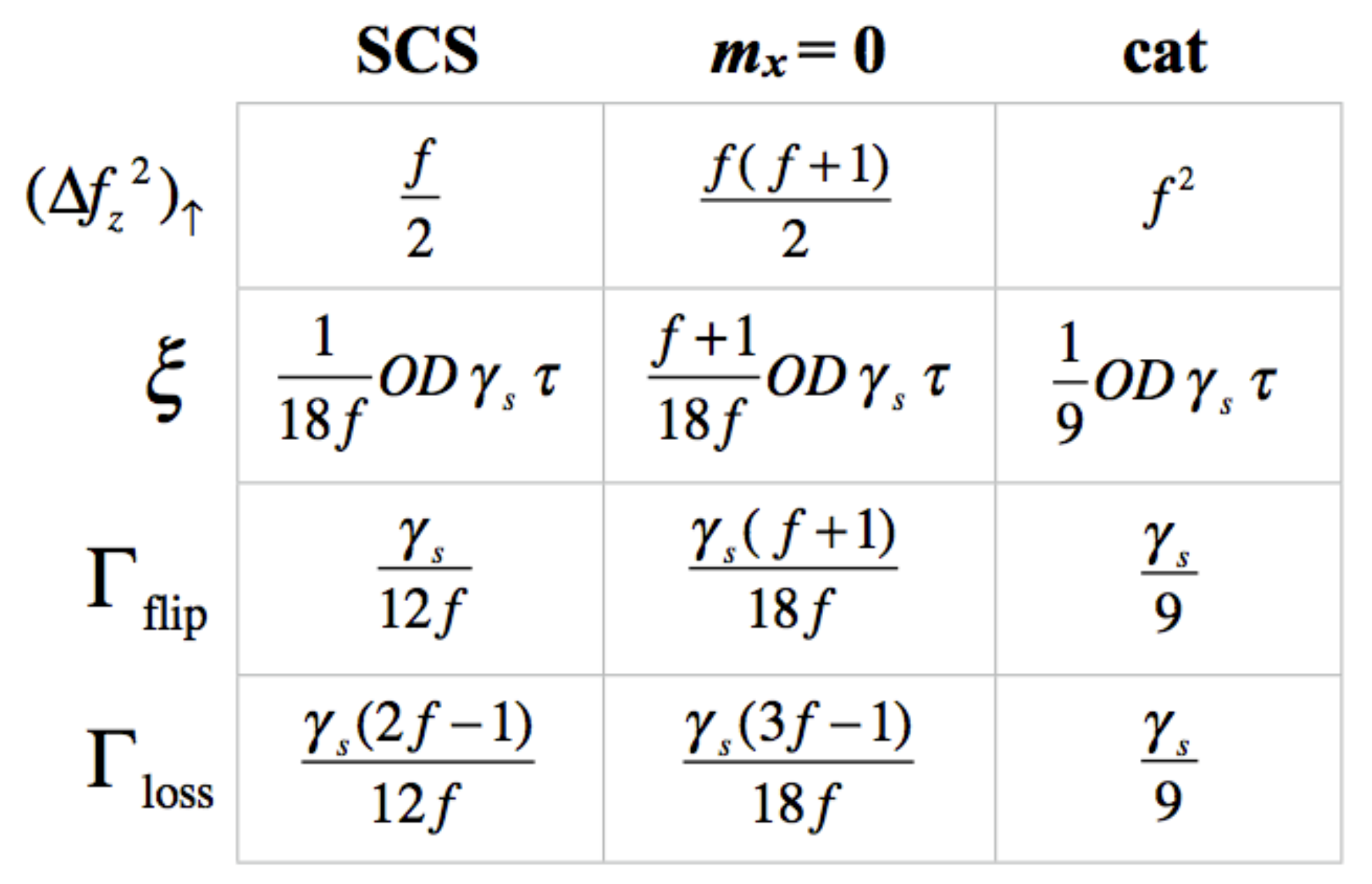}
\caption{The variance of the fiducial state, the collective spin coupling constant, the spin flip rate of the fiducial state and the total loss rate from the $f$ manifold for the SCS, $m_x=0$ and cat state preparations. Note that the spin flip and loss rates for the cat state are for $f\geq 1$. When $f=1/2$, the spin flip and loss rates for the cat are identical to those of the SCS. The spin flip and loss rates are calculated with the master equation in the rotating frame given in \erf{eq::MasterRotating}.}\label{fig::StateStats}
\end{figure}

 \begin{figure}
 \centering
\includegraphics[scale=.39]{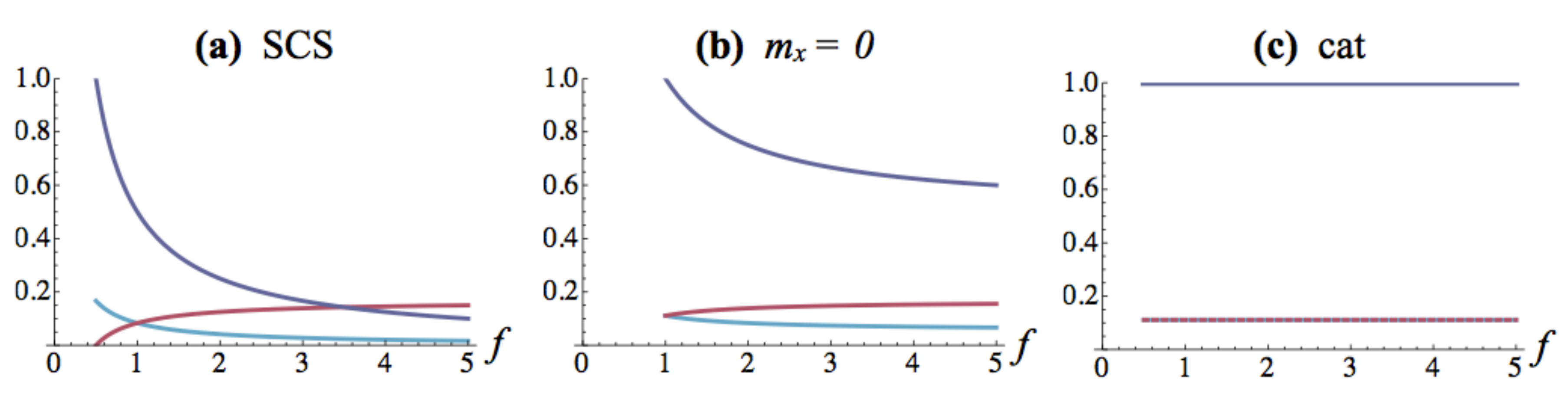}
\caption{Dependence of relevant quantities for the SCS and $m_x=0$ state preparations upon $f$. Plotted are the quantities $9\xi(OD\,\gamma_s\tau)^{-1}$ (purple), $\Gamma_\text{flip}\gamma_s^{-1}$ (light blue) and $\Gamma_\text{loss}\gamma_s^{-1}$ (red).}\label{fig::StateStatPlots}
\end{figure}

To guide our intuition,  the table in Fig. \ref{fig::StateStats} lists quantities relevant to the performance of each state preparation. These quantities - the variance of the fiducial state, the collective spin coupling constant, the spin flip rate of the fiducial state and the total loss rate of atoms from the $f$ manifold - are essential to interpreting the numerical results. Recall that the squeezing generated by the Faraday interaction increases with the collective spin coupling constant, $\xi$, which is proportional to the projection noise fluctuations of the fiducial state, $(\Delta f_z^2)_\downarrow$. Spin flips, the most damaging optical pumping processes, cause the mean spin to decay and inject noise back into the system, thereby increasing the variance of the collective spin.   While loss can result in an even greater decay of the mean spin, it is less damaging to spin squeezing because it also causes the collective variance to decay. Each of these quantities depend not only on the state preparation, but on the size of the hyperfine spin, $f$. Figure \ref{fig::StateStatPlots} provides a visual depiction of how these quantities vary with $f$. 

\begin{figure}[H]
 \centering
\includegraphics[scale=.6]{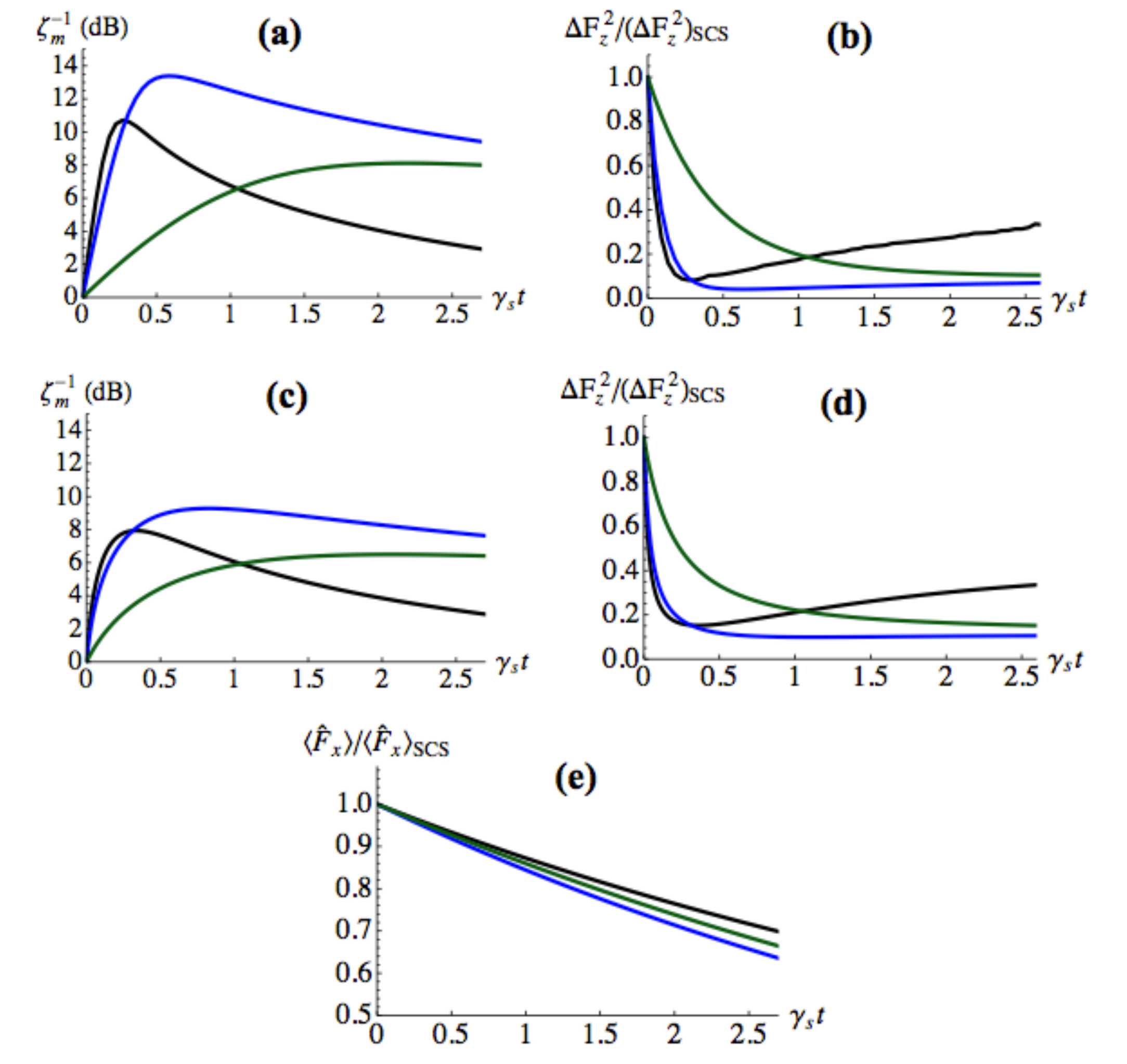}
\caption{Performance of the state preparations for $f=4$, SCS (green), cat (black) and $m_x=0$ (blue). For post-processing, a partial isometry mapping each state preparation to the SCS preparation was applied to the ensemble. Plot (a) shows the squeezing generated by each state preparation for the phase matching protocol. Plot (b) shows the corresponding variance, normalized by the variance of the spin coherent state. Plot (c) shows the squeezing generated by each state preparation for the QND measurement protocol with the corresponding variance, normalized by the variance of the spin coherent state shown in plot (d). For both squeezing protocols, the decay of the mean spin normalized by the mean spin of the spin coherent state is shown in plot (e). }\label{fig::4Plots}
\end{figure}

We first examine the performance of the state preparations for $f=4$, corresponding to the larger ground hyperfine manifold of cesium.  Figures \ref{fig::4Plots} (a) and (b) show the squeezing and the reduction in the collective variance generated by the double-pass phase matching protocol combined with a partial isometry to the SCS preparation. The same quantities for the QND measurement protocol combined with a partial isometry to the SCS preparation are plotted in (c) and (d). Also shown in Fig.  \ref{fig::4Plots} (e) is the decay of the mean spin, which is governed by optical pumping alone. Unsurprisingly, phase matching outperforms the QND measurement protocol. Our primary interest is the performance of the state preparations, however. The cat state has the largest fiducial state projection noise, but also the largest rate of spin flips. In addition, for this preparation, there is no transfer of coherence to mitigate the effect of spin flips. Consequently,  the protocols rapidly generate squeezing on the cat preparation, but this squeezing also decays at a rapid rate. The increase in $\Delta F_z^2$ for the cat preparation in plots (b) and (d) at longer times demonstrates the deleterious effect of spin flips. In contrast, the collective variances of the SCS and $m_x=0$ preparations appear to asymptote; the excess noise injection is balanced by squeezing and loss. Interestingly, the cat preparation's large rate of spin flips relative to loss make its mean spin the most robust to optical pumping. While a ``lost" atom contributes nothing to the mean spin, an atom that undergoes a spin flip from the fiducial to the coupled state contributes $\expect{\hat{f}_x}_{\downarrow'}$, where $\ket{\downarrow'}$ is the state to which the coupled state is mapped by the post-processing partial isometry. For a partial isometry to the SCS preparation, $\expect{\hat{f}_x}_{\downarrow'}=f-2$, meaning that a spin flip event reduces the mean spin by 2 as opposed to $f$ in the case of loss. Although the $m_x=0$ preparation has a smaller enhancement of the fiducial state projection noise compared to the cat, its reduced rate of spin flips make it more robust to optical pumping. Accordingly, the $m_x=0$ preparation outperforms both the cat and the SCS preparation, which has the smallest initial projection noise variance of the three state preparations. For the phase matching and QND protocols, the $m_x=0$ preparation achieves a peak squeezing of 13.4 dB and 9.3 dB, respectively. Curiously, the  $m_x=0$ preparation has a large rate of loss, which causes its mean spin to decay the fastest due to optical pumping. The enhanced squeezing of the collective variance for the $m_x=0$ preparation compensates for this, however. The SCS preparation has the smallest fiducial state projection noise and spin flip rate. As a consequence, the SCS preparation is much more robust to optical  pumping, but its peak squeezing is significantly smaller than that of the cat or $m_x=0$ because the entangling effect of the Faraday interaction is weaker.   

\begin{figure}[H]
 \centering
\includegraphics[scale=.6]{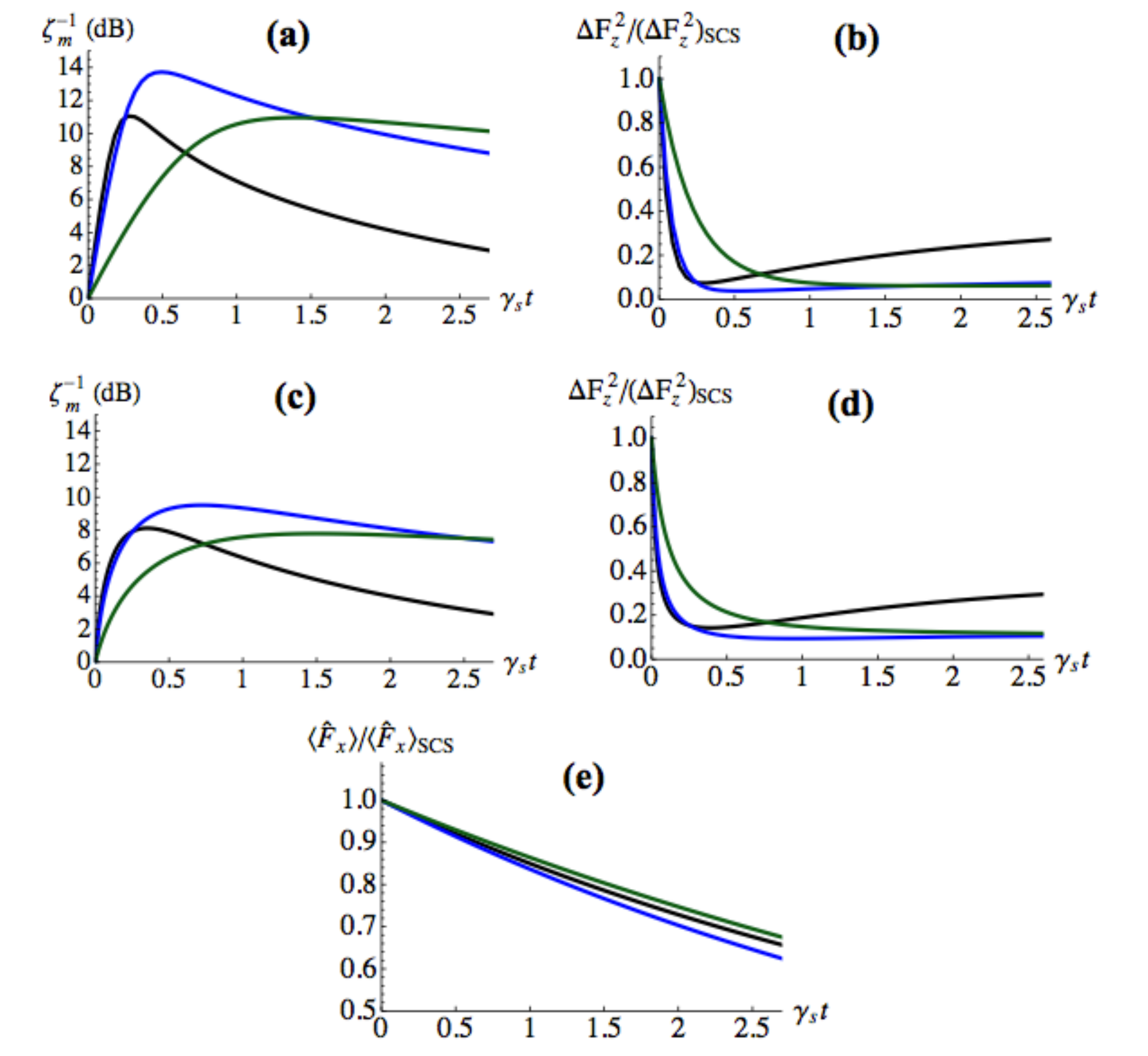}
\caption{Performance of the state preparations for $f=2$, SCS (green), cat (black) and $m_x=0$ (blue). For post-processing, a partial isometry mapping each state preparation to the SCS preparation was applied to the ensemble. Plot (a) shows the squeezing generated by each state preparation for the phase matching protocol. The corresponding variance, normalized by the variance of the spin coherent state, is plotted in (b). Plot (c) shows the squeezing generated by each state preparation for the QND measurement protocol with the corresponding variance, normalized by the variance of the spin coherent state shown in plot (d). For both squeezing protocols, the decay of the mean spin normalized by the mean spin of the spin coherent state is shown in plot (e).}\label{fig::2Plots}
\end{figure}

To illustrate the role of the spin size in determining squeezing, Fig. \ref{fig::2Plots} replicates the same plots in  Fig. \ref{fig::4Plots} with $f=2$ instead of $f=4$. As in the $f=4$ case, the $m_x=0$ preparation outperforms both the SCS and the cat preparations, attaining a peak squeezing of 13.7 dB for the double-pass phase matching protocol and 9.5 dB for the QND protocol. Unlike the $f=4$ case, the performances of the SCS and the cat preparations are comparable for $f=2$. For the phase matching protocol, the SCS and cat preparations achieve nearly identical peak squeezing values of 11.0 dB and 11.1dB, respectively, though the peak value is reached much faster for the cat state. For the QND measurement protocol, the SCS and cat preparations reach 7.8 dB and 8.1dB, respectively. The improved performance of the SCS preparation relative to the cat can be explained by the behavior of the mean spin. Note that the mean spin of the SCS preparation in  Fig. \ref{fig::2Plots} (e) decays the least, followed by the cat and the $m_x=0$ state. The order is different in the $f=4$ case, in which the mean spin of the cat decays the least, followed by the SCS and the $m_x=0$ state. Whereas the loss rate of the SCS preparation was substantially higher than the cat state preparation for $f=4$, their loss rates are nearly equal when $f=2$. The spin flip rate for the cat at $f=2$ is much larger, however. This causes the mean spin of the cat state to decay faster. Consequently, the squeezing produced by the SCS preparation is much improved when $f=2$. Despite this, the SCS preparation is still the lowest performer.  Also similar to the $f=4$ case, the cat state preparation is the least robust to optical pumping with its squeezing decaying the quickest. The SCS is the most robust, followed by the $m_x=0$ state. For both $f=4$ and $f=2$, the $m_x=0$ state preparation strikes the most optimal balance between enhanced squeezing due to the variance of the fiducial state and robustness to optical pumping. 

\subsection{Effect of Partial Isometries}
To convert the interatomic entanglement generated by the Faraday interaction into metrologically relevant squeezing, we have focused on internal spin control via two partial isometries. The first partial isometry, which maps an arbitrary state preparation to the SCS preparation, creates squeezing that depends upon interatomic entanglement alone, since $\zeta_m^{\uparrow_{SCS}}=1$. The second partial isometry maps an arbitrary state preparation to the Yurke or half-integer Yurke preparation, for which $\zeta_m^{\uparrow_{\text{y}}}<1$ and $\zeta_m^{\uparrow_{\text{hy}}}<1$. Because the Yurke states are squeezed, the spin squeezing generated by these partial isometries depends on internal spin squeezing as well as interatomic entanglement. 

\begin{figure}
 \centering
\includegraphics[scale=.4]{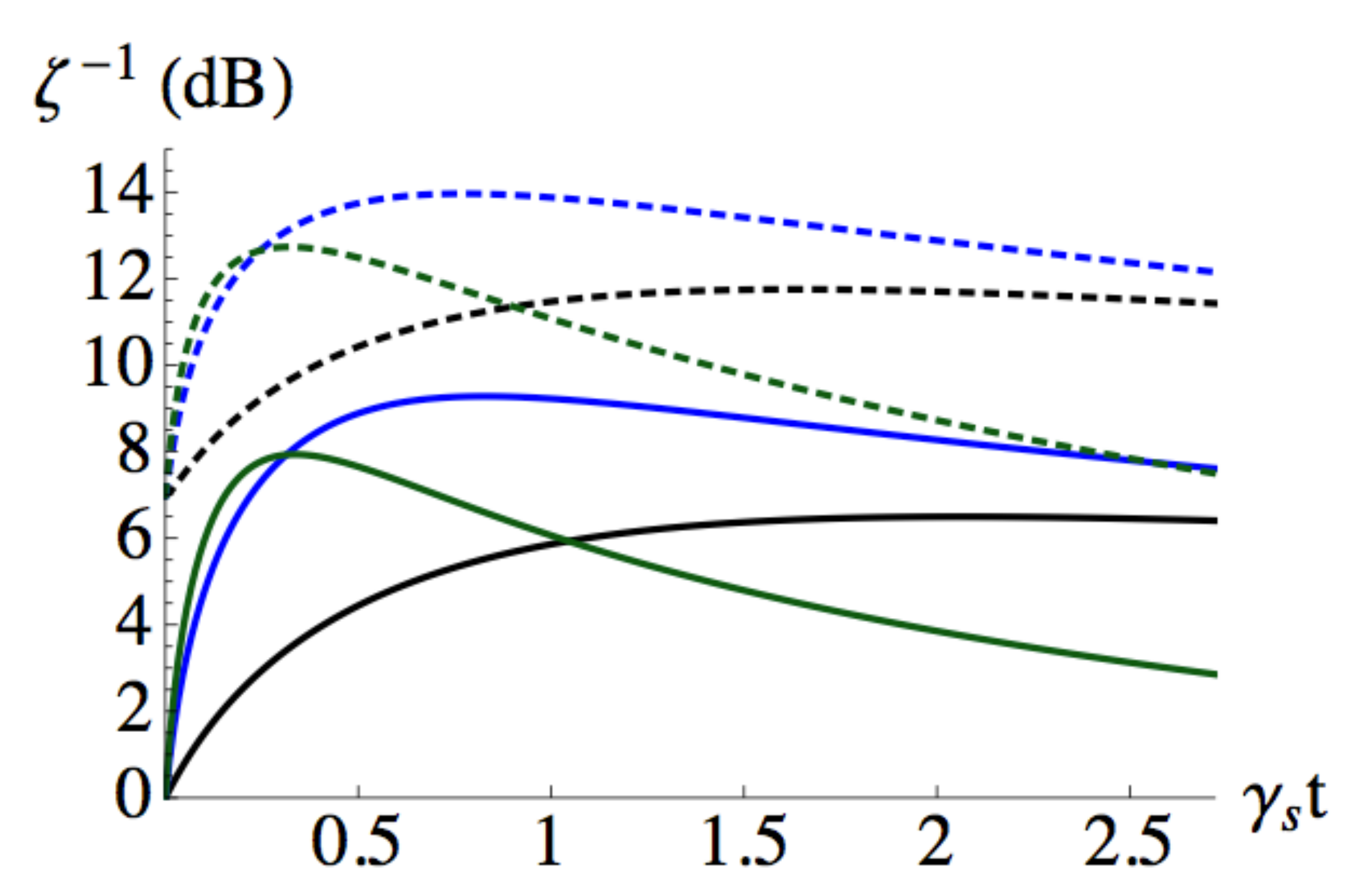}
\caption{The performance of the state preparations SCS (green), cat (black) and $m_x=0$ (blue) for $f=4$ under the QND measurement protocol with optical pumping. Squeezing is plotted versus time in units of scattering rate. Here, two different partial isometries are used to convert interatomic entanglement into spin squeezing, a map to the SCS preparation (solid), a map to the Yurke preparation (dashed).  }\label{fig::PICompare}
\end{figure}

Fig. \ref{fig::PICompare} examines the effect of internal spin squeezing on the collective spin squeezing achieved by the different state preparations. For $f=4$, Fig. \ref{fig::PICompare} shows the squeezing generated by QND measurement combined with either a partial isometry to the SCS preparation or a partial isometry to the Yurke preparation. This plot is generated using the formulas in Eqs.  (\ref{eq::SCSSqParamOP}) and (\ref{eq::YurkeSqParamOP}), which give the squeezing that results from a Faraday-effect squeezing protocol combined with a partial isometry in the presence of optical pumping. The parameter $\alpha$ in \erf{eq::YurkeSqParamOP}, which determines the squeezing of the Yurke state, is chosen by a numerical optimization. In absence of decoherence, the partial isometry to the Yurke preparation produces a multiplicative enhancement of the squeezing parameter by $(f+1)^{-1}$ as compared to the SCS preparation. When the inverse of the squeezing parameter is plotted in dB for $f=4$, as in Fig. \ref{fig::PICompare}, this enhancement appears as an increase in the amount of squeezing by 7.0 dB.
The dashed lines, which correspond to the Yurke partial isometry, are translated upward by nearly 7.0 dB at small times. The upward translation decreases, however, as time progresses and the state decoheres due to optical pumping. Nonetheless, as Fig. \ref{fig::PICompare} attests, collective spin squeezing is enhanced substantially by the internal spin squeezing resulting from the partial isometry to the Yurke preparation.

\subsection{Scaling of Squeezing with $f$}
\begin{figure}[H]
 \centering
\includegraphics[scale=.55]{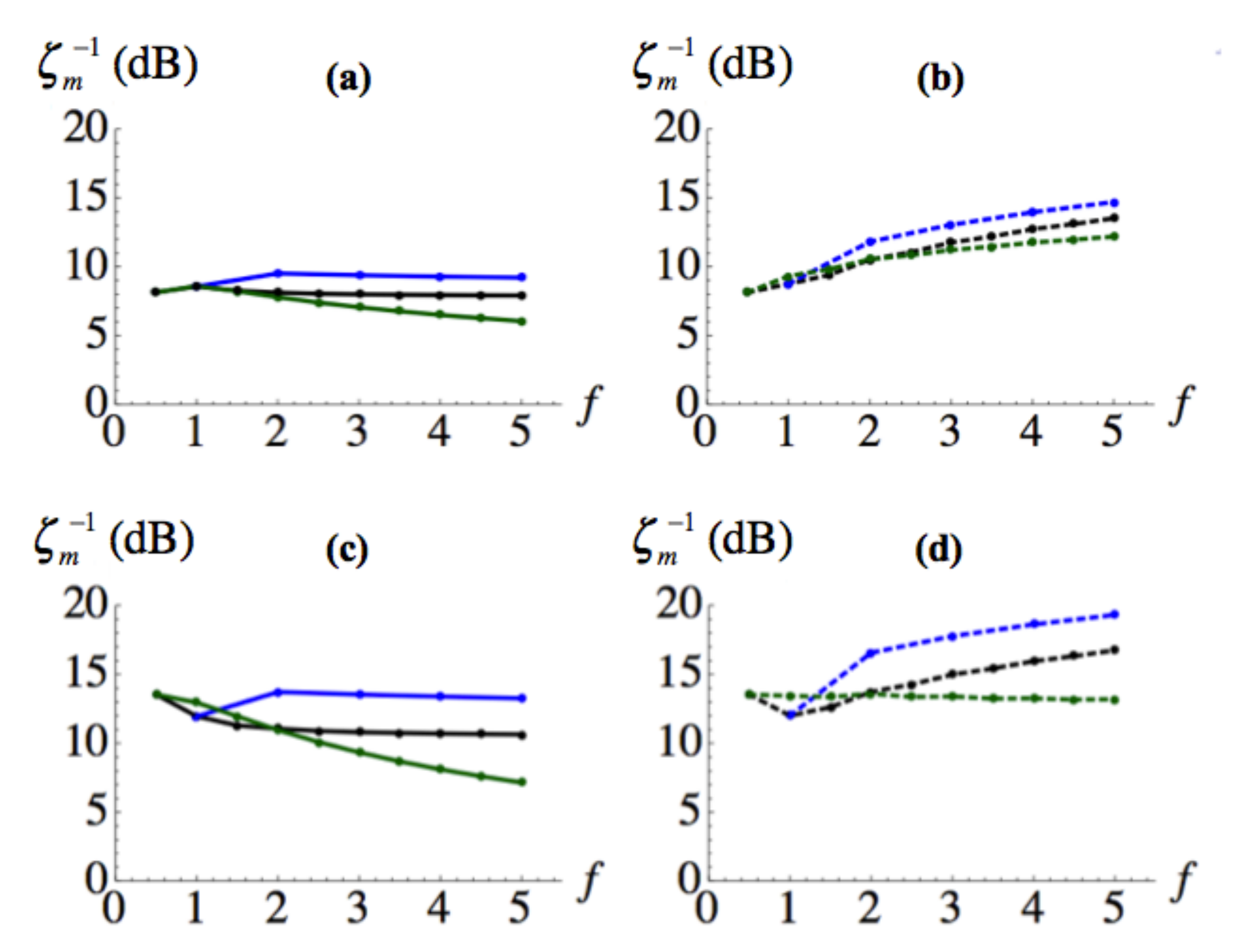}
\caption{Peak squeezing vs. spin size for SCS (green), cat (black) and $m_x=0$ (blue). These plots show the performance of the QND and phase matching squeezing protocols, as quantified by the inverse of the metrological squeezing parameter, versus the size of the hyperfine spin $f$. The topmost plots show the performance of the QND squeezing protocol with (a) a partial isometry to the SCS preparation and (b) a partial isometry to the Yurke preparation for integer $f$ and a partial isometry to the half-integer Yurke preparation for half-integer $f$. The bottom plots show the performance of the phase matching protocol with (c) a partial isometry to the SCS preparation and (d) a partial isometry to the Yurke preparation for integer $f$ and a partial isometry to the half-integer Yurke preparation for half-integer $f$.}\label{fig::fPlots}
\end{figure}

Figure \ref{fig::fPlots} further examines the dependence of squeezing on the spin size, $f$. Plotted in Fig. \ref{fig::fPlots} is the peak squeezing generated by the QND and phase matching protocols for partial isometries to both the SCS and Yurke preparations. We focus first on Fig. \ref{fig::fPlots} (a) and (c), which show the peak squeezing of the QND and phase matching protocols on different state preparations with a partial isometry to the SCS preparation. Plots (a) and (c) exhibit similar behavior, except at small values of $f$. Perhaps most significantly, both plots attain their maximum value of squeezing at $f=2$, not $f=1/2$, which one might naively expect given that the collective spin coupling constant decreases  monotonically with increasing $f$. The $m_x=0$ preparation performs the best for both the QND and phase matching protocols, peaking at $f=2$ and then gradually declining due to the decreased collective spin coupling constant. The $m_x=0$ preparation likely peaks at $f=2$ because it is the smallest spin for which there is a transfer of coherence, giving the ensemble additional robustness to decoherence. A smaller spin is favorable since the collective spin coupling constant decreases with increasing spin. For both protocols, the performance of the $SCS$ preparation falls off the most rapidly, beginning at $f=1$. This occurs because the collective spin coupling constant of the SCS preparation, which is proportional to $1/f$, has the greatest decline with increasing $f$. Although the spin flip rate of the $SCS$ preparation also decreases with $1/f$, this does not compensate for loss in the coherent interaction strength. For the cat preparation, the collective spin coupling constant, the rate of loss, and the rate of spin flips are all constant with $f$. This is reflected by the relative stability of the cat preparation after $f=1$.

Fig. \ref{fig::fPlots} (b) and (d) depict the dependence of the peak squeezing generated by the QND measurement and phase matching protocols on different state preparations combined with internal spin squeezing. The internal spin squeezing is generated by a partial isometry to the Yurke preparations. The comparative performances of the state preparations in plots (b) and (d) is the same as for (a) and (c). A significant difference between these plots, however, is that spin squeezing largely improves as $f$ increases in plots (b) and (d).  Although the collective spin coupling constant decreases with increasing $f$, internal spin squeezing generates a greater amount of collective spin squeezing for atoms with larger $f$. Internal spin squeezing more than compensates for the loss in coherent interaction strength as $f$ increases.

\subsection{Effect of the Transfer State}
 \begin{figure}
 \centering
\includegraphics[scale=.48]{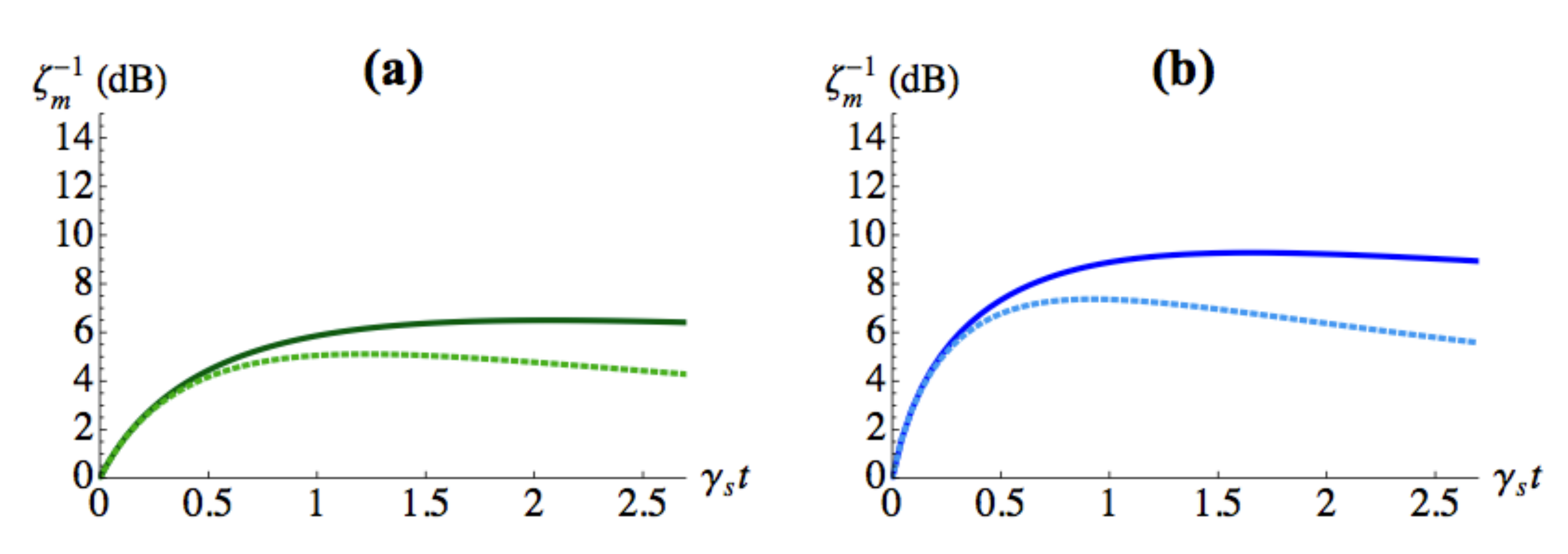}
\caption{Performance of state preparations with and without the transfer state for the QND measurement protocol and $f=4$. A partial isometry mapping to the SCS preparation is applied to the $m_x=0$ preparation. Plot (a) shows the squeezing generated by the SCS state preparation for $f=4$ when the transfer state is preserved (green solid) and when the transfer state is eliminated via internal spin control (light green dashed). Preserving the transfer state improves spin squeezing by 1.4 dB. Plot (b) shows the squeezing generated by the $m_x=0$ state preparation for $f=4$ when the transfer state is preserved (blue solid) and when the transfer state is eliminated (light blue dashed). For the $m_x=0$ state preparation, spin squeezing is improved by 1.9 dB when the transfer state is preserved.}\label{fig::transfer}
\end{figure}
Sec. \ref{sec::TransferState} introduced the counterintuitive idea that preserving the transfer state, rather than removing it by mapping it to the other ground hyperfine manifold with internal spin control, can improve spin squeezing in the presence of optical pumping. This idea is explored in Fig. \ref{fig::transfer}, which compares the squeezing generated by two state preparations under QND measurement and optical pumping when the transfer state is preserved and when it is eliminated. For $f=4$, the plots in Fig. \ref{fig::transfer} feature the SCS preparation and the $m_x=0$ preparation with a final partial isometry to the SCS preparation.  From Fig. \ref{fig::transfer}, it is evident that preserving the transfer state has a significant effect upon the achievable spin squeezing. In the case of the SCS preparation, shown in Fig. \ref{fig::transfer} (a), preserving the transfer state improves peak squeezing by 1.4 dB. For the $m_x=0$ preparation in Fig. \ref{fig::transfer} (b), preserving the transfer state results in a 1.9 dB improvement of spin squeezing. In addition to increasing the magnitude of the peak squeezing, the presence of the transfer state also reduces the rate at which spin squeezing decays. Overall, retaining the transfer state greatly improves the performance of both state preparations in the presence of optical pumping.

\chapter{Optimal Spin States for QND Squeezing}\label{sec::Beyond}
Thus far, the techniques we have used to determine the achievable spin squeezing in the atomic ensemble have been suited only to particular fiducial states. In Chapters \ref{sec::ALinterface} and \ref{sec::OpticalPumping}, we derived the coupled and transfer states from the fiducial state by making the assumption that $\expect{\hat{f}_z}_\uparrow=\expect{\hat{f}_z}_\downarrow=\expect{\hat{f}_z}_\wr=0$. 
The modified version of the multilevel Holstein-Primakoff approximation developed in Chapter \ref{sec::ModHPCovar} was applicable to ensembles prepared in fiducial states that satisfied the condition in \erf{eq::ModeCondition}, ensuring that pairwise coherences involving the fiducial and transfer states did not develop. While the SCS, cat, and $m_x=0$ state preparations satisfy both  $\expect{\hat{f}_z}_\uparrow=\expect{\hat{f}_z}_\downarrow=\expect{\hat{f}_z}_\wr=0$ and \erf{eq::ModeCondition}, this is not necessarily true of an arbitrary fiducial state. In this chapter we develop techniques to model the effects of squeezing and optical pumping for any fiducial state. 

The techniques of this chapter rely on expressing a squeezing protocol in differential form. The double pass squeezing protocols, such as phase matching, are not readily converted to differential form. Because the only way to model these protocols is as a series of updates upon the covariance matrix of a Gaussian state, we must rely upon the modified multilevel Holstein-Primakoff approximation of the previous chapter. A relatively simple differential form exists for the QND measurement protocol, however. The effects of QND measurement and optical pumping can, thus, be modeled more easily on any fiducial state. We utilize this method to perform a numerical search over the space of all fiducial states in order to find the state preparation that maximizes squeezing in the presence of optical pumping. 

\section{Coupled and Transfer States for Arbitrary\\ Fiducial States}
Determining squeezing when the ensemble is prepared in an arbitrary fiducial state requires that we generalize the definitions of the coupled and transfer states to the case where $\expect{\hat{f}_z}_\uparrow$, $\expect{\hat{f}_z}_\downarrow$ and $\expect{\hat{f}_z}_\wr$ are not necessarily zero. In Sec. \ref{sec::MultiHPEnsemble}, we derived the coupled state by considering the effect of the Faraday interaction on a single atom prepared in the fiducial state. Here, we do the same thing, but relax the assumption that $\expect{\hat{f}_z}_\uparrow=0$. For $\chi<<1$, 
\begin{align}\label{eq::CoupledNo0Mean}
e^{-i\chi\hat{S}_3\hat{f}_z}\ket{\uparrow}&\approx(\mathbb{I}-i\chi\hat{S}_3\hat{f}_z)\ket{\uparrow}\\\notag
&=\ket{\uparrow}-i\chi\hat{S}_3\expect{\hat{f}_z}_\uparrow\ket{\uparrow}
-i\chi\hat{S}_3(\hat{f}_z-\expect{\hat{f}_z}_\uparrow)\ket{\uparrow}.
\end{align}
The Faraday interaction maps the fiducial state back to itself and to an orthogonal state, proportional to $(\hat{f}_z-\expect{\hat{f}_z}_\uparrow)\ket{\uparrow}=\Delta\hat{f}_z\ket{\uparrow}$. This orthogonal state is, by definition, the coupled state.  From \erf{eq::CoupledNo0Mean}, the coupled state is given by
\begin{align}\label{eq::CoupledDefNo0}
\ket{\downarrow}=&\frac{\Delta\hat{f}_z\ket{\uparrow}}{\sqrt{(\Delta f_z^2)_\uparrow}}\\\notag
=&\frac{\Delta\hat{f}_z\ket{\uparrow}}{v(\uparrow)/\sqrt{2}}.
\end{align}
The transfer state arises by considering the effect of the Faraday interaction on the coupled state, which depends on
\begin{align}\label{eq::TransferNo0Mean}
\hat{f}_z\ket{\downarrow}=\expect{\hat{f}_z}_\downarrow\ket{\downarrow}+\sqrt{(\Delta f_z^2)_\uparrow}\ket{\uparrow}+
\Big((\hat{f}_z-\expect{\hat{f}}_\downarrow)\ket{\downarrow}-\sqrt{(\Delta f_z^2)_\uparrow}\ket{\uparrow}\Big).
\end{align}
The Faraday interaction maps the coupled state to a superposition of itself,  the fiducial state and another state orthogonal to both the coupled and the fiducial states, proportional to $(\hat{f}_z-\expect{\hat{f}}_\downarrow)\ket{\downarrow}-\sqrt{(\Delta f_z^2)_\uparrow}\ket{\uparrow}$. This orthogonal state is, by definition, the transfer state. From \erf{eq::TransferNo0Mean}, the transfer state is given by
\begin{align}\label{TransferStateNo0}
\ket{\wr}&=\frac{1}{\sqrt{(\Delta f_z^2)_\downarrow-(\Delta f_z^2)_\uparrow}}\left(\Delta\hat{f}_z\ket{\downarrow}-\sqrt{(\Delta f_z^2)_\uparrow}\ket{\uparrow}\right)\\\notag
&=\frac{1}{w(\uparrow)}\left(\sqrt{2}\Delta\hat{f}_z\ket{\downarrow}-v(\uparrow)\ket{\uparrow}\right).
\end{align}
From Eqs. (\ref{eq::CoupledDefNo0}) and (\ref{TransferStateNo0}), we see that the definitions of the coupled and transfer states are readily generalized to the case of nonzero $\expect{\hat{f}_z}_\uparrow$ and $\expect{\hat{f}_z}_\downarrow$.

Using the generalized definitions of the coupled and transfer states, the internal spin component $\hat{f}_z$ can be expressed in the embedded qutrit basis as
\begin{align}\label{eq::fzno0}
\hat{f}_z=&\sqrt{(\Delta f_z^2)_\uparrow}\left(\ket{\uparrow}\bra{\downarrow}+\ket{\downarrow}\bra{\uparrow}\right)+\expect{\hat{f}_z}_\uparrow\ket{\uparrow}\bra{\uparrow}+\expect{\hat{f}_z}_\downarrow\ket{\downarrow}\bra{\downarrow}
\\\notag&+\sqrt{(\Delta f_z^2)_\downarrow-(\Delta f_z^2)_\uparrow}\left(\ket{\downarrow}\bra{\wr}+\ket{\wr}\bra{\downarrow}\right)+\expect{\hat{f}_z}_\wr\ket{\wr}\bra{\wr}.
\end{align}
Note that the second term contains coherences between the fiducial and the coupled state and the fourth term contains coherences between the coupled and transfer state, both of which are responsible for generating negative correlations in spin squeezed states. Treating the expectation values $\expect{\hat{f}_z}_\uparrow$,  $\expect{\hat{f}_z}_\downarrow$ and  $\expect{\hat{f}_z}_\wr$ as nonzero contributes additional terms to $\hat{f}_z$, but preserves the essential coherences.

\subsection{Collective Spin Variance}
In Chapter \ref{sec::ModHPCovar}, we determined the collective spin variance $\Delta F_z^2$ by tracking the evolution of covariances between the collective pseudospin operators. Here, we show that the exact same thing is possible with the more general definitions of the coupled and transfer states. Due to the presence of the nonzero means in this expression, the collective spin depends upon both the collective pseudospins and the populations. From \erf{eq::fzno0},
\begin{align}
\hat{F}_z=v(\uparrow)\hat{X}_{\downarrow\uparrow}+w(\uparrow)\hat{X}_{\wr\downarrow}
+\expect{\hat{f}_z}_\uparrow N_\uparrow+\expect{\hat{f}_z}_\downarrow N_\downarrow+\expect{\hat{f}_z}_\wr N_\wr.
\end{align}
Because the populations are treated as c-numbers, however, they cancel from the uncertainty of the collective spin,
\begin{align}
\Delta\hat{F}_z=\hat{F}_z-\expect{\hat{F}_z}\approx v(\uparrow)\Delta\hat{X}_{\downarrow\uparrow}+w(\uparrow)\Delta\hat{X}_{\wr\downarrow}.
\end{align}
The collective spin variance,
\begin{align}
\Delta F_z^2=v(\uparrow)^2\Delta X_{\downarrow\uparrow}^2+
2v(\uparrow)w(\uparrow)\expect{\Delta\hat{X}_{\downarrow\uparrow}\Delta\hat{X}_{\wr\downarrow}}_S+
w(\uparrow)^2\Delta X_{\wr\downarrow}^2,
\end{align}
takes the exact same form as \erf{eq::FzVarNewObs}. 

\subsection{Faraday Interaction}
Because the Faraday interaction depends upon the collective spin $\hat{F}_z$, it takes a different form when the means $\expect{\hat{f}_z}_\uparrow$, $\expect{\hat{f}_z}_\downarrow$, and $\expect{\hat{f}_z}_\wr$ are nonzero. In terms of the collective pseudospins and populations of the ensemble,
\begin{align}\label{eq::HPfaraday}
\hat{H}=&\frac{\hbar\chi}{\Delta t}\hat{S}_3\Big(v(\uparrow)\hat{X}_{\downarrow\uparrow}+w(\uparrow)\hat{X}_{\wr\downarrow}\Big)\\\notag&+\frac{\hbar\chi}{\Delta t}\hat{S}_3\Big(\expect{\hat{f}_z}_\uparrow N_\uparrow+\expect{\hat{f}_z}_\downarrow N_\downarrow+\expect{\hat{f}_z}_\wr N_\wr\Big).
\end{align}
The first term in the expression is the familiar analogue of the Faraday interaction in the phase plane of the ensemble. The final term in this expression is absent when we assume that $\expect{\hat{f}_z}_\uparrow=\expect{\hat{f}_z}_\downarrow=\expect{\hat{f}_z}_\wr=0$. Even when these expectations are nonzero, this term has little effect on the state of the ensemble or the entanglement between the light and atoms. Because the populations are treated as c-numbers, the final term commutes with all ensemble observables and, thus, has no influence on the ensemble dynamics. For the light, this term generates a rotation of the light's polarization in the Poincar\'{e} sphere about $\hat{S}_3$. Because there is no projection noise contribution from the populations, the fluctuations in polarization induced by the ensemble on the light are unchanged by the presence of the final term. This term, therefore, has no role in creating entanglement between the light and ensemble. The final term also has no influence over preexisting entanglement. In the phase plane picture, the final term generates a rotation of the light's polarization, or equivalently a translation in phase space, that affects only the first order moments of observables. Because this term does not alter the variances or covariances,  the entanglement between the light and ensemble is unaffected. The second term can, therefore, be discarded without any impact on spin squeezing. The Faraday interaction takes the same form whether or not $\expect{\hat{f}_z}_\uparrow$, $\expect{\hat{f}_z}_\downarrow$, and $\expect{\hat{f}_z}_\wr$ are zero, producing identical dynamics.

\section{Differential Form of QND Measurement}
Previously in Sections \ref{sec::QNDmeas}, and \ref{sec::GaussSim}, we formulated the squeezing by QND measurement protocol through a series of updates on the covariance matrix of the light and ensemble. As previously noted, this treatment is only valid for fiducial states that satisfy \erf{eq::ModeCondition}. The dynamics of the ensemble state under QND measurement can also be written in the form of a stochastic master equation (SME) that is valid for any fiducial state. Combined with the generalizations of the coupled and transfer states from the previous section, this formalism can treat an ensemble prepared in any fiducial state. 

Under continuous measurement of $\hat{F}_z$, the SME describing the evolution of the ensemble state is given by 
\begin{align}\label{eq::planeSME}
d\hat{\rho}\big|_{\text{QND}}=\sqrt{\frac{\kappa}{4}}\mathcal{H}(\hat{\rho})dW+\frac{\kappa}{8}\mathcal{L}(\hat{\rho})dt,
\end{align}
where $\kappa=\chi^2\dot{N_L}$ is the measurement strength. Measurement backaction on the ensemble is taken into account by the superoperator 
\begin{align}
\mathcal{H}(\hat{\rho})=\hat{F}_z\hat{\rho}+\hat{\rho}\hat{F}_z-2\expect{\hat{F}_z}\hat{\rho}.
\end{align}
The Lindblad dissipator,
\begin{align}
\mathcal{L}(\hat{\rho})=[\hat{F}_z,[\hat{\rho},\hat{F}_z]],
\end{align}
describes the decoherence of the ensemble from light that goes unmeasured after interacting with the atoms. 

From the SME, we can determine the evolution of a covariance $\expect{\Delta\hat{O}\Delta\hat{A}}_S$ under continuous QND measurement of $\hat{F}_z$.
We first examine the evolution of the covariance due to measurement backaction,
\begin{align}\label{eq::backaction}
d\expect{\Delta\hat{O}\Delta\hat{A}}_S\big|_{\mathcal{H}}=&\frac{\sqrt{\kappa}}{4}\expect{\mathcal{H}(\hat{O}\hat{A}+\hat{A}\hat{O})}dW\\\notag
&-\sqrt{\frac{\kappa}{4}}\expect{\mathcal{H}(\hat{O})}\expect{\hat{A}}dW-\sqrt{\frac{\kappa}{4}}\expect{\hat{O}}\expect{\mathcal{H}(\hat{A})}dW\\\notag
&-\frac{\kappa}{4}\expect{\mathcal{H}(\hat{O})}\expect{\mathcal{H}(\hat{A})}dt.
\end{align}
Note that by the rules of It$\overline{\text{o}}$ calculus, differentials must be taken to second order \cite{JacSte06}. When $\hat{F}_z$ is written in the operator basis of collective pseudo-spins and populations in Eqs. (\ref{eq::Xdownup}) through (\ref{eq::PopWr}), all terms involving the populations vanish from this expression since they are treated as c-numbers. When $\hat{O}$ and $\hat{A}$ are collective pseudo spins, the first term in \erf{eq::backaction} contains third order moments of the collective pseudo spins. For an ensemble initially prepared in $\ket{\uparrow}^{\otimes N_A}$, the collective pseudo spins $\hat{X}_{\downarrow\uparrow}$, $\hat{Y}_{\downarrow\uparrow}$, $\hat{X}_{\uparrow\wr}$ and $\hat{Y}_{\uparrow\wr}$ are Gaussianly distributed. The collective pseudo spins $\hat{X}_{\wr\downarrow}$ and $\hat{Y}_{\wr\downarrow}$ are approximately Gaussianly distributed after population accumulates in the coupled state. Because the third order moments contain only Gaussian operators, they can be decomposed in terms of first and second order moments as
\begin{align}\label{eq::3rdGaussDecomp}
\frac{1}{6}\sum_{\text{perm}}\expect{\hat{O}\hat{A}\hat{Q}}=&\expect{\Delta\hat{O}\Delta\hat{Q}}_S\expect{\hat{A}}+
\expect{\Delta\hat{A}\Delta\hat{Q}}_S\expect{\hat{O}}+\expect{\Delta\hat{O}\Delta\hat{A}}_S\expect{\hat{Q}}
\\\notag&+\expect{\hat{O}}\expect{\hat{A}}\expect{\hat{Q}}.
\end{align}
The sum in the left hand side of this expression is taken over all permutations of $\hat{O}$, $\hat{A}$ and $\hat{Q}$. When the third order moments are decomposed under the Gaussian approximation, the equation of motion for the covariance reduces to the relatively simple expression
\begin{align}\label{eq::finalEoMQND7}
d\expect{\Delta\hat{O}\Delta\hat{A}}_S\big|_{\mathcal{H}}=&-\kappa\left(v(\uparrow)\expect{\Delta\hat{X}_{\downarrow\uparrow}\Delta\hat{O}}_S
+w(\uparrow)\expect{\Delta\hat{X}_{\wr\downarrow}\Delta\hat{O}}_S\right)\\\notag&
\times\left(v(\uparrow)\expect{\Delta\hat{X}_{\downarrow\uparrow}\Delta\hat{A}}_S
+w(\uparrow)\expect{\Delta\hat{X}_{\wr\downarrow}\Delta\hat{A}}_S\right)dt.
\end{align}
This differential equation nonlinearly couples the covariances of the collective pseudo-spins.

We next turn our attention to the evolution of the covariance due to the dissipative term in the SME,
\begin{align}
d\expect{\Delta\hat{O}\Delta\hat{A}}_S\big|_{\mathcal{L}}=&\frac{\kappa}{16}\expect{\mathcal{L}(\hat{O}\hat{A}+\hat{A}\hat{O})}dt-\frac{\kappa}{8}\expect{\mathcal{L}(\hat{O})}\expect{\hat{A}}dt\\\notag&-\frac{\kappa}{8}\expect{\hat{O}}\expect{\mathcal{L}(\hat{A})}dt\\\notag
=&\frac{\kappa}{8}\expect{\{[\hat{F}_z,\hat{O}],[\hat{A},\hat{F}_z]\}}dt-\frac{\kappa}{8}\expect{\Delta\hat{O}\Delta[\hat{F}_z,[\hat{A},\hat{F}_z]]}dt\\\notag
&-\frac{\kappa}{8}\expect{\Delta[\hat{F}_z,[\hat{O},\hat{F}_z]]\Delta\hat{A}}dt
\end{align}
When $\hat{O}$ and $\hat{A}$ are collective pseudospins, the final two terms in this expression involve either commutators or covariances of populations. Because the populations are treated as c-numbers, these terms are negligible. 

To obtain the full evolution of the covariance under QND measurement, we combine the contributions from measurement backaction and dissipation,
\begin{align}
d\expect{\Delta\hat{O}\Delta\hat{A}}_S\big|_{\text{QND}}=&d\expect{\Delta\hat{O}\Delta\hat{A}}_S\big|_{\mathcal{H}}+d\expect{\Delta\hat{O}\Delta\hat{A}}_S\big|_{\mathcal{L}}\\\notag
=&-\kappa\left(v(\uparrow)\expect{\Delta\hat{X}_{\downarrow\uparrow}\Delta\hat{O}}_S
+w(\uparrow)\expect{\Delta\hat{X}_{\wr\downarrow}\Delta\hat{O}}_S\right)\\\notag&
\times\left(v(\uparrow)\expect{\Delta\hat{X}_{\downarrow\uparrow}\Delta\hat{A}}_S
+w(\uparrow)\expect{\Delta\hat{X}_{\wr\downarrow}\Delta\hat{A}}_S\right)dt\\\notag
&+\frac{\kappa}{8}v(\uparrow)^2\expect{\{[\hat{X}_{\downarrow\uparrow},\hat{O}],[\hat{A},\hat{X}_{\downarrow\uparrow}]\}}dt\\\notag&
+\frac{\kappa}{8}v(\uparrow)w(\uparrow)\expect{\{[\hat{X}_{\wr\downarrow},\hat{O}],[\hat{A},\hat{X}_{\downarrow\uparrow}]\}}dt\\\notag
&+\frac{\kappa}{8}v(\uparrow)w(\uparrow)\expect{\{[\hat{X}_{\downarrow\uparrow},\hat{O}],[\hat{A},\hat{X}_{\wr\downarrow}]\}}dt\\\notag&+\frac{\kappa}{8}w(\uparrow)^2\expect{\{[\hat{X}_{\wr\downarrow},\hat{O}],[\hat{A},\hat{X}_{\wr\downarrow}]\}}dt.
\end{align}
For the complete evolution of the covariance, we must also include decoherence due to optical pumping, 
\begin{align}\label{eq::Covariance QNDop}
\frac{d}{dt}\expect{\Delta\hat{O}\Delta\hat{A}}_S=\frac{d}{dt}\expect{\Delta\hat{O}\Delta\hat{A}}_S\big|_{\text{QND}}
+\frac{d}{dt}\expect{\Delta\hat{O}\Delta\hat{A}}_S\big|_{\text{op}},
\end{align}
where the equation of motion for the covariances under optical pumping was derived previously in \erf{eq::CovarEvol2}. When the dynamics due to QND measurement and optical pumping are combined, we obtain a closed set of differential equations that couple the covariances between all pairs of collective pseudospins with the populations.

\section{Optimal Fiducial State}
The numerical results in Sec. \ref{sec::HPResults} demonstrate that the choice of fiducial state has a dramatic impact on the performance of Faraday-based squeezing protocols. Both the coherent squeezing and decoherence of the ensemble are highly dependent on the choice of fiducial state. Of the state preparations studied in Sec. \ref{sec::HPResults} in the presence of optical pumping, the $m_x=0$ preparation generated the most squeezing. A natural question is what fiducial state maximizes squeezing in the presence of optical pumping?  Through the equations of motion for the covariances derived in the previous section, we can utilize a numerical search to find the fiducial state that maximizes the squeezing when the ensemble is subject to QND measurement and optical pumping.

\subsection{Optimization}
In the optimization, we seek to minimize the peak squeezing parameter over all possible fiducial states. For an ensemble of spin-$f$ atoms, we parametrize the fiducial state in terms of $2(2f+1)$ real numbers as $\ket{\uparrow}=\sum_{m_z=-f}^{f}(p_{2m_z-1}+ip_{2m_z})\ket{f,m_z}$, where $\sum_{m_z=-f}^{f}(p_{2m_z-1}^2+p_{2m_z}^2)=1$. The choice of fiducial state generates a unique set of coupled differential equations describing the evolution of the variances and populations. The evolution of the covariances under QND measurement and optical pumping is given by \erf{eq::Covariance QNDop}. There are 21 equations of motion describing the covariances, corresponding to each pair of collective pseudospins plus the variances of each collecective pseudospin. Because QND measurement negligibly effects the populations, we consider only the evolution of the populations due to optical pumping, which is given in \erf{eq::FirstOrderPopulations2}. For each fiducial state, the set of differential equations giving the evolution of the covariances and populations can be solved to determine the squeezing parameter after post-processing. In the case where the fiducial state does not support a transfer state, such as for $\ket{\uparrow_{\text{cat}}}$, the squeezing parameter is given by the expression in \erf{eq::noTransfer}, which is then minimized over all $t$ to determine the peak squeezing. When the fiducial state does support a transfer state, both squeezing parameters in Eqs. (\ref{eq::unnormSqParam2}) and (\ref{eq::noTransfer}) are minimized over all $t$ with the smaller of the two representing the peak squeezing. This ensures that the transfer state is only preserved if it is beneficial to spin squeezing.

For integer and half integer spins from $f=1$ to $f=5$, we minimize the squeezing parameter over all fiducial states and time using an interior point algorithm in MATLAB \cite{InteriorPt}. For the post-processing partial isometry, we choose the map to the SCS preparation. The exact form of the squeezing parameter that is minimized is given in \erf{eq::SCSSqParamOP}. In the case that the transfer state is discarded,  \erf{eq::SCSSqParamOP} is minimized with $\Delta\hat{X}_{\wr\downarrow}^{2}\rightarrow N_\downarrow/2$, $\expect{\Delta\hat{X}_{\downarrow\uparrow}\Delta\hat{X}_{\wr\downarrow}}_S\rightarrow0$ and $N_\wr\rightarrow 0$. 

Because the landscape of the squeezing parameter as a function of time and the fiducial state has not been studied, little is known about the presence of local minima or saddle points. The existence of numerous local minima is likely, as the convergence of the algorithm is highly dependent upon the initial fiducial state and time seeded to the interior point algorithm. To compensate for this, over 100 randomly chosen fiducial states are selected for each $f$ and seeded to the interior point algorithm. The minimal value of the squeezing parameter that results from this large set of initial seeds is taken to be the optimal solution. It should be emphasized that because so little about the optimization landscape is known, we cannot assert that the minimum squeezing parameters found by our algorithm are absolute minima.

\subsection{Results}

\begin{figure}
\centering
\includegraphics[scale=.47]{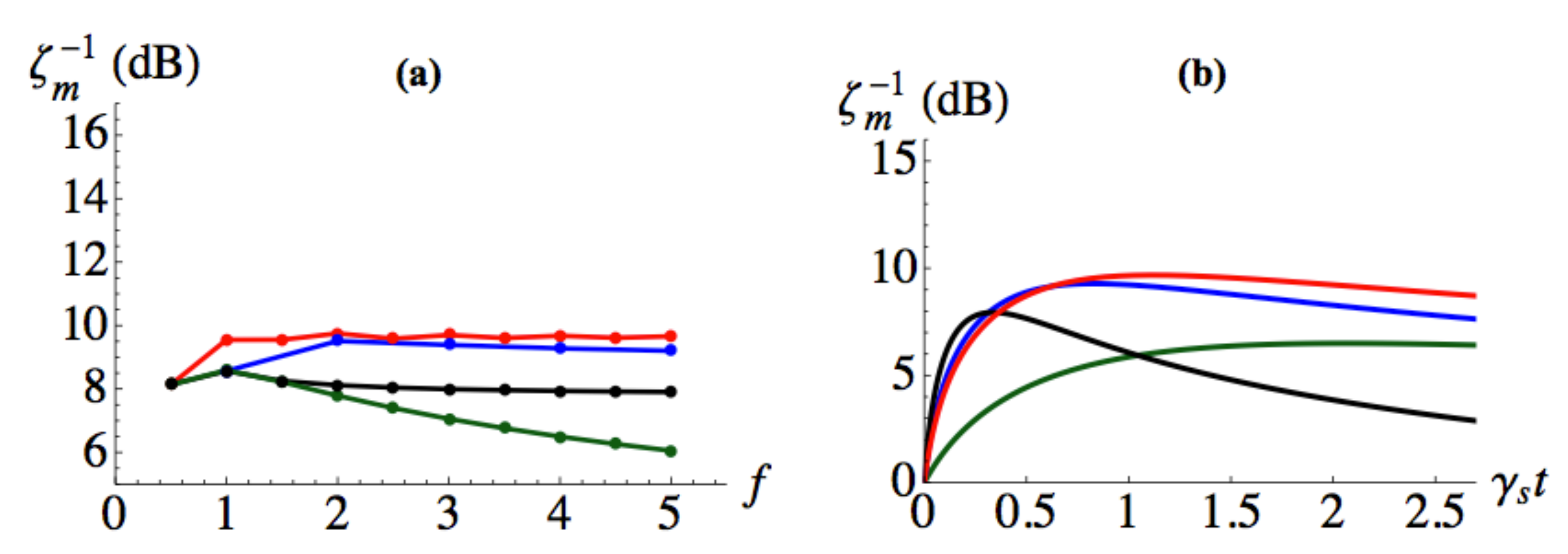}
\caption{Performance of the fiducial states found by numerical optimization compared to the other state preparations. Plot (a) shows the peak squeezing generated by various state preparations for each $f$. Plotted are the state preparations found by numerical optimization for each $f$ (red), the SCS preparation (green), the cat preparation (black) and the $m_x=0$ (blue). Plot (b) shows the performance of the state preparations for $f=4$ versus time.} \label{fig::Optf}
\end{figure}

The fiducial states found by the numerical optimization outperform the state preparations we have considered thus far, as shown in Fig. \ref{fig::Optf}. Figure \ref{fig::Optf} (a) depicts the peak squeezing determined by the numerical optimization for each $f$ along with the peak squeezing generated by the SCS, cat, and $m_x=0$ state preparations. The numerical optimization significantly outperforms the state preparations for smaller $f$, but for $f\geq 2$ it performs only slightly better than $m_x=0$.  Figure \ref{fig::Optf} (b) shows the squeezing generated by the fiducial state found by the numerical optimization for $f=4$ along with the state preparations. While the numerical optimization only improves upon the $m_x=0$ preparation by .3 dB, the fiducial state determined by numerical optimization appears more robust to optical pumping with its squeezing decaying more slowly with time. 

\begin{figure}
 \centering
\includegraphics[scale=.4]{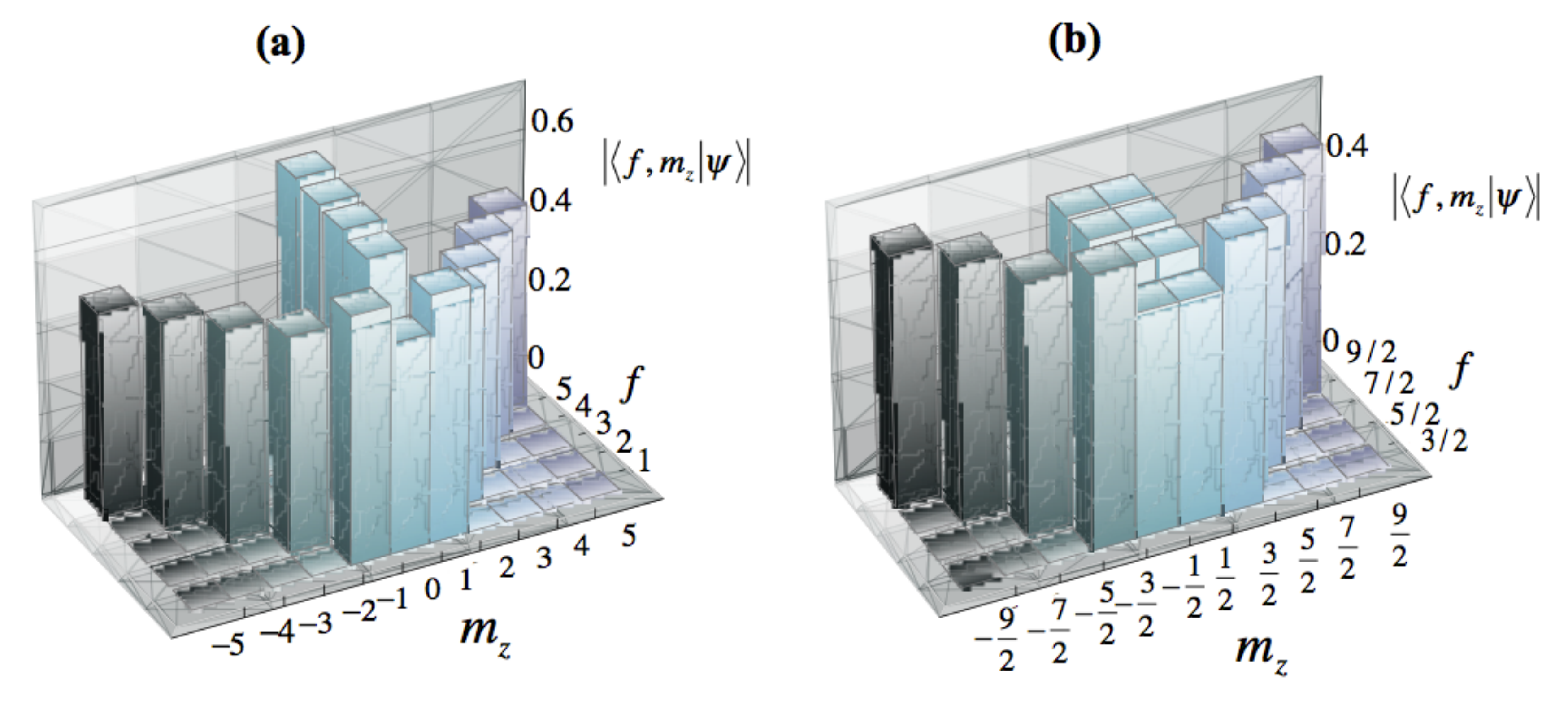}
\caption{Fiducial states found by numerical optimization for (a) integer $f$ and (b) half integer $f$. Each bar in position $(f, m_z)$ corresponds to the $\hat{f}_z$ eigenstate $\ket{f,\,m_z}$. The $z$-axis gives the weight, $|\bra{f,\,m_z}\psi\rangle|$, on each eigenstate for a fiducial state $\ket{\psi}$.}\label{fig::OptStates}
\end{figure}

We now examine the fiducial states found by the numerical optimization, which generate the squeezing depicted in Fig. \ref{fig::Optf}. Figure \ref{fig::OptStates} graphically represents the fiducial states that minimize the squeezing parameter for both integer and half-integer $f$. Interestingly, the fiducial states found through the numerical optimization all take a similar form. For integer $f$, the fiducial states are approximately of the form 
 \begin{align}
\ket{\uparrow_\text{int}}=c_f\ket{f,m_z=f}+c_0\ket{f,m_z=0}+c_{f}\ket{f,m_z=-f}. 
 \end{align}
For half-integer $f$, the fiducial states are approximately of the form 
 \begin{align}
\ket{\uparrow_\text{half}}=&c_f\ket{f,m_z=f}+c_{1/2}\ket{f,m_z=1/2}\\\notag&+c_{1/2}\ket{f,m_z=-1/2}+c_{f}\ket{f,m_z=-f}.
 \end{align}
 Here, $c_f$, $c_0$ and $c_{1/2}$ are real constants. 
 
 In both cases, the fiducial states are the superposition between a state with a large variance in $\hat{f}_z$ and one with less or no variance in $\hat{f}_z$.  For integer $f$, these states are $\ket{\uparrow_\text{cat}}$ and $\ket{f,m_z=0}$, the superposition of which yield a variance $(\Delta f_z^2)_{\uparrow_{\text{int}}}=2|c_f|^2f^2$. For half-integer $f$, the superposition consists of $\ket{\uparrow_\text{cat}}$ and $(\ket{f,m_z=1/2}\\+\ket{f,m_z=-1/2})/\sqrt{2}$, resulting in a variance of $(\Delta f_z^2)_{\uparrow_{\text{half}}}=2|c_f|^2f^2+|c_{1/2}|^2/2$. This superposition is likely caused by competition between the coherent squeezing interaction and the spin flip rate. The strength of the coherent squeezing interaction is governed by the collective spin coupling constant, given in \erf{eq::2ndXI}, which is proportional to $(\Delta f_z^2)_\uparrow$. From the master equation in \erf{eq::MasterRotating}, the spin flip rate also depends upon $(\Delta f_z^2)_\uparrow$,
 \begin{align}
 \Gamma_{\text{flip}}&=\gamma_s\bra{\downarrow}\mathcal{D}\left(\ket{\uparrow}\bra{\uparrow}\right)\ket{\downarrow}\\\notag
 &=\frac{g_f\gamma_s}{9}\left((\Delta f_z^2)_\uparrow+\frac{1}{2}|\bra{\uparrow}\hat{f}_y\ket{\downarrow}|^2
 +\frac{1}{2}|\bra{\uparrow}\hat{f}_x\ket{\downarrow}|^2\right).
  \end{align}
 Both the coherent squeezing interaction and spin flip rate increase with the variance of the fiducial state in $\hat{f}_z$. The superposition between states with large and small variances in $\hat{f}_z$ seem to balance the coherent and incoherent dynamics. The state with large variance boosts squeezing, while the state with smaller variance minimizes decoherence due to spin flips.
 
\begin{figure}
\centering
\includegraphics[scale=.49]{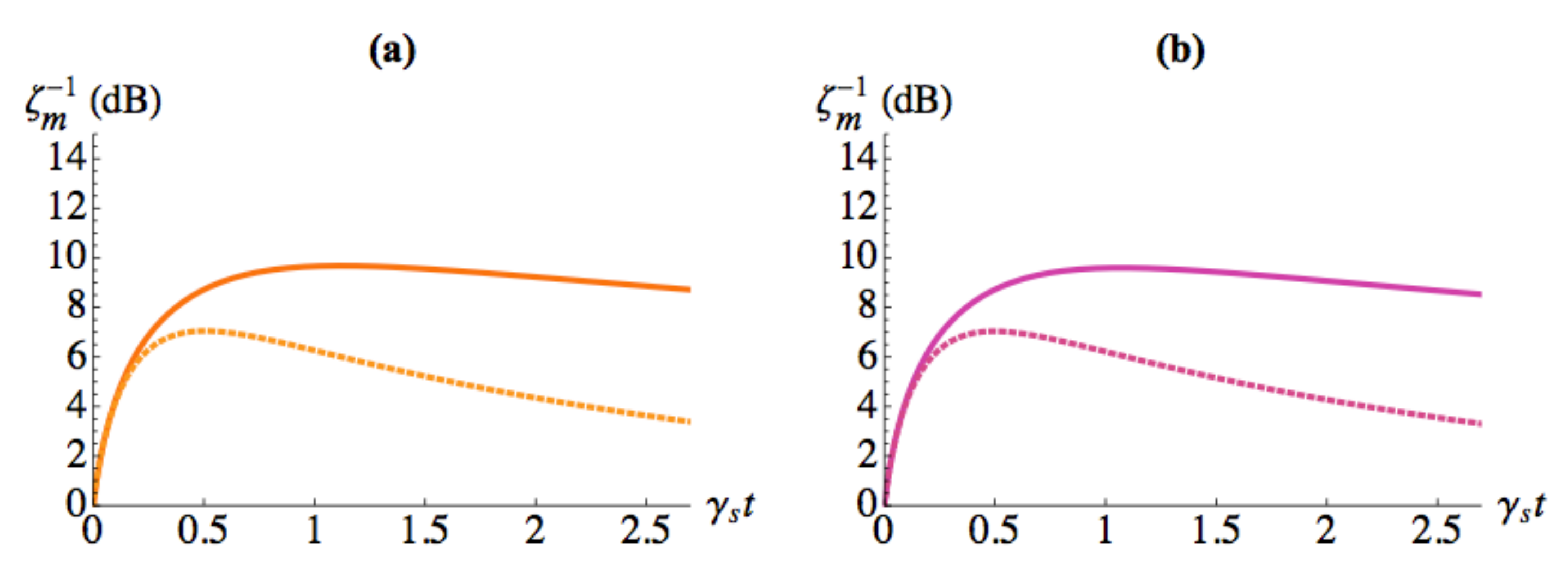}
\caption{Performance of fiducial states found by numerical optimization for (a) $f=4$ and (b) $f=7/2$. The squeezing generated by the state preparations is plotted versus time with the transfer state preserved (solid) and with the transfer state eliminated (dashed).}\label{fig::OptTransfer}
\end{figure}
 
Another interesting property of these fiducial states is the existence of a beneficial transfer state. For simplicity, we focus on the case where $f$ is an integer. From \erf{eq::coupledDef}, the coupled state is
\begin{align}
\ket{\downarrow_{\text{int}}}=\frac{1}{\sqrt{2}|c_f|}\left(c_f\ket{f,m_z=f}-c_f\ket{f,m_z=-f}\right).
\end{align}
The coupled state enables us to deduce the transfer state from \erf{CoherenceState}, 
\begin{align}
\ket{\wr_{\text{int}}}=&\frac{c_f|c_0|}{\sqrt{2}|c_f|}\left(\ket{f,m_z=f}+\ket{f,m_z=-f}\right)\\\notag
&-\frac{\sqrt{2}|c_f|c_0}{|c_0|}\ket{f,m_z=0}.
\end{align}
Recall that the transfer state is beneficial to spin squeezing if the conditions in Eqs. (\ref{eq::TofUp}) and (\ref{eq::up2squiggle}) are satisfied by the fiducial, coupled and coherence states. For the master equation in the rotating frame, the first condition is equivalent to
\begin{align}\label{eq::NewT}
T(\uparrow)=\text{Re}[\bra{\wr}\mathcal{D}\left(\ket{\downarrow}\bra{\uparrow}\right)\ket{\downarrow} >0.
\end{align}
This ensures that negative correlations are preserved due to the entanglement involving the transfer and the coupled states. The second condition is equivalent to
\begin{align}\label{eq::NewN}
N(\uparrow)=\bra{\wr}\mathcal{D}(\ket{\uparrow}\bra{\uparrow})\ket{\wr}=0.
\end{align}
This ensures the transfer state does not contribute to the noise injection. For $f>1$, the states $\ket{\uparrow_\text{int}}$, $\ket{\downarrow_\text{int}}$ and $\ket{\wr_\text{int}}$ satisfy these conditions, demonstrating a beneficial transfer of coherence. Note that when $c_0=0$, $\ket{\uparrow_\text{int}}$ is a cat state with no transfer of coherence. The presence of a term with weight on the eigenstate $\ket{f,m_z=0}$ ensures that a beneficial transfer state exists, making $\ket{\uparrow_\text{int}}$ more robust to decoherence than the cat state. 

For  $f>3/2$, $\ket{\uparrow_\text{half}}$, $\ket{\downarrow_\text{half}}$ and $\ket{\wr_\text{half}}$ satisfy conditions (\ref{eq::NewT}) and (\ref{eq::NewN}), implying that a transfer of coherence likewise exists in the half-integer case. The impact of the transfer state on the squeezing generated by both $\ket{\uparrow_\text{int}}$ and $\ket{\uparrow_\text{half}}$ is shown in Fig. \ref{fig::OptTransfer} for (a) $f=4$ and (b) $f=7/2$. Preserving the transfer state increases the squeezing produced by both preparations by 2.6 dB.  The presence of the transfer state additionally slows the decay of the squeezing due to optical pumping. Transfers of coherence are a significant factor in the performance of the fiducial states found through numerical optimization.

\chapter{Squeezing with Paraxial Beams}\label{paraxial}
In the previous discussion and in most work on spin squeezing involving free space atomic ensembles, the light is assumed to be a plane wave that couples identically to all atoms. While it is possible to create plane-like waves in a laboratory when the beam area is large compared to the spatial extent of the ensemble, this results in poor mode matching between their respective radiation patterns. Because generating interatomic entanglement requires the atoms to be as indistinguishable as possible, mode matching is essential for maximizing spin squeezing. In this chapter, we replace plane waves with focused paraxial beams and derive equations of motion for the  ensemble observables. Because paraxial beams, unlike plane waves, are spatially inhomogeneous, the geometries of both the atomic ensemble and probe beam are crucial factors in determining the degree of atom-light coupling. As we have seen in the preceding chapters, the peak squeezing that we can achieve depends on the balance between coherent squeezing and decoherence.  Since decoherence is also spatially inhomogeneous, the optimal geometries for squeezing depend both on spatial mode matching and optimizing this balance. Using our formalism, we find the optimal ensemble and beam geometries for maximal mode matching and generation of spin squeezing.

\section{Paraxial Beams}
In a realistic implementation of spin squeezing in a free space atomic ensemble, the light emerging from the probe laser is well approximated as a Gaussian beam in the TEM$_{00}$ spatial mode. The intensity of a beam in the TEM$_{00}$ mode is spatially varying and proportional to the square of the mode function,
\begin{align}\label{eq::TEM00mode}
u_{00}(\mathbf{r}_\bot,z)= \frac{w_0}{w(z)}e^{-\frac{\left|\mathbf{r}_\bot\right|^2}{[w(z)]^2}}e^{\frac{ik_0 \left|\mathbf{r}_\bot\right|^2}{2 R(z)}}e^{-i\Phi(z)}.
\end{align}
Here, the beam waist, radius of curvature, and Gouy phase are given by
\begin{align}
w(z) &= w_0 \sqrt{1+(z/z_R)^2} ,  \\
R(z) &= z\left(1+(z_R/z)^2\right)\;\;\;\text{and}   \\
\Phi(z) &= \tan^{-1}(z/z_R), 
\end{align}
for Rayleigh range $z_R \equiv k_0 w_0^2/2$ and minimum beam waist $w_0$. 

The field of the probe beam in the TEM$_{00}$ mode satisfies the paraxial wave equation. For a spatially elongated ensemble, the light coherently scattered by the atoms is also  paraxial. With this motivation, we partition the electric field scattered by the ensemble into two components,
	\begin{align} \label{Eq::ModeDecomp}
		\hat{\mathbf{E}}_{\text{scat}}^{(+)}(\mathbf{r}, t) = \hat{\mathbf{E}}_{\rm para}^{(+)}(\mathbf{r}, t) + \hat{\mathbf{E}}_{\rm diff}^{(+)}(\mathbf{r}, t),
	\end{align}
where $\hat{\mathbf{E}}_{\rm para}^{(+)}(\mathbf{r}, t)$ is the field of the coherently scattered paraxial light and $\hat{\mathbf{E}}_{\rm diff}^{(+)}(\mathbf{r}, t)$ is the field of the scattered non-paraxial light, which includes light spontaneously emitted by the atoms. Solutions to the paraxial wave equation can be decomposed into sums of orthogonal Gaussian mode functions, $u_{pl} (\mbf{r})$. Employing such a decomposition, the positive frequency component of the paraxial field becomes
	\begin{equation}\label{eq::PosFreqComp}
		\hat{\mbf{E}}^{(+)}_{\text{para}}(\mbf{r},t) = \sum_{pl, a=x, y} \sqrt{\frac{2 \pi \hbar \omega_0}{c A}}\, \mbf{e}_a \, \hat{a}_{pl, a}(z,\tau) \,  u_{pl}(\mbf{r},z)  e^{i(k_0 z - \omega_0 t)},
	\end{equation}
where $\tau = t-z/c$ is the retarded time along the propagation direction, $A$ is the transverse beam area and $a$ denotes the beam's polarization. As illustrated by the positive frequency component, each mode function describes the field amplitude of an orthogonal transverse spatial mode, which we denote by the indices $pl$. The operator $\hat{a}_{pl,a}(z_i,t)$ represents the annihilation of a photon in the transverse spatial mode $pl$ and the polarization mode $a$.

One of the most valuable features of the mode functions $u_{pl} (\mbf{r})$ is that they form a orthogonal basis over functions of the transverse coordinate, $\mbf{r}_\perp$, at a fixed $z$. Here, we have partitioned the spatial coordinate $\mathbf{r}$ into $\mathbf{r}_\bot$ and $z$, where $z$ is the longitudinal coordinate parallel to the beam's axis of propagation. The mode functions are chosen to be dimensionless, obeying the orthogonality and completeness relations
\begin{align} \label{Eq::TransverseOrthogonality}
\int d^2 \mbf{r}_\perp u^*_{pl} (\mbf{r}_\perp , z) u_{p'l'} &(\mbf{r}_\perp , z) = A \, \delta_{p, p'} \delta_{l, l'}
\end{align}
and
\begin{align}\label{Eq::TransverseCompSameZ}
\sum_{p,l} u_{pl}(\mbf{r}_\perp , z) u^*_{pl}(\mbf{r}_\perp' , z)  &=  A \, \delta^{(2)}(\mbf{r}_\perp-\mbf{r}_\perp').
\end{align} 
These conditions imply that any function of $\mbf{r}_\perp$ at fixed $z$ can be expressed as a sum of mode functions. This property will prove vital in deriving the equations of motion for the system.

\section{Three-Dimensional Atomic Ensemble}
The spatial distribution of the ensemble is critical in determining properties of the atom-light interface. As we will later discuss, it is the geometry of the ensemble that determines the decomposition of the coherently scattered field into transverse spatial modes. Additionally, because the intensity of the probe varies throughout the ensemble, properties such as the strength of the Faraday interaction, the rate of squeezing and the rate of optical pumping are all position dependent. Whereas in the plane wave case we can ignore the position degrees of freedom of the atoms, in the paraxial case we must associate to each atom $i$ a spatial coordinate $\mathbf{r}_i=(\mbf{r}_{\bot i},z_i)$. To simplify calculations, we will often approximate the discrete distribution of atoms in the ensemble as a continuous density, $\eta(\mathbf{r})$.  For instance, in the simulations presented in Sec. \ref{sec::MultimodeResults}, we treat the density of the ensemble as a cylindrically symmetric Gaussian cloud with 
	\begin{align} \label{Eq::AtomicDistribution}
		\eta(\mathbf{r}) = \eta_0 \exp \left( - 2\frac{\rho^2}{\sigma_\perp^2}  - 2\frac{z^2}{\sigma_z^2} \right).
	\end{align}
In this expression, $\sigma_\perp^2$ and $\sigma_z^2$ are the transverse and longitudinal $1/e^2$ variances and $\eta_0$ is the peak density at the center of the cloud. This approximates the density of a large cloud of cold atoms confined in a dipole trap. As in the classical propagation of a wave through a gas, it is this continuous density that is responsible for the index of refraction; the density fluctuations associated with the discrete positions are responsible for diffuse scattering.

\section{The Multimode Faraday Interaction}\label{sec::MMFaraday}
As in the case of the Faraday interaction with plane waves, the collective spin of the atomic ensemble causes the polarization of the light to rotate from $x$ to $y$. Unlike the plane wave case, however, the ensemble radiates the light into a superposition of spatial modes outside the TEM$_{00}$ mode of the probe. These effects are described by the multimode Faraday interaction \cite{Baragiola14}
\begin{align}\label{Eq::multiFaraday}
 		\hat{H} \!= \! -i  \frac{\hbar \sqrt{\kappa}}{2} \! \sum_{i,p,l} \! \Big[ & \beta^*_{pl}(\mbf{r}_{\perp i}, z_i ) \hat{a}_{pl,y}(z_i,t) \!- \! \mbox{h.c.} \Big] \! \hat{f}_z^{(i)}, 
\end{align}
where $\beta_{pl}(\mbf{r}_\perp, z)$ is a product of mode functions given by
\begin{align} \label{Eq::Beta}
		\beta_{pl}(\mbf{r}_\perp, z)  \equiv u^*_{pl}(\mbf{r}_\perp, z) u_{00}(\mbf{r}_\perp, z). 
\end{align}
This Hamiltonian corresponds to annihilation of a photon in the fundamental mode with x-polarization and creation in another paraxial mode with y-polarization.  The final result
follows from a Holstein-Primakoff approximation treating the fundamental mode as a macroscopically occupied c-number.  Equation (\ref{Eq::multiFaraday}) is summed over all transverse modes $pl$ and atoms in the ensemble, $i$, with spatial coordinates $(\mbf{r}_{\perp i}, z_i )$. Note that rather than being proportional to the collective spin operator $\hat{F}_z$, like the plane wave Faraday interaction in \erf{eq::FaradayDef}, the multimode Faraday interaction depends upon a sum of the internal spin operators weighted by the functions $\beta_{pl}(\mbf{r}_\perp, z)$. These weighted sums, referred to as spin waves,  are given by
\begin{align}\label{eq::SpinWaveDef}
\hat{F}_z^{pl}=\sum_{i} \beta_{pl}(\mbf{r}_{\perp i}, z_i)\hat{f}_z^{(i)}.
\end{align}
Each spin wave is the effective collective spin of atoms absorbing photons in the TEM$_{00}$ spatial mode of the probe beam and radiating into the transverse mode $pl$. A particularly important spin wave is the so-called fundamental spin wave, $\hat{F}_z^{00}$, which represents the effective collective spin of the atoms that re-radiate into the TEM$_{00}$ mode of the probe.

As in the plane wave case, squeezing of the ensemble can be created by QND measurement of the light. The presence of transverse spatial modes adds subtlety to this process, however. At the plane of the polarimeter, the probe light in the TEM${_{00}}$ mode interferes with the coherently scattered light, which is in a superposition of transverse spatial modes.  The probe light destructively interferes with all spatial mode components of $\hat{\mathbf{E}}_{\rm para}^{(+)}(\mathbf{r}, t)$, except for $pl=00$. Consequently, the detector only measures scattered light in the TEM$_{00}$ mode. This light carries information about the atomic ensemble just as it did in the plane wave case. The polarimetry signal takes the form,
\begin{equation}  \label{Eq::XplOut}
\hat{X}_{00}(\Delta t) =  \hat{X}_{00}(0) +\sqrt{\frac{\kappa}{2}} \hat{F}^{00}_z(0) .
\end{equation}
Because the TEM$_{00}$ mode only couples to the fundamental spin wave, the signal depends upon $\hat{F}^{00}_z$ alone. As in the plane wave case, a measurement of $\hat{X}_{00}$ extracts information about the fundamental spin wave and creates squeezing by measurement backaction. The strength of the measurement backaction depends on the entanglement between the light and ensemble created by the Faraday interaction. Analogous to the plane wave case discussed in Sec. \ref{sec::EntangleFaraday}, the entanglement between the light and ensemble increases with the projection noise fluctuations of the fundamental spin wave. Thus, we define a paraxial collective spin coupling constant analogous to $\xi$ from \erf{eq::2ndXI}, 
\begin{equation}\label{Eq::CouplingStrength}
\xi_{\text{para}} = \frac{\int_0^{\Delta t}dt(\Delta X_{00}^2)_{PN}}{\int_0^{\Delta t}dt(\Delta X_{00}^2)_{SN}}= \left( \Delta F_z^{00}(0)\right)^2  \kappa \Delta t.  
\end{equation}	
Here, $(\Delta X_{00}^2)_{PN}$ and $(\Delta X_{00}^2)_{SN}$ are the projection noise variance and shot noise variance of the signal. In the definition of the paraxial collective spin coupling constant, these signal variances in the detector are integrated over a time $\Delta t$. As in the plane wave case, backaction is maximized when the projection noise of the ensemble dominates over the shot noise of the light.

From the definition of the paraxial collective spin coupling constant, measurement backaction increases with the initial projection noise fluctuations of the fundamental spin wave. In a departure from the plane wave case, these fluctuations depends not just upon the fiducial state, but upon the probe beam and the density of the atomic ensemble. For an ensemble with each atom prepared in the fiducial state $\ket{\uparrow}$,
\begin{align}
(\Delta F_z^{00}(0))^2&=\sum_{i,j}\beta_{00}(\mbf{r}_i)\beta_{00}(\mbf{r}_j)\bra{\uparrow_i\uparrow_j}\Delta\hat{f}_z^{(i)}\Delta\hat{f}_z^{(j)}\ket{\uparrow_i\uparrow_j}\\&=\sum_i\beta_{00}(\mbf{r}_i)^2(\Delta f_z^2)_\uparrow.
\end{align}
By treating the atomic ensemble as a continuous density distribution, this expression becomes
\begin{align}
(\Delta F_z^{00}(0))^2=\int d^3\mbf{r} \, \eta (\mbf{r})\beta_{00}(\mbf{r})^2(\Delta f_z^2)_\uparrow=N_{\text{eff}}^{00\,(2)}(\Delta f_z^2)_\uparrow.
\end{align}
We refer to $N_{\text{eff}}^{00\,(2)} $ as an effective atom number in the fundamental mode. 

The general expression for the $K$th effective atom number in the transverse mode $pl$ is given by
\begin{align}
N_{\text{eff}}^{pl\,(K)} =\int d^3\mbf{r} \, \eta (\mbf{r})\beta_{pl}(\mbf{r})^K.
\end{align}
The effective atom numbers are associated with different physical quantities. The initial variances of the spin waves are related to the $K=2$ effective atom numbers, while the means are related to $K=1$. The mean of the spin wave operator 
$\hat{O}^{pl}=\sum_i\beta_{pl}(\mbf{r}_i)\hat{o}^{(i)}$ is given by
\begin{align}\label{eq::MeanSpinWaveN1}
\expect{\hat{O}^{pl}}=\int d^3\mbf{r} \, \eta (\mbf{r})\beta_{pl}(\mbf{r})\expect{\hat{o}}=N_{\text{eff}}^{pl\,(1)}\expect{\hat{o}}.
\end{align}

Because the collective spin coupling constant is proportional to $N_{\text{eff}}^{00\,(2)}$, this effective atom number has special significance. The parameter $N_{\text{eff}}^{00\,(2)}$ quantifies the effective number of atoms that are radiating light back into the probe mode. When $N_{\text{eff}}^{00\,(2)}$ is large, the ensemble is well mode matched and the atoms are less distinguishable, leading to enhanced interatomic entanglement and spin squeezing. In a parallel sense, the effective atom numbers $N_{\text{eff}}^{pl\,(2)}$ for $pl\neq 00$ quantify the strength of the scattered field in modes outside of the TEM$_{00}$ mode of the probe. Having large values of $N_{\text{eff}}^{pl\,(2)}$ for $pl\neq 00$ is a symptom of poor mode matching. 
In analogy with the plane wave case, we define the effective optical density in terms of $N_{\text{eff}}^{00\,(2)}$ as
\begin{align}\label{eq::ODeff}
O\!D_{\text{eff}}=N_{\text{eff}}^{00\,(2)}\frac{\sigma_0}{A}.
\end{align}
This quantifies the strength of the mode matched  coupling between the ensemble and probe. Note that the values of all effective atom numbers, and thus $O\!D_{\text{eff}}$, are governed by the density of the atomic ensemble and the geometry of the probe.

\section{Paraxial Spin Squeezing Parameter}
For a Gaussian probe beam in the TEM$_{00}$ mode, QND measurement creates squeezing in the associated spin wave $\hat{F}_z^{00}$.  Because we are interested in utilizing this spin wave squeezing for applications in metrology, we quantify its strength through the angular resolution of a magnetometer, $\Delta\phi$, introduced in Sec. \ref{sec::QuantSqueezing}. In a situation analogous to the plane wave case,  the ensemble is initially polarized along $x$. We wish to deduce the magnitude of a rotation, $\phi$, about $y$ by measuring the collective spin of the ensemble along $z$. This can be accomplished by the Faraday effect, i.e. shining a linearly polarized probe through the ensemble and measuring the rotation angle of the polarization. For a plane wave probe propagating in the $z$ direction, the rotation of the light's polarization is proportional to $\hat{F}_z$ by the Faraday interaction.  For a Gaussian probe in the TEM$_{00}$ mode propagating along $z$, the rotation of the light's polarization is proportional to the fundamental spin wave, $\hat{F}_z^{00}$. The precision with which we can determine $\phi$, thus, depends on the uncertainty $\Delta F_z^{00}$.  The precision also depends on  a mean spin wave component orthogonal to $z$, analogous to $\expect{\hat{F}_x}$ in the plane wave case. For an ensemble initially polarized along $x$, the signal is the mean spin of the effective number of atoms addressed by the probe light in the TEM$_{00}$ mode, 
\begin{align}
\expect{\hat{F}_x^{00}}=\sum_i\beta_{00}(\mbf{r}_i)\expect{\hat{f}_x^{(i)}}=N_{\text{eff}}^{00\,(1)}\expect{\hat{f}_x^{(i)}}.
\end{align}
Similar to Sec. \ref{sec::QuantSqueezing}, the complete expression for the angular resolution of the magnetometer is
\begin{align}\label{angRes00}
\Delta\phi=\frac{\Delta F_z^{00}}{\expect{\hat{F}_x^{00}}}.
\end{align}
For an ensemble prepared in a spin coherent state, the resolution is
\begin{align}
\Delta\phi_{SCS}=\frac{1}{N_{\text{eff}}^{00\,(1)}}\sqrt{\frac{N_{\text{eff}}^{00\,(2)}}{2f}}.
\end{align}
The Wineland squeezing parameter compares the performance of the state we wish quantify with the performance of a spin coherent state. For the paraxial case, the Wineland squeezing parameter becomes
\begin{equation} \label{Eq::SqueezingParam}
		\zeta_{\text{para}} \equiv \left(\frac{\Delta \phi}{\Delta \phi_{\text{SCS}}}\right)^2 = 2f \frac{ \big(N^{(1)}_\eff \big)^2 }{N_\eff^{(2)}} \frac{\left(\Delta F_z^{00}\right)^2}{\expect{\hat{F}_x^{00}}^2}.
	\end{equation}
This expression quantifies the degree to which QND measurement improves the angular resolution of a magnetometer over an initially spin coherent ensemble. Note that unlike the squeezing parameter for the plane wave case in \erf{eq::SqParameter}, the paraxial squeezing parameter depends on the geometry of the ensemble as well as the state of the collective spin due to the presence of the different effective atom numbers.

\section{Equation of Motion for Spin Waves}
To determine the paraxial squeezing parameter, we must track the evolution of the spin waves. We follow a procedure similar to Chapters \ref{sec::ModHPCovar} and \ref{sec::Beyond}, deriving a set of coupled differential equations that describe the behavior of the ensemble observables or spin waves, in this case. The spin waves evolve due to both optical pumping of the ensemble and continuous QND measurement.   For simplicity, in the text we present only the equations of motion for the spin waves in the $f=1/2$ case, where the ensemble is prepared in a SCS. The fully generalized case for $f\geq1/2$ and arbitrary fiducial states is given in Appendix \ref{sec::fSpinWaves}.

As mentioned previously, decoherence from optical pumping occurs when light from the probe beam is diffusely scattered. The optical pumping induced by a paraxial beam on the ensemble acts locally on each atom. It is described by the master equation 
\begin{align}\label{eq::FullMasterParaxial}
\frac{d\hat{\rho}}{dt}\Big|_{\text{op}}=\sum_i\gamma_s(\mathbf{r}_i)\mathcal{D}^{(i)}(\hat{\rho}),
\end{align}
where $\mathcal{D}^{(i)}$ is the superoperator given in \erf{eq::MasterD} in the rotating frame defined by the bias magnetic field along the $z$-axis. This master equation is almost identical to the master equation defined in \erf{eq::MasterRotating} for the plane wave case, except that the photon scattering rate is not uniform throughout the atomic ensemble. This is a consequence of the spatially inhomogeneous intensity of the probe, $I_{00}(\mathbf{r})$. The photon scattering rate, which increases with the intensity of the probe, is given by
\begin{align} \label{Eq::LocalScatRate}
		\gamma_s(\mathbf{r}) = I_{00}(\mathbf{r}) \frac{\sigma_0}{\hbar \omega} \frac{\Gamma^2}{4 \Delta^2} = \gamma_{0}\beta_{00}(\mathbf{r}),
	\end{align}
where $\gamma_0$ is the peak scattering rate at the ensemble's center. As a result of the inhomogeneous probe, atoms at positions in the ensemble with different intensities undergo different rates of optical pumping.

Continuous QND measurement of the fundamental spin wave is described by the stochastic master equation,
\begin{align} \label{Eq::HomodyneSME}
d \hat{\rho}&= \sqrt{ \frac{\kappa}{4} } \mathcal{H}_{00}(\hat{\rho}) \, dW + \frac{\kappa}{4} \sum_{p,l} \mathcal{L}_{pl}(\hat{\rho}) \, dt,
\end{align}
which takes a form similar to the SME in the plane wave case in \erf{eq::planeSME}. A detailed derivation of the paraxial SME is provided in Ref. \cite{Baragiola14}. The effect of measurement backaction on the ensemble is taken into account by the superoperator $\mathcal{H}_{00}(\hat{\rho}) $, where
\begin{align}  \label{Eq::HSuperoperator}
		\mathcal{H}_{00}(\hat{\rho}) = \hat{F}^{00}_z \hat{\rho} + \hat{\rho} \hat{F}^{00 \dag}_z - \text{Tr}((\hat{F}^{00}_z + \hat{F}^{00\dag}_z) \hat{\rho} ) \hat{\rho}.
	\end{align}
Because the fundamental spin wave is the measured observable, $\mathcal{H}_{00}$ depends upon $\hat{F}^{00}_z$ rather than $\hat{F}_z$, as in the plane wave case.  The Lindblad superoperator, 
	\begin{align} \label{Eq::LSuperoperator}
		\mathcal{L}_{pl}(\hat{\rho}) = \hat{F}_z^{pl} \hat{\rho} \hat{F}_z^{pl\dag} - \frac{1}{2} \hat{F}_z^{pl\dag} \hat{F}_z^{pl} \hat{\rho}  - \frac{1}{2} \hat{\rho} \hat{F}_z^{pl\dag} \hat{F}_z^{pl},
	\end{align}
describes decoherence of the ensemble arising from coherent scattering of the light into all transverse spatial modes. In the case of $pl\neq 00$, the light is unmeasured and carries away information about the ensemble state.

\subsection{Evolution of the Mean Spin Waves}
We first consider the evolution of $\expect{\hat{F}_x^{pl}}$, the mean of a spin wave in spatial mode $pl$, defined as 
\begin{align}\label{MeanFxpl}
\hat{F}_x^{pl} = \sum_i  \beta_{pl}(\mbf{r}_i) \hat{f}_x^{(i)}. 
\end{align}
Because collective scattering and measurement backaction negligibly affect the dynamics of the mean spin, to good approximation the evolution of $\expect{\hat{F}_x^{pl}}$ is dominated by optical pumping.  The spatially varying nature of the photon scattering rate makes deriving the equations of motion for the ensemble observables slightly more complicated than in Sec. \ref{sec::covUpdate}. The master equation describing the evolution of the density operator, $\hat{\rho}^{(i)}$, of a single atom takes the form
\begin{align}\label{eq::ParaxRhoi}
\frac{d\hat{\rho}^{(i)}}{dt}\Big|_{\text{op}}=\gamma_s(\mathbf{r}_i)\mathcal{D}^{(i)}(\hat{\rho}).
\end{align}
From this master equation, we can compute the equation of motion for the spin component $\hat{f}_x^{(i)}$ of a single atom,
\begin{align}\label{eq::fxParax}
\frac{d}{dt}\hat{f}_x^{(i)}\Big|_{\text{op}}=\gamma_s(\mathbf{r}_i)\mathcal{D}^{(i)}(\hat{f}_x^{(i)}).
\end{align}
For $f=1/2$, the superoperator in this expression simplifies to ${\mathcal{D}_i}(\hat{f}_x^{(i)}) =-\hat{f}_x^{(i)}/3$.  From \erf{MeanFxpl}, we obtain the equation of motion for the spin wave,
\begin{align}\label{Eq::SimplifiedMeanFx1}
\frac{d}{dt}\expect{\hat{F}_x^{pl}}=-\frac{\gamma_0}{3}\sum_i  \beta_{00}(\mbf{r}_i) \beta_{pl}(\mbf{r}_i)\expect{\hat{f}_x^{(i)}}.
\end{align} 
By decomposing $\beta_{00}(\mbf{r}) \beta_{pl}(\mbf{r})$ in terms of orthogonal mode functions, the right hand side of \erf{Eq::SimplifiedMeanFx1} can be expressed as a sum of spin wave operators. In terms of the mode functions,
	\begin{align}\notag
		 \beta_{00}(\mbf{r}_\perp,z)  \beta_{pl}(\mbf{r}_\perp,z)& =|u_{00}(\mbf{r}_\perp,z)|^2u_{pl}^*(\mbf{r}_\perp,z)u_{00}(\mbf{r}_\perp,z) \\
		&= \sum_{p',l'} c^{pl}_{p'l'}(z) \beta_{p'l'}(\mbf{r}_\perp,z), \label{Eq::ProjCoeff_proto}
\end{align}
where we have made use of the orthogonality and completeness conditions in Eqs. (\ref{Eq::TransverseOrthogonality}) and (\ref{Eq::TransverseCompSameZ}) to define projection coefficients,
	\begin{align} \label{Eq::ProjCoeff2}
		c^{pl}_{p'l'}(z)  \equiv  \frac{1}{A} \int d^2 \mathbf{r}_\perp \left[ u_{00}(\mathbf{r}_\perp, z)\right]^2 u^*_{pl}(\mathbf{r}_\perp, z) u_{p'l'}(\mathbf{r}_\perp, z).
	\end{align}
Because the projection coefficients depend on the longitudinal coordinate, $z$, we coarse grain the ensemble into longitudinal slices along $z$. When restricted to a coarse-grained slice $k$ of thickness $\delta z$ centered at longitudinal coordinate $z_k$, \erf{Eq::SimplifiedMeanFx1} becomes
\begin{align}\label{Eq::SimplifiedMeanFx2}
\frac{d}{dt}\expect{\hat{F}_x^{pl}(z_k)}=-\frac{\gamma_0}{3}\sum_{i_k}  \beta_{00}(\mbf{r}_{i_k}) \beta_{pl}(\mbf{r}_{i_k})\expect{\hat{f}_x^{(i_k)}},
\end{align} 	
where $i_k$ is an index over all atoms in slice $k$.	By performing the projection in \erf{Eq::ProjCoeff_proto}, we obtain an infinite hierarchy of differential equations that couple mean spin waves in a given slice to one another,
\begin{align}\label{Eq::ZkSliceFx}
\frac{d}{dt}\expect{\hat{F}_x^{pl}(z_k)}=-\frac{\gamma_0}{3}\sum_{p',l'}c^{pl}_{p'l'}(z_k)\expect{\hat{F}_x^{p'l'}(z_k)}.
\end{align} 

An approximate solution to \erf{Eq::ZkSliceFx} is found for each slice by choosing a width, $\delta z$, and truncating the number of spin waves at some index $p_{max}, \,l_{max}$.  The result is a finite system of coupled differential equations describing mean spins, $\expect{\hat{F}_x^{pl}(z_k)}$, in each slice $k$, where $0\leq l\leq l_{max}$ and $0\leq p\leq p_{max}$. Solving this system of coupled differential equations requires the initial conditions of the mean spin waves in each slice. Using $\expect{\hat{f}_x(0)}=1/2$ for the initial SCS state of the ensemble,
	\begin{align}\label{Eq::meanSlice}
		\expect{\hat{F}_x^{pl}(z_k, t=0)} = \sum_{i_k}\beta_{pl}(\mbf{r}_{i_k})\expect{\hat{f}_x^{(i_k)}(0)}  =\frac{1}{2}\sum_{i_k}\beta_{pl}(\mbf{r}_{i_k}),
	\end{align} 
where $i_k$ is an index  over all atoms in slice $k$. For an average atomic density, $\eta(\mathbf{r})$, the sum becomes an integral, 
	\begin{align}\label{Eq::meanSlice2}
		\expect{\hat{F}_x^{pl}(z_k, t=0)} = \frac{\delta z}{2}\int d^2\mathbf{r} \,\eta(\mathbf{r},z_k)\beta_{pl}(\mathbf{r},z_k).
	\end{align} 
After solving the system of coupled differential equations, summing over the solutions at each slice gives the mean of the fundamental spin wave,
\begin{align}\label{Eq::FundSpinwave}
\expect{\hat{F}_x^{00}(t)}=\sum_{k}\expect{\hat{F}_x^{00}(z_k,t)}.
\end{align} 
Equation (\ref{Eq::FundSpinwave}) is the fundamental mean spin in the definition of the paraxial squeezing parameter. 

\subsection{Evolution of the Spin Wave Variances}
To solve for the variance of the fundamental spin wave, we follow a similar procedure. We start with the covariance between spin waves on different transverse modes $pl$ and $p'l'$,
	\begin{align} \label{Eq::GenCovariance}
		\expect{\Delta\hat{F}_z^{pl}\Delta\hat{F}_z^{p'l'}} =\expect{\hat{F}_z^{pl}\hat{F}_z^{p'l'}}-\expect{\hat{F}_z^{pl}}\expect{\hat{F}_z^{p'l'}} .
		\end{align}
Unlike the mean spin, we cannot neglect the effects of continuous QND measurement. In this section, we solve for the equation of motion of the covariance under both optical pumping and continuous measurement.  	

We first consider optical pumping. 
Decomposing the spin waves in \erf{Eq::GenCovariance} via \erf{eq::SpinWaveDef} shows that the covariance is the sum of both single atom and pairwise atomic correlations functions,
\begin{align} \label{eq::CovarCorrFuncts}
\expect{\Delta\hat{F}_z^{pl}\Delta\hat{F}_z^{p'l'}} =
&\sum_i\beta_{pl}(\mathbf{r}_i)\beta_{p'l'}(\mathbf{r}_i)\expect{\Delta\hat{f}_z^{(i)\;2}}\\\notag
&+\sum_{i\neq j}\beta_{pl}(\mathbf{r}_i)\beta_{p'l'}(\mathbf{r}_j)\expect{\Delta\hat{f}_z^{(i)}\Delta\hat{f}_z^{(j)}}.
\end{align}
The equation of motion for the single atom correlation function, $\expect{\Delta\hat{f}_z^{(i)\;2}}$, can be determined through the master equation in \erf{eq::ParaxRhoi}, which gives the evolution of a single atom. From \erf{eq::ParaxRhoi},
\begin{align}\label{eq::1stOrderCorrSW}
\frac{d}{dt}\sum_i\expect{\Delta\hat{f}_z^{(i)\;2}}\big|_{\text{op}}=\sum_{i}  \gamma_s(\mbf{r}_i) \Big\{  \big\langle\mathcal{D}^{(i)} \big( \hat{f}_z^{(i)2} \big) \big\rangle  -2 \big\langle \mathcal{D}^{(i)} \big( \hat{f}_z^{(i)} \big) \big\rangle \big\langle \hat{f}_z^{(i)} \big\rangle  \Big\}. 
\end{align}
Deriving the equation of motion for the pairwise correlation function, $\expect{\Delta\hat{f}_z^{(i)}\Delta\hat{f}_z^{(j)}}_{i\neq j}$, in \erf{eq::CovarCorrFuncts} requires the master equation describing the evolution of any two atoms in the ensemble. This master equation is given by
\begin{align}\label{Eq::rhoij2}
	\frac{d}{dt}\hat{\rho}^{(i,j)}\Big|_{\rm op} =\gamma_s(\mbf{r}_i)\mathcal{D}^{(i)}(\hat{\rho}^{(i,j)})+ \gamma_s(\mbf{r}_j) \mathcal{D}^{(j)}(\hat{\rho}^{(i,j)}).
\end{align}
From \erf{Eq::rhoij2},
\begin{align} \label{eq::correlationDecay2}
&\frac{d}{dt}\sum_{i\neq j} \big\langle \Delta \hat{f}_z^{(i)}\Delta \hat{f}_z^{(j)}   \big\rangle \big|_{\rm op} =\\\notag
		& \;\;\;\;\; \sum_{i\neq j} \! \Big\{ \! \gamma_s(\mbf{r}_i) \big\langle \Delta\mathcal{D}_i \big[ \hat{f}_z^{(i)} \big]\Delta \hat{f}_z^{(j)}\big\rangle  \! +\!  \gamma_s(\mbf{r}_j) \big\langle \Delta \hat{f}_z^{(i)}\Delta \mathcal{D}_j \big[\hat{f}_z^{(j)}\big] \big\rangle \! \Big\}.
\end{align}

Combining Eqs. (\ref{eq::CovarCorrFuncts}), (\ref{eq::1stOrderCorrSW}) and (\ref{eq::correlationDecay2}) yields the equation of motion for the covariance,
\begin{align} \label{eq::CovarEvolParax}
\frac{d}{dt}&\expect{\Delta\hat{F}_z^{pl}\Delta\hat{F}_z^{p'l'}}\big|_{\text{op}} =\\\notag
&\sum_i\beta_{pl}(\mathbf{r}_i)\beta_{p'l'}(\mathbf{r}_i) \gamma_s(\mbf{r}_i) \big\{  \big\langle\mathcal{D}^{(i)} \big( \hat{f}_z^{(i)2} \big) \big\rangle  -2 \big\langle \mathcal{D}^{(i)} \big( \hat{f}_z^{(i)} \big) \big\rangle \big\langle \hat{f}_z^{(i)} \big\rangle  \big\} 
\\\notag
&+\sum_{i\neq j}\beta_{pl}(\mathbf{r}_i)\beta_{p'l'}(\mathbf{r}_j)\big\{ \! \gamma_s(\mbf{r}_i) \big\langle \Delta\mathcal{D}^{(i)} \big( \hat{f}_z^{(i)} \big)\Delta \hat{f}_z^{(j)}\big\rangle  \! +\!  \gamma_s(\mbf{r}_j) \big\langle \Delta \hat{f}_z^{(i)}\Delta \mathcal{D}^{(j)} \big(\hat{f}_z^{(j)}\big) \big\rangle \! \big\}.
\end{align}
After some algebra, this expression becomes
\begin{align} \label{eq::CovarEvolParax23}
\frac{d}{dt}&\expect{\Delta\hat{F}_z^{pl}\Delta\hat{F}_z^{p'l'}}\big|_{\text{op}} =\\\notag&\gamma_s\!\sum_{i, j}\beta_{pl}(\mathbf{r}_i)\beta_{p'l'}(\mathbf{r}_j)\big\{ \! \beta_{00}(\mbf{r}_i) \big\langle \Delta\mathcal{D}^{(i)} \big( \hat{f}_z^{(i)} \big)\Delta \hat{f}_z^{(j)}\big\rangle  \! +\!  \beta_{00}(\mbf{r}_j) \big\langle \Delta \hat{f}_z^{(i)}\Delta \mathcal{D}^{(j)} \big(\hat{f}_z^{(j)}\big) \big\rangle \! \big\}\\\notag
&+\gamma_s\sum_i\beta_{pl}(\mathbf{r}_i)\beta_{p'l'}(\mathbf{r}_i)\beta_{00}(\mathbf{r}_i)\big\{  \big\langle\mathcal{D}^{(i)} \big( \hat{f}_z^{(i)2} \big) \big\rangle  -\big\langle \{\mathcal{D}^{(i)} \big( \hat{f}_z^{(i)} \big), \hat{f}_z^{(i)}\} \big\rangle \big\}. 
\end{align}
For $f=1/2$, the terms in this equation of motion can be simplified by ${\mathcal{D}^{(i)}}(\hat{f}_z^{(i)})= -2\hat{f}_z^{(i)}/9$ and ${\mathcal{D}^{(i)}}(\hat{f}_z^{(i)\, 2})= {\mathcal{D}^{(i)}}(\mathbb{I}^{(i)}/4)=0$, where the latter equality is the consequence of ${\mathcal{D}^{(i)}}$ being trace preserving when $f=1/2$. Equation (\ref{eq::CovarEvolParax23}) becomes
\begin{align} \label{eq::CovarEvolParax2}
\frac{d}{dt}&\expect{\Delta\hat{F}_z^{pl}\Delta\hat{F}_z^{p'l'}}\big|_{\text{op}} =\\\notag&-\frac{2\gamma_s}{9}\sum_{i}\big( \beta_{pl}(\mathbf{r}_i)\beta_{00}(\mbf{r}_i)\expect{\Delta\hat{f}_z^{(i)}\Delta\hat{F}_z^{p'l'}} 
+ \beta_{p'l'}(\mathbf{r}_i)\beta_{00}(\mbf{r}_i)\expect{\Delta\hat{F}_z^{pl}\Delta\hat{f}_z^{(i)}}  \big)
\\\notag
&+\frac{\gamma_s}{9}\sum_i\beta_{pl}(\mathbf{r}_i)\beta_{p'l'}(\mathbf{r}_i)\beta_{00}(\mathbf{r}_i).
\end{align}
By coarse graining each spin wave into longitudinal slices in the same manner as the mean spin, we can use the projection coefficients in \erf{Eq::ProjCoeff2}. When the spin wave $\hat{F}_z^{pl}$ is restricted to slice $k$ and the spin wave $\hat{F}_z^{p'l'}$ is restricted to slice $k'$, the equation of motion becomes 
\begin{align} \label{eq::CovarEvolParax3}
\frac{d}{dt}\expect{\Delta\hat{F}_z^{pl}(z_k)&\Delta\hat{F}_z^{p'l'}(z_{k'})}\big|_{\text{op}} =\\\notag&-\frac{2\gamma_s}{9}\sum_{i_k} \beta_{pl}(\mathbf{r}_{i_k})\beta_{00}(\mbf{r}_{i_k})\expect{\Delta\hat{f}_z^{(i_k)}\Delta\hat{F}_z^{p'l'}(z_{k'})} 
\\\notag&-\frac{2\gamma_s}{9}\sum_{i_{k'}} \beta_{p'l'}(\mathbf{r}_{i_{k'}})\beta_{00}(\mbf{r}_{i_{k'}})\expect{\Delta\hat{F}_z^{pl}(z_k)\Delta\hat{f}_z^{({i_{k'}})}}  \big)
\\\notag
&+\delta_{k,k'}\frac{\gamma_s}{9}\sum_i\beta_{pl}(\mathbf{r}_i)\beta_{p'l'}(\mathbf{r}_i)\beta_{00}(\mathbf{r}_i)\end{align}
By performing the projection in \erf{Eq::ProjCoeff_proto}, we obtain
\begin{align}\label{eq::CovarEvolParaxfinal}
		 &\frac{d}{dt}  \expect{\Delta\hat{F}_z^{pl}(z_k)  \Delta\hat{F}_z^{p'l'}(z_{k'})}\big|_{\text{op}}=\\\notag 		
		 &-\!\frac{2\gamma_s}{9}\!\sum_{p''l''}\Big[ c^{pl}_{p''l''}(z_k)\expect{\Delta\hat{F}_z^{p''l''}(z_k)\Delta\hat{F}_z^{p'l'}(z_{k'})} \!+\!c_{p''l''}^{p'l'}(z_{k'})\expect{\Delta\hat{F}_z^{pl}(z_k)\Delta\hat{F}_z^{p''l''}(z_{k'})}\Big] \\\notag&+\frac{\gamma_s}{9}N_{p'l'}^{pl}(z_k)\delta_{k,k'}. \nonumber
	\end{align}
In the equation of motion, $N_{p'l'}^{pl}(z_k)$ arises from the sum in the final term of \erf{eq::CovarEvolParax3}, which can be expressed as an integral over the density of the atomic cloud,
\begin{align}
N_{p'l'}^{pl}(z_k)= \delta z \int d^2\mathbf{r}\eta(\mathbf{r},z_k)\beta_{00}(\mathbf{r},z_k)\beta_{pl}(\mathbf{r},z_k)\beta_{p'l'}(\mathbf{r},z_k).
\end{align}
Note that for the fundamental mode, $p,p',l,l'=0$ and $N_{p'l'}^{pl}(z_k)$ is $N^{(3)}_{\text{eff}}$ in slice $z_k$. 

We now turn our attention to the evolution of the covariances under continuous QND measurement. As demonstrated by \erf{eq::CovarEvolParaxfinal}, optical pumping couples the covariances $\expect{\Delta\hat{F}_z^{pl}(z_k)\Delta\hat{F}_z^{p'l'}(z_{k'})}$ between spin waves in slices $z_k$ and $z_{k'}$ to one another. From the SME in \erf{Eq::HomodyneSME}, we can find the equations of motion for these covariances as the fundamental spin wave is measured.  The SME includes decoherence from collective scattering into transverse modes other than the fundamental mode, which is described by the map $\mathcal{L}_{pl}$ in \erf{Eq::LSuperoperator}. Note that this map has no effect on the covariances since the spin waves $\hat{F}_z^{pl}$ commute with one another. The evolution of the covariances, thus, depends entirely on measurement backaction, which is described by the map $\mathcal{H}_{00}$ in \erf{Eq::HSuperoperator}. The equation of motion for the covariances is given by
\begin{align} 
		 d\expect{ \Delta&\hat{F}_z^{pl}(z_k)   \Delta\hat{F}_z^{p'l'}(z_{k'})}\Big|_{QND} =\label{Eq::CovHomodyne}  \\\notag
		&\sqrt{ \frac{\kappa}{4} } \bigg\{  \big\langle \mathcal{H}_{00}[\hat{F}_z^{pl}(z_k)\hat{F}_z^{p'l'}(z_{k'})]\big\rangle- \big\langle\mathcal{H}_{00}[\hat{F}_z^{pl}(z_k)]\big\rangle \big\langle\hat{F}_z^{p'l'}(z_{k'})\big\rangle \\\notag
& - \big\langle\hat{F}_z^{pl}(z_k)\big\rangle \big\langle\mathcal{H}_{00}[\hat{F}_z^{p'l'}(z_{k'})]\big\rangle
 \bigg\} dW -\frac{\kappa}{4}\big\langle\mathcal{H}_{00}[\hat{F}_z^{pl}(z_k)]\big\rangle\big\langle\mathcal{H}_{00}[\hat{F}_z^{p'l'}(z_{k'})]\big\rangle dt.
	\end{align} 
The final term in the equation of motion arises  from the rule of It\={o} calculus that differentials must be taken to second order \cite{JacSte06}, i.e. $d(XY) = (dX) Y + X (dY) + (dX)(dY)$. The map $\mathcal{H}_{00}$ couples the first and second order moments of the spin waves to higher order moments. For the initial SCS along $x$ and during its subsequent evolution, the spin waves $\hat{F}_z^{pl}$ are Gaussian distributed, both over the entire cloud and within each coarse-grained slice $k$. The third order moments of the spin waves can, therefore, be expressed in terms of first and second order moments through the relation \cite{JacSte06} 
\begin{align}
\expect{\hat{X}\hat{Y}\hat{Z}} =& \expect{\Delta\hat{X}\Delta\hat{Y}}\expect{\hat{Z}} + \expect{\Delta\hat{X}\Delta\hat{Z}}\expect{\hat{Y}} + \expect{\Delta\hat{Y}\Delta\hat{Z}}\expect{\hat{X}} \\\notag&+\expect{\hat{X}}\expect{\hat{Y}}\expect{\hat{Z}}.
\end{align}
The decomposition in \erf{eq::3rdGaussDecomp} reduces to this expression when the operators commute.
In this regime, all stochastic terms in \erf{Eq::CovHomodyne} cancel, leaving the deterministic equation: 
	\begin{align}\label{Eq::CovarianceBackaction}
		\frac{d}{dt}\expect{\Delta\hat{F}_z^{pl}(z_k)&  \Delta\hat{F}_z^{p'l'} (z_{k'})}\Big|_{QND} = -\kappa\big\langle\Delta\hat{F}_z^{pl}(z_k)\Delta\hat{F}_z^{00}\big\rangle\big\langle\Delta\hat{F}_z^{p'l'}(z_{k'})\Delta\hat{F}_z^{00}\big\rangle \\\label{Eq::CovarianceBackaction2}
		&=-\kappa\sum_{k'',k'''}\big\langle\Delta\hat{F}_z^{pl}(z_k)\Delta\hat{F}_z^{00}(z_{k''})\big\rangle\big\langle\Delta\hat{F}_z^{p'l'}(z_{k'})\Delta\hat{F}_z^{00}(z_{k'''})\big\rangle. 
	\end{align}
These dynamics, which arise from continuous polarimetry measurements, serve to generate the correlations that produce spin squeezing.  

	By combining \erf{eq::CovarEvolParaxfinal} and \erf{Eq::CovarianceBackaction2}, we obtain the full equation of motion for the covariances under both optical pumping and continuous measurement,
\begin{align}
\frac{d}{dt}\expect{&\Delta\hat{F}_z^{pl}(z_k)  \Delta\hat{F}_z^{p'l'} (z_{k'})}=
\frac{d}{dt}\expect{\Delta\hat{F}_z^{pl}(z_k)  \Delta\hat{F}_z^{p'l'} (z_{k'})}\Big|_{\text{op}}\\\notag&\;\;\;\;\;\;\;\;\;\;\;\;\;\;\;\;\;\;\;\;\;\;\;\;\;\;\;\;\;\;\;\;\;\;\;\;\;+
\frac{d}{dt}\expect{\Delta\hat{F}_z^{pl}(z_k)  \Delta\hat{F}_z^{p'l'} (z_{k'})}\Big|_{QND}\\\label{eq::FullEOMCOV}
= &-\kappa\sum_{k'',k'''}\big\langle\Delta\hat{F}_z^{pl}(z_k)\Delta\hat{F}_z^{00}(z_{k''})\big\rangle\big\langle\Delta\hat{F}_z^{p'l'}(z_{k'})\Delta\hat{F}_z^{00}(z_{k'''})\big\rangle\\\notag&\!-\!\frac{2\gamma_s}{9}\!\sum_{p''l''}\!\!\Big[ c^{pl}_{p''l''}\!(z_k)\!\expect{\Delta\hat{F}_z^{p''l''}\!(z_k)\Delta\hat{F}_z^{p'l'}(z_{k'})}\!+\!c_{p''l''}^{p'l'}(z_{k'})\!\expect{\Delta\hat{F}_z^{pl}\!(z_k)\Delta\hat{F}_z^{p''l''}\!(z_{k'})}\Big] \\\notag &+\frac{\gamma_s}{9}N_{p'l'}^{pl}(z_k)\delta_{k,k'}. 
\end{align}
This is an infinite set of differential equations that nonlinearly couples all covariances between spin waves in slices $k$ and $k'$ to one another. 

As in the case of the mean spin waves, the solution to this set of equations is approximated by truncating \erf{eq::FullEOMCOV} at some $p_{\text{max}}$ and  $l_{\text{max}}$. Solving the resulting finite set of differential equations requires the initial values of the covariances, which are given by
	\begin{align}\notag
		\expect{\Delta\hat{F}_z^{pl}(z_k)&\Delta\hat{F}_z^{p'l'}(z_{k'})}(0)= \!\! \int \!\!d^2\mathbf{r} \, \eta(\mathbf{r},z_k) \beta_{pl}(\mathbf{r},z_k)\beta_{p'l'}(\mathbf{r},z_k)\expect{\hat{f}_z^{(i_{k})}(0) \hat{f}_z^{(i_{k'})}(0)} \\
		&=  \delta_{k,k'}\frac{\delta z}{4}\int d^2\mathbf{r} \, \eta(\mathbf{r},z_k) \beta_{pl}(\mathbf{r},z_k)\beta_{p'l'}(\mathbf{r},z_k).
	\end{align}
The second equality follows because $\expect{\hat{f}_z^{(i_{k})}(0) \hat{f}_z^{(i_{k'})}(0)} =\delta_{k,k'}/4$ for the initial SCS of the ensemble. With these initial conditions and the equations of motion, we can solve for the evolution of all covariances under both QND measurement and decoherence by optical pumping. In particular, we can solve for the covariances between the fundamental spin waves in different slices. Summing these covariances over all slices $k$ and $k'$ yields the variance of the fundamental spin wave,
\begin{align}
(\Delta F_z^{00})^2=\sum_{k,k'}\expect{\Delta\hat{F}_z^{00}(z_k)\Delta\hat{F}_z^{00}(z_{k'})}.
\end{align}
From the variance of the fundamental spin wave, we can determine the paraxial squeezing parameter.

\section{Results}\label{sec::MultimodeResults}
In this section, we use the equations of motion for the spin waves to determine the optimal geometries of the atomic ensemble and probe beam for both mode matching and spin squeezing. In an experimental setting, the probe beam is tuned by varying the beam waist, $\omega_0$. Because the Rayleigh range increases with the square of the beam waist, $\omega_0$ determines both the longitudinal and transverse extent of the ensemble that lies within the region of high field intensity. We consider an experimental implementation in which the atoms are cooled and confined in a crossed beam dipole trap. In this case, the geometry of the atomic ensemble is well approximated by the continuous Gaussian density distribution in \erf{Eq::AtomicDistribution}. Adjusting the angle between the trapping beams creates atomic ensembles with different transverse and longitudinal $1/e^2$ variances, which we denote by $\sigma_\perp^2$ and $\sigma_z^2$. The geometry of the ensemble is described concisely with the aspect ratio, defined as $AR=\sigma_z/\sigma_\perp$. 

For different geometries of the probe beam and ensemble, we numerically solve the equations of motion for the mean spin waves and the spin wave covariances to obtain the squeezing parameter. The infinite set of differential equations describing the evolution of the ensemble observables are truncated at some  
transverse mode $p_{\text{max}},\,l_{\text{max}}$ once convergence in the set of numerical solutions is achieved. To define the region of maximum mode matching, we also solve for $OD_{\text{eff}}$ for different probe and ensemble geometries. The $OD_{\text{eff}}$ is determined through the formula in \erf{eq::ODeff}, where the beam area $A=\pi\omega_0^2/2$ for the Gaussian probe. Note that unlike the spin squeezing, the effective optical density is a purely geometric quantity.

\subsection{Ensemble Geometry and Optical Pumping}\label{sec::EnGeo}
\begin{figure}[H]
\centering
\includegraphics[scale=.37]{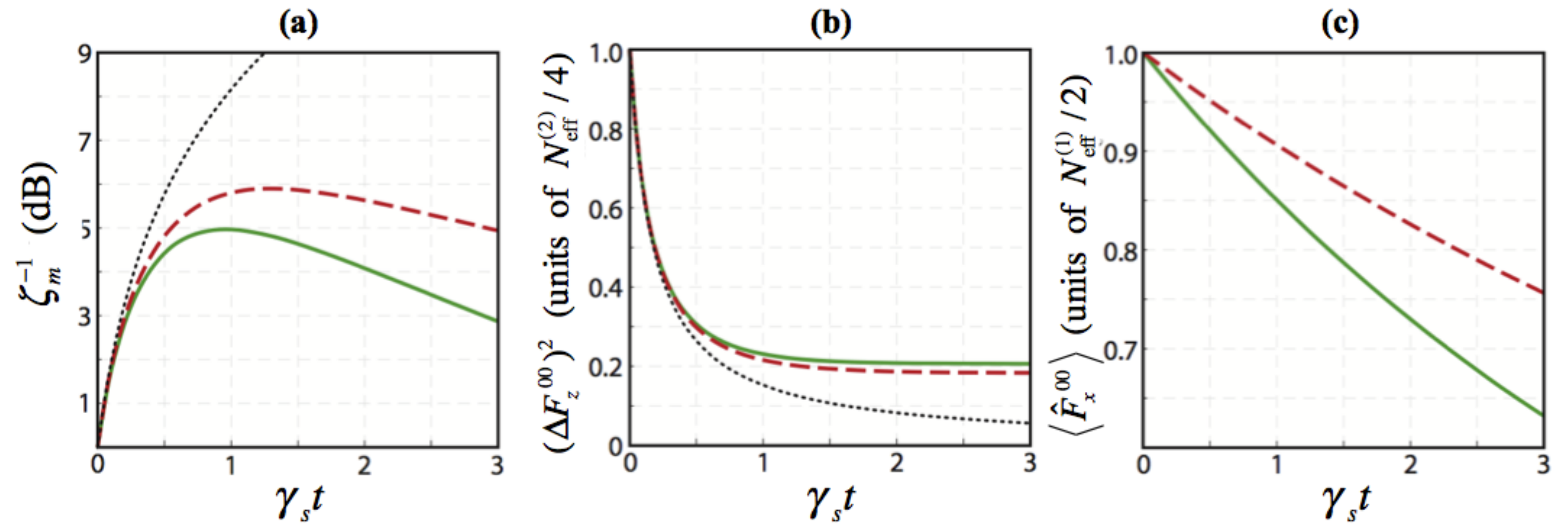}
\caption{Ensemble observables as a function of time for different ensemble geometries, fixed $OD_{\text{eff}}=50$ and fixed beam waist $\omega_0=20 \mu\text{m}$. (a) Squeezing versus time for an ensemble with $AR=0.1$ (green solid line) and $AR=316$ (red dashed line). Also shown are the dynamics of the squeezing without decoherence (black dotted line). (b) The fundamental spin wave variance, normalized by $N_{\text{eff}}^{(2)}/4$, versus time for $AR=0.1$ (green solid line), $AR=316$ (red dashed line) and without decoherence (black dotted line). (c) Fundamental mean spin, normalized by $N_{\text{eff}}^{(1)}/2$,  versus time for $AR=0.1$ (green solid line) and $AR=316$ (red dashed line).   }\label{fig::ODeff50}
\end{figure}

First, we examine how the geometry of the ensemble affects decoherence due to optical pumping. We focus on the case where the ensemble is prepared in an SCS for $f=1/2$.  Figure \ref{fig::ODeff50} shows the dynamics of various spin wave observables for two different ensemble geometries with fixed $OD_{\text{eff}}=50$ and $\omega_0=20\mu\text{m}$. To obtain a fixed optical density at different cloud geometries, we vary $\eta_0$, the peak density of the Gaussian density function in \erf{Eq::AtomicDistribution}. Because the effective optical density is held constant,  the strength of the coherent squeezing interaction is identical for each of the ensemble geometries. The effect of decoherence on the different ensemble geometries solely accounts for the discrepancies in the behavior of the observables depicted in Fig. \ref{fig::ODeff50}. Figure \ref{fig::ODeff50} (a) shows the squeezing generated for an ensemble with a ``pancake" geometry ($AR=.1$) and an ensemble with a ``pencil" geometry ($AR=316$).  In this plot, the pencil geometry generates substantially more squeezing than the pancake. For reference, the squeezing generated by QND measurement in the absence of decoherence is also depicted. Because the magnitude of this squeezing depends only upon the size of the fixed $OD_{\text{eff}}$, it represents the squeezing generated by either ensemble geometry in absence of decoherence. By comparing the performance of both geometries to the solution without decoherence, it is evident that the pancake geometry is far more susceptible to decoherence due to optical pumping.

To explain the relative robustness of the pencil geometry to optical pumping as compared to the pancake, we also examine the dynamics of the fundamental spin wave variance and mean spin wave for these geometries. Fig. \ref{fig::ODeff50} (b) shows that the behavior of the fundamental spin wave variance for the pancake and pencil geometries is comparable. There is a substantial difference in the decay of the fundamental mean spin wave for these geometries, however, as shown in Fig. \ref{fig::ODeff50} (c). The faster decay of the mean spin for the pancake geometry explains its poor performance compared to the pencil. The mean spin of the pencil decays more favorably because a larger fraction of the atoms are spread out longitudinally, far from the beam waist. Because of the reduced probe intensity farther away from the beam waist, these atoms have a lower rate of optical pumping. To achieve the same $OD_\text{eff}$ in the pancake geometry, the atoms must be concentrated at the waist, where they are more likely to undergo optical pumping. Consequently, the pancake is much more susceptible to optical pumping.

\subsection{Optimal Ensemble and Beam Geometries for Fixed Atom Number }
\begin{figure}[H]
\centering
\includegraphics[scale=.43]{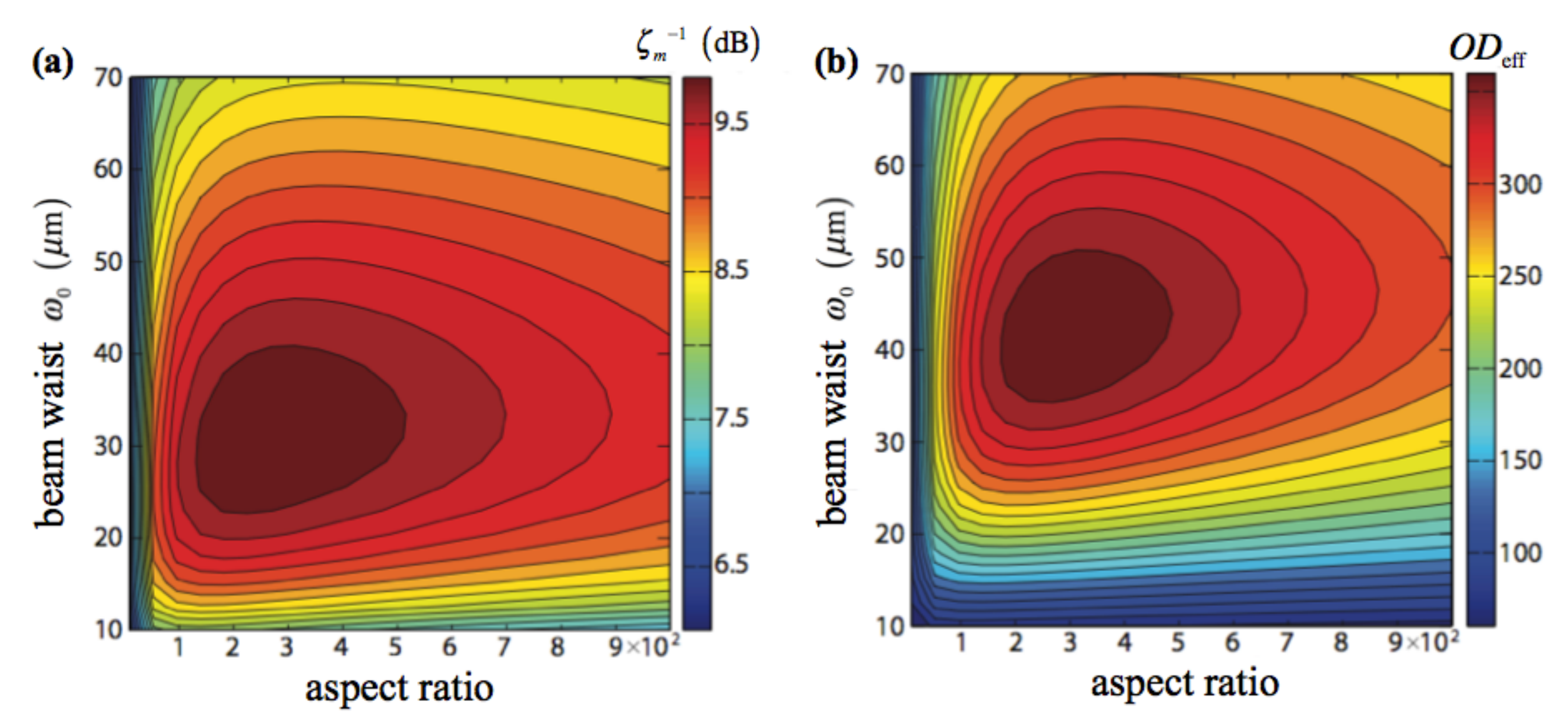}
\caption{Contours of (a) peak squeezing for $f=1/2$ and (b) $OD_\text{eff}$ as a function of the aspect ratio of the ensemble and the beam waist of the probe. The ensemble is prepared in a SCS. Note that the effective optical density depends only upon the ensemble and probe geometries, making it independent of the spin size $f$. For both plots, the volume of the cloud and total atom number, $N_A=9.84\times10^6$, are held constant. The maximum peak squeezing achieved in (a) is $\zeta_m^{-1}=10.0$ dB, occurring at $AR=256$ and $\omega_0=31\mu m$.}\label{fig::Multimode}
\end{figure}

In the previous section, we analyzed the effect of the ensemble geometry on the squeezing of the fundamental spin wave. We now investigate optimal geometries of both the ensemble and the beam for a fixed atom number, peak intensity and ensemble volume. The number of atoms and peak intensity are held constant at $N_A=9.8\times 10^6$ and $\eta_0=5\times 10^{11} \text{cm}^{-3}$ , respectively. Figure \ref{fig::Multimode} (a) shows contours of peak squeezing as a function of the aspect ratio and the beam waist for a $f=1/2$ ensemble initially prepared in an SCS. The maximum value of the peak squeezing, $\zeta_{m}^{-1} = 10.0$ dB, occurs at AR $= 256$ at a beam waist of $w^{\rm}_0=31$ $\mu$m. The ensemble geometry is a pencil at the maximum peak squeezing with its length extending over several Rayleigh ranges, $\sigma_z/z^{\rm opt}_R = 2.42$.  The transverse width of the cloud at the maximum is slightly larger than the beam waist with $\sigma_\perp/w^{\rm opt}_0 = 1.09$. 

To understand the optimal region for squeezing, we plot similar contours of $OD_\text{eff}$ as a function of the aspect ratio and the beam waist in Fig. \ref{fig::Multimode} (b).  Comparison of Figs. \ref{fig::Multimode} (a) and (b) shows that the maximum peak squeezing occurs in a region of high effective optical density, but does not perfectly coincide with the maximum of $OD_{\text{eff}}$. Because the effective optical density quantifies the strength of the coherent squeezing interaction, this discrepancy is due to decoherence. The maximum peak squeezing occurs in a region where the beam waist is smaller than in the region of maximum $OD_{\text{eff}}$. 
Maximum squeezing occurs at smaller beam waists because the region of the beam with greatest intensity, the Rayleigh range, is smaller. Because the scattering rate $\gamma_s(\mathbf{r})$ is proportional to the local intensity, atoms outside the Rayleigh range experience a decreased rate of optical pumping.  Although a smaller Rayleigh range implies a decreased $OD_{\text{eff}}$, the reduction of the decoherence rate dominates in this regime. This is a direct analogy to  Sec. \ref{sec::EnGeo}, in which pencil-shaped clouds were more robust to decay due to a large number of atoms farther away from the beam waist. 

The equations of motion for spin wave observables of ensembles with $f\geq 1/2$ are derived in Appendix \ref{sec::fSpinWaves}. In Fig. \ref{fig::Multimode4}, we show a preliminary result related to the higher spin case. This figure plots contours of peak squeezing as a function of the aspect ratio and the beam waist for a $f=4$ ensemble initially prepared in an SCS. Like the $f=1/2$ case described above, the number of atoms and the peak intensity are held constant at $N_A=9.8\times 10^6$ and $\eta_0=5\times 10^{11} \text{cm}^{-3}$ , respectively. The volume of the cloud is also fixed. For $f=4$, the maximum peak squeezing is $\zeta_m^{-1}=7.8$ dB, which is smaller than the maximum when $f=1/2$. This occurs because the coupling strength between the light and ensemble, quantified by $\xi_{\text{parax}}$ in \erf{Eq::CouplingStrength}, decreases with increasing $f$.
The maximum for $f=4$ occurs at $AR=300$, which is a pencil, similar to the optimal geometry for $f=1/2$. Also like $f=1/2$, the beam waist of the maximum squeezing, $\omega_0=28\mu m$, is smaller than the beam waist of maximum $OD_{\text{eff}}$.

\begin{figure}[H]
\centering
\includegraphics[scale=.4]{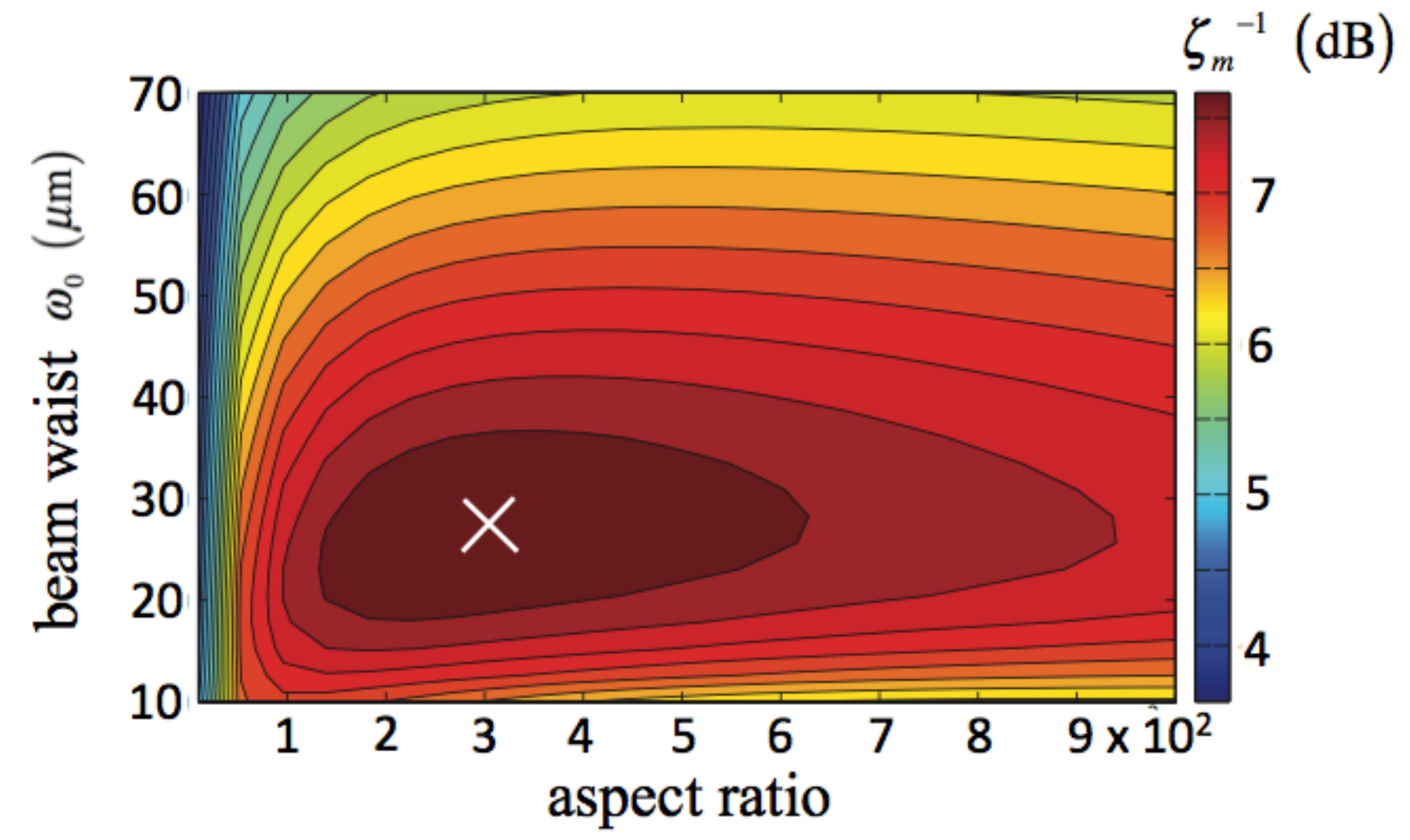}
\caption{Peak squeezing for an $f=4$ ensemble prepared in an SCS.  Contours of the peak squeezing are plotted as a function of the aspect ratio of the ensemble and the beam waist of the probe. Like the contour plots in Fig. \ref{fig::Multimode}, the volume of the cloud and total atom number, $N_A=9.84\times10^6$, are held constant. The maximum peak squeezing is $\zeta_m^{-1}=7.8$ dB, occurring at $AR=300$ and $\omega_0=28\mu m$, as indicated by the ``x". }\label{fig::Multimode4}
\end{figure}

\chapter{Conclusion and Outlook}\label{conclusion}
This dissertation explores how internal spin control affects spin squeezing and decoherence in large ensembles of alkali atoms with hyperfine spin $f$. While most studies of spin squeezing in atomic ensembles have been restricted to the $f=1/2$ case, we demonstrate that higher spin atoms offer substantial advantages. First, the number of internal degrees of freedom is greater, offering numerous options for internal spin control. By using internal spin control to prepare the ensemble in states with larger projection noise, we can enhance the entangling power of the Faraday interaction when $f>1/2$. Post-processing via internal spin control converts this increased interatomic entanglement into metrologically relevant spin squeezing. Post-processing can also squeeze the internal spin of the atoms, producing gains in spin squeezing that increase with $f$. Higher spin atoms also offer advantages due to their robustness to decoherence. Transfers of coherence can preserve interatomic entanglement under optical pumping  when $f>1/2$. Harmful optical pumping processes, such as spin flips, are also suppressed at larger $f$.  The initial state preparation of the ensemble determines the enhancement in the entangling power of the Faraday interaction and the susceptibility of the ensemble to optical pumping. For the appropriate choice of a fiducial state, higher spin atoms outperform $f=1/2$, even in absence of internal spin squeezing. The peak squeezing achieved by the optimization protocol in Chapter \ref{sec::Beyond} occurs for $f=2$, rather than $f=1/2$.

In this dissertation, we have also introduced new ways of modeling dissipative dynamics in large atomic ensembles. Even for large $f$, we have shown that restricting the hyperfine spin of each atom to an embedded qutrit captures the ensemble dynamics relevant to spin squeezing and optical pumping.  For certain fiducial states, we have shown that the multilevel Holstein-Primakoff approximation can be modified to accommodate dissipative dynamics. For these fiducial states, the ensemble of embedded qutrits becomes a Gaussian state on two effective bosonic modes. We formulate optical pumping as a update on the covariance matrix of the Gaussian ensemble state. Using the covariance matrix update formalism, optical pumping can be easily combined with coherent squeezing dynamics. For any fiducial state, the combined effects of squeezing by QND measurement and optical pumping can be modeled on the ensemble through a Stochastic master equation. Using the SME, we derive a system of differential equations describing the ensemble observables relevant to spin squeezing. The squeezing parameter can by optimized through these differential equations to reveal fiducial states that maximize spin squeezing. 

These computational methods  can also be extended to the case of a three-dimensional atomic ensemble interacting with a paraxial probe beam. This enables us to obtain parameter regimes that maximize mode matching and spin squeezing in the presence of optical pumping. This is the first treatment of optical pumping in a three dimensional atom-light interface. 

\section{Future Directions}
There are several future directions for the research presented in this dissertation. A natural project to pursue relates to Chapter \ref{sec::Beyond}, in which we found fiducial states that optimize the squeezing generated by QND measurement. Because we have not studied the landscape of the squeezing parameter as a function of time and the fiducial state, we cannot assert that the states we obtain via numerical search are global optima. A detailed study of the optimization landscape of the squeezing parameter can determine whether these states are global optima and, if not, the states that are. Extending the optimization procedure to squeezing protocols besides QND measurement would also be useful. The optimization procedure requires expressing a squeezing protocol in differential form, which is not accomplished as easily for the double pass protocols. An expression for the double pass squeezing protocols in differential form would also be valuable for the three-dimensional atom light interface presented in Chapter \ref{paraxial}, since we solve for the evolution of the spin wave observables using a truncated set of differential equations.

Another possible extension of this work concerns generating spin squeezing suited for particular metrological applications, such as atomic clocks. The Wineland squeezing parameter that we have used to quantify squeezing throughout this text was originally proposed in the context of Ramsey spectroscopy for atomic clocks. The precision of an atomic clock is dictated by the fluctuations in the measured frequency, $\Delta\omega=\Delta J_z/\expect{\hat{J}_x}$. Here, $\mathbf{J}$ is a collective spin composed of two level systems or qubits with energy splitting $\hbar\omega_0$, where $\omega_0$ is the clock frequency. When the collective spin is composed of qudits, $\Delta\omega^2$ is equivalent to the angular resolution $\Delta\phi^2$ presented in \erf{AngRes}, which determines the squeezing parameter. Atomic clocks composed of alkali ensembles typically treat each atom as a qubit formed by the clock states $\ket{f_\pm,m=0}$, where $f_\pm$ are the angular momenta of the ground hyperfine manifolds. The frequency resolution can be improved by squeezing the collective spin $\mathbf{J}$, which is composed of the ``clock qubits".  Many protocols create squeezing in atomic clocks by probing on the clock transition of the atoms, $\ket{f_-,m=0}\rightarrow\ket{f_+,m=0}$ \cite{Oblak05,SchleierSmith09,CHen14}. Using internal spin control, we can prepare each atom in the ensemble in any fiducial state $\ket{\uparrow}$ in the $f$ manifold.  A squeezing protocol based on the Faraday interaction can create squeezing in the ensemble of embedded qubits consisting of the fiducial state and coupled state, $\ket{\downarrow}$. Internal spin control can then be used to map the fiducial and coupled states to the clock states, $\ket{\uparrow}\rightarrow\ket{f_+,m=0}$ and $\ket{\downarrow}\rightarrow\ket{f_-,m=0}$. For some choice of fiducial state, can this protocol generate more squeezing than probing directly on the clock transition? This is another avenue for investigation.

Lastly, the continuous tomography protocol described Refs. \cite{Silberfarb05} and \cite{Riofrio11} can reconstruct the spin state of an alkali atom prepared in one of the ground hyperfine manifolds. This is accomplished by evolving an ensemble of identical alkalis through a set of informationally complete observables by internal spin control while the ensemble simultaneously undergoes QND measurement. Rather than reconstructing the state of a single alkali atom $i$, finding a way to reconstruct the second order correlation functions
\begin{align}\label{corr9}
\expect{\Delta\hat{f}_z^{(i)}\Delta\hat{f}_z^{(j)}}_{i\neq j} 
\end{align}
would offer a novel way of measuring spin squeezing and other phenomena related to interatomic entanglement in the ensemble. Because \erf{corr9} depends upon covariances between the qutrit operators in Sec. \ref{sec::NewObs}, the differential methods presented in this dissertation might be a natural way of reconstructing the second order correlation function. Tracking the evolution of the covariances under QND measurement, control and optical pumping requires that we use a basis of fiducial, coupled and transfer states that evolve with the internal spin control Hamiltonian. This ensures that the populations remain approximately c-numbers and that the Gaussian approximation used to derive the equation of motion for the covariances under QND measurement in \erf{eq::finalEoMQND7} remains valid. The internal spin control Hamiltonian, however, cannot be expressed in the basis of fiducial, coupled and transfer states for $f>1$. Finding the complete evolution of the covariances requires some adaptation of the procedure.

\chapter*{Appendices}
\addcontentsline{toc}{chapter}{Appendices}
\appendix
\include{ch_appendix}

\chapter{Optical Pumping Update Matrices}\label{sec::MgammaUps}
In Sec. \ref{sec::covUpdate}, we derived the updates on the covariance matrix and populations of the atomic ensemble describing the effects of optical pumping. Here, we list the specific updates for the SCS, cat and $m_x=0$ preparations that were used in the numerical simulations presented in Sec. \ref{sec::HPResults}.

For the SCS, the update on the populations is
\begin{align}
J_\gamma=\mathbb{I}+\gamma_s\Delta t\left(\begin{matrix}-\frac{1}{6}&\frac{1}{12f}\\
\frac{1}{12f}&-\frac{3f^2+2f-1}{18f^2}\end{matrix}\right)
\end{align}
and the update on the covariance matrix is given by
\begin{align}
&M_{A\gamma}=\mathbb{I}\;+\\\notag
&\small{\gamma_s\Delta t\left(\begin{matrix}-\frac{6f+1}{36f}&0&\frac{\sqrt{f(2f-1)}}{12f^2}&0\\
0&-\frac{2f+1}{12f}&0&\frac{\sqrt{f(2f-1)}}{12f^2}\\
\frac{\sqrt{f(2f-1)}}{12f^2}&0&-\frac{6f^2+4f-3}{36f^2}&0\\
0&\frac{\sqrt{f(2f-1)}}{12f^2}&0&-\frac{6f^2+8f-5}{36f^2}\end{matrix}\right)}
\end{align}
and
\begin{align}
N_{A\gamma}=\gamma_s\Delta tN_{\uparrow_{SCS}}\left(\begin{matrix}
\frac{8f+5}{72f}&0&-\frac{\sqrt{f(2f-1)}}{24f^2}&0\\
0&\frac{4f+6}{36f}&0&-\frac{\sqrt{f(2f-1)}}{24f^2}\\
-\frac{\sqrt{f(2f-1)}}{24f^2}&0&\frac{1}{24f}&0\\
0&-\frac{\sqrt{f(2f-1)}}{24f^2}&0&\frac{1}{24f}
\end{matrix}\right)\\
+\gamma_s\Delta tN_{\downarrow_{SCS}}\left(\begin{matrix}
\frac{8f^2-3f+4}{72f^2}&0&-\frac{\sqrt{f(2f-1)}}{12f^2}&0\\
0&\frac{2f^2+f+1}{18f^2}&0&-\frac{\sqrt{f(2f-1)}}{12f^2}\\
-\frac{\sqrt{f(2f-1)}}{12f^2}&0&\frac{6f^2+10f-7}{72f^2}&0\\
0&-\frac{\sqrt{f(2f-1)}}{12f^2}&0&\frac{6f^2+18f-11}{72f^2}
\end{matrix}\right).
\end{align}

For the cat, the populations evolve via
\begin{align}
J_\gamma=\mathbb{I}+\gamma_s\Delta t\left(\begin{matrix}-\frac{2}{9}&\frac{1}{9}\\
\frac{1}{9} &-\frac{2}{9}\end{matrix}\right)
\end{align}
and the update on the covariance matrix is given by the matrices
\begin{align}
M_{A\gamma}=\mathbb{I}+\gamma_s\Delta t\left(\begin{matrix}-\frac{1}{9}&0\\0&-\frac{1}{3}\end{matrix}\right)
\end{align}
and
\begin{align}
N_{A\gamma}=\gamma_s\Delta t(N_{\uparrow_\text{cat}}+N_{\downarrow_\text{cat}})\left(\begin{matrix}\frac{1}{18}&0\\
0&\frac{5}{18}\end{matrix}\right).
\end{align}

For the $m_x=0$ preparation, the update on the populations is
\begin{align}
J_\gamma=\mathbb{I}+\gamma_s\Delta t\left(\begin{matrix}-\frac{2}{9} & \frac{f+1}{18f} \\ \frac{f+1}{18f} &-\frac{2}{9}\end{matrix}\right).
\end{align}
The update on the covariance matrix is given by the maps
\begin{align}
&M_{A\gamma}=\mathbb{I}+
\gamma_s\Delta t\times\\\notag
&\tiny{\left(\begin{matrix}-\frac{3f-1}{18f}&0&\frac{1}{18f^2}\sqrt{\frac{f(f+1)(f-1)(f+2)}{2}}&0\\
0&-\frac{5f+1}{18f}&0&\frac{1}{18f^2}\sqrt{\frac{f(f+1)(f-1)(f+2)}{2}}\\
\frac{1}{18f^2}\sqrt{\frac{f(f+1)(f-1)(f+2)}{2}}&0&-\frac{7f^2-f+2}{36f^2}&0\\
0&\frac{1}{18f^2}\sqrt{\frac{f(f+1)(f-1)(f+2)}{2}}&0&-\frac{9f^2+f-2}{36f^2}\end{matrix}\right)}
\end{align}
and
\begin{align}
&N_{A\gamma}=
\gamma_s\Delta tN_{\uparrow_{0}}\times\\\notag&\tiny{\left(\!\begin{matrix}
\frac{3f-1}{36f}&0&-\frac{1}{36f^2}\sqrt{\frac{f(f+1)(f-1)(f+2)}{2}}&0\\
0&\frac{7f+3}{36f}&0&-\frac{1}{36f^2}\sqrt{\frac{f(f+1)(f-1)(f+2)}{2}}\\
-\frac{1}{36f^2}\sqrt{\frac{f(f+1)(f-1)(f+2)}{2}}&0&\frac{f+1}{36f}&0\\
0&-\frac{1}{36f^2}\sqrt{\frac{f(f+1)(f-1)(f+2)}{2}}&0&\frac{f+1}{36f}
\end{matrix}\!\right)}\\\notag\\\notag&+\gamma_s\Delta tN_{\downarrow_{0}}\times\\\notag&\tiny{\left(\!\begin{matrix}
\frac{3f-1}{36f}&0&-\frac{1}{18f^2}\sqrt{\frac{f(f+1)(f-1)(f+2)}{2}}&0\\
0&\frac{7f+3}{36f}&0&-\frac{1}{18f^2}\sqrt{\frac{f(f+1)(f-1)(f+2)}{2}}\\
-\frac{1}{18f^2}\sqrt{\frac{f(f+1)(f-1)(f+2)}{2}}&0&\frac{7f^2-f+2}{72f^2}&0\\
0&-\frac{1}{18f^2}\sqrt{\frac{f(f+1)(f-1)(f+2)}{2}}&0&\frac{11f^2+3f-6}{72f^2}
\end{matrix}\!\right)}.
\end{align}

\chapter{Spin Wave Equations of Motion}\label{sec::fSpinWaves}
Presented in the text were the equations of motion for the means and covariances of the spin waves in the case of $f=1/2$. Here, we generalize to the case of $f>1/2$ by deriving equations of motion for the means and covariances of the ``pseudo spin waves" and ``effective populations". These are the spin wave analogues of the operator basis defined in Eqs. (\ref{eq::Xdownup}) through (\ref{eq::PopWr}).  Through the equations of motion for the first and second order moments of this spin wave operator basis, we can solve for the variance and mean of the fundamental spin wave.

\section{First Order Spin Wave Operators}
To begin, we turn our attention to the evolution of the first order spin waves under optical pumping. The master equation describing the evolution of the density operator, $\hat{\rho}^{(i)}$, of a single atom takes the form
\begin{align}\label{eq::ParaxRhoiAppendix}
\frac{d\hat{\rho}^{(i)}}{dt}\Big|_{\text{op}}=\gamma_s(\mathbf{r}_i)\mathcal{D}^{(i)}(\hat{\rho}).
\end{align}
From \erf{eq::ParaxRhoiAppendix}, the equation of motion for the expectation value of a first order spin wave operator $\hat{O}^{pl}=\sum_i\beta_{pl}(\mathbf{r}_i)\hat{o}^{(i)}$ under optical pumping is
\begin{align}\label{eq::InitWaveY}
\frac{d}{dt}\expect{\hat{O}^{pl}}\Big|_{\text{op}}=\gamma_s\sum_i\beta_{00}(\mathbf{r}_i)\beta_{pl}(\mathbf{r}_i)
\expect{\mathcal{D}^{(i)}(\hat{o}^{(i)}}).
\end{align}
As in the $f=1/2$ case, we can make use of \erf{Eq::ProjCoeff_proto} and the projection coefficients in \erf{Eq::ProjCoeff2} to write this expression as
\begin{align}\label{eq::2WaveY}
\frac{d}{dt}\expect{\hat{O}^{pl}}\Big|_{\text{op}}=\gamma_s\sum_{i,p',l'}c^{pl}_{p'l'}(z_i) \beta_{p'l'}(\mbf{r}_{\perp i},z_i)
\expect{\mathcal{D}^{(i)}(\hat{o}^{(i)}}).
\end{align}
Because the projection coefficients depend upon the longitudinal coordinate $z$, the right hand side of this expression cannot be written in terms of spin wave operators like the left hand side. 

To place the right hand and left hand sides of \erf{eq::2WaveY} on similar footing, we again coarse grain the ensemble into longitudinal slices of thickness $\delta z$ centered at coordinates $z_k$. Corresponding to each slice is a coarse-grained spin wave operator, defined as
\begin{align}
\hat{O}^{pl}(z_k)=\sum_{i_k} \beta_{pl}(\mbf{r}_{\perp i_k},z_k)\hat{o}^{(i_k)},
\end{align}
where the index $i_k$ is summed over each atom contained in the slice $k$ centered at $z_k$. Because of the small longitudinal width of the slices, the mode functions at each atom $i_k$ are approximately $u_{pl}(\mbf{r}_{\perp i_k},z_k)$. When each side of the equation is projected onto the slice $k$,  \erf{eq::2WaveY} becomes 
\begin{align}\label{eq::3WaveY}
\frac{d}{dt}\expect{\hat{O}^{pl}(z_k)}\Big|_{\text{op}}=\gamma_s\,c^{pl}_{p'l'}(z_k)\sum_{i_k,\,p',l'} \beta_{p'l'}(\mbf{r}_{\perp i_k},z_k)\expect{\mathcal{D}^{(i_k)}(\hat{o}^{(i_k)}}).
\end{align}
Note that the right hand side of this expression is a sum of spin wave operators on different transverse modes, but all in the same slice $k$. This equation of motion for a spin wave operator in slice $k$ can be used to solve for first order moments of any spin wave. The spin wave is retrieved from the coarse-grained operators by summing over all slices of the ensemble,
\begin{align}
\expect{\hat{O}^{pl}}=\sum_k\expect{\hat{O}^{pl}(z_k)}.
\end{align}

We now apply the equation of motion for a coarse-grained spin wave operator to solve for the mean spin wave $\expect{\hat{F}_x^{00}}$, which appears in the definition of the paraxial squeezing parameter. We first define the effective populations, 
\begin{align}
N_\psi^{pl}=\sum_i\beta_{pl}(\mbf{r}_i)\hat{n}_\psi^{(i)},
\end{align}
where $\hat{n}_\psi=\ket{\psi}\bra{\psi}$ for $\psi\in\{\uparrow,\downarrow, \wr\}$. These parameters represent the effective number of atoms in the internal state $\ket{\psi}$ that are radiating into the transverse mode $pl$. Like the populations in the plane wave case, the effective populations have small variances compared to the other ensemble observables and can be treated as c-numbers. Similar to the plane wave case, the mean spin wave can be written in terms of the effective populations in the fiducial, coupled and coherence states as
\begin{align}\label{eq::FundMeanSpin}
\expect{\hat{F}_x^{00}}=N_\uparrow^{00}\expect{\hat{f}_x}_\uparrow+N_\downarrow^{00}\expect{\hat{f}_x}_\downarrow
+N_\wr^{00}\expect{\hat{f}_x}_\wr.
\end{align}

To find the evolution of the effective populations, we replace $\hat{O}^{pl}(z_k)$ in \erf{eq::3WaveY} with the effective population $N_\psi^{pl}(z_k)$ in slice $k$. Utilizing the qutrit operator basis defined in Eqs. (\ref{eq::qutritOp1}) through (\ref{eq::qutritOp9}), the evolution of the effective population in slice $k$ becomes
\begin{align}\label{eq::CGNup}
\frac{d}{dt}N_\psi^{pl}(z_k)\Big|_{\text{op}}=\gamma_s\,c^{pl}_{p'l'}(z_k)\sum_{p',l'}[&\text{Tr}(\mathcal{D}(\hat{n}_\psi)\hat{n}_\uparrow)N_\uparrow^{p'l'}(z_k)
\\\notag+\text{Tr}(\mathcal{D}(\hat{n}_\psi)\hat{n}_\downarrow)N_\downarrow^{p'l'}(z_k)
&+\text{Tr}(\mathcal{D}(\hat{n}_\psi)\hat{n}_\wr)N_\wr^{p'l'}(z_k)].
\end{align}
As in the plane wave case, we have discarded projections onto the basis elements $\hat{x}_{\downarrow\uparrow}$, $\hat{y}_{\downarrow\uparrow}$, $\hat{x}_{\wr\downarrow}$, $\hat{y}_{\wr\downarrow}$, $\hat{x}_{\uparrow\wr}$ and $\hat{y}_{\uparrow\wr}$, since the means of the corresponding ``collective pseudo spin waves" are negligible compared to the effective populations.  Together, the equations of motion for $N_\uparrow^{pl}(z_k)$, $N_\downarrow^{pl}(z_k)$ and $N_\wr^{pl}(z_k)$ form a set of differential equations that couple effective populations in slice $k$ to one another. Because an effective population in the transverse mode $pl$ is coupled to effective populations in all other transverse modes $p'l'$, the number of differential equations is infinite. The projection coefficients $c^{pl}_{p'l'}(z_k)$ decrease as the difference grows between $pl$ and $p'l'$, however. Because we are ultimately concerned with the behavior of spin waves in the fundamental mode, we can truncate the set of differential equations at some finite $p'l'$ once convergence is achieved in the observables for which $pl=00$. 

Similar to the populations in the plane wave case, the effective populations are negligibly affected by QND measurement. Therefore, the equation of motion due to optical pumping in \erf{eq::CGNup} provides a complete description of the effective population dynamics. After truncating these equations at a finite transverse mode $p'l'$, solving them requires only the initial conditions for the coarse grained effective populations. For an initial ensemble state with each atom prepared in $\ket{\uparrow}$, these values take a form similar to \erf{Eq::meanSlice2},
\begin{align}\label{eq::effectPopInit} 
N_\psi^{pl}(z_k,t=0)&=\sum_{i_k}\beta_{pl}(\mbf{r}_{i_k})\bra{\uparrow}\hat{n}_\psi\ket{\uparrow}\\&=\delta z\int d^2\mathbf{r}_\bot \eta(\mathbf{r}_\bot ,z_k)\beta_{pl}(\mathbf{r}_\bot ,z_k)\bra{\uparrow}\hat{n}_\psi\ket{\uparrow}.
\end{align}
The last equality in this expression arises from treating the ensemble as a continuous density distribution. Note that, with the exception of $N_\uparrow^{pl}(z_k)$, all effective populations in slice $k$ are initially zero at $t=0$. The initial conditions enable us to
numerically solve the set of differential equations for the effective populations in every slice. We determine the effective populations in the fundamental mode by summing over all longitudinal slices, $N_\psi^{00}(t)=\sum_kN_\psi^{00}(z_k,t)$. The mean spin, $\expect{\hat{F}_x^{00}}$, can then be obtained via \erf{eq::FundMeanSpin}.

\section{Second Order Spin Wave Operators}
We next consider the evolution of second order covariances of the spin wave operators.  In particular, we focus upon the dynamics of the fundamental spin wave variance, $(\Delta F_z^{00})^2$, which appears in the paraxial squeezing parameter. Like the effective populations, we decompose $(\Delta F_z^{00})^2$ into spin waves in longitudinal slices,
\begin{align}\label{eq::Fz00decompCG}
(\Delta F_z^{00})^2=\sum_{k,k'}\expect{\Delta\hat{F}_z^{00}(z_k)\Delta\hat{F}_z^{00}(z_{k'})}.
\end{align}
Similar to $\hat{F}_z$ in the plane wave case, the fundamental spin wave can be written as the sum of ``collective pseudo spin waves",
\begin{align}
\hat{F}_z^{00}(z_k)=v(\uparrow)\hat{X}_{\downarrow\uparrow}^{00}(z_k)+w(\uparrow)\hat{X}_{\wr\downarrow}^{00}(z_k).
\end{align}
Here, $v(\uparrow)$ and $w(\uparrow)$ are given in Eqs. (\ref{eq::vDef}) and (\ref{eq::wDef}). For arbitrary spatial modes, the collective pseudo spin waves take the form
\begin{align}
\hat{X}_{\downarrow\uparrow}^{pl}=\sum_i\beta_{pl}(\mathbf{r}_i)\hat{x}_{\downarrow\uparrow}^{(i)}\\
\hat{Y}_{\downarrow\uparrow}^{pl}=\sum_i\beta_{pl}(\mathbf{r}_i)\hat{y}_{\downarrow\uparrow}^{(i)}\\
\hat{X}_{\wr\downarrow}^{pl}=\sum_i\beta_{pl}(\mathbf{r}_i)\hat{x}_{\wr\downarrow}^{(i)}\\
\hat{Y}_{\wr\downarrow}^{pl}=\sum_i\beta_{pl}(\mathbf{r}_i)\hat{y}_{\wr\downarrow}^{(i)}\\
\hat{X}_{\uparrow\wr}^{pl}=\sum_i\beta_{pl}(\mathbf{r}_i)\hat{x}_{\uparrow\wr}^{(i)}
\end{align}
and
\begin{align}
\hat{Y}_{\uparrow\wr}^{pl}=\sum_i\beta_{pl}(\mathbf{r}_i)\hat{y}_{\uparrow\wr}^{(i)},
\end{align}
where $\hat{x}_{\downarrow\uparrow}$, $\hat{y}_{\downarrow\uparrow}$, $\hat{x}_{\wr\downarrow}$, $\hat{y}_{\wr\downarrow}$, $\hat{x}_{\uparrow\wr}$ and $\hat{y}_{\uparrow\wr}$ are the qutrit basis operators defined in Eqs. (\ref{eq::qutritOp1}) through (\ref{eq::qutritOp9}). In terms of the pseudo spin waves, each covariance on the right hand side of \erf{eq::Fz00decompCG} becomes,
\begin{align}
\big\langle\Delta\hat{F}_z^{00}(z_k)&\Delta\hat{F}_z^{00}(z_{k'})\big\rangle= v(\uparrow)^2\expect{\Delta\hat{X}_{\downarrow\uparrow}^{00}(z_k)\Delta\hat{X}_{\downarrow\uparrow}^{00}(z_{k'})}_S\\\notag
+&v(\uparrow)w(\uparrow)\left(\expect{\Delta\hat{X}_{\downarrow\uparrow}^{00}(z_k)\Delta\hat{X}_{\wr\downarrow}^{00}(z_{k'})}_S+
\expect{\Delta\hat{X}_{\wr\downarrow}^{00}(z_{k})\Delta\hat{X}_{\downarrow\uparrow}^{00}(z_{k'})}_S\right)\\\notag
+& w(\uparrow)^2\expect{\Delta\hat{X}_{\wr\downarrow}^{00}(z_k)\Delta\hat{X}_{\wr\downarrow}^{00}(z_{k'})}_S.
\end{align}
Solving for the variance of the fundamental spin wave becomes a matter of determining the variances and covariances between collective pseudo spin waves. We, thus, seek to derive the equations of motion for the covariances $\expect{\Delta\hat{X}_\theta^{pl} (z_k)\Delta\hat{X}_{\theta'}^{p'l'} (z_{k'})}_S$, where $\theta,\theta'\in\{\downarrow\uparrow,\,\wr\!\downarrow\}$.

Determining the evolution of the covariances under optical pumping requires the equation of motion for the density operator of any two atoms $i$ and $j$,  
\begin{align}\label{Eq::rhoij}
	\frac{d}{dt}\hat{\rho}^{(i,j)}\Big|_{ \text{op}} =\gamma_s(\mbf{r}_i)\mathcal{D}_i(\hat{\rho}^{(i,j)})+ \gamma_s(\mbf{r}_j) \mathcal{D}_j(\hat{\rho}^{(i,j)}).
\end{align}
This expression arises from restricting the sum in the full master equation in \erf{eq::FullMasterParaxial} to two indices. Using the equations of motion for $\hat{\rho}^{(i,j)}$ and $\hat{\rho}^{(i)}$ in \erf{eq::ParaxRhoi} and \erf{Eq::rhoij2}, we determine the dynamics of the covariance $\expect{\Delta\hat{X}_\theta^{pl}  (z_k)\Delta\hat{X}_{\theta'}^{p'l'} (z_{k'})}_S$ through a procedure similar to Sec. \ref{sec::2ndOrderOP}. The equation of motion for the quadrature covariance becomes
\begin{align}\label{Eq::YTcoarse}
\frac{d}{dt}\expect{\Delta\hat{X}_{\theta}^{pl}(z_k)&\Delta\hat{X}_{\theta'}^{p'l'}(z_{k'})}_S\Big|_{\text{op}}=\\\notag
&\gamma_s\sum_{i_k}\beta_{00}(\mbf{r}_{i_k})\beta_{pl}(\mbf{r}_{i_k})
\expect{\Delta\mathcal{D}^{(i_k)}(\hat{x}_\theta^{(i_k)})\Delta\hat{X}_{\theta'}^{p'l'}(z_{k'})}_S\\\notag
&+\gamma_s\sum_{i_{k'}}\beta_{00}(\mbf{r}_{i_{k'}})\beta_{p'l'}(\mbf{r}_{i_{k'}})
\expect{\Delta\hat{X}_\theta^{pl}(z_k)\Delta\mathcal{D}^{(i_{k'})}(\hat{x}_{\theta'}^{(i_{k'})})}_S\\\notag
&+\delta_{k,k'}\gamma_s\sum_{i_k}\beta_{00}(\mbf{r}_{i_k})\beta_{pl}(\mbf{r}_{i_k})
\beta_{p'l'}(\mbf{r}_{i_k})\expect{\mathcal{N}(\hat{x}_\theta^{(i_k)},\hat{x}_{\theta'}^{(i_k)})}.
\end{align}
The first two terms in this expression depend upon covariances between pseudospin wave operators.  The third term, depending upon first order spin wave operators, is the paraxial analogue to the noise term in the plane wave case, given in \erf{eq::noiseSuperOp}. 

The equation of motion for the covariance can be simplified by utilizing the basis of Gaussian mode functions.   We first simplify the noise term through the mode function decomposition
\begin{align}
\beta_{00}(\mbf{r}_{\bot},z)\beta_{pl}(\mbf{r}_{\bot},z)\beta_{p'l'}(\mbf{r}_{\bot},z)=
\sum_{p''l''}G^{plp'l'}_{p''l''}(z)\beta_{p''l''}(\mbf{r}_{\bot},z)
\end{align}
where the projection coefficient  $G^{plp'l'}_{p''l''}(z)$ is defined as
\begin{align}
G^{plp'l'}_{p''l''}(z)=\int d^2\mbf{r}_{\bot}u_{p''l''}(\mbf{r}_{\bot},z)u_{00}(\mbf{r}_{\bot},z)
\beta_{pl}(\mbf{r}_{\bot},z)\beta_{p'l'}(\mbf{r}_{\bot},z).
\end{align}
By substituting the projection coefficients $G^{plp'l'}_{p''l''}(z)$ and $C^{pl}_{p'l'}(z)$ from \erf{Eq::ProjCoeff2}  into the equation of motion, we obtain
\begin{align}
\frac{d}{dt}\expect{\Delta\hat{X}_\theta^{pl}&(z_k)\Delta\hat{X}_{\theta'}^{p'l'}(z_{k'})}_S\Big|_\text{op}=\\\notag
&\gamma_s\!\!\!\!\sum_{i_k,\,p'',l''}C_{p''l''}^{pl}(z_k)\beta_{p''l''}(\mbf{r}_{\perp i_k},z_k)
\expect{\Delta\mathcal{D}^{(i_k)}(\hat{x}_\theta^{(i_k)})\Delta\hat{X}_{\theta'}^{p'l'}(z_{k'})}_S\\\notag
+&\gamma_s\!\!\!\!\sum_{i_{k'},\,p'',l''}C_{p''l''}^{p'l'}(z_{k'})\beta_{p''l''}(\mbf{r}_{\perp i_{k'}},z_{k'})
\expect{\Delta\hat{X}_\theta^{pl}(z_{k})\Delta\mathcal{D}^{(i_{k'})}(\hat{x}_{\theta'}^{\,(i_{k'})})}_S\\\notag
+&\gamma_s\,\delta_{k',k}\sum_{i_k,\,p'',l''}G_{p''l''}^{plp'l'}(z_k)\beta_{p''l''}(\mbf{r}_{\perp i_{k}},z_{k})
\expect{\mathcal{N}(\hat{x}_\theta^{(i_k)},\hat{x}_{\theta'}^{(i_k)})}.
\end{align}

Expressing the spin wave covariances in terms of the collective pseudo spin waves us to again take advantage of the qutrit operator basis in Eqs. (\ref{eq::qutritOp1}) through (\ref{eq::qutritOp9}). By decomposing the internal spin operators in this basis, we can rewrite the equation of motion in terms of collective pseudo spin waves and effective populations,
\begin{align}\label{eq::CGcovarFinal}
\frac{d}{dt}\expect{\Delta\hat{X}_\theta^{pl}&(z_k)\Delta\hat{X}_{\theta'}^{p'l'}(z_{k'})}_S\Big|_\text{op}=\\\notag
&\gamma_s\sum_{p'',\,l''}\sum_{\hat{X}\in \mathcal{S}}C_{p''l''}^{pl}(z_k)\;\text{Tr}\left(\mathcal{D}(\hat{x}_\theta)\hat{x}\right)
\expect{\Delta\hat{X}^{p''l''}\!(z_k)\Delta\hat{X}_{\theta'}^{p'l'}(z_{k'})}_S\\\notag
+&\gamma_s\sum_{p'',\,l''}\sum_{\hat{X}\in \mathcal{S}}C_{p''l''}^{p'l'}(z_{k'})\;\text{Tr}\left(\mathcal{D}(\hat{x}_{\theta'}\,)\hat{x}\right)
\expect{\Delta\hat{X}_\theta^{pl}(z_{k})\Delta\hat{X}^{p''l''}(z_{k'})}_S\\\notag
+&\gamma_s\,\delta_{k',k}\sum_{p'',\,l''}\sum_{\psi\in\{\uparrow,\,\downarrow,\,\wr\}}G_{p''l''}^{plp'l'}(z_k)\;
\text{Tr}\left(\mathcal{N}(\hat{x}_\theta,\hat{x}_{\theta'})\,\hat{n}_\psi\right)N_\psi^{p''l''}(z_k),
\end{align}
where $\hat{X}^{pl}=\sum_{i}\beta_{pl}(\mbf{r})\hat{x}^{(i)}$ and $\mathcal{S}=\{\hat{X}_{\downarrow\uparrow},\hat{Y}_{\downarrow\uparrow},
\hat{X}_{\wr\downarrow},\hat{Y}_{\wr\downarrow},\hat{X}_{\uparrow\wr},\hat{Y}_{\uparrow\wr}\}$.  Similar to the equation of motion for the covariance in the plane wave case, given in \erf{eq::CovarEvol2}, the operator $\hat{X}$ only takes the values of the pseudo spin waves, since the effective populations are approximated as c-numbers. As in the plane wave case, the noise term can be written entirely in terms of effective populations in slice $k$, since the means of the collective pseudo spin waves are negligible by comparison. Note that because the noise term is a first order operator, it is nonzero only in the case when the collective pseudo spin waves $\hat{X}_\theta^{pl}(z_k)$ and $\hat{X}_{\theta'}^{p'l'}(z_{k'})$ are defined on the same slice, i.e. $k=k'$.

When the ensemble is prepared in the SCS, cat, or $m_x=0$ state preparations, the equation of motion for the covariance $\expect{\Delta\hat{X}_\theta^{pl}(z_k)\Delta\hat{X}_{\theta'}^{p'l'}(z_{k'})}_S\,$ simplifies substantially. For these state preparations,
$\text{Tr}\left(\mathcal{D}(\hat{x}_\theta)\hat{x}\right)=0\,$ when $\hat{X}\in\{\hat{Y}_{\downarrow\uparrow},\hat{Y}_{\wr\downarrow},\hat{X}_{\uparrow\wr},\hat{Y}_{\uparrow\wr}\}$. The covariances $\expect{\Delta\hat{X}_\theta^{pl}(z_k)\Delta\hat{X}_{\theta'}^{p'l'}(z_{k'})}_S$ are, thus, coupled only to other covariances of the form  
$\expect{\Delta\hat{X}_\theta^{pl}(z_k)\Delta\hat{X}_{\theta'}^{p'l'}(z_{k'})}_S$. The equation of motion becomes,
\begin{align}\label{eq::CGcovarFinal}
\frac{d}{dt}&\expect{\Delta\hat{X}_\theta^{pl}(z_k)\Delta\hat{X}_{\theta'}^{p'l'}(z_{k'})}_S\Big|_\text{op}=
\\\notag&\;\;\;\;\gamma_s\sum_{p'',\,l''}C_{p''l''}^{pl}(z_k)(\text{Tr}(\mathcal{D}(\hat{x}_\theta)\hat{x}_{\downarrow\uparrow})
\expect{\Delta\hat{X}_{\downarrow\uparrow}^{p''l''}\!(z_k)\Delta\hat{X}_{\theta'}^{p'l'}\!(z_{k'})}_S\\\notag&\;\;\;\;\;\;\;\;\;\;\;\;\;\;\;\;\;\;\;\;\;\;\;\;\;\;\;\;\;+
\text{Tr}(\mathcal{D}(\hat{x}_\theta)\hat{x}_{\wr\downarrow})
\expect{\Delta\hat{X}_{\wr\downarrow}^{p''l''}\!(z_k)\Delta\hat{X}_{\theta'}^{p'l'}\!(z_{k'})}_S)\\\notag
&+\gamma_s\sum_{p'',\,l''}C_{p''l''}^{p'l'}(z_{k'})(\text{Tr}(\mathcal{D}(\hat{x}_{\theta'})\hat{x}_{\downarrow\uparrow})
\expect{\Delta\hat{X}_\theta^{pl}\!(z_k)\Delta\hat{X}_{\downarrow\uparrow}^{p''l''}\!(z_{k'})}_S\\\notag&\;\;\;\;\;\;\;\;\;\;\;\;\;\;\;\;\;\;\;\;\;\;\;\;\;\;\;\;\;+\text{Tr}(\mathcal{D}(\hat{x}_{\theta'})\hat{x}_{\wr\downarrow})
\expect{\Delta\hat{X}_\theta^{pl}\!(z_k)\Delta\hat{X}_{\wr\downarrow}^{p''l''}\!(z_{k'})}_S)\\\notag
&+\gamma_s\,\delta_{k',k}\sum_{p'',\,l''}\sum_{\psi\in\{\uparrow,\,\downarrow,\,\wr\}}G_{p''l''}^{plp'l'}(z_k)\;
\text{Tr}(\mathcal{N}(\hat{x}_\theta,\hat{x}_{\theta'})\hat{n}_\psi)N_\psi^{p''l''}(z_k).
\end{align}
This equation couples all covariances $\expect{\Delta\hat{X}_\theta^{pl}(z_k)\Delta\hat{X}_{\theta'}^{p'l'}(z_{k'})}_S$ in slices $k$ and $k'$ to one another. In the event that $k=k'$, the covariances are also coupled to the effective populations in slice $k$. Because each covariance is coupled to infinitely many covariances in the transverse modes $p''l''$, the equation of motion must be truncated at a finite $p''l''$ like the equation of motion for the effective populations in  \erf{eq::CGNup}. After this step, the covariances $\expect{\Delta\hat{X}_\theta^{pl}(z_k)\Delta\hat{X}_{\theta'}^{p'l'}(z_{k'})}_S$ are related to the effective populations and eachother by a set of closed differential equations.

We now examine the evolution of the covariances  $\expect{\Delta\hat{X}_\theta^{pl}(z_k)\Delta\hat{X}_{\theta'}^{p'l'}(z_{k'})}_S$ under QND measurement. When the ensemble is prepared in the SCS, cat, or $m_x=0\,$ state preparations, the collective pseudospin waves $\hat{X}_{\downarrow\uparrow}^{pl}(z_k)$ and $\hat{X}_{\wr\downarrow}^{pl}(z_k)$ approximately commute with the spin waves along $z$, as shown in Sec. \ref{sec::EffModes}.  As a result, these collective pseudospin waves are unaffected by the Lindblad dissipator in \erf{Eq::LSuperoperator}. Because of this, we need only consider the effect of measurement backaction upon the covariances, which is given by the superoperator $\mathcal{H}_{00}$ in \erf{Eq::HSuperoperator}. The evolution of the covariances under QND measurement is 
\begin{align}\label{eq::HCovar}
d\expect{\Delta\hat{X}_\theta^{pl}(z_k)&\Delta\hat{X}_{\theta'}^{p'l'}(z_{k'})}_S\Big|_{\text{QND}}=\\\notag&\sqrt{\frac{\kappa}{16}}\!\left\langle\mathcal{H}_{00}\!\left(\!\Delta\hat{X}_\theta^{pl}(z_k)\Delta\hat{X}_{\theta'}^{p'l'}(z_{k'})+\Delta\hat{X}_{\theta'}^{p'l'}(z_{k'})\Delta\hat{X}_\theta^{pl}(z_k)\!\right)\right\rangle\\\notag
&-\sqrt{\frac{\kappa}{4}}\expect{\mathcal{H}_{00}(\hat{X}_\theta^{pl}(z_k))}\expect{\hat{X}_{\theta'}^{p'l'}(z_{k'})}\\\notag
&-\sqrt{\frac{\kappa}{4}}\expect{\hat{X}_\theta^{pl}(z_k)}\expect{\mathcal{H}_{00}(\hat{X}_{\theta'}^{p'l'}(z_{k'}))}\\\notag
&-\frac{\kappa}{4}\expect{\mathcal{H}_{00}(\hat{X}_\theta^{pl}(z_k))}\expect{\mathcal{H}_{00}(\hat{X}_{\theta'}^{p'l'}(z_{k'}))}.
\end{align}
Note that by the rules of It\={o} calculus differentials are taken to second order. When the ensemble is prepared in one of the state preparations discussed in Sec. \ref{sec::StatePreps}, the spin wave $\hat{F}_z^{00}$ is approximately Gaussian distributed along with the collective pseudospin waves $\hat{X}_{\downarrow\uparrow}^{pl}(z_k)$ and $\hat{X}_{\wr\downarrow}^{pl}(z_k)$. The third order moments arising in \erf{eq::HCovar} can, thus, be decomposed in terms of means and covariances via 
\begin{align}
\frac{1}{6}\sum_{\text{perm}}\expect{\hat{O}\hat{A}\hat{Q}}=&\expect{\Delta\hat{O}\Delta\hat{Q}}_S\expect{\hat{A}}+
\expect{\Delta\hat{A}\Delta\hat{Q}}_S\expect{\hat{O}}+\expect{\Delta\hat{O}\Delta\hat{A}}_S\expect{\hat{Q}}
\\\notag&+\expect{\hat{O}}\expect{\hat{A}}\expect{\hat{Q}},
\end{align}
where the sum on the left is taken over all permutations of Gaussian operators $\hat{O}$, $\hat{A}$ and $\hat{Q}$. Under the Gaussian approximation, \erf{eq::HCovar} reduces to the relatively simple expression
\begin{align}\label{eq::QND00}
d\expect{\Delta\hat{X}_\theta^{pl}(z_k)\Delta\hat{X}_{\theta'}^{p'l'}(z_{k'})}_S\big|_{\text{QND}}\!=-&\kappa\expect{\Delta\hat{F}_z^{00}\Delta\hat{X}_\theta^{pl}(z_k)}_S\\\notag
&\times\expect{\Delta\hat{F}_z^{00}\Delta\hat{X}_{\theta'}^{p'l'}(z_{k'})}_S\,dt.
\end{align}
After writing $\Delta\hat{F}_z^{00}$ in terms of spin waves, this expression becomes
\begin{align}\label{eq::finalEoMQND}
d\expect{\Delta\hat{X}_\theta^{pl}&(z_k)\Delta\hat{X}_{\theta'}^{p'l'}(z_{k'})}_S\big|_{\text{QND}}\!=\!\\\notag-\kappa\sum_{k'',k'''}&\left(v(\uparrow)\expect{\Delta\hat{X}_{\downarrow\uparrow}^{00}(z_{k''})\Delta\hat{X}_\theta^{pl}(z_k)}_S
+w(\uparrow)\expect{\Delta\hat{X}_{\wr\downarrow}^{00}(z_{k''})\Delta\hat{X}_\theta^{pl}(z_k)}_S\right)\\\notag
\times&\left(v(\uparrow)\expect{\Delta\hat{X}_{\downarrow\uparrow}^{00}(z_{k'''})\Delta\hat{X}_{\theta'}^{p'l'}(z_{k'})}_S
+w(\uparrow)\expect{\Delta\hat{X}_{\wr\downarrow}^{00}(z_{k'''})\Delta\hat{X}_{\theta'}^{p'l'}(z_{k'})}_S\right).
\end{align}
a nonlinear differential equation that couples covariances between collective pseudospin waves in every slice with one another.

The full equation of motion for the covariances is comprised of the contributions from QND measurement in \erf{eq::finalEoMQND} and  optical pumping in \erf{eq::CGcovarFinal} ,
\begin{align}
\frac{d}{dt}\expect{\Delta\hat{X}_\theta^{pl}(z_k)\Delta\hat{X}_{\theta'}^{p'l'}(z_{k'})}_S=&
\frac{d}{dt}\expect{\Delta\hat{X}^{pl}(z_k)\Delta\hat{X}_{\theta'}^{p'l'}(z_{k'})}_S\Big|_\text{QND}\\\notag&+
\frac{d}{dt}\expect{\Delta\hat{X}_\theta^{pl}(z_k)\Delta\hat{X}_{\theta'}^{p'l'}(z_{k'})}_S\Big|_\text{op}.
\end{align}
When combined with the equation of motion for the effective populations in  \erf{eq::CGNup}, we obtain a closed set of differential equations coupling the covariances and the effective populations. Along with the initial conditions of the coarse-grained effective populations in \erf{eq::effectPopInit}, the initial conditions of the collective pseudospin wave covariances in all slices are required to solve this set. The initial conditions for the coarse-grained covariances are given by
\begin{align}
\expect{\Delta\hat{X}_\theta^{pl}&(z_k)\Delta\hat{X}_{\theta'}^{p'l'}(z_{k'})}(t=0)\\\notag
&=\delta_{k,k'}\frac{\delta z}{2}\int d^2\mbf{r}_\bot\beta_{pl}(\mbf{r},z_k)\beta_{p'l'}(\mbf{r},z_k)\bra{\uparrow}
(\Delta\hat{x}_\theta\Delta\hat{x}_{\theta'}+\Delta\hat{x}_{\theta'}\Delta\hat{x}_\theta)\ket{\uparrow}.
\end{align}

The coarse grained covariances and effective populations enable us to solve for the fundamental spin wave variance and the fundamental mean spin  that determine the paraxial squeezing parameter.  As in the plane wave case, the spin wave squeezing resulting from QND measurement can be enhanced by post-processing. Applying the post-processing partial isometry $\hat{U}_{\uparrow'}$, introduced in \erf{UArb}, transforms the fundamental spin wave variance as
\begin{align}
(\Delta &F_z^{00})^2= v(\uparrow')^2\sum_{k,k'}\expect{\Delta\hat{X}_{\downarrow\uparrow}^{00}(z_k)\Delta\hat{X}_{\downarrow\uparrow}^{00}(z_{k'})}_S\\\notag
+&v(\uparrow')w(\uparrow')\sum_{k,k'}\left(\expect{\Delta\hat{X}_{\downarrow\uparrow}^{00}(z_k)\Delta\hat{X}_{\wr\downarrow}^{00}(z_{k'})}_S+
\expect{\Delta\hat{X}_{\wr\downarrow}^{00}(z_{k})\Delta\hat{X}_{\downarrow\uparrow}^{00}(z_{k'})}_S\right)\\\notag
+& w(\uparrow')^2\sum_{k,k'}\expect{\Delta\hat{X}_{\wr\downarrow}^{00}(z_k)\Delta\hat{X}_{\wr\downarrow}^{00}(z_{k'})}_S
\end{align}
and the fundamental mean spin as
\begin{align}
\expect{\hat{F}_x^{00}}=\sum_{k}\left(N_\uparrow^{00}(z_k)\expect{\hat{f}_x}_{\uparrow'}
+N_\downarrow^{00}(z_k)\expect{\hat{f}_x}_{\downarrow'}
+N_\wr^{00}(z_k)\expect{\hat{f}_x}_{\wr'}\right).
\end{align}
By substituting these this quantities into the definition for the paraxial squeezing parameter in \erf{Eq::SqueezingParam}, we can quantify the effectiveness of squeezing by QND measurement on this inhomogeneous system with $f>1/2$.

\addcontentsline{toc}{chapter}{{\bf bibliography}}
\bibliographystyle{ieeetr}
 \bibliography{SpinfReferences}

\begin{thebibliography}{10}

\bibitem{Fleischhauer2002}
M.~Fleischhauer and M.~D. Lukin {\em Phys. Rev. A}, vol.~65, p.~022314, 2002.

\bibitem{Polzik2004}
B.~Julsgaard, J.~Sherson, J.~I. Cirac, J.~Fiuraek, and E.~S. Polzik {\em
  Nature}, vol.~432, p.~482, 2004.

\bibitem{Kimble2008}
K.~S. Choi, H.~Deng, J.~Laurat, and H.~J. Kimble {\em Nature}, vol.~452, p.~67,
  2008.

\bibitem{DLCZ}
L.~M. Duan, M.~D. Lukin, J.~Cirac, and P.~Zoller {\em Nature}, vol.~414,
  p.~413, 2001.

\bibitem{Kuzmich2004}
D.~N. Matsukevich and A.~Kuzmich {\em Science}, vol.~306, p.~663, 2004.

\bibitem{ContinuousVar}
S.~L. Braunstein and P.~van Loock {\em Rev. Mod. Phys.}, vol.~77, p.~513, 2005.

\bibitem{KuzBig00}
A.~Kuzmich, L.~Mandel, and N.~P. Bigelow {\em Phys. Rev. Lett.}, vol.~85,
  p.~1594, 2000.

\bibitem{appel09}
J.~Appel, P.~J. Windpassinger, D.~Oblak, U.~B. Hoff, N.~Kjaergaard, and E.~S.
  Polzik {\em Proc. Nat. Acad. Sci. USA}, vol.~106, p.~10960, 2009.

\bibitem{Takano2009}
T.~Takano, M.~Fuyama, R.~Namiki, and Y.~Takahashi {\em Phys. Rev. Lett.},
  vol.~102, p.~033601, 2009.

\bibitem{VulSqueezingClock}
I.~D. Leroux, M.~H. Schleier-Smith, and V.~Vuletic {\em Phys. Rev. Lett.},
  vol.~104, p.~250801, 2010.

\bibitem{Koschorreck2010}
M.~Koschorreck, M.~Napolitano, B.~Dubost, and M.~W. Mitchell {\em Phys. Rev.
  Lett.}, vol.~105, p.~093602, 2010.

\bibitem{TakTak05}
M.~Takeuchi, S.~Ichihara, T.~Takano, M.~Kumakura, T.~Yabuzaki, and Y.~Takahashi
  {\em Phys. Rev. Lett.}, vol.~94, p.~023003, 2005.

\bibitem{Wineland94}
D.~J. Wineland, J.~J. Bollinger, W.~M. Itano, F.~L. Moore, and D.~J. Heinzen
  {\em Phys. Rev. A}, vol.~46, p.~6797, 1992.

\bibitem{KosMitSq}
M.~Koschorreck, M.~Napolitano, B.~Dubost, and M.~W. Mitchell {\em Phys. Rev.
  Lett.}, vol.~104, p.~093602, 2010.

\bibitem{BudkerSq}
S.~M. Rochester, M.~P. Ledbetter, T.~Zigdon, A.~D. Wilson-Gordon, and D.~Budker
  {\em Phys. Rev. A}, vol.~85, p.~022125, 2012.

\bibitem{ASmith13}
A.~Smith, B.~E. Anderson, H.~Sosa-Martinez, C.~A. Riofrio, I.~H. Deutsch, and
  P.~S. Jessen {\em Phys. Rev. Lett.}, vol.~111, p.~170502, 2013.

\bibitem{MerkelControlPRA}
S.~T. Merkel, P.~S. Jessen, and I.~H. Deutsch {\em Phys. Rev. A}, vol.~78,
  p.~023404, 2008.

\bibitem{DeuJes09}
I.~H. Deutsch and P.~S. Jessen {\em Opt. Comm.}, vol.~283, p.~681, 2009.

\bibitem{KitagawaUeda93}
M.~Kitagawa and M.~Ueda {\em Phys. Rev. A}, vol.~47, p.~5138, 1993.

\bibitem{Chaudhury07}
S.~Chaudhury, S.~Merkel, T.~Herr, A.~Silberfarb, I.~H. Deutsch, and P.~S.
  Jessen {\em Phys. Rev. Lett.}, vol.~99, p.~163002, 2007.

\bibitem{Fernholz08}
T.~Fernholz, H.~Krauter, K.~Jensen, J.~F. Sherson, A.~S. S$\o$rensen, and E.~S.
  Polzik {\em Phys. Rev. Lett.}, vol.~101, p.~073601, 2008.

\bibitem{Stoler71}
D.~Stoler {\em Phys. Rev D}, vol.~1, p.~3217, 1970.

\bibitem{Lu72}
E.~Y.~C. Lu {\em Lett. Nuovo Cimento}, vol.~3, p.~585, 1972.

\bibitem{Hollenhorst79}
J.~N. Hollenhorst {\em Phys. Rev. D}, vol.~19, p.~1669, 1979.

\bibitem{HP}
T.~Holstein and H.~Primakoff {\em Physical Review}, vol.~58, p.~1098, 1940.

\bibitem{MultilevelHP}
Z.~Kurucz and K.~M$\o$lmer {\em Phys. Rev. A}, vol.~81, p.~032314, 2010.

\bibitem{Smith03}
G.~A. Smith, S.~Chaudhury, A.~Silberfarb, I.~H. Deutsch, and P.~S. Jessen {\em
  Phys. Rev. Lett.}, vol.~93, p.~163602, 2003.

\bibitem{Baragiola14}
B.~Q. Baragiola, L.~M. Norris, E.~Monta$\tilde{\text{n}}$o, P.~G. Mickelson,
  P.~S. Jessen, and I.~H. Deutsch {\em Phys. Rev. A}, vol.~89, p.~033850, 2014.

\bibitem{Giedke02}
G.~Giedke and J.~I. Cirac {\em Phys. Rev. A}, vol.~66, p.~032316, 2002.

\bibitem{PlenioEisert}
J.~Eisert and M.~B. Plenio {\em Int. J. Quantum Inform.}, vol.~01, p.~479,
  2003.

\bibitem{Wang07}
X.~B. Wang, T.~Hiroshima, A.~Tomita, and M.~Hayashi {\em Phys. Rep.}, vol.~448,
  2007.

\bibitem{Weedbrook12}
C.~Weedbrook, S.~Pirandola, R.~Garc$\acute{\text{i}}$a-Patr$\acute{\text{o}}$n,
  N.~J. Cerf, T.~C. Ralph, J.~H. Shapiro, and S.~Lloyd {\em Rev. Mod. Phys.},
  vol.~84, p.~621, 2012.

\bibitem{Adesso14}
G.~Adesso, S.~Ragy, and A.~R. Lee {\em Open Syst. Inf. Dyn.}, vol.~21,
  p.~1440001, 2014.

\bibitem{SymplecticGeo}
M.~de~Gosson, {\em Symplectic Geometry and Quantum Mechanics}.
\newblock Birkh\"{a}user Verlag, 2006.

\bibitem{KuzMan98}
A.~Kuzmich, N.~P. Bigelow, and L.~Mandel {\em Europhys. Lett.}, vol.~42,
  p.~481, 1998.

\bibitem{TraDeu10}
C.~M. Trail, P.~S. Jessen, and I.~H. Deutsch {\em Phys. Rev. Lett.}, vol.~105,
  p.~193602, 2010.

\bibitem{Scully91}
M.~O. Scully, B.~Englert, and H.~Walther {\em Nature}, vol.~351, p.~111, 1991.

\bibitem{NorDeu12}
L.~M. Norris, C.~M. Trail, P.~S. Jessen, and I.~H. Deutsch {\em Phys. Rev.
  Lett.}, vol.~109, p.~173603, 2012.

\bibitem{Combes05}
J.~Combes and H.~M. Wiseman {\em J. Opt. B}, vol.~7, p.~14, 2005.

\bibitem{YurkeState}
B.~Yurke {\em Phys. Rev. Lett.}, vol.~56, p.~1515, 1986.

\bibitem{CohenTannoudji75}
C.~Cohen-Tannoudji, ``Atoms in strong resonant fields,'' in {\em Frontiers in
  laser spectroscopy} (R.~B. . S. H. .~S. Liberman, ed.), Session XXXVIII,
  p.~1, Les Houches, North-Holland, 1975.

\bibitem{MadMol}
L.~B. Madsen and K.~M$\o$lmer {\em Phys. Rev. A}, vol.~70, p.~052324, 2004.

\bibitem{JacSte06}
K.~Jacobs and D.~A. Steck {\em Contemp. Phys.}, vol.~47, p.~279, 2006.

\bibitem{InteriorPt}
A.~Forsgren, P.~E. Gill, and M.~H. Wright {\em SIAM}, vol.~44, p.~525, 2002.

\bibitem{Oblak05}
D.~Oblak, P.~G. Petrov, C.~L.~G. Alzar, W.~Tittel, A.~K. Vershovski, J.~K.
  Mikkelsen, J.~L. S$\o$rensen, and E.~S. Polzik {\em Phys. Rev. A}, vol.~71,
  p.~043807, 2005.

\bibitem{SchleierSmith09}
R.~C$\hat{\text{o}}$t$\acute{\text{e}}$, P.~L. Gould, M.~Rozman, and W.~W.
  Smith, eds., {\em Spin Squeezing on an Atomic Clock Transition}, Proceedings
  of the XXI. International Conference on Atomic Physics, World Scientific,
  2009.

\bibitem{CHen14}
Z.~Chen, J.~G. Bohnet, J.~M. Weiner, K.~C. Cox, and J.~K. Thompson {\em Phys.
  Rev. A}, vol.~89, p.~043837, 2014.

\bibitem{Silberfarb05}
A.~Silberfarb, P.~S. Jessen, and I.~H. Deutsch {\em Phys. Rev. Lett.}, vol.~95,
  p.~030402, 2005.

\bibitem{Riofrio11}
C.~A. Riofrio, P.~S. Jessen, and I.~H. Deutsch {\em J. Phys. B}, vol.~44,
  p.~154007, 2011.

\end{thebibliography}

%

\end{document}